\documentclass[11pt,twoside] {report}

\usepackage{fancyheadings}
\usepackage{times}

\usepackage{epsfig}
\catcode`@=11
\def\@citex[#1]#2{\if@filesw\immediate\write\@auxout{\string\citation{#2}}\fi
  \def\@citea{}\@cite{\@for\@citeb:=#2\do
    {\@citea\def\@citea{,\penalty\@m}\@ifundefined
      {b@\@citeb}{{\bf ?}\@warning
       {Citation `\@citeb' on page \thepage \space undefined}}%
\hbox{\csname b@\@citeb\endcsname}}}{#1}}

\def\citer{\@ifnextchar [{\@tempswatrue\@citexr}{\@tempswafalse\@citexr[]}}

\def\@citexr[#1]#2{\if@filesw\immediate\write\@auxout{\string\citation{#2}}\fi
  \def\@citea{}\@cite{\@for\@citeb:=#2\do
    {\@citea\def\@citea{--\penalty\@m}\@ifundefined
       {b@\@citeb}{{\bf ?}\@warning
       {Citation `\@citeb' on page \thepage \space undefined}}%
\hbox{\csname b@\@citeb\endcsname}}}{#1}}
\catcode`@=12
\newcommand{\dis}{\displaystyle}

\def\cseff{C_7^{\mbox{eff}}}
\def\cneff{C_9^{\mbox{eff}}}
\def\s{\hat{s}}
\def\u{\hat{u}}
\def\z{v \cdot \hat{q}}
\def\cnt{C_9^{\mbox{eff}} \mp C_{10}}
\def\ml{\hat{m}_l}
\def\ms{\hat{m}_s}

\def\mc{\hat{m}_c}
\def\mb{\frac{M_B}{{m_b}^3}}
\def\lo{\hat{\lambda}_1}
\def\lt{\hat{\lambda}_2}
\def\loo{\lambda_1}
\def\lto{\lambda_2}
\def\q{\hat{q}}
\def\bxsll{$B \rightarrow X_s \ell^+ \ell^- $ }
\def\absvcb{\left| V_{cb} \right|}

\def\bsll{$b \rightarrow s \ell^+ \ell^- $ }

\def\bxsg{$B \rightarrow X_s \gamma $ }

\def\mb{\frac{m_B}{{m_b}^3}}

\begin{document}
\setlength{\baselineskip}{24pt}
% {\baselineskip}{8mm} is desy default
\setlength{\baselineskip}{6mm}

\input{qm.cmd}

\begin{titlepage}

\begin{center}

\vspace*{1cm}

\huge \bf \sc

Improved QCD perturbative contributions and power corrections in radiative 
and semileptonic rare $B$ decays 

\vfill

\large

\vfill

 Dissertation \\
 zur Erlangung des Doktorgrades\\
 des Fachbereichs Physik\\
 der Universit\"at Hamburg\\

\vfill

 vorgelegt von \\
 Gudrun Hiller \\
 aus Hamburg\\

\vfill

 Hamburg\\
 1998

\vspace*{1cm}

\end{center}
\newpage
\thispagestyle{empty}

\vspace*{14cm}
\vfill
\small 
% die naechsten Zeilen auskommentieren zur Druchversion

\begin{tabular}{ll}
\begin{minipage}[t]{6.0cm}
Gutachter der Dissertation:
\end{minipage} 
& 
\begin{minipage}[t]{5.0cm}
Prof.~Dr.~A.~Ali\\
Prof.~Dr.~G.~Kramer\\
\end{minipage} \\
\bigskip

\begin{minipage}[t]{6.0cm}
Gutachter der Disputation:
\end{minipage} 
& 
\begin{minipage}[t]{5.0cm}
Prof.~Dr.~A.~Ali\\
Prof.~Dr.~J.~Bartels\\
\end{minipage} \\
\bigskip

\begin{minipage}[t]{6.0cm}
Datum der Disputation:
\end{minipage} 
& 
\begin{minipage}[t]{5.0cm}
16.~Juni 1998
\end{minipage} \\
\bigskip

\begin{minipage}[t]{6.0cm}
Sprecher des Fachbereichs Physik \\
und Vorsitzender des \\
Promotionsausschusses:
\end{minipage} 
& 
\begin{minipage}[t]{5.0cm}
${}$ \\
${}$ \\
Prof.~Dr.~B.~Kramer\\
\end{minipage} 
\end{tabular}

\newpage
\mbox{ }\\\vfill
\thispagestyle{empty}
\begin{center}
\begin{minipage}[t]{12cm}
{\begin{center}\bf{Abstract}\end{center}}
We calculate the leading power corrections to the decay rates,
distributions and hadronic spectral moments in rare inclusive \bxsll
decays in the standard model, using heavy quark expansion (HQE) in $(1/m_b)$
and a phenomenological model implementing the Fermi motion effects of the
$b$-quark bound in the $B$-hadron.
We include next-to-leading order perturbative
QCD corrections and work out the dependences of the spectra, decay rates and 
hadronic moments on the model parameters in
either HQE and the Fermi motion model.
In the latter, we take into account long-distance effects via
$B \to X_s +(J/\psi, \psi^\prime,...) \to X_s \ell^+ \ell^- $
with a vector meson dominance ansatz and
study the influence of kinematical cuts in the dilepton and hadronic 
invariant masses on branching ratios, hadron spectra and hadronic moments.

We present leading logarithmic QCD corrections to the 
$b \rightarrow s \gamma\gamma$ amplitude.
The QCD perturbative improved 
$B_{s}\rightarrow\gamma\gamma$  branching ratio is given in the standard 
model including our estimate of long-distance effects via
$B_s \to \phi \gamma \to \g \g$ and
$B_s \to \phi \psi \to \phi \g \to \g \g $ decays.
The uncertainties due to the 
renormalization scale and the parameters of the HQE inspired bound state model
are worked out.
  \\
{\begin{center}\bf{Zusammenfassung}\end{center}}
Wir berechnen im Standardmodell mit der Entwicklung schwerer Quarks (ESQ)
Kor\-rekturen in $(1/m_b)$ zu
Zerfallsbreiten, Verteilungen und 
hadronischen Momenten des seltenen, inklusiven Zerfalls \bxsll.
Dieselbe Analyse haben wir im ph\"anomenologischen Fermibewegungsmodell
durchgef\"uhrt, welches die Effekte
des im $B$-Meson gebundenen $b$-Quarks beschreibt.
In beiden Methoden zeigen wir unter Einbeziehung nicht f\"uhrender 
QCD--Korrekturen die Abh\"angigkeit der Spektren, Brei\-ten und 
Momente von den Modellparametern.
Mit einem Vektormesondominanzansatz modellieren wir
im Fermibewegungsmodell die langreichweitigen Beitr\"age der Charmoniumresonanzen, die durch $B \to X_s +(J/\psi, \psi^\prime,...) \to X_s \ell^+ \ell^- $
entstehen.
Wir studieren den Einflu\ss{} kinematischer Schnitte in der invarianten Masse 
des Leptonpaares und des hadronischen Endzustandes auf 
Verzweigungsverh\"altnisse, hadronische Verteilungen und Momente des
Zerfalls \bxsll.

Ferner pr\"asentieren wir f\"uhrende logarithmische QCD--Korrekturen in der
$b \to s \g \g$ Amplitude.
Wir geben das QCD--verbesserte Verzweigungsverh\"altnis von
$B_{s}\rightarrow\gamma\gamma$ im Standardmodell unter Ber\"ucksichtigung 
unserer Absch\"atzung langreichweitiger 
Beitr\"age durch die Zerfallsketten
$B_s \to \phi \gamma \to \g \g$ und
$B_s \to \phi \psi \to \phi \g \to \g \g $ an.
Wir sch\"atzen die Unsicherheiten durch die Renormierungsskala und
des $B_s$ Bindungszustandsmodelles, welches
der Theorie schwerer Quarks entlehnt ist, ab.
\end{minipage}
\end{center}

\vfill
\mbox{ }

\newpage
\pagestyle{empty}
\mbox{}

\newpage
\pagestyle{empty}
\setcounter{page}{1}
%\tableofcontents

\end{titlepage}

\tableofcontents
\cleardoublepage
\pagestyle{fancyplain}
\chapter{Introduction}

%{\it "Respect all. Trust few. Fear none."}

Rare $B$ decays probe the flavour sector of the 
standard model (SM) \cite{GSW} and perturbative and non-perturbative aspects 
of QCD.
Since the first measurement of rare radiative $B \to K^* \g$ decays  by the
CLEO collaboration \cite{CLEOkstar}
this area of particle physics has received much experimental 
\cite{skwarnicki97} and theoretical \cite{aliapctp97} attention.
Flavour changing neutral current (FCNC) induced decays are of particular 
interest, since in the SM they are governed by loop effects and depend 
sensitively on virtual particles like, e.g., the top quark and Cabibbo 
Kobayashi Maskawa (CKM) matrix elements \cite{CKM}.

The theory of the effective Hamiltonian 
%\cite{effhamali,effhamburas,Ciuchini,misiakE,grinstein}
\citer{effhamali,grinstein} (see section \ref{sec:effham} for a  discussion) 
$\ch_{eff} \sim \sum C_i O_i$ enables a 
description of low energy weak processes in terms of
short-distance (Wilson) coefficients $C_i$, 
which can be calculated perturbatively.
The (new) vertices $O_i$, which are absent in the full Lagrangian, 
are obtained by integrating out the heavy particles
($W,t,\phi$ in the SM ) from the full theory. 
Their coupling strength is given by
the $C_i$, which characterize the short-distance dynamics of 
the underlying theory.

The Wilson coefficient $\cseff$ corresponds to the effective $bs\g$ vertex. 
Its modulus $|\cseff|$ is constrained by the measurement of the inclusive 
$B \to X_s \g$ branching 
ratio at CLEO \cite{CLEO95inkl}.
The CLEO result is in agreement with the present theoretical SM prediction in
\bxsg decay and moreover, can exclude large parameter spaces of 
non-standard models.

The transition $b \to s \ell^+ \ell^{-}$ with $\ell=e,\mu,\tau$ involves
besides the electromagnetic penguin $b \to s \g^* \to s \ell^+ \ell^{-}$ also
electroweak penguins $b \to s Z^{0 *} \to s \ell^+ \ell^{-}$ and boxes.
They give rise to two additional Wilson coefficients in the semileptonic 
decay \bxsll, $C_9$ and $C_{10}$.
A model independent fit of the short-distance coefficients 
$C_9,C_{10}$ and $\cseff$ is possible from the following 
three observables \cite{agm94}: 
The \bxsg branching ratio
${\cal{B}}(B \to X_s \gamma)$, the (partly integrated)
invariant dilepton mass spectrum  and the 
Forward-Backward (FB) asymmetry \cite{amm91} in \bxsll decays.
They involve independent combinations of the Wilson coefficients,
which allows the determination
of sign and magnitude of $\cseff,C_9$ and $C_{10}$ from data.

The presence of charmonium resonances in the decay \bxsll
complicates this analysis. The $c \bar{c}$ states appear via 
$B \to X_s +(J/\psi, \psi^\prime,...) \to X_s \ell^+ \ell^- $ and
can be taken into account in a 
phenomenological vector meson dominance (VMD) ansatz, which is assumed to 
hold near the $J/\psi,\psi',\dots$ peaks \cite{amm91}.
Hence, kinematical cuts on the dilepton mass $q^2$ have to be imposed to 
remove the dominant  
resonance contributions and to disentangle the short-distance information from
the long-distant one, e.g., $q^2 < m_{J/\psi}^2-\delta$. Here $\delta$ is 
an experimental cut-off parameter and typically of order a few hundred MeV.
The restriction to certain regions of phase space
explains the use of {\it partly} integrated spectra.
The net effect of the resonances is an additional error on the 
distributions and hinders the determination of the short-distance 
coefficients $C_i$ from experimental data.
Hence, they must be evaluated by taking into account theoretical dispersion
by using different phenomenological models \cite{AH98-3}
or further experimental input.

The bound state nature of heavy hadrons can be explored within the
$B$-system.
Non-perturbative power corrections in $(\Lambda_{QCD}/m_b)^n$ are 
systematically obtained with the heavy quark expansion (HQE) technique  
\cite{georgi},
parametrizing distributions, decay rates etc.\ in terms of higher order matrix 
elements, denoted by $\lambda_1,\lambda_2$ for $n=2$.
We recall the HQE relation
$m_B=m_b+\bar{\Lambda}-(\lambda_1+3 \lambda_2)/2 m_b +  {\cal{O}}(1/m_b^2)$ 
between the physical $B$-meson and the $b$-quark mass, where $\bar{\Lambda}$
accounts for the ``binding energy".
The HQE method has been applied to semileptonic charged current 
$B \to X \ell \nu_\ell$ \cite{MW} and FCNC \bxsg \cite{FLSold} 
decays, and it is known that inclusive spectra are not entirely calculable 
with the HQE approach 
%\cite{MW,AHHM97,Bigietal2,falketal1,JR94}.
\cite{MW}, \citer{AHHM97,JR94}.
$B$-meson wave function effects have been estimated in 
$B \to X_{u,c} \ell \nu_\ell$ \cite{aliqcd,greubrey} and \bxsg 
\cite{effhamali,ag1} decays with a phenomenological Fermi motion model (FM) 
\cite{aliqcd}.

We present here a detailed analysis of inclusive \bxsll decays in the SM
with $\ell=e,\mu$, (since we neglected lepton masses in our calculation the 
result cannot be used in the $\tau$ case), following similar studies
for charged current $B \to X \ell \nu_\ell$ and radiative \bxsg decays.

This thesis contains the following points \cite{AHHM97,AH98-1,AH98-2,AH98-3}:
\begin{itemize}
\item We calculate the $1/m_b^2$ power corrections using HQE techniques 
in \bxsll decays in the dilepton invariant mass distribution. 
This corrects an earlier calculation
\cite{FLSold} and has been confirmed recently \cite{buchallaisidorirey}.
We find that the HQE  breaks down near the high $q^2$ end-point, 
hence the spectrum cannot be used in this region. 

\item Alternatively, we study $B$-meson wave function effects with a
Gaussian FM.
We present the dilepton invariant mass distribution and the FB asymmetry and
investigate the dependence on the FM model parameters.

\item We include $c \bar{c}$ resonance effects in \bxsll spectra with the 
help of a VMD model and present the distributions including
next-to-leading order perturbative QCD corrections.

\item We present $1/m_b$ power corrections in the HQE approach 
and ${\cal{O}}(\alpha_s)$ perturbative
corrections in the hadron spectra and hadronic spectral moments in \bxsll
decays.

\item The explicit dependences of the lowest moments of the hadronic energy
$E_H$ and the hadronic invariant mass $S_H$,
$\langle E_H^n \rangle$, $\langle S_H^n \rangle$ for $n=1,2$
on the non-perturbative HQE parameters are worked out.
We find that the first two moments of the hadronic invariant mass 
in \bxsll decay are sensitive
to $\bar{\Lambda}$ and $\lambda_1$ .

\item We complement this profile of hadron spectra and moments in
\bxsll decays by an analysis in the FM. The hadronic energy spectrum 
is found to be stable against a variation of the FM parameters, however, the 
hadronic invariant mass distribution depends significantly on them.

%\item The hadronic spectral moments in \bxsll decay in the FM and the HQE 
%approach are found 
%to be very similar for equivalent values of their parameters.

\item We incorporate the charmonium resonances by means of the VMD model
into the analysis of hadron spectra and moments in \bxsll decays. 
The broadening of the resonances in the FM in the hadron spectra is worked out.
Also the $c \bar{c}$ resonances turn out to be important in the moments.

\item We investigate the resulting uncertainties in spectra and moments in
\bxsll decays from different parametrizations \cite{KS96,LSW97,AHHM97} of the 
resonant and non-resonant $c\bar{c}$ contributions.

\item We work out hadron spectra, spectral moments and branching ratios 
in the FM  with kinematical cuts, as used in the CLEO analysis in their search 
for the decays \bxsll, $\ell=e,\mu$ \cite{cleobsll97}. 
They imposed two kinds of cuts:
one in the dilepton invariant mass to exclude the main bulk of the 
$J/\psi,\psi',\dots$ resonances and another one in the hadronic invariant mass
$S_H < 3.24 \, {\mbox{GeV}}^2$, to suppress the $B \bar{B}$ background.
Our study of the \bxsll branching ratios with cuts is of direct use 
to estimate the efficiency of the remaining signal.
\end{itemize}

To  summarize, we shall present spectra, decay rates and hadronic
spectral moments in inclusive rare \bxsll decays in the standard model.
The dilepton mass spectra and the FB asymmetry presented here are of use to
extract the short-distance coefficients. We show that the uncertainties lying 
in different parametrizations of the charmonium resonance effects are not the 
dominant ones.
Hadron spectra and moments in
\bxsll decays can be used to test HQE and FM and/or to determine 
their parameters.
We point out that the HQE, where it is valid, and the FM show very similar 
behaviour.
Moreover, some of their parameters can be related.
In particular, the moments of the hadronic invariant mass in \bxsll decays 
provide a complementary constraint on the non-perturbative HQE parameters 
$\bar{\Lambda},\lambda_1$ to the one in charged current
semileptonic $B \to X \ell \nu_\ell$ decays \cite{gremm}. These decays
can be used for a precise determination of these parameters.
A related issue is the question of universality of the FM/HQE parameters
for $b \to q$ transitions with final quarks $q=u,d,s$ or $q=c$.
This remains to be tested.
Finally, relating partly integrated hadron spectra of \bxsll to
semileptonic $B \to X_u \ell \nu_\ell$ decays, we expect the cancelation of 
uncertainties resulting from bound state effects, thus this offers the possibility of a precise determination of $V_{ub}$.

Further, we study the exclusive mode $B_s \to \g \g$,
which in the SM has a branching ratio in reach of future $B$-facilities
\cite{skwarnicki97}.
We improve earlier analyses 
%\cite{yao,simma,herrlichkalinowski}
\citer{yao,herrlichkalinowski} by including
leading logarithmic QCD corrections to the short-distance
$b \to s \g \g$ amplitudes.
We use the same effective Hamiltonian \cite{effhamali,effhamburas} 
as for $b \to s \g$, which, as we will show, 
contains a complete set of operators for both decays.
The $B_s \to \g \g$ decay rate gets enhanced under renormalization, 
like the \bxsg decay rate \cite{effhamali}.
Likewise, we obtain
a strong dependence on the renormalization scale of the 
$B_s \to \g \g$ observables,
the branching ratio and the CP ratio \cite{aliev}, the latter resulting
from CP-odd and CP-even parts in the FCNC 2-photon amplitude.
Moreover, modeling $B_s$ bound state effects in a HQE inspired approach,
we avoid the constituent quark mass value $m_s \sim m_K$, as used in
previous analyses.

In our analysis of $B_s \to \g \g$ decay we take into account 
long-distance (LD) effects via intermediate (neutral) vector mesons.
Especially the contribution due to $B_s \to \phi \g $ and subsequent decay 
$\phi \to \g$ is estimated and found to sizeably reduce the
$B_s \to \g \g$ branching ratio. 
We use QCD sum rules to evaluate the $B_s \to \phi \g$
form factor and include the contribution from the gluon condensate. 
The VMD model is employed for the $\phi$-meson photon conversion. 

Finally, we perform a VMD based calculation of the decay
$B_s \to \phi \psi \to \phi \g \to \g \g $, where we abbreviate
$\psi \equiv (J/\psi, \psi^\prime,...)$.
We compare the LD-contribution to 
$B_{s}\rightarrow\gamma\gamma$ decay resulting from intermediate
$\psi$ production
with the one obtained by the
interaction of the virtual charm loop with soft gluons \cite{Ruckl}.
We find that both amplitudes are in good agreement within the accuracy of the 
calculation. 
The contribution due to intermediate charmonium 
resonances changes the $B_s \to \g \g$ branching ratio
including the intermediate $\phi$ contribution by less than $1 \%$.

Our work in $B_s \to \g \g$ decays \cite{gudi,hilleriltanpsi} can be 
summarized as follows:
\begin{itemize}
\item We present leading logarithmic QCD corrections to the
short-distance amplitude in $b\to s \g \g$ decay.
\item We model $B_s$ bound state effects in a HQE inspired approach, in 
contrast to the constituent quark model, which is used in the literature.
\item We estimate the LD effect due to the decay chain
$B_s \to \phi \gamma \to \g \g$, using QCD sum rules and VMD.
\item We estimate the LD contribution through
$B_s \to \phi \psi \to \phi \g \to \g \g $ using VMD.
\end{itemize}  

We present the branching ratio and the CP ratio for
$B_s \to \g \g$ decay in the SM, taking into account
improved perturbative ${\cal{O}}(\alpha_s)$ contributions and long-distance 
effects via intermediate vector mesons. Uncertainties  resulting from the 
renormalization scale and the bound state parameters are worked out.

\subsubsection{Organization of the work}
An introduction to rare $B$ decays and the methods used is given in
Chapter \ref{chap:rare}, where we
discuss the effective Hamiltonian theory, rare radiative \bxsg decays
and long-distance methods in inclusive $B$ decays (HQET, FM and VMD).
Chapter \ref{chap:bsll} is based on refs. \cite{AHHM97,AH98-1,AH98-2,AH98-3}.
Here we investigate inclusive \bxsll decays in the SM. 
We present branching ratios and various spectra, furthermore 
hadronic spectral moments are estimated.
In chapter \ref{chap:bsgg} the exclusive channel $B_s \to \g \g$ is analysed 
\cite{gudi,hilleriltanpsi}.
SM based branching ratios and CP ratios in  $B_s \to \g \g$ decay are given
and their uncertainties are worked out.
Finally, chapter \ref{chap:out} contains a summary and an outlook.
Input parameters, Feynman rules and utilities are collected in 
appendix \ref{app:generalities}.
The power corrections to the structure functions of the hadron tensor in 
\bxsll decays are given in appendix \ref{app:dilepton}, together with 
auxiliary functions and the FM double differential Dalitz distribution in the 
context of the dilepton mass spectrum and the FB asymmetry.
Appendix \ref{app:hadron} contains analytic
expressions used in the derivation of hadron spectra and hadronic spectral 
moment in \bxsll decays.

\clearpage
\chapter{Rare $B$ Decays: Motivation and Methods \label{chap:rare}}

%{\it I smiled sadly, for a love I cannot obey...\\
%Oh how I sighed when they asked if I knew his name}\\
%"Lady Stardust",D. Bowie

In this chapter we outline the flavour structure of the standard model (SM).
We discuss the CKM mixing matrix and motivate the importance of studying 
flavour changing neutral current (FCNC) $b \to s$ transitions. 
We introduce the necessary tool to include
QCD perturbative corrections in weak decays, the effective 
Hamiltonian theory. 
As an application of the former we discuss $B \to X_s \gamma$ decay as the 
most prominent example of a rare $B$ decay.
Finally non-perturbative methods like the
heavy quark expansion technique, the Fermi motion model and vector meson 
dominance are sketched.

\section{The Flavour Sector in the Standard Model}

In the quark sector of the SM, there are six quarks organized in 3 families.
The left-handed quarks are put into weak isospin $SU(2)_L$ doublets 
\begin{eqnarray}
{q_{up} \choose q_{down}^{\prime}}_{i=1,2,3}=
{u \choose d^{\prime} }_L, \; {c \choose s^{\prime} }_L, \; 
{t \choose b^{\prime} }_L \; ,
\end{eqnarray}
and the corresponding 
right-handed fields transform as singlets under $SU(2)_L$.
Under the weak interaction an up-quark (with $Q_{u}=2/3 e$) can decay into a
down-quark (with $Q_{d}=-1/3 e$) and a $W^+$ boson.
This charged current is given as
\begin{equation}
J_{\mu}^{CC}=
 \frac{e}{ \sqrt{2} \sin \theta_W }
 \left( \bar{u}, \bar{c}, \bar{t} \right)_L
\gamma_\mu V_{\mbox{\footnotesize{ CKM}}}
\left( \begin{array}{c} d\\ s \\ b \end{array} \right)_L \; ,
\end{equation}
where the subscript $L =(1-\gamma_5)/2$ denotes the left-handed projector and 
reflects the $V-A$ structure of $J_{\mu}^{CC}$ in the SM.
Here the weak mixing (Weinberg-)angle $\theta_W$ is a 
parameter of the SM, which is measured with high accuracy \cite{PDG}.
The so-called Cabibbo Kobayashi Maskawa (CKM) matrix $V_{CKM}$ \cite{CKM} 
describes the mixing between different quark flavours. 
It contains the angles describing the rotation between the eigen vectors of 
the weak interaction $(q^{\prime})$ and the mass eigen states $(q)$
\begin{equation}
\left( \begin{array}{c} d^{\prime}\\ s^{\prime} \\ b^{\prime} \end{array} 
\right)=
V_{CKM}  \left( \begin{array}{c} d\\ s \\ b \end{array} \right)\; .
\end{equation}
Symbolically, $V_{CKM}$ can be written as
\begin{equation}
\label{eq:ckm}
V_{\mbox{\footnotesize CKM}} \equiv \left( \begin{array}{lll}
V_{ud} & V_{us} & V_{ub}\\
V_{cd} & V_{cs} & V_{cb}\\
V_{td} & V_{ts} & V_{tb}
\end{array} \right).
\end{equation}
In general all the entries are complex numbers, only restricted by
unitarity $V_{CKM} V_{CKM}^{\dagger}=1$.
They are parameters of the SM and can only be obtained from an 
experiment.
Note that only three independent real parameters and one phase 
are left after imposing the unitarity condition. 
Some parametrizations of $V_{CKM}$ can be seen in ref.~\cite{PDG}.

A useful parametrization of the CKM matrix has been proposed by Wolfenstein 
\cite{Wolfenstein}
 \begin{equation}
V_{\mbox{\footnotesize Wolfenstein}} = \left( \begin{array}{lll}
1-\frac{1}{2} \lambda^2 & \lambda &
A\lambda^3 (\rho - i \eta) \\
- \lambda & 1-\frac{1}{2} \lambda^2 
 & A \lambda^2 \\
A\lambda^3 (1-\rho-i \eta) & -A \lambda^2 & 1
\end{array} \right) +{\cal{O}}(\lambda^4) \; .
\label{eq:wolfenstein}
\end{equation}
The parameters $A,\lambda,\rho$ and the phase $\eta$ are real numbers.
$\lambda$ is related to the Cabibbo angle through
$\lambda=\sin \theta_C$ \cite{PDG}, 
which describes the quark mixing with 4 quark flavours. 
Since $\lambda \simeq 0.221$,
the relative sizes of the matrix elements in eq.~(\ref{eq:ckm})
can be read off from eq.~(\ref{eq:wolfenstein}). 
As can be seen, the diagonal entries are close to unity and the more 
off-diagonal they are, the smaller is the value of their corresponding 
matrix elements.
The parameter $A$ has been determined from the decays $b \to c \ell \nu_\ell$
and $B \to D^* \ell \nu_\ell$, yielding
$A=0.81 \pm 0.07$. The measurement of the ratio $|V_{ub}/V_{cb}|=0.08 \pm 0.02$ yields $\sqrt{\rho^2+\eta^2}=0.36 \pm 0.09$.
Likewise the mass difference $\triangle M_d \equiv M(B_d^{(1)})-M(B_d^{(2)})
\simeq 0.46 \, (ps)^{-1}$ constrains the combination
$\sqrt{(1-\rho)^2+\eta^2}$.
The observed CP-asymmetry parameter $\epsilon_K=2.26 \cdot 10^{-3}$ constrains
$\rho$ and $\eta$.
The precise determination of the parameters $\rho$ and $\eta$ is a high and 
important 
goal, since it corresponds to two important questions:
\begin{itemize}
\item Does CP hold in the SM ??
A non zero phase $\eta\not=0$ in the CKM matrix directly leads to CP 
violating effects.
\item The unitarity of the CKM matrix can be used to write down relations 
between its elements 
$\sum_{j=1}^{3} V_{ij} V^{\dagger}_{j k}$ $ =\delta_{ik}, \; i,k=1,2,3$. 
There are 6 orthogonality equations possible ($i\not=k$), and each
can be represented graphically as a triangle, a so-called unitarity 
triangle (UT) \cite{jarlskog88}.
The sides and angles of such an UT can be constrained by different types of 
experiments.
For {\it the} UT given by the relation
\begin{eqnarray}
V_{ud} V_{td}^* + V_{us} V_{ts}^* + V_{ub} V_{tb}^* = 0 \; ,
%V_{ub} + V_{td}^* = V_{us} V_{cb} \; ,
\end{eqnarray}
there are 3 scenarios possible, which at present are not excluded 
experimentally and are a sign for {\it new physics}:
1. the UT does not close, i. e., $\sum_{i=1}^3 \alpha_i \not=0$, where
$\alpha_i$ denotes the three angles of the triangle.
2. $\sum_{i=1}^3 \alpha_i =0$, but the values of the $\alpha_i$ are outside of 
their SM ranges determined by another type of experiment
3. $\sum_{i=1}^3 \alpha_i =0$, but the values of the angles are inconsistent 
with the measured sides of the triangle.
\end{itemize}
In the literature special unitarity triangles are discussed.
A recent review over the present status on the CKM matrix and {\it the} 
unitarity triangle is given by \cite{aliapctp97}.

\subsection{Flavour changing neutral currents and why do we look for them in 
the $B$-system ?}

In the SM, the neutral current mediated through the gauge bosons
$Z^0, \gamma, g$ does not change flavour. Therefore, the so-called 
Flavour changing neutral currents (FCNC) do not appear at tree level and
are described by loop effects.
The subject of the present work is an analysis of such rare 
(FCNC mediated) $B$ decays in the SM.
The quarks are grouped into {\it light} $(u,d,s)$ and
{\it heavy} $(c,b,t)$ ones in the sense, that the mass of a heavy quark is
 much larger than the typical scale of the strong interaction, 
$\Lambda_{QCD} \sim 200 \, {\mbox{MeV}}$.
The sixth quark, the top, is too heavy to build 
bound states because it decays too fast.
The special role of the $b$-quark is that it is the heaviest one 
building hadrons.
We will not discuss the ``double" heavy $B_c$ and concentrate on 
$B \equiv (\bar{b} q)$ meson transitions with light $q=u,d,s$.
Since the $b$-quark is heavy, the $B$-system is well suited for a clean 
extraction of the underlying short-distance dynamics.
Unlike the $K$-system, long-distance effects are expected to play 
a subdominant role in $B$ decays except where such effects are present in a 
resonant form.

The motivation to investigate
$b\to s (d)$ transitions is to improve the knowledge of the
CKM matrix elements and to study loop effects. 
For the latter the interest is large, since there is no tree level FCNC decay 
possible in the SM. 
The leading loops give the leading contribution and they are 
sensitive to the masses and other properties of the internal virtual 
particles like e.~g. the top. 
They can be heavy and therefore can be studied in a rare $B$ decay at energies 
which are much lower than the ones necessary for a direct production of such 
particles. 
The idea is to compare the SM based prediction for a rare $B$ decay with an
experiment. A possible deviation gives a hint not only for the existence, but
also for the structure of the ``new physics" beyond the SM.

Further the $B$-system can be used as a testing ground for QCD, to check 
perturbative and non-perturbative methods. One example is the decay 
$B \to X_s \gamma$, which can be described in the lowest order at parton level
through $b \to s \gamma$. As a 2-body decay, the photon energy in the $b$-quark
rest frame is fixed: $E_\gamma=(m_b^2-m_s^2)/2 m_b$ for an on-shell $\gamma$.
A possible non trivial spectrum 
%which has been already measured by cite{CLEOspectrum}
can result from gluon bremsstrahlung $b \to s \gamma g$ and/or a 
non-perturbative mechanism, which is responsible for the motion of 
the $b$-quark inside the meson thus boosting the distribution.
Such a bound state effect can be incorporated with e.g., the Fermi motion
model, see section \ref{subsec:FM} for a brief discussion.

Some rare $B$ decays have already been detected.
The channel $B\to K^{\ast} \gamma$ has been measured from the CLEO 
collaboration some time ago \cite{CLEOkstar}, however the most
prominent example of a rare $B$ decay is the inclusive $B\to X_s \gamma$ 
\cite{CLEO95inkl}, 
where $X_s$ is any hadronic state with strangeness $s=1$ and $B$ is a mixture
of $B^\pm$ and $B^0 (\bar{B}^0)$.
The branching ratios are found to be
\begin{eqnarray}
{\cal{B}}(B\to K^{\ast} \gamma)^{\mbox{{\small{CLEO}}}}&=&
4.2 \pm 0.8 \pm 0.6 \cdot 10^{-5} \; , \\ 
{\cal{B}}(B\to X_s \gamma)^{\mbox{{\small{CLEO}}}}&=&
2.32 \pm 0.57 \pm 0.35 \cdot 10^{-4} \; .
\label{eq:bsgexp}
\end{eqnarray}

Also the ALEPH collaboration has presented a preliminary result \cite{alephbsg}
\begin{eqnarray}
{\cal{B}}(h_b\to X_s \gamma)^{{\mbox{{\small{ALEPH}}  }}}&=&
3.29 \pm 0.71 \pm 0.68 \cdot 10^{-4} \; ,
\label{eq:aleph}
\end{eqnarray}
where $h_b$ is any $b$ flavoured hadron originating from $Z^0$ decays,
$Z^0 \to h_b X$.

The calculation of the exclusive mode introduces large theoretical 
uncertainties due to the  hadronic matrix elements. 
The inclusive decay is under better control, leading to the 
following result in the recently completed NLO calculation \cite{nlogreub},
\cite{nlomisiak}
\begin{eqnarray}
{\cal{B}}(B\to X_s \gamma)_{NLO}
=3.50 \pm 0.33 \cdot 10^{-4} \; .
\label{eq:bsgnlo}
\end{eqnarray}
Comparing eq.~({\ref{eq:bsgexp}}) and  eq.~({\ref{eq:bsgnlo}}), 
the CLEO measurement is found to be $2 \sigma$ away from the theory, but
the SM cannot be ruled out.
One has to look for other decay modes, since improving the theoretical 
accuracy in $B\to X_s \gamma$ decay seems not at hand. 
After displaying the methods developed in
$B \to X_s \gamma$, two other rare $B$ decays, \bxsll with $\ell=e, \mu$ and
the exclusive decay $B_s \to \gamma \gamma$ will be discussed in 
chapter \ref{chap:bsll} and chapter \ref{chap:bsgg} of this work, respectively.
Both candidates have branching ratios $\sim 10^{-6}$ which are in reach of 
future $B$ experiments. 
The aim of this thesis is to analyse these decays within the framework of 
the SM and to present up to date predictions for measurable 
quantities (branching ratios, distributions, asymmetries, etc) 
as accurately as possible with the present available techniques.
\begin{figure}[htb]
\vskip -0.8truein
\centerline{\epsfysize=10in
{\epsffile{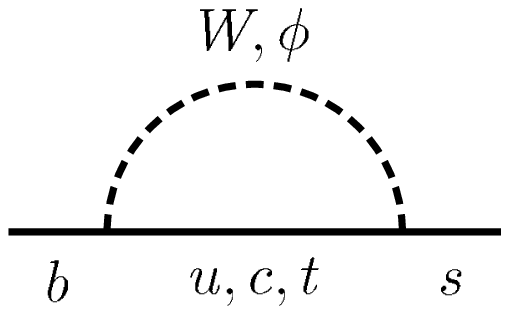}}}
\vskip -7.8truein
\caption{ \it A FCNC $b \to s$ diagram.}
\label{fig:bsg}
\end{figure}  

A typical diagram for $b \to s $ is displayed in Fig.~\ref{fig:bsg}
from where the CKM couplings can be directly read off.
The amplitude $T$ is the sum over all internal up-quarks
\begin{eqnarray}
T= \sum_{i=u,c,t} \lambda_i  T_i \; \; ; \; \;   
\lambda_i \equiv V_{i b} V_{i s}^{\ast} \; .
\end{eqnarray}
Using the CKM unitarity $\sum_{i=u,c,t} \lambda_i=0$ and the smallness of
$V_{ub}$  yielding $\lambda_u \ll \lambda_t$, we arrive at 
\begin{eqnarray}
T=\lambda_t (T_t-T_c)+\lambda_u (T_u-T_c) \simeq \lambda_t (T_t-T_c)
\end{eqnarray}
for a $b \to s$ amplitude in the SM.
In the $D$-system the FCNC transition rates ($c \to u$) are much more 
suppressed due to an inbuilt GIM mechanism \cite{GIM}. Here we have 
\begin{eqnarray}
T_{c \to u}& = &\sum_{i=d,s,b}   V_{c i} V_{u i}^{\ast} T_i \nonumber   \\
&=& V_{c b} V_{u b}^{\ast} (T_b-T_s)+ V_{c d} V_{u d}^{\ast}(T_d-T_s) \; ,
\end{eqnarray}
in which the first term is CKM suppressed and the second one GIM suppressed 
since $m_s^2-m_d^2 \ll m_W^2$.
The SM rates in the charm sector for decays such as 
$D^0 \to \gamma \gamma$, $D^0 \to \ell^+ \ell^-$ are out of reach for present 
experiments. If one nevertheless finds something in the rare charm sector, 
it would be a direct hint for the desired physics beyond the SM.

\section{The Effective Hamiltonian Theory \label{sec:effham}}

As a weak decay under the presence of the strong interaction, rare $B$ decays
require special techniques, to be treated economically.
The main tool to calculate such rare $B$ decays is the effective Hamiltonian
theory. It is a two step program, starting with an operator product 
expansion (OPE) and performing
a renormalization group equation (RGE) analysis afterwards.
The necessary machinery has been developed over the last years, see
%\cite{grinstein,Ciuchini,effhamali,effhamburas,misiakE,buchallaburasreview} 
\citer{effhamali,grinstein}, \cite{buchallaburasreview}
and references therein.

\begin{figure}[htb]
\vskip -1.0truein
\centerline{\epsfysize=10in
{\epsffile{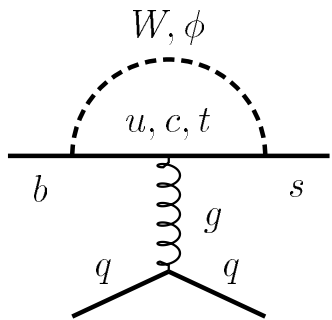}}}
\vskip -7.3truein
\caption{ \it A gluonic penguin diagram.}
\label{fig:penguin}
\end{figure}

The derivation starts as follows:
If the kinematics of the decay are of the kind that the masses of the internal particles $m_i$ are much larger than the external momenta $p$
$m_i^2 \gg p^2$, then 
the heavy particles 
%$(W,t,\phi,\ldots)$ 
can be {\it integrated out}. This concept takes a concrete form with 
the functional integral formalism. 
It means that the heavy particles  are removed
as dynamical degrees of freedom  from the theory, hence their fields do not 
appear in the (effective) Lagrangian anymore. 
Their residual effect lies in the generated effective vertices.
In this way an effective low energy theory can be constructed from a full 
theory like the SM.
A well known example is the four-Fermi interaction, where the 
$W$-boson propagator is made local for $q^2 \ll m_W^2$
($q$ denotes the momentum transfer through  the $W$):
\begin{eqnarray}
-i \frac{g_{\mu \nu}}{q^2-m_W^2} \to 
i g_{\mu \nu} (\frac{1}{m_W^2} + \frac{q^2}{m_W^4} +\dots \,) \; ,
\end{eqnarray}
where the ellipses denote terms of higher order in $1/m_W$.
\footnote{
We remark here that the original way was reversed:
The main historical step was to extrapolate the observed low energy 4-Fermi 
theory in nuclear $\beta$-decay to a dynamical theory of the weak interaction
with heavy particle exchange.}
Performing an OPE for QCD and electroweak 
interactions,
the effective Hamiltonian for a FCNC $b\to s \gamma$ transition in the SM
can be obtained by integrating out $W,t,\phi$.
Up to ${\cal{O}}(\frac{1}{m_W^4})$ it is given as:\\
\begin{eqnarray}
{\cal{H}}_{eff}(b\to s \gamma)=-4 \frac{G_{F}}{\sqrt{2}} \lambda_t 
\sum_{i=1}^{8} C_{i}(\mu) O_{i}(\mu) \; ,
\label{eq:heff}
\end{eqnarray}
where the weak couplings $g_W=\frac{e}{ \sin{\theta_W}}$ are collected in the 
Fermi constant $G_F$
\begin{eqnarray}
\frac{G_F}{\sqrt{2}}&=&\frac{g_W^2}{8 m_W^2} \; , \\
G_F &=&1.16639 \cdot 10^{-5} \; {\mbox{GeV}}^{-2} \; .
\end{eqnarray}
The on-shell operator basis is chosen to be \cite{effhamburas,effhamali}
\begin{eqnarray}
\label{eq:operatorbasis}
 O_1 &=& (\bar{s}_{L \alpha} \gamma_\mu b_{L \alpha})
               (\bar{c}_{L \beta} \gamma^\mu c_{L \beta}), \nonumber   \\
 O_2 &=& (\bar{s}_{L \alpha} \gamma_\mu b_{L \beta})
               (\bar{c}_{L \beta} \gamma^\mu c_{L \alpha}), \nonumber   \\
 O_3 &=& (\bar{s}_{L \alpha} \gamma_\mu b_{L \alpha})
               \sum_{q=u,d,s,c,b}
               (\bar{q}_{L \beta} \gamma^\mu q_{L \beta}), \nonumber   \\
 O_4 &=& (\bar{s}_{L \alpha} \gamma_\mu b_{L \beta})
                \sum_{q=u,d,s,c,b}
               (\bar{q}_{L \beta} \gamma^\mu q_{L \alpha}), \nonumber   \\
 O_5 &=& (\bar{s}_{L \alpha} \gamma_\mu b_{L \alpha})
               \sum_{q=u,d,s,c,b}
               (\bar{q}_{R \beta} \gamma^\mu q_{R \beta}), \nonumber   \\
 O_6 &=& (\bar{s}_{L \alpha} \gamma_\mu b_{L \beta})
                \sum_{q=u,d,s,c,b}
               (\bar{q}_{R \beta} \gamma^\mu q_{R \alpha}), \nonumber   \\  
 O_7 &=& \frac{e}{16 \pi^2}
          \bar{s}_{\alpha} \sigma_{\mu \nu} (m_b R + m_s L) b_{\alpha}
                F^{\mu \nu},    \nonumber                               \\
 O_8 &=& \frac{g}{16 \pi^2}
    \bar{s}_{\alpha} T_{\alpha \beta}^a \sigma_{\mu \nu} (m_b R + m_s L)  
          b_{\beta} G^{a \mu \nu} \; ,
%  O_9 &=& \frac{e^2}{16 \pi^2} \bar{s}_\alpha \gamma^{\mu} L b_\alpha
%  \bar{\ell} \gamma_{\mu} \ell , \nonumber\\
%  O_{10} &=& \frac{e^2}{16 \pi^2} \bar{s}_\alpha \gamma^{\mu} L
%     b_\alpha \bar{\ell} \gamma_{\mu}\gamma_5 \ell ~ \nonumber.
\end{eqnarray}
where $L(R)=1/2(1\mp \gamma_5)$, 
$\sigma_{\mu \nu}=\frac{i}{2} [\gamma_{\mu}, \gamma_{\nu}]$ and 
$\alpha,\beta$ are $SU(3)$ colour indices. $T^a, \, a=1 \dots 8$ are the
generators of QCD, some of their identities can be seen appendix \ref{app:qcd}.
Here $F^{\mu \nu}, \, G^{a \mu \nu}$ denote the electromagnetic and 
chromomagnetic field strength tensor, respectively.
As can be seen from the operator basis, only degrees of freedom which are 
light compared to the heavy integrated out fields ($W,t,\phi$), 
remain in the theory.
The basis given above contains four-quark operators $O_{1 \dots 6}$, which 
differ by colour and left-right structure. 
Among them, the current-current operators $O_1$ and $O_2$ are the 
dominant four-Fermi operators.
A typical diagram generating the so-called gluonic penguins $O_{3 \dots 6}$  
is displayed in Fig.~\ref{fig:penguin}.
The operators $O_7$ and $O_8$
are effective $b \to s \gamma$, $b \to s g$ vertices, respectively.
All operators have dimension 6.
For $b \to s \ell^+ \ell^{-}$ transitions the basis 
eq.~(\ref{eq:operatorbasis})
should be complemented by two additional operators containing dileptons. They
are discussed together with their corresponding Wilson coefficients in 
chapter \ref{chap:bsll}.

The coupling strength of the introduced effective vertices $O_i$ is given by
the (c-numbers) Wilson coefficients $C_{i}(\mu)$.
Their values at a large scale $\mu=m_W$ are obtained from a
``matching" of the effective with the full theory.
In the SM, the $C_{i}(m_W)$ read as follows \cite{inamilim} 
\begin{eqnarray}
\label{eq:CimW}
C_{1,3 \dots 6}(m_W)&=&0  \; , \\
C_2(m_W)&=&1  \; , \\
C_7(m_W)&=&\frac{3 x^3-2 x^2}{4(x-1)^4} \ln x+
\frac{-8x^3-5 x^2+7 x}{24 (x-1)^3} \; , \\
C_8(m_W)&=&\frac{-3 x^2}{4(x-1)^4}\ln x+
\frac{-x^3+5 x^2+2 x}{8 (x-1)^3} \; ,
\label{eq:CimWend}
\end{eqnarray}
with $x=m_t^2/m_W^2$.
It is convenient to define {\it effective} coefficients
$C_{7,8}^{{\mbox{eff}}}(\mu)$ of the operators $O_7$ and $O_8$. They
contain renormalization scheme dependent contributions from the 
four-quark operators $O_{1\ldots 6}$ in ${\cal H}_{eff}$ to the effective 
vertices in $b \rightarrow s \gamma$ and $b \rightarrow s g$, respectively.
In the NDR scheme
{\footnote{We recall that in the naive dimensional regularization (NDR) scheme
the $\gamma_5$ matrix is total anti-commuting, 
i. e. $\{\gamma_5,\gamma_\mu \}=0$, thus $L \gamma_\mu=\gamma_\mu R$.}}
, which will be used throughout this work, they are written as
\cite{effhamburas}
\begin{eqnarray}
\cseff(\mu)&=&C_7(\mu)+Q_d C_5(\mu)+Q_d N_c C_6(\mu) \; , \\
C_8^{{\mbox{eff}}}(\mu)&=&C_8(\mu)+ C_5(\mu)  \; .
\end{eqnarray}
Here $N_c$ denotes the number of colours, $N_c=3$ for QCD.
\begin{figure}[htb]
\vskip -1.0truein
\centerline{\epsfysize=10in
{\epsffile{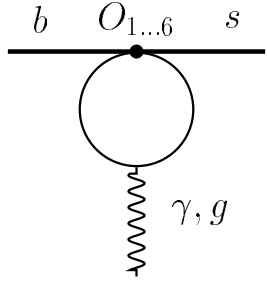}}}
\vskip -7.3truein
\caption{ \it The diagram contributing to the one loop $b \to s \gamma$,
$b \to s g$ matrix element, respectively.}
\label{fig:c78eff}
\end{figure}  
The above expressions can be found from evaluating the diagram shown in 
Fig.~\ref{fig:c78eff}.
Contributions from $O_{1 \dots 4}$, which correspond to an
$\gamma_\mu L \otimes \gamma_\mu L $ like insertion, vanish for an on-shell 
photon, gluon, respectively.
The Feynman rules consistent with these definitions are given in 
appendix \ref{app:feynrules}.

\subsection{QCD improved $\alpha_s$ corrections}

Our aim is now to include perturbative QCD corrections in the framework of 
the effective Hamiltonian theory.
This can be done by writing down the
renormalization group equation for the Wilson coefficients
{\footnote{with $C_i=C_i(\mu,g)$ we have equivalently
$\mu \frac{{\mbox{d}}}{{\mbox{d}}\mu} C_i=
(\mu \frac{\partial}{\partial \mu} +\mu  \frac{{\mbox{d}} g}{{\mbox{d}}\mu}
\frac{\partial}{\partial g}) C_i $.}}
\begin{eqnarray}
\label{eq:rge}
\mu \frac{{\mbox{d}}}{{\mbox{d}}\mu} C_i(\mu)=\gamma_{j i} C_j(\mu) \; ,
\end{eqnarray}
where $\gamma$ denotes the anomalous dimension matrix, i.e., in general the 
operators mix under renormalization.
Solving this equation yields
the running of the couplings $C_i(\mu)$ under QCD 
from the large matching scale (here $\mu=m_W$) down to the low scale 
$\mu \approx m_b$, 
which is the relevant one for $b$-decays.
Eq.~(\ref{eq:rge}) can be solved in perturbation theory $g^2=4 \pi \alpha_s$:
\begin{eqnarray}
\gamma_{j i}&=&\frac{g^2}{16 \pi^2} \gamma_{j i}^{(0)} 
+ (\frac{g^2}{16 \pi^2})^2 \gamma_{j i}^{(1)} + \ldots \; , \\
C_i(\mu)&=&C_i(\mu)^{(0)}+\frac{g^2}{16 \pi^2} C_i(\mu)^{(1)}  + \ldots \; .
\end{eqnarray}
The initial values of the above RGE are the $C_i(m_W)$, which in 
the lowest order in the SM are given in eq.~(\ref{eq:CimW}-\ref{eq:CimWend}).

Let us for the moment concentrate on the special case that $\gamma$ is a 
number. Then the lowest order solution is given by
\begin{eqnarray}
\label{eq:LLOG}
%C(\mu)&=&(\frac{\alpha_s(m_W)}{\alpha_s(\mu)})^
%{\frac{\gamma^{(0)}}{2 \beta_0}} C(m_W) \; , 
C(\mu)&=&\eta^{\frac{\gamma^{(0)}}{2 \beta_0}} C(m_W) \; , \\
\eta&=& \frac{\alpha_s(m_W)}{\alpha_s(\mu)} \; ,
\end{eqnarray}
which can be easily checked by substituting it into eq.~(\ref{eq:rge}).
In the derivation we have used the QCD $\beta$ function, 
which describes the running of the strong coupling:
\begin{eqnarray}
\label{eq:qcdbeta}
\beta(g)= \mu \frac{d}{d \mu} g =-g (\frac{g^2}{16 \pi^2} \beta_0 
+ (\frac{g^2}{16 \pi^2})^2 \beta_1) + \ldots \; , 
\end{eqnarray}
with its lowest order solution
\begin{eqnarray}
\label{eq:alphaslo}
\frac{\alpha_s(m_W)}{\alpha_s(\mu)}=\frac{1}
{1+\beta_0 \frac{\alpha_s(\mu)}{4 \pi} \ln(\frac{m_W^2}{\mu^2})} \; .
\end{eqnarray}
We see that our obtained solution eq.~(\ref{eq:LLOG}) contains all powers of 
$\alpha_s(\mu) \ln(\frac{\mu}{m_W})$.
It is called leading logarithmic (LLog) approximation and
is an improvement of the conventional perturbation theory.
In general such a QCD improved solution contains all large 
logarithms like $n=0,1, \dots$ (here with $\mu = m_b$)
\begin{eqnarray}
\alpha^n_s(m_b) \ln^m(\frac{m_b}{m_W}) \; ,
\end{eqnarray} 
where  $m=n$ corresponds to LLog.
A calculation including the next to lowest order terms is called
next to leading order (NLO) and would result in a summation of all terms with
$m=n-1$ and so on.
In the following we use the 2-loop expression for $\alpha_s(\mu)$
which can be always cast into the form
\begin{eqnarray}
\label{eq:2loop}
\alpha_s(\mu)&=&\frac{4 \pi}{\beta_0 \ln(\mu^2/\Lambda_{QCD}^2)}
 \left[1- \frac{\beta_1 \ln \ln(\mu^2/\Lambda_{QCD}^2)}
{\beta_0^2 \ln(\mu^2/\Lambda_{QCD}^2)} \right]\; .
\end{eqnarray}
With $N_f=5$ {\it active} flavours (note that we integrated out the top)
and $SU(N_c=3)$ the values of the coefficients of the $\beta$ function are 
\begin{eqnarray}
 \beta_0 =\frac{23}{3} \; ,  \hspace{1cm} \beta_1 =\frac{116}{3} \; .
\end{eqnarray}
They are given in appendix \ref{app:qcd} for arbitrary $N_c$ and $N_f$.
The strong scale parameter $\Lambda_{QCD} \equiv \Lambda_{QCD}^{(N_f=5)}$ is 
chosen to reproduce the
measured value of $\alpha_s(\mu)$ at the $Z^0$ pole.
For $\alpha_s(m_Z)=0.112,0.117,0.122$ we have
$\Lambda_{QCD}^{(5)}=160,214,280$ MeV, corresponding to the values of the 
input parameters listed in appendix \ref{app:input}.

We recall that in LLog the calculation of the anomalous 
dimension and the matching conditions at lowest order, 
$\gamma^{(0)},C_i^{(0)}(m_W)$ is required.
In NLO a further evaluation of higher order diagrams yielding
$\gamma^{(1)},C_i^{(1)}(m_W)$ is necessary and in addition the hadronic 
matrix elements $\langle O_i \rangle$ have also to be known in 
${\cal{O}}(\alpha_s)$.

In a general theory and also in the one relevant for rare radiative $b$ 
decays given in eq.~(\ref{eq:heff}), the operators mix and the matrix $\gamma$ 
has to diagonalized.
In the latter case the $(8\times 8)$ matrix $\gamma^{(0)}$ has been obtained 
by \cite{Ciuchini,misiakE} and the running of the $C_i(\mu)$ in LLog 
approximation cannot be given analytically.
The LLog solution for the Wilson coefficients ready for numerical analysis 
can be taken from \cite{burasmuenz}. We display the $C_i$ for different
values of the scale $\mu$ in Table \ref{LOWilson}. As can be seen, there is a 
strong dependence on the renormalization scale $\mu$, especially for 
$C_1$ and $\cseff$.
Other sources of uncertainty in the short-distance coefficients $C_i$
are the top mass and the value of $\alpha_s(m_Z)$. We keep them fixed to their 
central values given in appendix \ref{app:input}.
\begin{table}[h]
        \begin{center}
        \begin{tabular}{|l|r|r|r|r|}
        \hline
        \multicolumn{1}{|c|}{{\small{$C_i(\mu)$}}}&{{\small{$\mu=m_W$}}} & 
{{\small{$\mu=10$ GeV}}}&{{\small{$\mu=5$ GeV}}}&{{\small{$\mu=2.5$ GeV}}} \\
        \hline \hline
$C_1$       &  $0$     &$-0.161$  & $-0.240$ & $-0.347$  \\
$C_2$       &  $1$     &$1.064$   & $1.103$  & $1.161$\\
$C_3$       &  $0$     &$0.007$   & $0.011$  & $0.016$         \\
$C_4$       &  $0$     &$-0.017$  & $-0.025$ & $-0.035$ \\
$C_5$       &  $0$     &$0.005$   & $0.007$  & $0.010$  \\
$C_6$       &  $0$     &$-0.019$  & $-0.030$ & $-0.046$  \\   
$\cseff$ & $-0.196$ &$-0.277$  & $-0.311$ & $-0.353$  \\
$C^{{\mbox{eff}}}_8$ & $-0.098$ &$-0.134$  & $-0.148$ & $-0.164$\\
        \hline
        \end{tabular}
        \end{center}
\caption{ \it Leading order Wilson coefficients in the Standard Model as a 
function of the renormalization scale $\mu$.}
\label{LOWilson}
\end{table}

Here a comment about power counting in our effective theory is in order:
As can be seen from Fig.~\ref{fig:c78eff} with an external photon, the 
insertion of four-Fermi operators generates a contribution to $b\to s \gamma$,
which is also called a ``penguin".
It is a 1-loop diagram, but unlike ``normal" perturbation theory, of order
$\alpha_s^0$. To get the $\alpha_s^1$ contribution, one has to perform already
2 loops and so on.
That means, the calculation of the LO(NLO) anomalous dimension matrix was a 
2(3)-loop task.

A comprehensive discussion of weak decays beyond leading logarithms can 
be seen in ref.~\cite{buchallaburasreview}.
The main results of the NLO calculation in $B \to X_s \gamma$ decay 
will be given in section \ref{sec:bsgamma}.

The advantages of the effective theory 
compared to the full theory can be summarized as follows:
\begin{itemize}
\item The effective theory is the more appropriate way to
include QCD corrections. Large logarithms like $\ln(\mu/m_W)$ multiplied by 
powers of the strong coupling $\alpha_s(\mu)$, which spoil the
perturbation series in the full theory, can be resummed with the help of the 
RGE.
\item On the level of a generic amplitude $A=\langle {\cal{H}}_{eff} \rangle 
\sim \sum_i C_i(\mu) \langle O_i\rangle (\mu)$
the problem can be factorized into two parts: 
The short-distance (SD) information,
which can be calculated perturbatively, is encoded in the $C_i$, and it is 
independent of the external states, i.e. quarks or hadrons.
The long-distance (LD) contribution lies in the hadronic matrix elements.
Both are separated
by the renormalization scale $\mu$. Of course the complete physical answer 
should not depend on the scale $\mu$, truncating the perturbation 
series causes such a remaining dependence, which can be reduced only after 
including higher order terms or a full resummation of the theory.
\item As long as the basis is complete, the effective Hamiltonian theory 
enables one to write down a model independent analysis in terms of the SD 
coefficients $C_i$. This is true for SM {\it near} extensions like the 
two Higgs doublet model (2HDM) 
and the minimal supersymmetric model (MSSM). 
Here one can try to fit the $C_i$ from the data \cite{agm94}.
However, new physics scenarios like, e.g., the 
left-right symmetric model (LRM) require an extended operator 
set \cite{chomisiak,rizzo98,hewett98}.
Wilson coefficients in the 2HDM and in supersymmetry (SUSY) can be seen in 
ref.~\cite{grinstein90} and ref.~\cite{bsgsusy}, respectively.
\end{itemize}

\section{$B \rightarrow X_s \gamma$ in the Effective Hamiltonian Theory
\label{sec:bsgamma}}

The effective Hamiltonian theory displayed in the previous section is applied 
to $B \rightarrow X_s \gamma$ decay. Several groups have worked on the 
completion of the LLog calculation \cite{Ciuchini,misiakE}.
The anomalous dimension matrix at leading order $\gamma^{(0)}$
and the lowest order matching conditions 
(eq.~(\ref{eq:CimW}-\ref{eq:CimWend})) govern the 
running of the LLog Wilson coefficients, denoted in this and only this 
section by $C_i^{(0)}(\mu)$, to separate them from the NLO coefficients. 
We discuss the improvement of the theory in $B \rightarrow X_s \gamma$ 
obtained from NLO analysis. In the remainder of this work we treat the
Wilson coefficients $C_{i}, \, i=1, \dots 8$ in LLog approximation.

In the spectator model, the branching ratio for $B \rightarrow X_s \gamma$  
in LLog approximation can be written as 
\begin{eqnarray}
{\cal{B}}(B \to X_s \gamma) = 
{\cal{B}}_{sl} \frac{\Gamma(b \to s \gamma)}{\Gamma(b \to c e \bar{\nu}_e)}=
\frac{|\lambda_t|^2}{|V_{cb}|^2} \frac{6 \alpha}{\pi f(\mc)} 
|C_7^{(0) {\mbox{eff}}}(\mu)|^2 \; ,
\end{eqnarray}
where a normalization to the semileptonic decay $B \to X_c \ell \nu_{\ell}$
to reduce the uncertainty in the $b$-quark mass has been performed.
Here ${\cal B}_{sl}$ denotes the measured semileptonic branching ratio and
the phase space factor $f(\mc)$ with $\mc=m_c/m_b$ for 
$\Gamma (B \rightarrow X_c \ell \nu_{\ell})$ 
can be seen in  eq.~(\ref{eqn:fr}).

As the branching ratio for $B \to X_s \gamma$ is mainly driven by 
$C_7^{(0) {\mbox{eff}}}(\mu)$, several effects can be deduced:
\begin{itemize}
\item Including LLog QCD corrections enhance the branching ratio for 
$B \to X_s \gamma$ about a factor $2-3$, as can be seen in 
Table \ref{LOWilson} (here denoted by $C_i(\mu)$) and changing the scale from 
$\mu=m_W$ down to $\mu \sim m_b$.
\item While the sign of $C_7^{(0) {\mbox{eff}}}$ is fixed within the SM, i.e. 
negative, it can be plus or minus in possible extensions of the SM. 
A measurement of ${\cal{B}}(B \to X_s \gamma)$
alone is not sufficient to determine the sign of $C_7^{(0) {\mbox{eff}}}$, or 
in general, the sign of $C_7^{{\mbox{eff}}}$ resulting from possible higher 
order calculations.
\item The strong scale dependence of the value of $C_7^{(0) {\mbox{eff}}}(\mu)$
causes serious problems in the accuracy of the LLog prediction.
Varying the scale between $\frac{m_b}{2} \leq \mu \leq 2 m_b$ , results in an 
error in the branching ratio of $\pm 25 \%$ 
\cite{aligreub93}, \cite{effhamburas}.
\end{itemize}
Because of the last point the NLO calculation was required.
Several steps have been necessary for a complete NLO analysis.
Let us illustrate how the individual pieces look like:
At NLO, the matrix element for $b \to s \gamma$ renormalized around
$\mu=m_b$ can be written as \cite{effhamburas}:
\begin{eqnarray}
{\cal{M}}(b \to s \gamma)=-4 \frac{G_{F}}{\sqrt{2}} \lambda_t 
D \langle O_{7}(m_b) \rangle_{tree} \; ,
\label{eq:matbsg}
\end{eqnarray}
with
\begin{eqnarray}
D=\cseff(\mu)+\frac{\alpha_s(m_b)}{4 \pi} \sum_{i} \left(
C_i^{(0) \mbox{eff}}(\mu) \gamma_{i7}^{(0)} \ln \frac{m_b}{\mu}+
C_i^{(0)\mbox{eff}}(\mu) r_{i7} \right)  \; .
\label{eq:D}
\end{eqnarray}
The $r_{i7}$ are computed in ref.~\cite{nlogreub}.
They contain the bremsstrahlung corrections \cite{effhamali}, \cite{Pott}
$b \to s \gamma g$  and virtual corrections to the $O_7$ matrix element 
\cite{nlogreub}.
Especially the latter with an $O_2$ operator insertion demands
an involved 2-loop calculation, see Figs.~1-4 in \cite{nlogreub}, where the 
corresponding diagrams are shown.
It is consistent to keep the pieces in the parentheses in eq.~(\ref{eq:D}), 
which are multiplied by $\alpha_s(m_b)$, in LLog approximation.

Now $\cseff(\mu)$ has be be known at NLO precision,
\begin{eqnarray}
\cseff(\mu)=C_7^{(0)\mbox{eff}}(\mu)+
\frac{\alpha_s(m_b)}{4 \pi}C_7^{(1)\mbox{eff}}(\mu)  \; .
\end{eqnarray}
As this job consists out of two parts, 
the work has been done by two groups: The ${\cal{O}}(\alpha_s^2)$ anomalous 
dimension matrix was obtained in ref.~\cite{nlomisiak}, which required the 
calculation of 
the residue of a large number of 3-loop diagrams, describing the mixing 
between the four-Fermi operators $O_{1 \dots 6}$ and $O_{7,8}$.
The second part, the NLO matching at $\mu=m_W$ has been done in 
ref.~\cite{adelyao94} and confirmed in ref.~\cite{greubhurth}.
The NLO calculation reduces the $\mu={\cal{O}}(m_b)$ scale uncertainty 
in varying $\mu$ in the range $\frac{m_b}{2} \leq \mu \leq 2 m_b$
drastically to $\pm 4.3 \%$ \cite{buraskwiatpott}
and suggests for $B \to X_s \gamma$ a scale $\mu=\frac{m_b}{2}$ as an 
``effective" NLO calculation through
\begin{equation}
\Gamma(B \rightarrow X_s \gamma)_{LO} (\mu=\frac{m_b}{2})
\approx \Gamma (B \rightarrow X_s \gamma)_{NLO} \; .
\end{equation}

As a final remark on scale uncertainties it should be noted that in the 
foregoing the top quark and the $W$ have been integrated out at the same scale
$\mu = m_W$, which is an approximation to be tested.
It is justified by the fact that the difference between
$\alpha_s(m_W)$ and $\alpha_s(m_t)$ is much smaller than the one
between $\alpha_s(m_W)$ and $\alpha_s(m_b)$.
\footnote{Using eq.~(\ref{eq:2loop}) and the input parameters in 
Table~\ref{parameters}, we have
$\alpha_s(m_W)=0.12,\alpha_s(m_t)=0.11$ and $\alpha_s(m_b)=0.21$.}
The authors of \cite{buraskwiatpott} analysed the dependencies on both
the $W$ matching scale $\mu_{W}={\cal{O}}(m_W)$ and the one at which the 
running top mass is defined: $\bar{m}_{t}(\mu_t)$ and 
$\mu_{W} \not=\mu_{t}$.
Similar to the $m_b$ scale they allowed for $\mu_W,\mu_t$ a possible range:
$\frac{m_x}{2} \leq \mu_x \leq 2 m_{x}$  where $x=W,t$.
Their findings are that the $\mu_{W},\mu_{t}$ uncertainty is
much smaller 
(namely $\pm 1.1 \%,\pm 0.4 \%$  at $\mu \sim m_b$ in NLO, respectively) than 
the uncertainty in the scale around $m_b$ and therefore negligible.

\section{Long-Distance Effects in Inclusive $B$ Decays}

In this section we sketch the methods to treat the LD effects in 
inclusive $B$ decays. 
We have effects due to the confinement of the quarks in 
a bound state and due to resonances.
They will be explained more detailed in the following chapters when and where 
necessary.
For the evaluation of the exclusive channels we refer to chapter 
\ref{chap:bsgg} in which the rare mode $B_s \to \gamma \gamma$ is discussed,
especially section \ref{sec:qcdsumrule}.

There are mainly two different approaches to take into account the effects of 
the $B$-meson bound state, the heavy quark expansion (HQE) and the 
phenomenologically motivated Fermi motion model (FM).
While the former is a field theory in the framework of the heavy 
quark limit of QCD and has an interest of its own, 
the latter serves as a model of the data and has no intrinsic problems like 
end-point singularities etc. like HQE.
Both models have parameters which can be related to each other 
and as they are used as 
{\it inclusive} methods, no form factors appear in the amplitudes.
Inclusive decays are good from theoretical point of view 
and a challenge for experimentalists:
An inclusive final state $X$ is an average over a sufficiently high number
of exclusive single (resonances) and continuous multi body states with the 
same quantum numbers as $X$.
Inclusive decays involve the calculation of quark level processes.
The underlying assumption of quark-hadron duality requires a large and
dense enough populated phase space.
By means of a ``smearing" procedure, the singular behaviour in a local form is 
avoided and the differential spectra can be measured in a distribution sense.

Another type of LD effect beside the bound state effect mentioned above, is 
due to resonances. 
A $(q \bar{q})$ spin 1 state can hadronize from a virtual $q$-loop. 
The conversion of such a vector meson into a photon is described by the 
phenomenological vector meson dominance (VMD) model.

\subsection{The heavy quark expansion of QCD \label{sec:hqet}}

Consider a hadron containing one heavy quark $Q$ in the limit 
$m_Q \to \infty$. 
The other ingredients, light quarks and gluons, are seen 
as a light cloud around the heavy quark, 
sometimes also called the ``light brown muck" \cite{georgitasi}, which
exchanges small momenta of order $\Lambda_{QCD}$ with Q for 
which a perturbative expansion is not useful.
The parameter $\Lambda_{QCD}$ characterizes the soft 
hadronic interaction scale.
The heavy quark inside the bound state is treated as
a static source of gauge charge (colour and electric charge): 
Q is so heavy compared to $\Lambda_{QCD}$, 
that it does not recoil as a result of the soft 
exchanges, it sits at rest in the hadron rest frame.
This is the heavy quark limit of QCD.

New symmetries can be explored which are exact in the limit of an 
infinitely heavy quark:
The light degrees of freedom are insensitive to the mass, flavour and spin of 
the heavy quark!
This brings an enormous simplification of certain aspects of QCD, like the 
calculation of heavy quark matrix elements and hadron spectroscopy. 
However, for a firm phenomenological analysis
we need to go from the predictions of heavy quark symmetry in the strict
limit $m_Q \to \infty$ to a theory which provides a controlled expansion 
around this (academic) case. This can be done with the heavy quark expansion 
technique (HQET) in the limit 
$m_Q \gg \Lambda_{QCD}$. 
The necessary technology has been developed over the last decade and can be 
seen in a selection of papers \cite{georgi,MW} and references therein.
A nice review on the HQE technique is given in ref.~\cite{falklecture96}.

\subsubsection{$1/m_b$ power corrections}

Let us now switch to the system under consideration, the $B$-mesons.
Since the $b$-quark is heavy, i. e. 
$m_b \sim 4.8 {\mbox{GeV}} \gg \Lambda_{QCD} \sim 200 \, {\mbox{MeV}}$, 
the success of the spectator model in $B \equiv (\bar{b} q)$ meson decays
can be understood. Moreover, corrections in inverse powers of $m_b$ to this 
can be systematically obtained with the help of the HQET. 

The light degrees of freedom in the $B$-meson give rise to the parameter 
$\bar{\Lambda}$ which accounts for the binding 
energy of the bound state. 
In the limit of an infinitely heavy $b$-quark, i.~e. $m_b \to \infty$ we have
\begin{eqnarray}
m_B=m_b+\bar{\Lambda} \; ,
\end{eqnarray}
where $m_B$ denotes the $B$-meson mass.
Corrections to this can be calculated within the following 
set up of HQE:
The heavy $b$-quark momentum is written as $p_b=m_b v+k$, where 
$k$ is a  small residual momentum of order $\Lambda_{QCD}$. $v$ denotes 
the velocity of the meson 
with  momentum $P=m_{B} v$, which at rest is  $v=(1,0,0,0)$.
It follows that the relative movement between the heavy quark and the meson 
is suppressed by powers of $k/m_b$.
%%%%%%%%%%%%%
Performing an operator product expansion up to operators with canonical field 
dimension 5, the HQET mass relation modifies to \cite{Luke90}
\begin{eqnarray}
m_{B}=m_b +\bar{\Lambda}-\frac{1}{2 m_b} (\lambda_1+3 \lambda_2)+\dots \; ,
\label{hqetmass}
\end{eqnarray}
likewise for $B^*$ vector mesons
\begin{eqnarray}
m_{B^*} &=& m_b + \bar{\Lambda} -\frac{1}{2 m_b} (\loo -\lto)+ \dots \; ,
\end{eqnarray}
where the ellipses denote terms higher order in $1/m_b$.
In general, the next to leading power corrections in HQET are
parameterized in terms of these matrix elements of the kinetic energy and 
the magnetic moment operators $\lambda_1$ and $\lambda_2$, respectively.
We can get the value of $\lambda_2$ from spectroscopy
\begin{eqnarray}
\lambda_2 \simeq \frac{m_{B^{*}}^2 -m_B^2}{4}= 0.12 \,  {\mbox{GeV}}^2 \; .
\end{eqnarray}
The quantity $\loo$  is subject to a theoretical dispersion.
Its value has been determined from
QCD sum rules, yielding $\lambda_1=-(0.52 \pm 0.12)$ GeV$^2$ (Ball and
Braun in \cite{kineticsr}) and $\lambda_1=-(0.10 \pm 0.05)$ GeV$^2$
(Neubert \cite{neubertsr96}).
Further, the value for $\lambda_1$ has been extracted from an analysis of
data on semileptonic $B$ decays ($B \to X \ell \nu_{\ell}$), yielding 
$\lambda_1=-0.20 ~\mbox{GeV}^2$ with 
a corresponding value $\bar{\Lambda}=0.39 \, 
\mbox{GeV}$, as the two are correlated \cite{gremm}.
For a review on the spread in the present values of these non-perturbative 
parameters extracted from inclusive decay spectra, 
see Table \ref{tab:nonpertparam}, which is adopted from \cite{neubert98}.

Now there is an intrinsic difficulty in the relation eq.~(\ref{hqetmass}). 
One can ask for the meaning of the ``pole" mass $m_b$ and $\bar{\Lambda}$ ?
First of all, they are non-perturbative parameters and they add up in 
the combination given by eq.~(\ref{hqetmass}) to the physical $B$-meson mass.
However, while the sum is fixed, there is a scheme dependent 
``renormalon" ambiguity of order $\Lambda_{QCD}$ in both $m_b$ and 
$\bar{\Lambda}$, which cancels out in physically measurable quantities 
\cite{neubertsachr}.
Assuming universality, the parameters $m_b, \bar{\Lambda}$
determined by one experiment can be used to help the analysis of another decay,
provided that one uses the same renormalization scheme prescription.

The power corrections in \bxsg decay including $1/m_b^2$ terms have been 
calculated in ref.~\cite{FLSold}. The $1/m_b^3$ corrections have been recently 
reported in ref.~\cite{Bauer97}.
However, the use of the last calculation is limited by the fact, that the 
matrix elements of the higher dimensional operators are almost unknown.
In \bxsll decay the $1/m_b^2$ have been first calculated in ref.~\cite{FLSold}
(with massless $s$-quark) and corrected in ref.~\cite{AHHM97} with full $m_s$.
The latter has been confirmed recently in the massless $s$-quark case in 
ref.~\cite{buchallaisidorirey}. Details of the HQET calculation
are given in chapter \ref{chap:bsll}.
\begin{table}
\begin{center}
\begin{tabular}{|l|l|cc|}\hline
{}~Reference & ~Method & ~$\bar\Lambda$ [GeV]~ &
 ~$\lambda_1$ [GeV$^2$]~ \\
\hline
{}~Falk et al. \cite{FLSphenom}  & ~Hadron Spectrum~ &
 $\approx 0.45$ & $\approx -0.1$ \\
\hline
{}~Gremm et al. \cite{gremm}  & ~Lepton Spectrum &
 $0.39\pm 0.11$ & $-0.19\pm 0.10$ \\
{}~Chernyak \cite{chernyak96}  & ~$(\bar B\to X\,\ell\,\bar\nu)$ &
 $0.28\pm 0.04$ & $-0.14\pm 0.03$ \\
{}~Gremm, Stewart \cite{gremmstewart}  & & $0.33\pm 0.11$ &
 $-0.17\pm 0.10$ \\
\hline
{}~Li, Yu \cite{liyu}  & ~Photon Spectrum &
 $0.65_{-0.30}^{+0.42}$ & $-0.71^{+0.70}_{-1.16}$ \\
 & ~$(\bar B\to X_s\gamma)$ & & \\
\hline
\end{tabular}
\end{center}
\caption{ \it Determinations of the parameters $\bar\Lambda$ and $\lambda_1$
from inclusive decay spectra.}
\label{tab:nonpertparam}
\end{table}

\subsubsection{$1/m_c$ power corrections}
\begin{figure}[htb]
\vskip -1.0truein
\centerline{\epsfysize=10in
{\epsffile{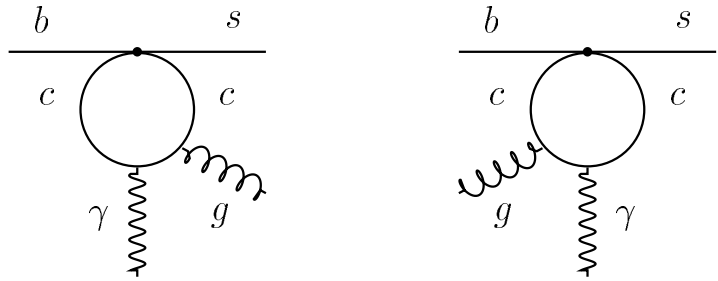}}}
\vskip -7.3truein
\caption{ \it The $b s \gamma g$-vertex at lowest order in QCD.}
\label{fig:1/mc}
\end{figure}

With $m_c \sim 1.4\mbox{GeV} \gg \Lambda_{QCD}$ the charm is still a heavy 
quark and also $1/m_c$ power corrections are subject of present $B$-physics.
A theoretically interesting structure, an effective $b s \gamma g$-vertex 
appears from the diagrams displayed in Fig.~\ref{fig:1/mc} \cite{Ruckl}. 
The amplitude of this operator can be expanded in $1/m_c$. 
The power corrections in  $\Lambda_{QCD}/m_c$ to the \bxsg decay rate have 
been calculated first in ref.~\cite{voloshin} and \cite{grantetal}, 
however with the wrong sign. This has been settled now in favor of 
\cite{grant,LRW97,buchallaisidorirey}. 
The resulting correction to the decay rate is found to be small:
\begin{eqnarray}
\frac{\delta \Gamma(B \to X_s \gamma)}{\Gamma(B \to X_s \gamma)}
=-\frac{C_2 \lambda_2}{9 \cseff m_c^2} \sim +0.03 \; .
\end{eqnarray}
The $1/m_c$ expansion is an alternative description of the virtual charm loop 
to the traditional vector meson dominance ansatz,
(see section \ref{subsec:vmd}, Fig.~\ref{fig:vmdconversion})
in regions of 
momentum transfer $q^2$ far away from the resonances.

Pros and cons of the HQE approach
\begin{itemize}
\item systematic expansion in $\Lambda_{QCD}/m_{c,b}$
\item well-defined limit of QCD
\item unknown matrix elements of higher order operators
\item expansion breaks down in some regions of the phase space
\end{itemize}
The last point above corresponds to an uncomfortable property of the HQE 
method: The development of end-point singularities in inclusive decay spectra
\cite{MW,AHHM97}, hence only quantities smeared over a 
sufficiently large phase space interval are calculable. 
Such a smearing is incorporated, e.g., in the Fermi motion
model \cite{aliqcd}, discussed in the next section.

\subsection{The Fermi motion model \label{subsec:FM}}

Another model to handle the effects of the bound state is the phenomenological
Fermi motion model (FM) \cite{aliqcd}. 
The FM is defined through the requirement that the $\bar{b}$-quark and the 
spectator quark $q$ four-momenta add up to the $B=(\bar{b} q)$-meson 
four-momentum.
In the rest frame of the $B$-meson the quarks 
fly back to back with momentum
$\vec{p}\equiv \vec{p_b}=-\vec{p_q}$. 
From energy conservation follows than that either the spectator or the 
$\bar{b}$ quark has to have a momentum dependent mass. 
In this work we choose $m_q$ to be a parameter of the model
and the $\bar{b}$ quark to have a $p$ dependent mass. It reads
\begin {eqnarray}
m_b^2(p) = {m_B}^2 + {m_q}^2 -2m_B \sqrt{p^2 + {m_q}^2} 
\quad ; \quad p = |\vec{p}|\, .
\end{eqnarray}
The next assumption is that the momentum $p$ obeys the Gaussian 
distribution function $\phi(p)$ weighted with the Fermi momentum $p_F$
\begin{equation}
\label{lett13}
 \phi(p)= \frac {4}{\sqrt{\pi}{p_F}^3} \exp (\frac {-p^2}{{p_F}^2}) \; ,
\end{equation}
with the normalization
$ \int_0^\infty \, dp \, p^2 \, \phi(p) = 1 $. 
The procedure to implement these wave function effects to a
general parton model distribution obtained in the $\bar{b}$ quark rest frame
is as follows: \\
1. replace the $\bar{b}$ quark mass by $m_b(p)$ \\
2. boost the distribution into the $B$-meson rest frame and\\ 
3. fold the result with the wave function given in eq.~(\ref{lett13}).

For subsequent
use in working out the normalization (decay widths) in the FM model, we
also define an {\it effective} $b$-quark mass by
\begin{equation}
\label{effbmass}
m_b^{{\mbox{eff}}}\equiv(\int_0^\infty  dp \, p^2 \, m_b(p)^5 
\phi(p))^{1/5}~.
\end{equation}

The two parameters of the FM model, the Fermi momentum $p_F$ and the spectator 
mass $m_q$ can be fitted from data, however, up to now this procedure has not 
been very conclusive as still large ranges of the parameters are possible.
The question appears here whether the FM parameters do depend on flavour,
i. e., are they universal for $B \to X_f+(\gamma,leptons)$ with 
$f=d,u,s,c$ transitions.
Further relations between the FM parameters $(p_F,m_q)$ 
and the HQET parameters $(\lambda_1, \bar{\Lambda})$ can be 
obtained. However, there is no analogue of the magnetic moment coupling 
$\lambda_2$ in the FM.

We will return to the FM in chapter {\ref{chap:bsll}} to model the
wave function effects in \bxsll decay.

\subsection{Vector meson dominance \label{subsec:vmd}} 

\begin{figure}[htb]
\vskip -1.0truein
\centerline{\epsfysize=10in
{\epsffile{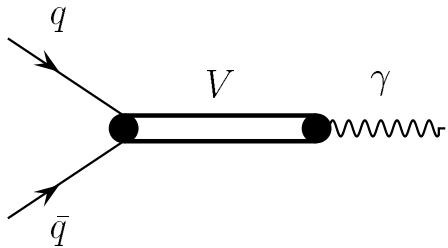}}}
\vskip -7.3truein
\caption{ \it The VMD conversion.}
\label{fig:vmdconversion}
\end{figure}

Vector meson dominance (VMD) provides a mechanism to convert a spin 1 meson, 
here generically denoted by $V \equiv (q \bar{q})$ 
into a photon \cite{sakurai}. 
The creation of a $ (q \bar{q})$
bound state from a virtual $q, \bar{q}$ pair 
and its subsequent conversion into a photon is displayed in 
Fig.~\ref{fig:vmdconversion}.
The intermediate vector meson propagator equals
$1/(m_V^2-q^2-i m_V \Gamma^{tot}_V)$,
where the imaginary parts takes the effect of a finite total width 
$\Gamma^{tot}_V$ of the $V$ into account.
The matrix element of the constituent current is defined as
\begin{eqnarray}
<0|\bar{q} \gamma_{\mu} q|V (q,\epsilon^V)>=
f_{V}(q^2) m_{V} \epsilon_{\mu}^V \; ,
\end{eqnarray}
and the photon interacts with the (neutral) bound state through the 
electromagnetic current 
$J_{\mu}^{em}=e Q_q \bar{q} \gamma_{\mu} q$ as:
\begin{eqnarray}
<0|J_{\mu}^{em}|V (q,\epsilon^V)>=
e Q_q f_{V}(q^2) m_{V} \epsilon_{\mu}^V \; .
\end{eqnarray}
Here $q,\epsilon^V$ are the momentum and the polarization vector of the
vector meson $V$, respectively and $Q_q$ denotes the charge of the quark $q$
in units of $e$.
VMD conversion means that while $V \to \gamma$ also 
$\epsilon_{\mu}^V \to \epsilon_{\mu}$, where 
$\epsilon_{\mu}$ is the polarization vector of the photon.
The form factor at $q^2=m_V^2$ can be obtained from measurement of the
leptonic width
\begin{eqnarray}
\Gamma(V \rightarrow e^{+}e^{-})=f^{2}_V 
\frac{Q_q^2 4 \pi\alpha^{2}}{3 m_V} \; .
\end{eqnarray}

What about  $f_V(q^2)$ at $q^2\not =m_V^2$ ?
For the calculation of the form factor at 
$0 \leq q^2 \leq m_V^2$, there exist different ways, which are neither 
straightforward nor unique.
a) First of all data on photo production $\gamma N\to V N$ at
$q^2=0$ can be used. They indicate a large suppression of $f_V(0)$ compared 
to  $f_V(m_V^2)$ only if the vector meson is heavy such as $V=J/\psi,\psi'$.
As in this case $q^2=m_{J/\psi}^2,m_{\psi'}^2$ to $q^2=0$ involves a large extrapolation in $q^2$. It is not expected to be so significant for
$V=\rho,\omega,\phi$ as $q^2=m_{\rho}^2$ etc. is not far away from $q^2=0$. 
This is what all the methods listed here share, that 
$f_V(q^2)$ decreases with decreasing $q^2$.
b) Often a single-pole form is assumed to extrapolate the form factor
to smaller values of the momentum transfer
\begin{eqnarray}
f_V(q^2)=\frac{f_V(0)}{1-q^{2}/m_{pole}^{2}} \; ,
\end{eqnarray}
where $m_{pole}$ corresponds to the masses of higher resonances of the $V$.
%This works well for light mesons where the extrapolation goes not too far 
%from $q^2 \sim 0$. 
c) The approach by \cite{terasaki} is based on a dispersion relation 
calculation. It yields an interpolation formula between $f_V(0)$
and $f_V(m_V^2)$ where both can be fixed by data. 
The situation for $q^2 >m_V^2$ is unclear.

%If the photon is on-shell, another VMD requirement is transversality.
%A gauge invariant amplitude has to be constructed for $V^{T} \to \gamma$,
%i. e. $\epsilon_{\mu}^{V T} \to \epsilon_{\mu}$,
%where the superscript $T$ denotes the transversal components of the 
%(massive) vector meson field.
%This is not a problem for
%light mesons, but needs further study for heavier ones like the charm 
%resonances, to obtain a transverse VMD amplitude.

The VMD mechanism has been applied to \bxsg decay to estimate the long 
distance contribution through $B \to X_s \psi^i \to X_s \gamma$ by
\cite{deshpande,pakvasa}.
Here $ \psi^i=J/\psi, \psi^{\prime}, \dots \psi^{(v)}$ are the known six
charmonium resonances, see \cite{PDG}. 
The  $\psi^i \rightarrow \gamma $ conversion requires the knowledge of the 
form factor at $q^2=0$ for an on-shell photon. It has been shown in ref.
\cite{deshpande}, that the methods a) and c) 
listed above yield a consistent suppression at $q^2=0$. 
The longitudinal degrees of freedom of the $\psi^i$ have been removed
using the procedure proposed by \cite{pakvasa}.
However, assumptions made remain as uncertainties in the calculation.

We will make extensive use of VMD in chapter \ref{chap:bsll}
and chapter \ref{chap:bsgg} to include long-distance effects
from intermediate vector mesons in \bxsll and 
$B_s \to \g \g$ decays, respectively.
However, in chapter \ref{chap:bsll} we estimate the resulting uncertainties 
which emerge from various theoretical approaches in implementing the 
$q^2$-dependence of the VMD-dominated amplitude.

\chapter{Inclusive \bxsll Decay  \label{chap:bsll}}

%{\it I'm a fountain of blood, in the shape of a girl...like a killer whale}\\
%"bachelorette", Bj\"ork

This chapter contains a comprehensive analysis of
\bxsll decay in the standard model (SM). 
We include QCD improved ${\cal{O}}(\alpha_s)$ corrections, use heavy quark 
expansion techniques (HQET) and apply the Fermi motion model (FM).
Further, the long-distance effects via intermediate $J/\psi, \psi^\prime,...$
resonances are taken into account with a vector meson dominance (VMD) ansatz.

\section{Introduction}
Flavour changing neutral current (FCNC) decays \bxsll and \bxsg are 
governed in the SM by loop effects.
They provide a sensitive probe of the flavour sector in the SM and search for 
physics beyond. In the context of rare $B$ decays the radiative mode \bxsg 
has been extensively discussed in chapter
\ref{chap:rare}.

In this chapter we address inclusive \bxsll decay with $\ell=e,\mu$. 
Since we are 
neglecting finite lepton masses we cannot apply our results to the $\tau$-case.
The $b \to s \ell^+ \ell^-$ transition has been studied earlier in the free 
quark model in 
%refs.~\cite{HWS87,Grinstein89,JW90} 
refs.~\citer{HWS87,JW90} in the lowest order in the SM context.
The NLO ${\cal{O}}(\alpha_s)$ improvement in the invariant dilepton mass 
distribution and the decay rate has been worked out in 
refs.~\cite{misiakE,burasmuenz}.
Leading $(1/m_b^2)$ power corrections in the HQET framework \cite{georgi,MW}
in the invariant 
dilepton spectrum in \bxsll decay have been reported in ref.~\cite{AHHM97}, 
correcting an earlier calculation \cite{FLSold}. 
This has been recently confirmed in ref.~\cite{buchallaisidorirey} for the 
massless $s$-quark case.
Another interesting quantity in \bxsll is the FB asymmetry \cite{amm91},
also known at $1/m_b^2$ \cite{AHHM97}.
It can be used together with the branching ratio of $B \to X_s \g$ 
and the dilepton spectrum in \bxsll for a model independent analysis of the
short-distance coefficients \cite{agm94} in the search for SUSY effects 
%\cite{choetal,Gotoetal97,Hewett96,hewett98}.
\citer{choetal,Hewett96}, \cite{hewett98}.
 The $1/m_b^2$ power corrections to the left-right
asymmetry \cite{LD96,BP97} have been presented
in \cite{BI98} correcting an earlier calculation of the same
\cite{BP97}. 
Both the FB asymmetry and the left-right asymmetry are defined in
section \ref{sec:asymmetries}.
The longitudinal
polarization of the lepton, $P_L$, in $B \to X_s \tau^+ \tau^-$ at the
partonic level has been worked out \cite{Hewettpol}; the other two
orthogonal polarization components $P_T$ (the component in the decay plane)
and $P_L$ (the component normal to the decay plane) were subsequently worked 
out in \cite{KS96}.
The ${\cal{O}}(1/m_c^2)$ correction to the dilepton invariant mass spectrum 
in \bxsll has also been 
calculated in \cite{chenrupaksavage}, however, the result differs in 
sign from the one in \cite{buchallaisidorirey}.
This controversy, which goes back to the corresponding one in \bxsg decay
(see section \ref{sec:hqet} for a discussion) has been settled in 
favor of \cite{buchallaisidorirey}.
It is known that inclusive decay spectra are not entirely calculable with HQET
%\cite{Bigietal2,falketal1,JR94,MW},
\citer{Bigietal2,JR94}, \cite{MW},
especially, the expansion in powers of $1/m_b$ diverges in the high
dilepton mass $q^2$ region in \bxsll decay \cite{AHHM97}.
An alternative approach to take into account $B$-meson bound state effects 
is the FM model \cite{aliqcd}.
In the FM a prediction for the entire $q^2$ range for the dilepton mass 
distribution and FB asymmetry in 
\bxsll decay has been given ref.~\cite{AHHM97}. The sensitivity of the 
distributions on the FM parameters is worked out there.
Long-distance effects due to intermediate 
$B \to X_s +(J/\psi, \psi^\prime,...) \to X_s \ell^+ \ell^- $ have been 
discussed in refs.~\cite{long,KS96,Ahmady96} and recently \cite{AH98-3}.
Hadron spectra and hadronic spectral moments are presented in refs.~ 
\cite{AH98-1,AH98-2,AH98-3} in both the HQE approach and the FM.

This chapter is divided into two parts.
The first one (this section up to and including section \ref{sec:LD1}) 
based on ref.~\cite{AHHM97}, contains an introduction to 
\bxsll decay, basic definitions and the ${\cal{O}}(\alpha_s)$ and $1/m_b^2$
corrected matrix element for $b \to s \ell^+ \ell^{-}$. Is is mainly devoted 
to the analysis of the dilepton invariant 
mass distribution and the FB asymmetry.
In doing that, we derive leading power corrections to the decay rates
and $q^2$ distributions in the decay \bxsll using heavy quark expansion 
(HQE) in $(1/m_b)$.
Further, wave function effects of the $b$-quark bound in the $B$-hadron are 
studied by us in the phenomenologically motivated Gaussian Fermi motion 
model.
Using this model for estimating the non-perturbative effects, 
we include the dominant long-distance (LD) contributions from the decays
$B \to X_s +(J/\psi, \psi^\prime,...) \to X_s \ell^+ \ell^- $.
Further, taking into account the next-to-leading order perturbative QCD 
corrections in $b \to s \ell^+ \ell^-$, we present the decay rates and 
distributions for the inclusive process \bxsll in the SM.

The second part, starting from section \ref{sec:hadronintro} 
complements the study of \bxsll decay and investigates hadron 
spectra and hadronic spectral moments.
It is mainly based on refs.~\cite{AH98-1,AH98-2,AH98-3}.
We compute the leading order (in $\alpha_s$) perturbative QCD and power 
($1/m_b^2)$ corrections 
to the hadronic invariant mass and hadron energy spectra in the decay $B 
\to X_s \ell^+ \ell^-$. The computations are carried out 
using HQET and a perturbative-QCD
improved Fermi motion model which takes into account $B$-meson bound state
effects. 
We also present results for the
first two hadronic moments $\langle S_H^n\rangle$ and $\langle 
E_H^n\rangle$, $n=1,2$, working out their sensitivity on the 
HQET and FM model parameters. 
In the FM, also the LD effects due to intermediate charmonium 
resonances are taken into account. We study uncertainties in the 
parametrization of the $c \bar{c}$ effects. 
Further, we investigated the effect of the experimental cuts,
used recently by the CLEO collaboration in searching for the decay 
\bxsll \cite{cleobsll97}, on the branching ratios, hadron spectra and 
hadronic invariant mass moments using the FM model.

\subsection{Kinematics \label{sec:kin}}

We start with the definition of the kinematics of \bxsll decay at parton level,
\begin{equation}
b (p_b) \to s (p_s) (+g (p_g))+\ell^{+} (p_{+})+\ell^{-}(p_{-}) ~,
\end{equation} 
where $g$ denotes a gluon from the $O(\alpha_s)$ correction
(see Fig.~\ref{fig:o9}). 
We define the momentum transfer to the lepton pair and the invariant 
mass of the dilepton system, respectively, as
\begin{eqnarray}
q &\equiv & p_{+}+p_{-} \; , \\
s &\equiv & q^2 \; .
\end{eqnarray}
The dimensionless variables with a
hat are related to the dimensionful variables by the scale $m_b$, 
the $b$-quark mass, e.g., 
\begin{eqnarray}
\s= \frac{s}{m_b^2} \; , ~~\ms=\frac{m_s}{m_b} \; ,
\end{eqnarray}
etc..
Further, we define a 4-vector $v$, which denotes the velocity of both the
$b$-quark and the $B$-meson, $p_b=m_b v$ and $p_B=m_B v$. We shall also
need the variable $u$ and the scaled variable $\u=\frac{u}{m_b^2}$, 
defined as:
\begin{eqnarray}
u  & \equiv& -(p_b-p_+)^2+(p_b-p_{-})^2 \; , \\
\u & =& 2 v \cdot (\hat{p}_{+}-\hat{p}_{-}) \; ,
\end{eqnarray}
and further the kinematical phase factor
\begin{eqnarray}
  u(s,m_s)&=&\sqrt{(s-(m_b + m_s )^2)
(s - (m_b -m_s )^2)} \; .
\label{kinvar}  
\end{eqnarray}
The scaled variables $\hat{s}$ and $\hat{u}$
in the decay \bsll are bounded as follows,
\begin{eqnarray}
& &-  \u(\s,\ms)< \u < + \u(\s,\ms)\; ,  \nonumber\\
\label{eq:boundaries}
\u(\s,\ms)  & = &
                 \sqrt{\left[ \s - (1 + \ms)^2 \right]
                 \left[ \s - (1 - \ms)^2 \right] } \; , \nonumber\\
& & 4 \ml^2< \s < (1-\ms)^2 \; .
\end{eqnarray}

\subsection{NLO-corrected amplitude for \bsll \label{sec:NLO}}

Next, the  explicit 
expressions for the matrix element and (partial) branching ratios in
the decays  \bsll are presented in terms of the Wilson coefficients of the 
effective 
Hamiltonian obtained by integrating out the top quark and the $W^\pm$ bosons,
\begin{equation}\label{heffbsll}
{\cal H}_{eff}(b \to s + \ell^+ \ell^-)
  = {\cal H}_{eff} (b \to s + \gamma) -\frac{4 G_F}{\sqrt{2}} V_{ts}^* V_{tb}
\left[ C_9(\mu)  O_9 +C_{10} O_{10} \right],
\end{equation}
where ${\cal H}_{eff}(b \to s +\gamma)$ together with the operators
$O_{1 \dots 8}$ and their corresponding Wilson coefficients $C_i(\mu)$ 
\cite{effhamburas,effhamali} can be seen in section \ref{sec:effham}.
The two additional operators involving the dileptons $O_9$ and $O_{10}$ 
are defined as:
\begin{eqnarray}
 O_9 &=& \frac{e^2}{16 \pi^2} \bar{s}_\alpha \gamma^{\mu} L b_\alpha
\bar{\ell} \gamma_{\mu} \ell \; , \nonumber\\
 O_{10} &=& \frac{e^2}{16 \pi^2} \bar{s}_\alpha \gamma^{\mu} L
b_\alpha \bar{\ell} \gamma_{\mu}\gamma_5 \ell  \; .
\end{eqnarray}
A usual, CKM unitarity has been used in factoring out the product $V_{ts}^\ast
V_{tb}$. 
Note that the chromomagnetic 
operator $O_8$ does not contribute to the decay \bxsll in the 
approximation which we use here. 
The Wilson coefficients are given in the
literature (see, for example, \cite{misiakE,burasmuenz}).
They depend, in general, on the renormalization scale $\mu$,
except for $C_{10}$. 
At leading logarithmic (LLog) approximation, we use the values of the $C_i$ 
given in Table \ref{wilson}.
\begin{table}[h]
        \begin{center}
        \begin{tabular}{|c|c|c|c|c|c|c|c|c|c|}
        \hline
        \multicolumn{1}{|c|}{ $C_1$}       & 
        \multicolumn{1}{|c|}{ $C_2$}       &
        \multicolumn{1}{|c|}{ $C_3$}       & 
        \multicolumn{1}{|c|}{ $C_4$}       &
        \multicolumn{1}{|c|}{ $C_5$}       & 
        \multicolumn{1}{|c|}{ $C_6$}       &
        \multicolumn{1}{|c|}{ $C_7^{\mbox{eff}}$}       & 
        \multicolumn{1}{|c|}{ $C_9$}       &
                \multicolumn{1}{|c|}{$C_{10}$} &
 \multicolumn{1}{|c|}{ $C^{(0)}$ }     \\
        \hline 
        $-0.240$ & $+1.103$ & $+0.011$ & $-0.025$ & $+0.007$ & $-0.030$ &
   $-0.311$ &   $+4.153$ &    $-4.546$    & $+0.381$     \\
        \hline
        \end{tabular}
        \end{center}
\caption{ \it Values of the Wilson coefficients used in the numerical
          calculations corresponding to the central values 
          of the parameters given in Table \protect\ref{parameters}.
Here, $C_7^{\mbox{eff}} \equiv C_7 -C_5/3 -C_6$, and for $C_9$ we use the 
NDR scheme and $C^{(0)} \equiv 3 C_1 + C_2 + 3 C_3 + C_4 + 3 C_5 + C_6$.} 
\label{wilson}
\end{table}

 With the help of the
effective Hamiltonian in eq.~(\ref{heffbsll})
 the matrix element for the decay \bsll can be factorized 
into a leptonic and a hadronic part as,
\begin{eqnarray}
        {\cal M (\mbox{\bsll})} & = & 
        \frac{G_F \alpha}{\sqrt{2} \pi} \, V_{ts}^\ast V_{tb} \, 
        \left[ \left( C_9^{\mbox{eff}} - C_{10} \right) 
                \left( \bar{s} \, \gamma_\mu \, L \, b \right)
                \left( \bar{\ell} \, \gamma^\mu \, L \, \ell \right) 
                \right. \nonumber \\
        & & \left.
                \; \; \; \; \; \; \; \; \;
                \; \; \; \; \; \; \; \; \;
                + \left( C_9^{\mbox{eff}} + C_{10} \right) 
                \left( \bar{s} \, \gamma_\mu \, L \, b \right)
                \left( \bar{\ell} \, \gamma^\mu \, R \, \ell \right)  
                \right. \nonumber \\
        & & \left. 
                \; \; \; \; \; \; \; \; \;
                \; \; \; \; \; \; \; \; \;
                - 2 C_7^{\mbox{eff}} \left( \bar{s} \, i \, \sigma_{\mu \nu} \, 
                \frac{q^\nu}{q^2} (m_s L + m_b R) \, b \right)  
                \left( \bar{\ell} \, \gamma^\mu \, \ell \right) 
                \right] \; ,
        \label{eqn:hamiltonian}
\end{eqnarray}
where we abbreviate $ C_9^{\mbox{eff}} \equiv  C_9^{\mbox{eff}}(\s)$.
We have kept the $s$-quark mass term in the matrix element explicitly
and this will be kept consistently in the calculation of power corrections
and phase space. The above matrix element can be written in a compact form,
\begin{equation}
        {\cal M (\mbox{\bsll})} =
        \frac{G_F \alpha}{\sqrt{2} \pi} \, V_{ts}^\ast V_{tb} \, 
        \left( {\Gamma^L}_\mu \, {L^L}^\mu 
        +  {\Gamma^R}_\mu \, {L^R}^\mu \right) \, ,
\end{equation}
with
\begin{eqnarray}
        {L^{L/R}}_\mu & \equiv & 
                \bar{\ell} \, \gamma_\mu \, L(R) \, \ell \, , \\
        {\Gamma^{L/R}}_\mu & \equiv & 
                \bar{s} \left[ 
                R \, \gamma_\mu 
                        \left( C_9^{\mbox{eff}} \mp C_{10} + 2 C_7^{\mbox{eff}} \, 
                        \frac{\hat{\slash{q}}}{\s} \right)
                + 2 \hat{m}_s \, C_7^{\mbox{eff}} \, \gamma_\mu \, 
                        \frac{\hat{\slash{q}}}{\s} L 
                \right] b \; .  
        \label{eqn:gammai}
\end{eqnarray}
where we have already used massless leptons in substituting
$-2 i \sigma_{\mu \nu} q^\nu=[\gamma_\mu,\slash{q}]$ by 
$2 \gamma_\mu \slash{q}$ in the term proportional to $\cseff$.

The effective Wilson coefficient $C_9^{\mbox{eff}}(\s)$ 
receives contributions from various pieces. The
resonant $c\bar{c}$ states also contribute to $C_9^{\mbox{eff}}(\s)$
and will be discussed in section \ref{sec:LD1}; hence
the contribution given below is just the perturbative part:
\begin{eqnarray}
C_9^{\mbox{eff}}(\s)=C_9 \eta(\s) + Y(\s) \, .
\end{eqnarray}
\begin{figure}[htb]
\vskip -0.4truein
\centerline{\epsfysize=7in
{\epsffile{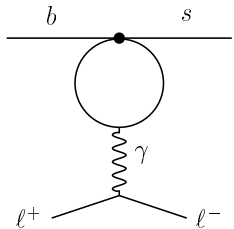}}}
\vskip -5.0truein
\caption[]{ \it The Feynman diagram responsible for the 
four-Fermi-operator contribution (depicted by the blob) to the operator $O_9$.}
\label{fig:ycharm}
\end{figure}
The function $Y(\hat{s})$ represents the one-loop matrix element of the
four-Fermi operators \cite{burasmuenz,misiakE}, see Fig.~\ref{fig:ycharm}. 
It is written as:
\begin{eqnarray}
\label{Ypert}
        Y(\s) & = & g(\mc,\s)
                \left(3 \, C_1 + C_2 + 3 \, C_3
                + C_4 + 3 \, C_5 + C_6 \right)
\nonumber \\
        & & - \frac{1}{2} g(1,\s)
                \left( 4 \, C_3 + 4 \, C_4 + 3 \,
                C_5 + C_6 \right) 
         - \frac{1}{2} g(0,\s) \left( C_3 +   
                3 \, C_4 \right) \nonumber \\
        & &     + \frac{2}{9} \left( 3 \, C_3 + C_4 +
                3 \, C_5 + C_6 \right) 
             - \xi \, \frac{4}{9} \left( 3 \, C_1 +
                C_2 - C_3 - 3 \, C_4 \right)
                \label{eqn:y} \; ,
\end{eqnarray}
 \begin{equation}
        \xi = \left\{
                \begin{array}{ll}
                        0       & \mbox{(NDR)}\; , \\
                        -1      & \mbox{(HV)}\; ,
                \end{array}
                \right.
\end{equation}
where (NDR) and (HV) correspond to the naive dimensional regularization
and the 't Hooft-Veltman schemes, respectively.
We recall that while $C_9$ is a renormalization scheme-dependent quantity, this
dependence cancels out with the corresponding one in the function $Y(\s)$
(the value of $\xi$, see above).
The function $g(z,\hat{s})$ includes the
quark-antiquark pair contribution
\cite{misiakE,burasmuenz}:
\begin{eqnarray}
\label{gpert}
g(z,\hat{s}) &=& -\frac{8}{9}\ln (\frac{m_b}{\mu})
 -\frac{8}{9} \ln z + \frac{8}{27} +\frac{4}{9}y
-\frac{2}{9}(2 + y) \sqrt{\vert 1-y \vert}\nonumber\\
&\times & \left[\Theta(1-y)(\ln\frac{1+\sqrt{1-y}}{1-\sqrt{1-y}} -i\pi )
+\Theta(y-1) 2 \arctan \frac{1}{\sqrt{y-1}} \right] ~, \\
g(0,\hat{s})& =& \frac{8}{27}-\frac{8}{9}\ln (\frac{m_b}{\mu})
              -\frac{4}{9}\ln \hat{s} + \frac{4}{9}i\pi ~,
\end{eqnarray}
where $y=4z^2/\hat{s}$.
As can be seen from the above equations, 
internal $b$-quarks $\sim g(1,\s)$, $c$-quarks $\sim g(\mc,\s)$ and 
light quarks $q$, (with $m_q=0$ for $q=u,d,s$) $\sim g(0,\s)$
contribute to the function $Y(\s)$; only the charm loop involves the dominant 
``current-current" operators $O_1$ and $O_2$.

The $O(\alpha_s)$ correction  \cite{jezkuhn} from the 
one-gluon exchange in the matrix element of $ O_9$ in the 
invariant dilepton mass $\s$ is represented by
\begin{eqnarray}
        \eta(\s)  =  1 + \frac{\alpha_s(\mu)}{\pi}
                \omega(\s) \; ,
\end{eqnarray}
where
\begin{eqnarray}
\omega(\hat{s}) &=& -\frac{2}{9}\pi^2 -\frac{4}{3}{\mbox Li}_2(\hat{s})
-\frac{2}{3}
\ln \hat{s} \ln(1-\hat{s}) -
\frac{5+4\hat{s}}{3(1+2\hat{s})}\ln(1-\hat{s})\nonumber\\
&-& \frac{2\hat{s}(1+\hat{s})(1-2\hat{s})}{3(1-\hat{s})^2(1+2\hat{s})}
\ln \hat{s} + \frac{5 + 9\hat{s} -6\hat{s}^2}{6(1-\hat{s})(1+2 \hat{s})} \; .
\label{omegahats}
\end{eqnarray}
Note that the function $\omega(\hat{s})$ is given with $m_s=0$.
The one-gluon correction to $O_9$ with respect to the final partonic energy
and the invariant mass will be presented below in section \ref{sec:pertqcd}.

\begin{figure}[htb]
\vskip -0.4truein
\centerline{\epsfysize=7in
{\epsffile{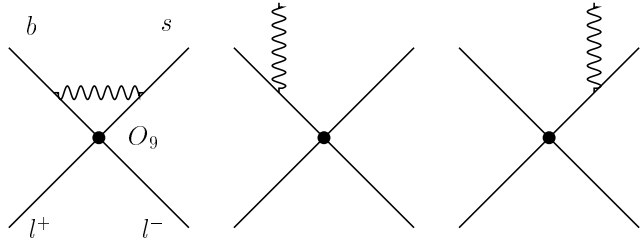}}}
\vskip -5.0truein
\caption[]{ \it Feynman diagrams contributing to the explicit
order $\alpha_s$ corrections of the operator $O_9$. 
Curly lines denote a gluon. Wave function corrections are not shown.}
\label{fig:o9}
\end{figure}

In the order we are working only $O_9$ is subject to $\alpha_s$ corrections 
since the renormalization 
group improved perturbation series for $C_9$ is 
${\cal{O}}(1/\alpha_s)+{\cal{O}}(1)+{\cal{O}}(\alpha_s)+ \dots$,
 due to the 
large logarithm in $C_9$ represented by ${\cal{O}}(1/\alpha_s)$
\cite{burasmuenz}.
The Feynman diagrams, which contribute to the matrix element of $O_9$ in 
$O(\alpha_s)$, corresponding to the virtual one-gluon and 
bremsstrahlung corrections, are shown in Fig.~\ref{fig:o9}.
%Hence, we are working in next-to-leading order, 
%but collect just leading logs. 

With the help of the above expressions, the differential
decay width becomes on using $p_{\pm}=(E_{\pm}, \mbox{\boldmath $p_{\pm}$})$, 
\begin{equation}
        {\rm d} \Gamma = \frac{1}{2 m_B} 
                \frac{{G_F}^2 \, \alpha^2}{2 \pi^2} 
                \left| V_{ts}^\ast V_{tb} \right|^2 
                \frac{{\rm d}^3 \mbox{\boldmath $p_+$}}{(2 \pi)^3 2 E_+} 
                \frac{{\rm d}^3 \mbox{\boldmath $p_-$}}{(2 \pi)^3 2 E_-} 
                \left( {W^L}_{\mu \nu} \, {L^L}^{\mu \nu} 
                +  {W^R}_{\mu \nu} \, {L^R}^{\mu \nu} \right) \, ,
\end{equation}
where $W_{\mu \nu}$ and $L_{\mu \nu}$ are the
hadronic and leptonic tensors, respectively.
The hadronic tensor $W_{\mu\nu}^{L/R}$
is related to the discontinuity in the forward scattering amplitude, 
denoted by
$T_{\mu \nu}^{L/R}$, through the relation $W_{\mu \nu} = 2 \, {\rm Im} \, 
T_{\mu \nu}$.  Transforming the integration variables 
 to $\hat{s}$, $\hat{u}$ and $v \cdot \hat{q}$, one can express the 
Dalitz distribution in \bsll (neglecting the lepton masses) as:  
\begin{equation}
        \frac{{\rm d} \Gamma}{{\rm d}\u \, {\rm d}\s \, {\rm d}(\z)} = 
                \frac{1}{2 \, m_B}
                \frac{{G_F}^2 \, \alpha^2}{2 \, \pi^2} 
                \frac{{m_b}^4}{256 \, \pi^4}
                \left| V_{ts}^\ast V_{tb} \right|^2 
                \, 2 \, {\rm Im} 
%                \int {\rm d}(\z) \,
                \left( {T^L}_{\mu \nu} \, {L^L}^{\mu \nu}
                +  {T^R}_{\mu \nu} \, {L^R}^{\mu \nu} \right) \, ,
        \label{eqn:dgds}
\end{equation}
with
\begin{eqnarray}
\label{eq:hadtensor}
        {T^{L/R}}_{\mu \nu} & \equiv & 
        i \, \int {\rm d}^4 y \, e^{-i \, \hat{q} \cdot y}      
        \left< B \left| {\rm T} \left\{ 
                {{\Gamma_1}^{L/R}_\mu} (y), 
                {\Gamma_2}^{L/R}_\nu (0) \right\} \right| B \right>\, , \\
        {L^{L/R}}^{\mu \nu} & \equiv & 
        \sum_{spin} 
        \left[ \bar{v}^{L/R}(p_+) \, \gamma^\mu \, u^{L/R}(p_-) \right]
        \left[ \bar{u}^{L/R}(p_-) \, \gamma^\nu \, v^{L/R}(p_+) \right]
                \nonumber \\
        & = & 2 \left[ {p_+}^\mu \, {p_-}^\nu + {p_-}^\mu \, {p_+}^\nu 
                - g^{\mu \nu} (p_+ \cdot p_-) 
                \mp i \epsilon^{\mu \nu \alpha \beta} \, 
                        {p_+}_\alpha \, {p_-}_\beta \right] \, , 
\end{eqnarray}
where ${{\Gamma_1}^{L/R}_\mu}^\dagger = {\Gamma_2}^{L/R}_\mu = 
\Gamma^{L/R}_\mu $, given in eq. (\ref{eqn:gammai}). The Dalitz distribution 
eq.~(\ref{eqn:dgds}) contains the explicit $O(\alpha_s)$-improvement, and 
the distributions in which we are principally interested in can be obtained
by straight-forward integrations.

 Using Lorentz decomposition, the tensor $T_{\mu \nu}$ can be expanded in 
terms of three structure functions
\footnote{We use the convention
$Tr(\gamma^{\mu} \gamma^{\nu} \gamma^{\alpha} \gamma^{\beta} \gamma_5)=
-4i\epsilon^{\mu \nu \alpha \beta}$, with $\epsilon^{0 1 2 3}=-1$.} , 
\begin{equation}
\label{eq:hadrontensor}
        T_{\mu \nu} = -T_1 \, g_{\mu \nu} + T_2 \, v_\mu \, v_\nu 
                + T_3 \, i \epsilon_{\mu \nu \alpha \beta} \, 
                        v^\alpha \, \hat{q}^\beta \, ,
\end{equation}
where the structure functions which do not contribute to the
amplitude in the limit of massless leptons have been neglected.
After contracting the hadronic and leptonic tensors, one finds
\begin{equation}
        {T^{L/R}}_{\mu \nu} \, {L^{L/R}}^{\mu \nu} = 
                {m_b}^2 \left\{ 2 \, \s \, {T_1}^{L/R} 
                + \left[ (\z)^2 - \frac{1}{4} \u^2 - \s \right] {T_2}^{L/R} 
                \mp \s \, \u \, {T_3}^{L/R} \right\} \, . 
        \label{eq:tlr}
\end{equation}
We remark here that the $T_3$ term will contribute to 
the FB asymmetry but not to the branching ratio or
the dilepton invariant mass spectrum in the decay \bxsll.

\subsubsection{Normalization}
It has become customary to express the branching ratio for \bxsll in terms
of the well-measured semileptonic branching ratio ${\cal B}_{sl}$
for the decays $B \to (X_c,X_u) \ell \nu_\ell$
according to
\begin{eqnarray}
d {\cal{B}}(B \to X_s \ell^+ \ell^-)={\cal B}_{sl} 
\frac{d \Gamma(B \to X_s \ell^+ \ell^-)}{\Gamma(B \to (X_c,X_u) \ell \nu_\ell) }\; .
\end{eqnarray} 
This fixes the normalization constant ${\cal B}_0$, which will be used 
throughout the following sections to be,
\begin{equation}
        {\cal B}_0 \equiv
                {\cal B}_{sl} \frac{3 \, \alpha^2}{16 \pi^2} \frac{
    {\vert V_{ts}^* V_{tb}\vert}^2}{\absvcb^2} \frac{1}{f(\mc) \kappa(\mc)}
                \; .
\label{eqn:seminorm}
\end{equation}
Here $f(\mc)$ is the phase space factor for 
$\Gamma (B \rightarrow X_c \ell \nu_{\ell})$
and the function $\kappa(\mc)$ accounts for both the $O(\alpha_s)$ QCD 
correction to the semileptonic decay  width \cite{CM78} and the leading order
$(1/m_b)^2$ power correction \cite{georgi}. 
They read as:
\begin{equation}
        f(\mc) = 1 - 8 \, \mc^2 + 8 \, \mc^6 - \mc^8 - 24 \, \mc^4 \, \ln \mc
        \label{eqn:fr}
\end{equation}
and 
\begin{equation}
\kappa(\mc) = 1 + \frac{\alpha_s(m_b)}{\pi} g(\mc)
    + \frac{h(\mc)}{2 m_b^2} \; ,
\end{equation}
where 
\begin{eqnarray}
g(\mc)&=& \frac{A_0(\mc)}{f(\mc)} \; , \\
h(\mc) &=& \lambda_1 + \frac{\lambda_2}{f(\mc)} \left[ -9 +24 \mc^2
-72\mc^4 + 72\mc^6 -15\mc^8 -72 \mc^4 \ln \mc \right]\; , 
\label{eqn:ghr}
\end{eqnarray}
and the analytic form of $A_0(\mc)$ can be seen in ref.~\cite{FLS}.
Note that the frequently used approximation
$g (z) \approx -\frac{2}{3} ((\pi^2-\frac{31}{4})(1-z)^2 + \frac{3}{2})$
holds within $1.4 \%$ accuracy in the range $0.2 \leq z\leq 0.4$. 
The equation
$g (z) = -1.671+2.04(z-0.3)-2.15(z-0.3)^2$
is accurate for $0.2 \leq z\leq 0.4$ to better than one per mille accuracy.

\subsection{Asymmetries in \bxsll Decay \label{sec:asymmetries}}

Besides the differential branching ratio, \bxsll decay offers other 
distributions (with different combinations of Wilson coefficients)
to be measured.
An interesting quantity is the Forward-Backward (FB) asymmetry
defined in \cite{amm91,agm94}
\begin{equation}
        \frac{{\rm d}{\cal A}(\s)}{{\rm d}\s} = \int_0^1
                \frac{{\rm d}^2 {\cal B}}{{\rm d}\s \, {\rm d}z} \, {\rm d}z
                - \int_{-1}^{0}
                \frac{{\rm d}^2 {\cal B}}{{\rm d}\s \, {\rm d}z} \, {\rm d}z
\label{eqn:fbasy}
        \, ,
\end{equation}
where $z \equiv \cos \theta$ is the angle of $\ell^+$ measured w.r.t. the
$b$-quark direction in the dilepton c.m. system.
From the experimental point of view, a more useful quantity is the
normalized FB asymmetry, obtained by normalizing $d{\cal A}/d\s$ with the 
dilepton mass distribution, $d{\cal B}/d\s$,
\begin{equation}
\frac{{\rm d} \overline{{\cal A}}}{{\rm d}\s} =  \frac{{\rm d}{\cal A}}{{\rm 
d}\s}/
 \frac{{\rm d}{\cal B}}{{\rm d}\s} \; .   
\end{equation}
The asymmetry $\bar{\ca}$, which we recall is defined in the dilepton 
c.m.s.~frame, is identical to the
energy asymmetry $A_{E}$ introduced in \cite{choetal}, 
as shown in ref.~\cite{AHHM97}.
It is defined in the $B$ rest frame as
\begin{equation}
A_E \equiv \frac{\left( N (E_- > E_+) - N (E_+ > E_-) \right)}{
\left( N (E_- > E_+) + N (E_+ > E_-) \right)} \; .
\end{equation}
Here $N(E_- > E_+)$ denotes the number of lepton pairs whose negatively
charged member is more energetic than its positive partner, where
$E_\pm$ denote the $\ell^\pm$ charged lepton energy in the $B$ rest frame. 
The FB asymmetry is odd under charge conjugation in contrary to 
the differential branching ratio, which is charge conjugation even. 
Both observables contain non overlapping 
information which together can be used to test the SM.

Another quantity is the left-right-asymmetry \cite{LD96,BP97,BI98},
defined as
\begin{eqnarray}
\frac{{\rm d}{\cal A}^{LR}}{{\rm d}\s}=\frac{{\rm d}{\cal B}^L}{{\rm d}\s}-
\frac{{\rm d}{\cal B}^R}{{\rm d}\s} \; ,
\end{eqnarray}
with ${\rm d}{\cal B}^{L}/ {\rm d} \s$ 
(${\rm d}{\cal B}^R /{\rm d}\s$) denoting the invariant dilepton mass 
distribution for \bxsll decay
into purely left-handed (right-handed) leptons.
We can obtain ${\rm d}{\cal B}^{L/R}/{\rm d} \s$ from the dilepton invariant 
mass distribution ${\rm d}{\cal B}/{\rm d}\s$ by the replacements
\begin{eqnarray}
\cneff \to \frac{\cneff \mp C_{10}}{2} \; ,~~ C_{10} \to \frac{C_{10} \mp \cneff}{2}\; ,~~ |\cseff|^2 \to \frac{1}{2}|\cseff|^2 \; . 
\end{eqnarray}
Measurement of these asymmetries provides additional information on the 
underlying short-distance physics.

\subsection{Leading power $(1/m_b)$ corrections in the decay  \bxsll}

We start with a discussion of the analyticity properties of the forward 
scattering amplitude $T_{\mu \nu}$.
They are determined by cuts, depending on the external states.
We consider real particle production in the inclusive decay \bxsll, 
thus we have $p_B=p_X+q$, 
where $p_B,p_X$ denotes the 4-momentum of the $B$-meson, 
final hadronic state $X_s$, respectively.
The hadronic invariant mass is in the range
\begin{eqnarray}
m_K^2 \leq p_X^2 \leq m_B^2 \; ,
\end{eqnarray}
and the physical cut runs along the real axis in the complex
$v \cdot q$ plane in the limits
\begin{eqnarray}
\sqrt{q^2} \leq v \cdot q \leq \frac{m_B^2+q^2-m_K^2}{2 m_B} \; .
\end{eqnarray} 
The phase space integration, which follows the above cut,
is over intermediate physical states and hence, depends on long-distance QCD.
Moreover, at the upper bound of the cut, where $p_X^2 \sim m_s^2$, resonances
are dominating.
In order to perform a reliable expansion in perturbative QCD, 
the contour of the integration has to be deformed in such a way that a) it 
encloses the cut and 
b) stays away from it by a distance large compared to $\Lambda_{QCD}$
(see Fig.~1 in \cite{MW} for the contour of integration).
The expansion is valid except in the corner of the Dalitz plot, where the hadronic invariant mass of the final state is small (the $s$-quark in 
Fig.~\ref{fig:hqetdiag} is almost on-shell.)
However, results of perturbative QCD are expected to be recovered after 
suitable smearing.

\subsubsection{Operator product expansion}

The next task is to expand the forward scattering amplitude 
$T_{\mu \nu}$ in eq.~(\ref{eq:hadrontensor}) in inverse powers of $m_b$. 
We employ HQE techniques which have been already
sketched in section \ref{sec:hqet}.
The leading term in this expansion, i.e., ${\cal O}(m_b^0)$ reproduces the 
parton model result. 
Let us describe how to get next to leading power corrections.
First we write the momentum of the heavy $b$-quark as 
$p_b{_\mu}=m_b v_\mu + k_\mu$, 
fix the four-velocity of the external $b$-quark field to be $v_\mu$
and treat the components of the ``residual momentum" $k_\mu$
of order $\Lambda_{QCD}$.
We obtain the condition $p_b^2=m_b^2+2 m_b v.k+k^2$, which yields 
$p_b^2=m_b^2 + {\cal{O}}(k/m_b)$.
The heavy quark remains almost on-shell under soft gluon exchange with the 
light degrees of freedom, there is no anti-quark generated and the
total $b$-number is conserved.

It is customary to define a field $h$ with fixed velocity $v$ through
\begin{equation}
h(x)=e ^{i m_b v.x} P_+ b(x) \; ,
\end{equation}
with inversion
\begin{equation}
\label{bhrel}
b(x) = e ^{i m_b v.x} \left[1+i \frac{\Slash{D}}{2 m_b} + ...\right] h(x) 
\; ,   
\label{btohx}
\end{equation}
and the projection operators $P_{\pm}=(1 \pm \slash{v})/2$.
For the Dirac field
$b(x)$ the following identities hold $P_+b(x)=b(x)$ and
$P_{-}b(x)=0$, which are corrected by terms of ${\cal{O}}(1/m_b)$.
Inserting this into the usual QCD Lagrangian 
${\cal{L}}=\bar{b} (i \Slash{D}-m_q) b$
we get the one in the HQET:
\begin{eqnarray}
\label{eq:Lhqet}
{\cal{L}}_{HQET}=\bar{h} i v . D h+ \delta {\cal{L}} \; ,
\end{eqnarray}
with $\delta {\cal{L}}$ containing the corrections in $1/m_b$:
\begin{eqnarray}
\delta {\cal{L}}=\frac{1}{2 m_b} \bar{h} (i D)^2 h-
\frac{1}{2 m_b} Z_1(\mu) \bar{h} (i v . D)^2 h+
\frac{1}{2 m_b} Z_2(\mu) \bar{h} \frac{-i}{2} \sigma^{\mu \nu}
                 G_{\mu \nu} h +{\cal{O}}(1/m_b^2)\; .
\label{eq:deltaL}
\end{eqnarray}
Here, $ G_{\mu \nu} \equiv[iD_{\mu},i D_{\nu}]$ 
denotes the gluon field strength tensor.
{\footnote{Note that here and only in this section we use this 
definition of the gluon field strength tensor $G=G^{FLS}$, following the 
conventions of 
ref.~\cite{FLSold}. It is not consistent with the usual one appearing in QCD 
text books, eq.~(\ref{eq:Gqcd}), denoted here by $G^{QCD}$. The two are 
related by $G^{FLS} = i g G^{QCD}$.}}
For definition of the renormalization constants $Z_{1,2}(\mu)$ we refer 
to \cite{FLSold} and references therein.
For the sake of completeness we give the Feynman rules in the HQET
in appendix \ref{app:hqetrules}. 
The matrix elements of the above higher dimensional operators are given as
 \begin{eqnarray}
        \left< B \left| \bar{h} \, (i \, D)^2 \, h \right| B\right>   
                & \equiv & 2 \, m_B \, \loo
                \, , \nonumber \\
        \left< B \left| \bar{h} \, \frac{-i}{2} \sigma^{\mu \nu}
                \, G_{\mu \nu} \, h \right| B \right>
                & \equiv & 6 \, m_B \, \lto
                \, ,
\label{hqetpar}
\end{eqnarray}
where $B$ denotes the pseudoscalar $B$-meson, see section \ref{sec:hqet} 
for a discussion of the values of the parameters $\lambda_1$ and $\lambda_2$.
The second term on the r.h.s. of eq.~(\ref{eq:deltaL}) vanishes by the 
lowest order equation of motion $i v . D h=0$.
The on-shell condition of the heavy quark is
$m_b^2=m_b^2+2 m_b v. k+k^2$. Neglecting the last term, we have the 
simple condition $v . k=0$.
This is equivalent to the lowest order EOM of a heavy quark, thus we have the
correspondence $k \leftrightarrow iD$.

\begin{figure}[htb]
\vskip -0.5truein
\centerline{\epsfysize=7in
{\epsffile{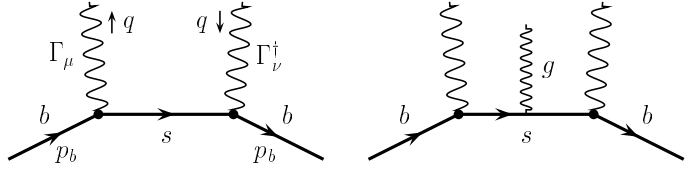}}}
\vskip -5.3truein
\caption{ \it The diagrams contributing to the operator product expansion.}
\label{fig:hqetdiag}
\end{figure}  

Suppressing the Lorentz indices for the time being, this operator product 
expansion (OPE) can be formally represented as:  
\begin{eqnarray}
        \int {\rm d}^4 y \, e^{-i \, \hat{q} \cdot y} 
        \left< B \left| {\rm T} \left\{ \Gamma_1 (y), 
        \Gamma_2 (0) \right\} \right| B \right> 
        & = & - \frac{1}{m_b} \left[ 
        \left< B \left| {\cal O}_0 \right| B \right> 
        + \frac{1}{2 \, m_b} \left< B \left| {\cal O}_1\right| B \right> 
        \right. 
        \nonumber \\
        & & \left.
        \, \, \, \, \, \, \,    
        + \frac{1}{4 \, {m_b}^2} \left< B \left| {\cal O}_2 \right| B \right>  
        + \cdots \right] \; .
\label{toproduct}
\end{eqnarray}
The expressions for the operators ${\cal O}_0, {\cal O}_1$ and 
${\cal O}_2$ have been first derived in ref.~\cite{FLSold}, 
which we have checked and confirm.
They are given as:
\begin{eqnarray}
{\cal O}_0& =& \frac{1}{x} \bar{b} \Gamma_1 (\slash{v} -\slash{\q} + 
\ms) \Gamma_2 b \; , \\
{\cal O}_1& =& \frac{2}{x} \bar{h} \Gamma_1 \gamma^{\alpha} \Gamma_2 i
D_{\alpha} h 
- \frac{4}{x^2}(v-\q)^{\alpha} \bar{h} \Gamma_1 (\slash{v} -\slash{\q} + 
\ms) \Gamma_2 i D_{\alpha} h \; ,
\end{eqnarray}
and
\begin{eqnarray}
{\cal O}_2 & = &{\cal O}_2^{(2)}+{\cal O}_2^{(g)}+{\cal O}_2^{(1)}
\; ,
\end{eqnarray}
where 
\begin{eqnarray}
{\cal O}_2^{(2)} & = &
\frac{16}{x^3} (v- \q)^{\alpha} (v- \q)^{\beta} \bar{h}
\Gamma_1 (\slash{v} -\slash{\q} + \ms) \Gamma_2 i D_{\alpha} i D_{\beta} h -
\frac{4}{x^2} \bar{h} \Gamma_1 (\slash{v} -\slash{\q} + \ms) \Gamma_2 (i D)^2 h
\nonumber \\
& - & \frac{4}{x^2}(v-\q)^{\beta} \bar{h} \Gamma_1 \gamma^{\alpha} \Gamma_2 
(i D_{\alpha} i D_{\beta}+ i D_{\beta} i D_{\alpha}) h \; , \\
{\cal O}_2^{(g)} & = & \frac{2}{x^2} \ms
 \bar{h} \Gamma_1 i \sigma_{\alpha \beta} \Gamma_2 G^{\alpha \beta} h
 + \frac{2}{x^2} i \epsilon^{\mu \lambda \alpha \beta}
 ( v- \q)_{\lambda}\bar{h}
\Gamma_1 \gamma_{\mu} \gamma_{5} \Gamma_2 G_{\alpha \beta} h \; , \\
{\cal O}_2^{(1)} & = &\frac{2}{x} \bar{h} (\gamma^{\beta} \Gamma_1 \gamma^{\alpha} \Gamma_2+
 \Gamma_1 \gamma^{\beta} \Gamma_2 \gamma^{\alpha})  i D_{\beta} i D_{\alpha} h
- \frac{4}{x^2}(v-\q)^{\alpha} \bar{h} \gamma^{\beta}
 \Gamma_1  (\slash{v} -\slash{\q} + \ms) \Gamma_2 i D_{\beta} i D_{\alpha} h
\nonumber \\ 
&-& \frac{4}{x^2}(v-\q)^{\alpha} \bar{h} \Gamma_1 
(\slash{v} -\slash{\q} + \ms) \Gamma_2 \gamma^{\beta} i D_{\alpha} i 
D_{\beta} h \; .
\end{eqnarray}
Here $x \equiv 1 + \s - 2 \, (\z) - \ms^2 + i \, \epsilon $.
The operator $O_3$ responsible for $1/m_b^3$ corrections can be seen in 
ref.~\cite{Bauer97}.

The above ${\cal{O}}_i$, $i=0,1,2$ 
are obtained by expanding the Feynman diagrams 
shown in Fig.~\ref{fig:hqetdiag}, which contributes to the time-ordered 
product on the l.h.s. of eq.~(\ref{toproduct}).
The diagram on the left is responsible for the operators
${\cal{O}}_0,{\cal{O}}_1,{\cal{O}}_2^{(1)}$ and ${\cal{O}}_2^{(2)}$.
To be definite, we write
the intermediate $s$-quark propagator 
using 4-momentum conservation $p_s=p_b-q=m_b v-q+k$ as
\begin{eqnarray}
i \frac{\slash{p}_s+m_s}{p_s^2-m_s^2+i \epsilon}=
i\frac{1}{m_b} 
\frac{\slash{v}-\slash{\q}+\slash{k}/m_b+\ms}
{x-2 \q \cdot k/m_b+2 v \cdot k/m_b+k^2/m_b^2} \; .
\end{eqnarray} 
and insert this into the diagrams Fig.~\ref{fig:hqetdiag}. 
Expanding as well the propagator as the Dirac field $b$, which  
sandwiches the amplitudes, as the normalization of the states in powers of
$k/m_b$, we obtain the desired OPE.
Note that the leading operator 
${\cal O}_0$ is defined in terms of the ``full" four-component field $b$.
The other two subleading operators ${\cal O}_1$ and ${\cal O}_2$ are, however,
written in terms of the two-component effective fields $h$.
In rewriting the operator ${\cal O}_1$ from $b \to h$ fields by means of
eq.~(\ref{btohx}), we obtain the operator ${\cal O}_2^{(1)}$.
Evaluation of the one-gluon diagram in Fig.~\ref{fig:hqetdiag}
results in the operator ${\cal O}_2^{(g)}$.

The results of the power corrections to the structure functions $T_i$, 
$i=1,2,3$ can be 
decomposed into the sum of various terms, denoted by $T_{i}^{(j)}$, 
which can be traced back to well 
defined pieces in the evaluation of the time-ordered product given above
\cite{AHHM97}:
\begin{equation}
T_{i}(v.\hat{q},\s) = \sum_{j=0,1,2,s,g,\delta} T_{i}^{(j)}(v.\hat{q}, \s)\,.
\label{Tijhqet}
\end{equation}
The expressions for $T_{i}^{(j)}(v.\hat{q},\s)$ calculated up to 
$O(m_B/m_b^3)$ are given in appendix \ref{app:Ti}.
They contain the parton model expressions $T_{i}^{(0)}(v.\hat{q},\s)$ and 
the power corrections in the HQE approach which depend on the two HQE-specific 
parameters $\loo$ and $\lto$  defined in eqs.~(\ref{hqetpar}).
Note that the $s$-quark mass terms are explicitly kept in
 $T_{i}^{(j)}(v.\hat{q},\s)$.
The origin  of the various terms in the expansion given in 
eq.~(\ref{Tijhqet}) can be specified, as follows:
\begin{eqnarray}
T_{i}^{(0)}(v.\hat{q}, \s)&=&
\langle B \vert {\cal{O}}_0 \vert B \rangle \; \; \; \mbox{for} \; \; \; 
\lambda_1=\lambda_2=0 \;  ,  \nonumber \\
T_{i}^{(s)}(v.\hat{q}, \s)&=&
\langle B \vert {\cal{O}}_0 \vert B \rangle-T_{i}^{(0)}(v.\hat{q}, \s)\;  , 
 \nonumber \\
T_{i}^{(j)}(v.\hat{q}, \s)&=&
\langle B \vert {\cal{O}}_2^{(j)} \vert B \rangle \; \; \; \mbox{for} \; \; \; j=1,2,g \;  ,  \nonumber \\
T_{i}^{(\delta)}(v.\hat{q}, \s)&=&
\langle B \vert {\cal{O}}_1 \vert B \rangle \; .
\label{eq:Tij}
\end{eqnarray}
In the leading order in $(1/m_b)$ the matrix element of ${\cal O}_1$ vanishes, 
but in the sub-leading order it receives a non-trivial contribution 
which can be calculated by using the equation of motion \cite{FLSold}.
The contributions 
$T_{i}^{(s)}$ arise from the matrix element of the scalar 
operator $\bar{b} b$, i.e use of eq.~(\ref{eq:bbnumber}) given below.
We recall that the scalar current can be written in terms of 
the vector current plus higher dimensional operators as \cite{georgi}
\begin{equation}
\bar{b} b = v_\mu \bar{b} \gamma^\mu b + \frac{1}{2m_b^2}
\bar{h} 
\left[ (iD)^2 - (v.iD)^2 - (i/2) \sigma^{\mu \nu} G_{\mu\nu} \right] h +...\; .
\end{equation}
With our normalization:
\begin{equation}
\langle B \vert \bar{b} \gamma_\mu b \vert B \rangle = 2 (p_B)_{\mu} \; ,
\end{equation}
it follows then
\begin{equation}
\langle B \vert \bar{b} b \vert B \rangle = 2 m_B (1+\frac{1}{2 m_b^2}
(\lambda_1+3 \lambda_2))+{\cal{O}}(m_B/m_b^3) \; .
\label{eq:bbnumber}
\end{equation}
Other possible Lorentz structures like $\gamma_5, \gamma_5 \gamma_\mu,
\gamma_\mu \gamma_\nu$ sandwiched between $\bar{b}$ and $b$ give zero after 
taking the $B$-meson matrix element.

{}From the expressions for $T_{i}^{(j)}$ given in appendix \ref{app:Ti},
we see that $T_{i}^{(0)} (i=1,2,3)$ are of order $m_B/m_b$
and the rest $T_{i}^{(1)}$, $T_{i}^{(\delta)}, T_{i}^{(2)}, T_{i}^{(s)}$ 
and  $ T_{i}^{(g)}$ are all of order ${m_B}\lambda_1/{ m_ {b} }^3$
or ${m_B}\lambda_2/{ m_ {b} }^3$.
Since the ratio $m_B/m_b = 1 + O(1/m_b)$, we note that the Dalitz 
distribution in \bxsll has linear corrections in $1/m_b$.

\section{Power Corrections to the Dilepton Invariant Mass Distribution
            and FB Asymmetry}

The integration in the complex plane $\z$ can be done using the relation
\begin{eqnarray}
{\rm Im} \frac{1}{x^n} 
\propto \frac{(-1)^{n-1}}{(n-1) !} \delta^{(n-1)} (1-2 \z+\s -\ms^2) \; .
\label{eq:imx}
\end{eqnarray}
Further, the integrand should be multiplied by the function 
$\theta(4 \z^2-4 \s^2- \u^2)$ 
responsible for the correct integration boundary. 
{\footnote{This
corresponds to $q^2_{max}=4 E_+ E_{-}$, with lepton energies
$E_{\pm}=v \cdot q/2 \pm u/(4 m_b)$.}}
The resulting double differential branching ratio in \bxsll can be expressed 
as,
\begin{eqnarray}
\frac{{\rm d}{\cal B}}{{\rm d}\s \, {\rm d}\u} & = & 
                {\cal B}_0 \, 
        \left( \left\{ 
        \left[ \left( 1 - \ms^2 \right)^2 - \s^2 - \u^2 
        - \frac{1}{3} \left( 2 \, \lo (-1 + 2 \ms^2 -\ms^4 - 2 \, \s + \s^2) 
\right. \right. \right. \right.
\nonumber \\
& &  \left. \left. \left. \left.
                + 3 \, \lt (-1 +6 \ms^2 -5 \ms^4 - 8 \, \s + 5 \, \s^2) 
        \right)
        \right] \left( |C_9^{\mbox{eff}}|^2 + |C_{10}|^2 \right)
        \right. \right.
        \nonumber \\
        & & \left. \left.
        + \, \left[ 4 \left( 
        1 - \ms^2 -\ms^4+\ms^6-8\ms^2 \s -\s^2 -\ms^2 \s^2+ \u^2 +\ms^2 \u^2 
              \right)  
        \right. \right.  \right. 
        \nonumber \\
        & & 
        - \frac{4}{3} \left( 2 \, \lo (-1 + \ms^2 +\ms^4 -\ms^6 + 2 \, \s +10 \ms^2 \s + \s^2 +\ms^2 \s^2)
\right.
\nonumber \\
& & \left. \left.
                + 3 \, \lt (3 +5 \ms^2 -3 \ms^4 -5 \ms^6 + 4 \, \s + 28 \ms^2 \s + 5 \, \s^2 + 5 \ms^2 \s^2) \right)
        \right] \frac{|C_7^{\mbox{eff}}|^2}{\s}
        \nonumber \\
        & &  
        - 8 \, \left[ \left( \s (1 + \ms^2) - (1 - \ms^2)^2 \right)
        + \frac{2}{3}  \lo (-1 +2 \ms^2 - \ms^4 + \s + \ms^2 \s) 
 \right.
\nonumber \\
& & \left.
                +  \lt (5 \ms^2 - 5 \ms^4 + 2 \s + 5 \ms^2 \s) 
        \right] Re(C_9^{\mbox{eff}}) \, C_7^{\mbox{eff}}
        \nonumber \\
        & & \left. \left.
        + 2 \, \left[ 2 +  \lo + 5 \, \lt  \right] 
                \u \, \s \, Re(C_9^{\mbox{eff}}) \, C_{10}
        \right. \right.
        \nonumber \\
        & & \left. \left.
        +  4 \, \left[ 2 \left( 1 + \ms^2 \right)  
        +  \lo (1+ \ms^2)+  \lt (3+ 5 \ms^2)  \right] \u \, Re(C_{10}) \, C_7^{\mbox{eff}}
        \right\} \theta \left[ {\u(\s,\ms)}^2 - \u^2 \right]
        \right.
        \nonumber \\
        & & \left. 
        - E_1 (\s, \u) \, \delta \left[ {\u(\s,\ms)}^2 - \u^2 \right] 
        - E_2 (\s, \u) \, \delta^\prime \left[ {\u(\s,\ms)}^2 - \u^2 \right] 
        \right) \, , 
        \label{eqn:dddw}
\end{eqnarray}
where $\lo=\lambda_{1}/m_{b}^2$ and $\lt=\lambda_{2}/m_{b}^2$.
The auxiliary functions $E_i (\s, \u)$ ($i = 1,2$), introduced  here
for ease of writing, are given explicitly in appendix \ref{app:Ei}.
The boundary of the Dalitz distribution is as usual determined by the
argument of the $\theta$-function and in the $(\u,\s)$-plane
it has been specified in eq.~(\ref{eq:boundaries}).
 The analytic 
form of the result (\ref{eqn:dddw}) is very similar to the corresponding 
double differential
distributions derived by Manohar and Wise in \cite{MW} for
the semileptonic decays $B \to (X_c,X_u) \ell \nu_\ell$. Further comparisons
with this work in the $V-A$ limit for the single differential  and 
integrated rates are given a little later at the end of this section.

Finally, after integrating over the variable $\hat{u}$, we derive the 
differential branching ratio in the scaled dilepton invariant mass 
for \bxsll , 
\begin{eqnarray}
        \frac{{\rm d}{\cal B}}{{\rm d}\s} & = & 2 \; {\cal B}_0 
                \left\{ 
                  \left[
                \frac{2}{3} \u(\s,\ms) ((1 - \ms^2)^2 + \s (1+\ms^2) -2 \s^2) 
        +       \frac{1}{3} (1 -4 \ms^2 + 6 \ms^4 -4 \ms^6 + \ms^8 -\s 
\right. \right.
                \nonumber \\
        & &   
+ \ms^2 \s +
\ms^4 \s - \ms^6 \s -3 \s^2 -2 \ms^2 \s^2 -3 \ms^4 \s^2 + 5 \s^3 +5 \ms^2 \s^3-2 \s^4 ) \frac{\lo}{ \u(\s,\ms)}  
\nonumber \\
  & &                   + \left( 1 -8 \ms^2 + 18 \ms^4 -16 \ms^6 + 5 \ms^8 -\s 
-3 \ms^2 \s + 9 \ms^4 \s -5 \ms^6 \s -15 \s^2 -18 \ms^2 \s^2 
\right.
\nonumber \\
& &
\left. \left.
-15 \ms^4 \s^2 + 25 \s^3
 + 25 \ms^2 \s^3 -10 \s^4 \right) \frac{\lt}{ \u(\s,\ms)}  
                        \right] 
                \left( |C_9^{\mbox{eff}}|^2 + |C_{10}|^2 \right)
                \nonumber \\
        &  &
            +   \left[
                 \frac{8}{3} \u(\s, \ms) (2 (1+\ms^2)(1-\ms^2)^2-
                (1+14 \ms^2 +\ms^4) \s -(1+\ms^2) \s^2 )
                \right.
                \nonumber \\
        & &  
          +     \frac{4}{3} (2 - 6 \ms^2 + 4 \ms^4 +4 \ms^6 -6 \ms^8+
 2 \ms^{10} -5 \s -12 \ms^2 \s + 34 \ms^4 \s -12 \ms^6 \s -5 \ms^8 \s + 3 \s^2
\nonumber \\
& & 
 + 29 \ms^2 \s^2 + 29 \ms^4 \s^2 
+3 \ms^6 \s^2+ \s^3 -10 \ms^2 \s^3 +\ms^4 \s^3-\s^4-\ms^2 \s^4)
  \frac{\lo}{ \u(\s,\ms)} + 4  \left(-6 + 2 \ms^2
\right.
\nonumber \\
& &
 + 20 \ms^4 -12 \ms^6 - 14 \ms^8 +10 \ms^{10} +   3 \s+   16 \ms^2 \s +
 62 \ms^4 \s  -56 \ms^6 \s -25 \ms^8 \s + 3 \s^2
\nonumber \\
& & \left. \left.
  + 73 \ms^2 \s^2 + 101 \ms^4 \s^2 +15 \ms^6 \s^2+ 5 \s^3-26 \ms^2 \s^3+
5 \ms^4 \s^3 -5 \s^4-5 \ms^2 \s^4 \right) \frac{\lt}{ \u(\s,\ms)}
                \right] \frac{|C_7^{\mbox{eff}}|^2}{\s}
                \nonumber \\
        & &     
         +      \left[
                8 \u(\s, \ms) ((1-\ms^2)^2-(1+\ms^2) \s)
                + 4 ( 1 - 2 \ms^2 +\ms^4  - \s-\ms^2 \s) \; \u(\s,\ms) \;  \lo 
\right.
\nonumber \\
 & &
     + 4 \left( -5 +30\ms^4-40 \ms^6 +15 \ms^8 -\s + 21 \ms^2 \s +
 25 \ms^4 \s -45 \ms^6 \s+ 13 \s^2 + 22 \ms^2 \s^2\right.
\nonumber \\
& & 
\left. \left. \left.
+45 \ms^4 \s^2 -7 \s^3-15 \ms^2 \s^3 \right)\frac{\lt}{ \u(\s,\ms)}    
                        \right] Re(C_9^{\mbox{eff}}) \, C_7^{\mbox{eff}} 
                \right\} \; . 
\label{eqn:dbds}
\end{eqnarray}
 The leading power corrected expression for the
FB-asymmetry ${\cal A}(\s)$ is:
\begin{eqnarray}
        \frac{{\rm d}{\cal A}(\s)}{{\rm d}\s} & = &
                - 2 \; {\cal B}_0
                \left\{
                \left[
                2 (\u(\s ,\ms))^2 \s
        + \frac{\s}{3} (3 -6 \ms^2 + 3 \ms^4 +2 \s -6 \ms^2 \s+3 \s^2) \lo
\right. \right.
\nonumber \\
&+ & \left. 
   \s \, (-9 -6 \ms^2 + 15 \ms^4 -14 \s -30 \ms^2 \s+ 15 \s^2) \, \lt 
\right]  
                         \, Re(C_9^{\mbox{eff}}) \, C_{10}
                \nonumber \\
        &+&
                 \left[ 4 (\u(\s ,\ms))^2 (1+\ms^2)
+ \frac{2}{3}\; (1+\ms^2) \;
 (3 -6 \ms^2 + 3 \ms^4 + 2 \s -6 \ms^2 \s + 3\s^2) \lo
\right. \\
&+ & \left. \left.
                 2 (-7 -3 \ms^2 -5 \ms^4
+15 \ms^6 -10 \s -24 \ms^2 \s-30 \ms^4 \s + 9 \s^2+ 15 \ms^2 \s^2)\,
 \lt \right]\,  Re(C_{10}) \, C_7^{\mbox{eff}}\,  \right\}  . \nonumber
\end{eqnarray}

The results derived for the $O(\alpha_s)$-improved and power-corrected 
Dalitz distribution, dilepton invariant mass, and FB-asymmetry in \bxsll 
are the principal new results in this section.
It is useful to write the 
corresponding expressions in the limit $m_s=0$. For the dilepton invariant
mass distribution, we get 
\begin{eqnarray}
        \frac{{\rm d}{\cal B}}{{\rm d}\s} & = & 2 \; {\cal B}_0 
                \left\{ 
                  \left[
                \frac{1}{3} (1-\s)^2 (1+2 \s) \; (2 + \lo) 
                + ( 1 - 15  \s^2 + 10 \s^3\right) \lt
                        \right] 
                \left( |C_9^{\mbox{eff}} |^2 + |C_{10}|^2 \right)
                \nonumber \\
        & & 
             +  \left[
                 \frac{4}{3} (1-\s)^2 (2+ \s) \; (2 + \lo)
                + 4  \left( -6 -3  \s + 5 \s^3 \right) \lt
                \right] \frac{|C_7^{\mbox{eff}}|^2}{\s}
                \nonumber \\
        & &     \left.
           +    \left[
                4 (1-\s)^2 (2+ \lo)
               + 4  \left( -5 -6  \s + 7 \s^2 \right) \lt   
                                \right] Re(C_9^{\mbox{eff}}) \, C_7^{\mbox{eff}} 
                \right\}\, .
\label{eqn:dbds0}
\end{eqnarray}

 The (unnormalized) FB asymmetry reads as,
\begin{eqnarray}
        \frac{{\rm d}{\cal A}}{{\rm d}\s} & = & 
                - 2 \; {\cal B}_0 
                \left\{
                \left[ 
                2 (1 - \s)^2 \s 
                + \frac{\s}{3} (3+2 \s +3 \s^2) \lo 
                +  \s \, (-9 -14 \s + 15 \s^2) \, \lt \right] 
                         \, Re(C_9^{\mbox{eff}} ) \, C_{10}
                \right. 
                \nonumber \\
        & &     \left. 
                + \left[ 4 (1-\s)^2   
                + \frac{2}{3} (3+ 2 \s + 3\s^2) \lo 
                + 2 (-7 -10 \s + 9 \s^2)\, \lt \right] 
\, Re(C_{10}) \, C_7^{\mbox{eff}} 
                 \,  \right\}\, .
\end{eqnarray}

Our result \cite{AHHM97} for the dilepton invariant mass distribution given 
in eq.~(\ref{eqn:dbds0}) has been confirmed recently by
\cite{buchallaisidorirey} in the $m_s=0$ limit and is in disagreement with an 
earlier publication \cite{FLSold}.
(The differences between the previous result eq.~(3.21) of the 
paper by \cite{FLSold} have been discussed at length in
 \cite{AHHM97}.)

\begin{figure}[htb]
\vskip -1.4truein
\centerline{\epsfysize=5in
{\epsffile{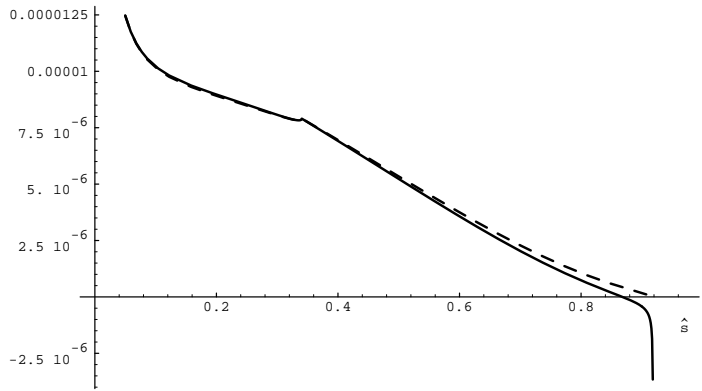}}}
\vskip -1.60truein
\caption[]{ \it 
Dilepton invariant mass spectrum ${\rm d}{\cal B} (B \to X_s e^+
e^-)/{\rm d} \hat{s}$ in the parton model (dashed curve) and with leading 
power corrections calculated in the
HQE approach (solid curve). The parameters used are given in
Table \ref{parameters}.}
\label{fig:hillerfig1}
\end{figure}

Concerning the invariant dilepton mass spectrum derived by us and given
in eq.~(\ref{eqn:dbds}),
we would like to make the following observations:
First, the leading order power corrections in the dilepton mass distribution 
are found to be small over a good part of the dilepton mass $\s$.
However, we find
that the power corrections become increasingly
large and negative as one approaches $\hat{s} \to \hat{s}^{max}$,
where $\s^{max}=(1-\ms)^2$.
Since the parton model spectrum falls steeply near the
end-point $\hat{s} \to \hat{s}^{max}$, this leads to the
uncomfortable result that the power corrected
dilepton mass distribution becomes negative for the high dilepton masses.
We show in Fig.~\ref{fig:hillerfig1} this distribution in the parton model 
and the HQE approach, using the central values of the parameters in Table 
\ref{parameters}.
Further, the power-corrected dilepton invariant mass
distribution retains the characteristic
$1/\hat{s}$ behaviour following from the one-photon exchange 
in the parton model.
We note that the correction proportional to the kinetic energy
term $\hat{\lambda}_1$ renormalizes the parton model invariant 
mass distribution multiplicatively by approximately the factor 
$(1 + \lambda_1/(2m_b^2))$, which is exact in the limit $m_s=0$ and no new 
functional dependence in $\hat{s}$ is introduced (moreover, this factor is
hardly different from 1). Hence,
the negative probability near the end-point is largely driven by the magnetic
moment term $\hat{\lambda}_2$.  

\begin{figure}[htb]
\vskip -1.4truein
\centerline{\epsfysize=5in
{\epsffile{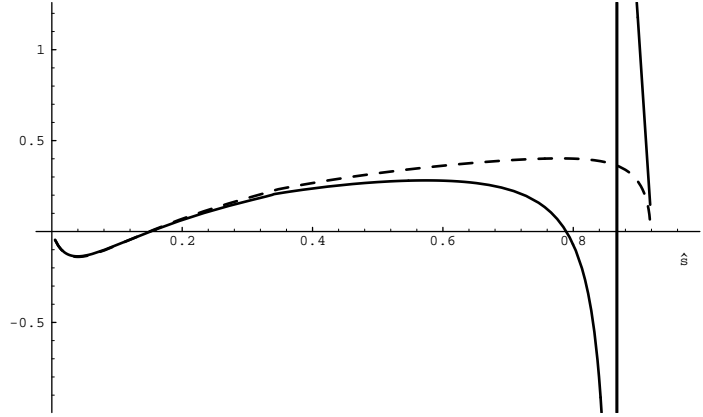}}}
\vskip -1.6truein   
\caption[]{\it
FB asymmetry (normalized) ${\rm d}\overline{{\cal A}} (B \to X_s e^+
e^-)/{\rm d} \hat{s}$ in the parton model (dashed curve) and with power 
corrections calculated in the
HQE approach (solid curve). The parameters used are given in
Table \ref{parameters}.}
\label{fig:hillerfig2}
\end{figure}

The  normalized FB asymmetry, $d\bar{\cal A}(\s)/d\s$,
in the HQE-approach and the parton model are shown in 
Fig.~\ref{fig:hillerfig2}. We find that this asymmetry is stable against 
leading order power corrections up to $\s \leq 0.6$, but the corrections
become increasingly large due to the unphysical
behaviour of the HQE-based dilepton mass distribution as $\s$ approaches 
$\s^{max}$ (see Fig.~\ref{fig:hillerfig1}).
Based on these investigations, we must conclude that the HQE-based approach
has a  restrictive kinematical domain for its validity. In particular,
it breaks down for the high dilepton invariant mass region in \bxsll.
  This behaviour of the dilepton mass spectrum in \bxsll is not unexpected,
as similar behaviours have been derived near the end-point of the lepton
energy spectra in the decays $B \to X \ell \nu_\ell$
in the HQE approach \cite{MW}.
 To stress these similarities, we show the 
power correction in the dilepton 
mass distribution as calculated in the HQE approach compared to the parton 
model through the ratio defined as:
\begin{equation}
 R^{\mbox{\small HQE}} (\s) \equiv \frac{{\rm d}{\cal 
B}/d\hat{s}(\mbox{HQE}) - 
{\rm d}{\cal B}/d\hat{s}(\mbox{Parton Model})}{ {\rm d}{\cal
B}/d\hat{s}(\mbox{Parton Model})} \; .
\label{ratiohqepm}
\end{equation}

\begin{figure}[htb]
\vskip -1.4truein
\centerline{\epsfysize=5in
{\epsffile{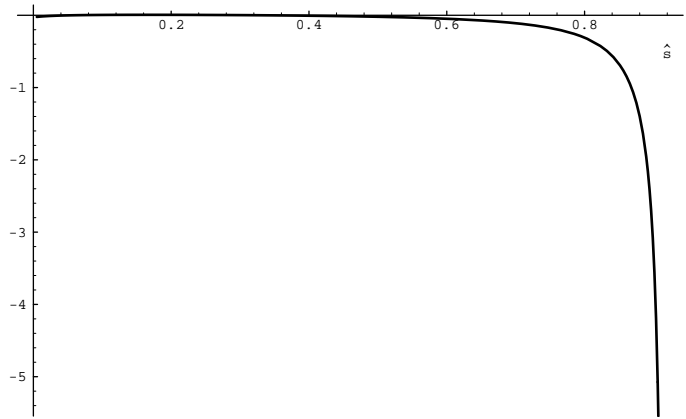}}}
\vskip -1.6truein
\caption[]{ \it
The correction factor $R^{HQE}(\s)$ (in percentage) as defined in
eq. (\protect\ref{ratiohqepm}) for the dilepton mass spectrum ${\rm 
d}{\cal B} (B \to X_s \ell^+\ell^-)/{\rm d} \hat{s}$.
The parameters used are given in Table \ref{parameters}.}
\label{fig:hillerfig3}
\end{figure}

 The correction factor  $R^{\mbox{HQE}} (\s)$ for \bxsll
shown in Fig.~\ref{fig:hillerfig3} is qualitatively similar to 
the corresponding factor in the lepton
energy spectrum in the decay $ B \to X_c \ell \nu_\ell$, given in Fig. 6
of \cite{MW}.
We note that we have been able to derive
the power corrected rate for the semileptonic decays $B \to X_c \ell
\nu_\ell$ obtained by Manohar and Wise in \cite{MW}.
\footnote{The HQE matrix elements in our convention and the MW ones
are related by
$\loo = - 2 {m_b}^2  K_b$,
$  3 \lto  =- 2 {m_b}^2 G_b$ and
$\loo + 3 \lto  =- 2 {m_b}^2 E_b=- 2 {m_b}^2 (K_b+G_b)$, likewise we have for 
the normalization of states $|B\rangle=\sqrt{2 m_B} |B\rangle^{MW}$.}
In doing this, we shall reduce the matrix element for the
decay \bxsll to the one encountered in
$B \to X_c \ell \nu_\ell$, obtained by the replacements ($V-A$ limit):
\begin{eqnarray} \label{eq:c9CC}
        C_9^{\mbox{eff}} & = & - C_{10} = \frac{1}{2}
                \, , \\
        C_7^{\mbox{eff}} & = & 0 
                \, , \\
        \left( \frac{G_F \, \alpha}{\sqrt{2} \, \pi}
                V_{ts}^\ast V_{tb}\right)
                & \rightarrow &
        \left( - \frac{4 \, G_F}{\sqrt{2}} V_{cb} \right)
                \, .  
\label{eq:CKMCC} 
\end{eqnarray}
This amounts to keeping only the charged current $V-A$ contribution in \bxsll
decays. 

Finally, since the HQE-improved expression for the decay rate cannot be given
analytically due to the Wilson coefficient $\cneff(\s)$ which is a 
complicated function of
$\s$, we give below the results in numerical form:
\begin{equation}
\Gamma^{\mbox{\small HQE}}=\Gamma^{b} (1 + c_1\hat{\lambda}_1 + c_2 
\hat{\lambda}_2 )\; ,
\label{gammahqet}
 \end{equation}
where $\Gamma^{b}$ is the parton model decay width for \bsll and the
coefficients depend on the input parameters. For the central values of
the parameters given in Table \ref{parameters}, they have the values
$c_1 =0.501$ and $c_2 = -7.425$. This leads to a reduction in the decay
width by $-4.1\%$, using the values of $\lambda_1$ and $\lambda_2$ given
in Table \ref{parameters}. Moreover, this reduction is mostly contributed by 
the $\lambda_2$-dependent term. We recall that the coefficient of the 
$\hat{\lambda}_1$ term $c_1$ is (almost) the same as in the semileptonic 
width $\Gamma(B \to X_u \ell \nu_\ell)$ obtained by Bigi et al \cite{georgi}
\begin{equation}
         \Gamma^{\mbox{\small HQE}}_{sl} = 
                \Gamma^b_{sl} \, \left( 1 + \frac{1}{2} \lo
                - \frac{9}{2} \, \lt \right) 
                \; , 
                \label{eqn:gmw}
\end{equation} 
where $\Gamma^b_{sl}$ is the parton model decay width. This points towards
the universality of this coefficient.
The coefficient of the $\hat{\lambda}_2$ term $c_2$ for \bsll decay is larger 
than the corresponding one in the semileptonic decay width.
Hence, the power corrections in $\Gamma(B \to X_u \ell \nu_\ell)$ and
$\Gamma(\mbox{\bxsll})$ are rather similar but not identical. 
\section{Dilepton Invariant Mass and the FB Asymmetry in the Fermi Motion 
Model \label{wavefunction}}
  In this section, we present our estimates of the non-perturbative
effects on the decay distributions in \bxsll. These effects are connected
with the bound state nature of the $B$-hadron and the
physical threshold in the \bxsll in the final state.
In order to implement these effects on the decay 
distributions in \bxsll, we resort to the Gaussian Fermi motion (FM) model 
\cite{aliqcd} introduced in section \ref{subsec:FM}.

 In the Fermi motion
model, the problem of negative probabilities encountered in the HQE approach
for the high dilepton masses near $s \to s_{max}$ is not present, which
motivates us to use
this model as a reasonable approximation of the non-perturbative effects  
in the entire dilepton mass range. The success of this model in describing
the inclusive lepton energy spectra in $B \to (X_c,X_u) \ell \nu_\ell$
and \bxsg strengthens this hope.

     In the decay \bxsll, the distribution
$d {\cal B}/d\hat{s}$ depends on the Lorentz-invariant variable $\hat{s}$ 
only. So, the Lorentz boost involved in the Fermi motion 
model (Doppler shift) leaves the dilepton mass distribution invariant. 
However, since the $b$-quark mass $m_b(p)$ is 
now a momentum-dependent quantity, this distribution is affected due to the
 difference ($m_b(p)-m_b)$ (mass defect), which 
rescales the variable $\hat{s}$ and hence smears the dilepton distribution
calculated in the parton model.
 For different choices of the model
parameters $(p_F,m_q)$ corresponding to the same effective $b$-quark mass,
$m_b^{\mbox{eff}}$ which is defined in eq.~(\ref{effbmass})
the dilepton mass distributions should be very similar
\cite{greubrey}, 
which indeed is the case as we have checked numerically but do not show
the resulting distributions here.

The situation with the FB asymmetry is, however, quite different. 
Being an angle-dependent quantity,  it is not
Lorentz-invariant and is sensitive to both the Doppler shift and the 
mass defect. We give in appendix \ref{app:dalitz}, 
the Dalitz distribution $d^2\Gamma(B \to
X_s \ell^+ \ell^-)/ds du$ in the Fermi motion model.

As we calculate the
branching ratio for the inclusive decay \bxsll in terms of the
semileptonic decay branching ratio ${\cal B} (B \to X\ell \nu_\ell)$, we
have to correct the normalization due to the variable $b$-quark mass in
both the decay rates. 
%Fixing $m_b$ but varying the model parameters $p_F$ and $m_q$ yields 
%variable effective (momentum-dependent) $b$-quark mass $m_b^{\mbox{eff}}$.
We recall that the decay widths for \bxsll and $B \to X
\ell \nu_\ell$ in this model are proportional to $(m_b^{\mbox{eff}})^5$
\cite{ag1,effhamali,aligreub93}. Hence the  decay widths for both the decays
individually are rather sensitive
to $m_b^{\mbox{eff}}$. This dependence largely (but not exactly) cancels
out in the branching ratio ${\cal B}(\mbox{\bxsll})$. Thus, varying
$m_b^{\mbox{eff}}$ in the range  $m_b^{\mbox{eff}}= 4.8 \pm 0.1$ GeV
results in $\Delta \Gamma(\mbox{\bxsll})/\Gamma = \pm 10.8\%$.
 However, the change in the
branching ratio itself is rather modest, namely
$\Delta {\cal B}(\mbox{\bxsll})/{\cal B} = \pm 2.3\%$.
This is rather similar to what we have obtained in the HQE approach.

The theoretical uncertainties in the branching ratios
for \bxsll from the 
perturbative part, such as the ones from the indeterminacy in the top 
quark mass,   
the QCD scale $\Lambda_{QCD}$ and the renormalization scale $\mu$,  
have been investigated in the literature \cite{misiakE,burasmuenz}.
 We have
recalculated them for the indicated ranges of the parameters in
Table \ref{parameters}. The resulting (SD) branching ratios and their 
present uncertainties are found to be:
\begin{eqnarray}
{\cal B}(B \to X_s e^+ e^-) &=& (8.4 \pm 1.9) \times 10^{-6} \; , \nonumber\\ 
{\cal B}(B \to X_s \mu^+ \mu^-)& =& (5.7 \pm 0.9) \times 10^{-6} \; , 
\nonumber\\ 
{\cal B}(B \to X_s \tau^+ \tau^-) &=& (2.6 \pm 0.4) \times 10^{-7} \; ,
\label{eq:Brnumbers} 
\end{eqnarray}
where in calculating the branching ratio ${\cal B}(B \to X_s \tau^+ \tau^-)$,
we have included the $\tau$-lepton mass terms in the matrix element
\cite{KS96}.
These uncertainties, typically $\pm 20\%$, are
much larger than the wave function-dependent uncertainties, and so the
theoretical accuracy of the SD-part
in the SM in these decays is not compromised by the non-perturbative 
effects.

\begin{figure}[t]
     \mbox{ }\hspace{-0.7cm}
     \begin{minipage}[t]{7.6cm}
     \mbox{ }\hfill\hspace{1cm}(a)\hfill\mbox{ }
     \epsfig{file=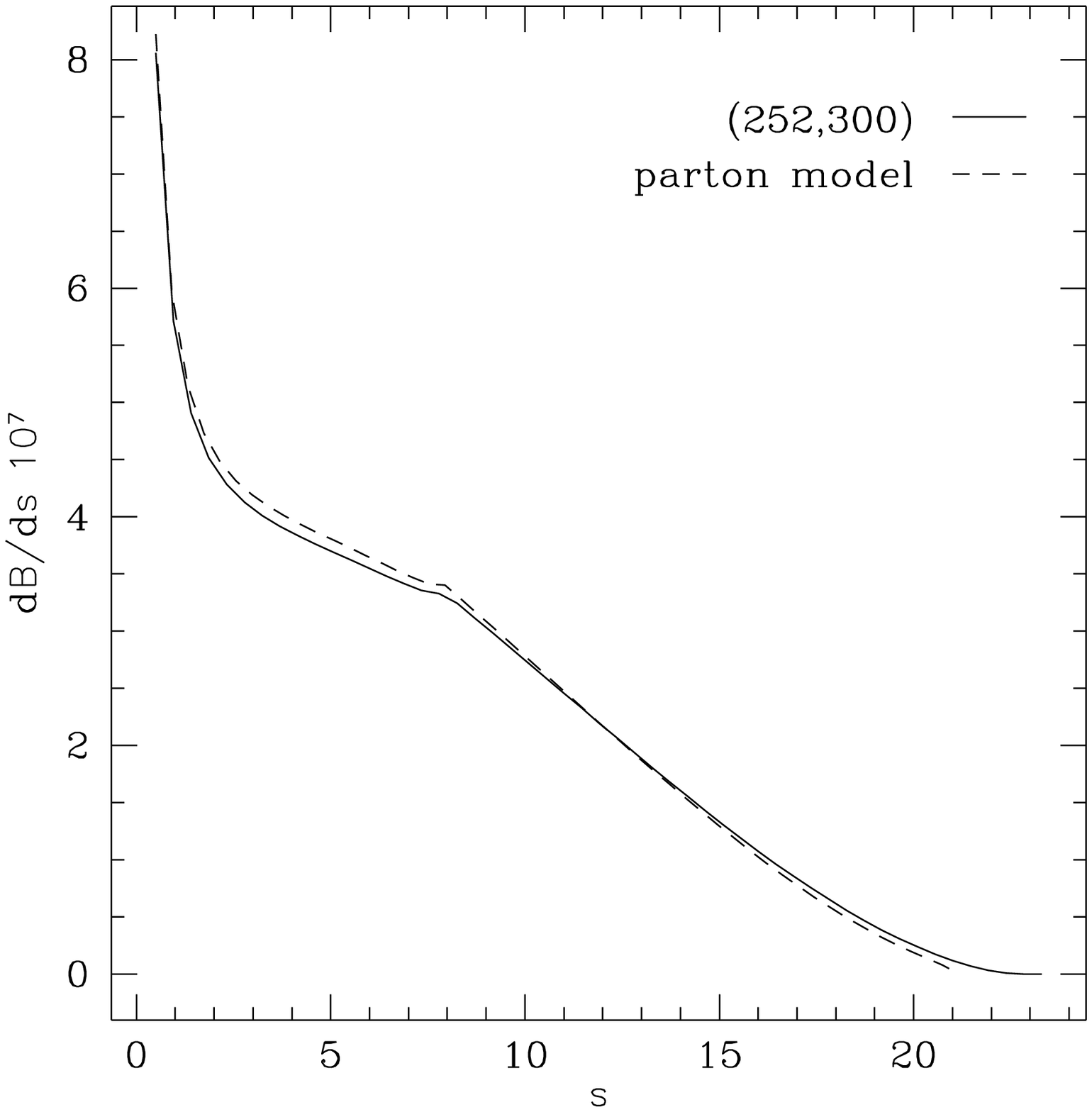,width=7.6cm}
     \end{minipage}
     \hspace{-0.4cm}
     \begin{minipage}[t]{6.3cm}
     \mbox{ }\hfill\hspace{3.1cm}(b)\hfill\mbox{ }
     \epsfig{file=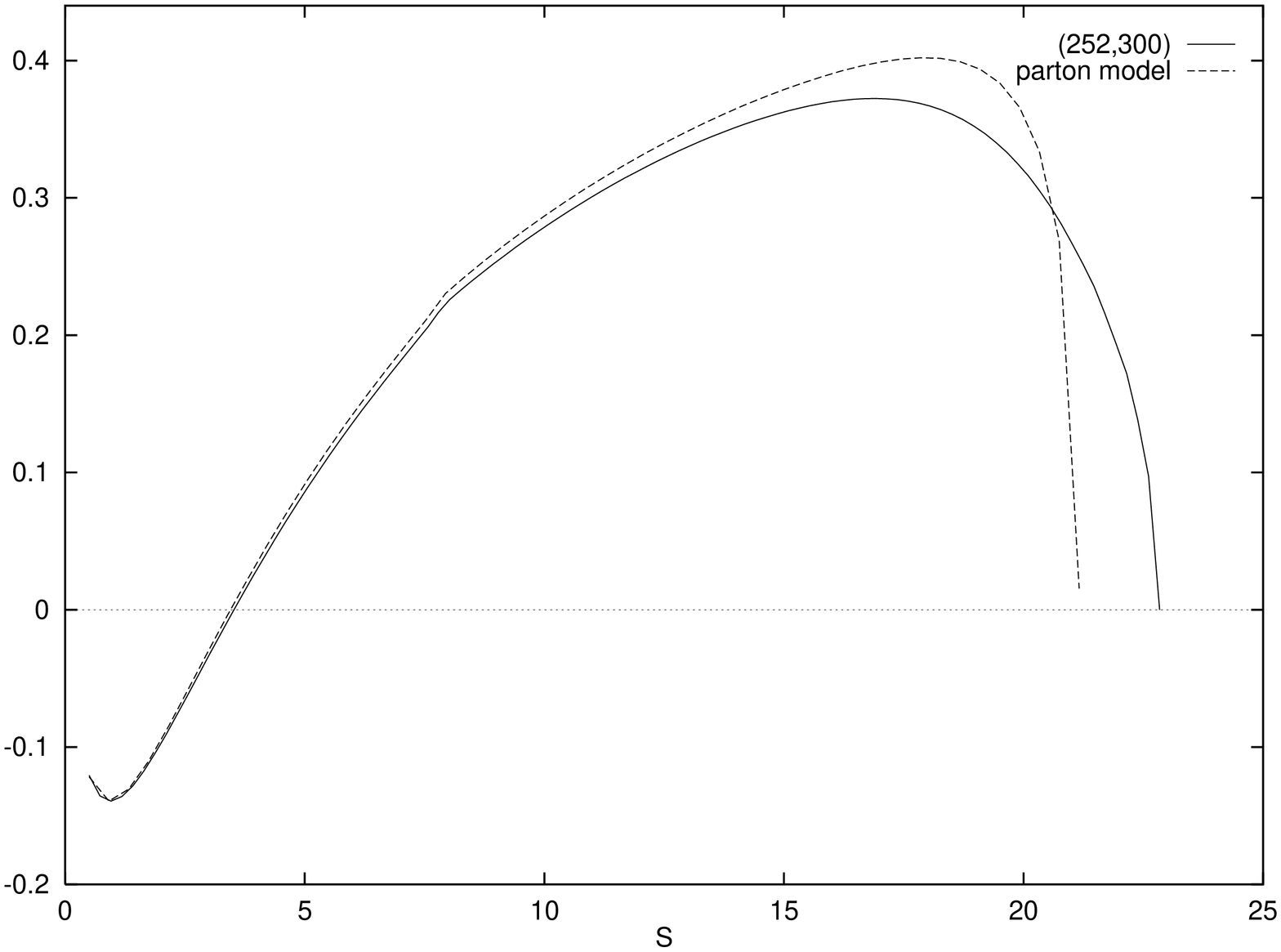,width=6.3cm,angle=270}
     \end{minipage}  
     \caption{\it 
Differential branching ratio $d {\cal B}/ds$ for $B \to X_s \ell^+ \ell^-$ 
(a) and normalized differential FB asymmetry $d\overline{{\cal A}}(s)/ds$ (b)
in the SM including the next-to-leading order QCD corrections.
The dashed curve corresponds to the parton model with the parameters given in 
Table \ref{parameters} and the
solid curve results from the Fermi motion model with the model parameters
$(p_F,m_q)=(252,300)$ MeV, yielding an effective $b$-quark mass
 $m_b^{eff}=4.85$ GeV.}
 \label{fig:fermi1}
\end{figure}
\begin{figure}[t]
     \mbox{ }\hspace{-0.7cm}
     \begin{minipage}[t]{7.6cm}
     \mbox{ }\hfill\hspace{1cm}(a)\hfill\mbox{ }
     \epsfig{file=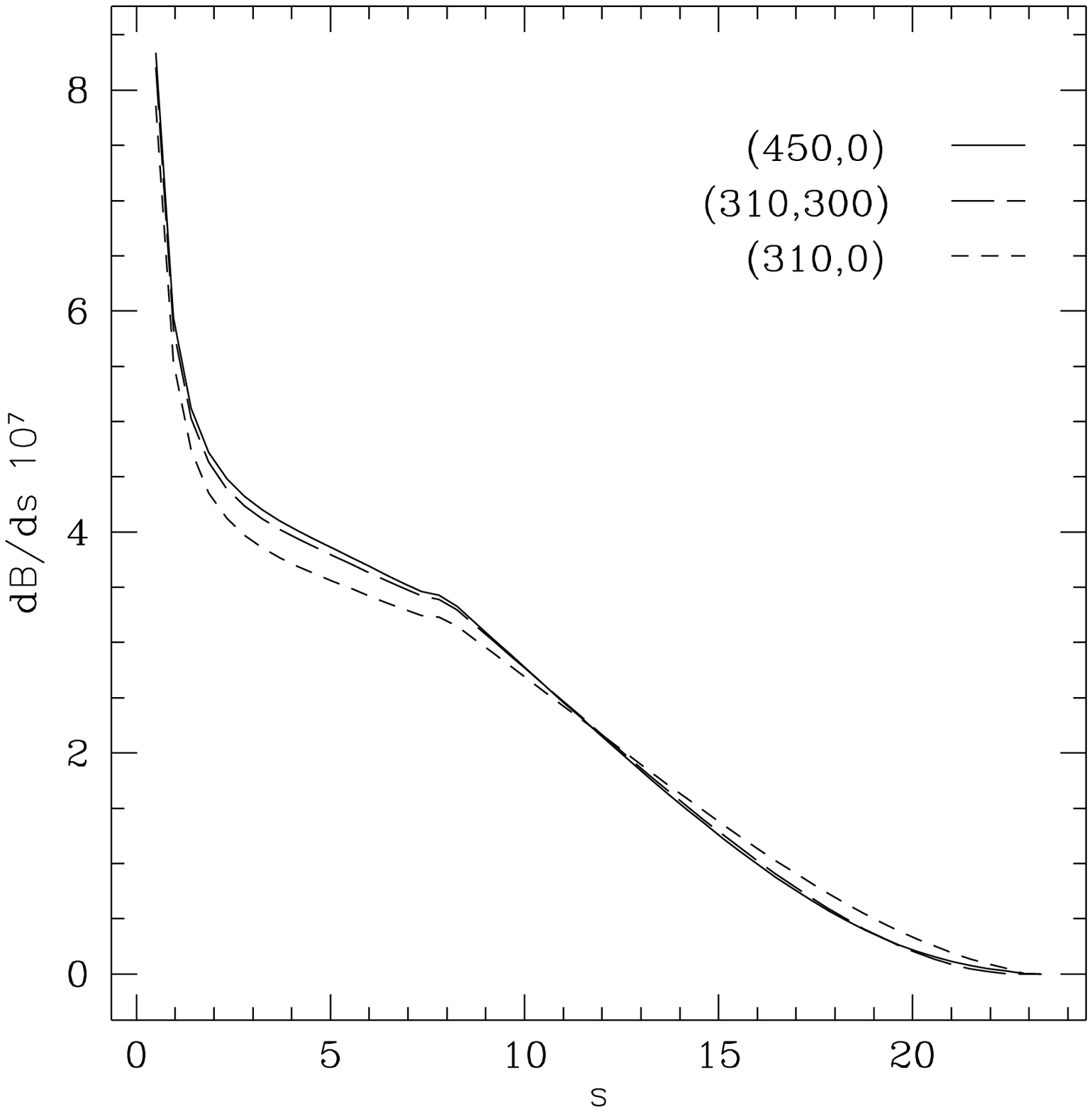,width=7.6cm}
     \end{minipage}
     \hspace{-0.4cm}
     \begin{minipage}[t]{6.3cm}
     \mbox{ }\hfill\hspace{3.1cm}(b)\hfill\mbox{ }
     \epsfig{file=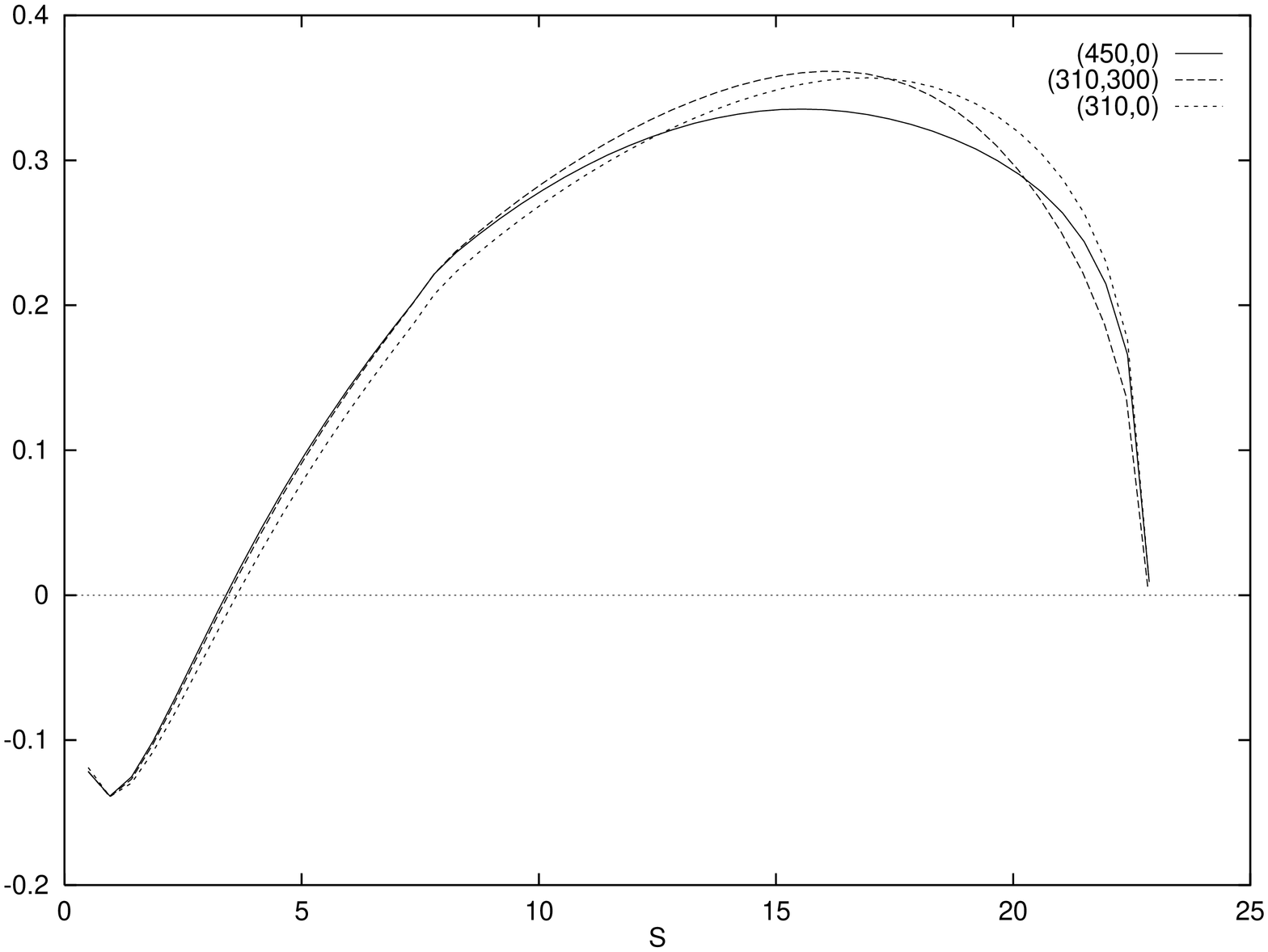,width=6.3cm,angle=270}
     \end{minipage}  
     \caption{\it 
Differential branching ratio $d {\cal B}/ds$ for
$B \to X_s \ell^+ \ell^-$ (a) and normalized differential FB asymmetry 
$d\overline{{\cal A}}(s)/ds$ (b)
using the Fermi motion model for three different pairs of the 
model parameters $(p_F,m_q)=(450,0)$ MeV (solid curve), $(310,300)$ MeV
(long dashed curve), and $(p_F,m_q)=(310,0)$ MeV (short dashed curve)
 yielding the effective $b$-quark masses
 $m_b^{eff} =4.76$ GeV, $4.80$ GeV, and $4.92$ GeV, respectively.}
\label{fig:fermi5}
\end{figure}
\begin{figure}[t]
     \mbox{ }\hspace{-0.7cm}
     \begin{minipage}[t]{7.6cm}
     \mbox{ }\hfill\hspace{1cm}(a)\hfill\mbox{ }
     \epsfig{file=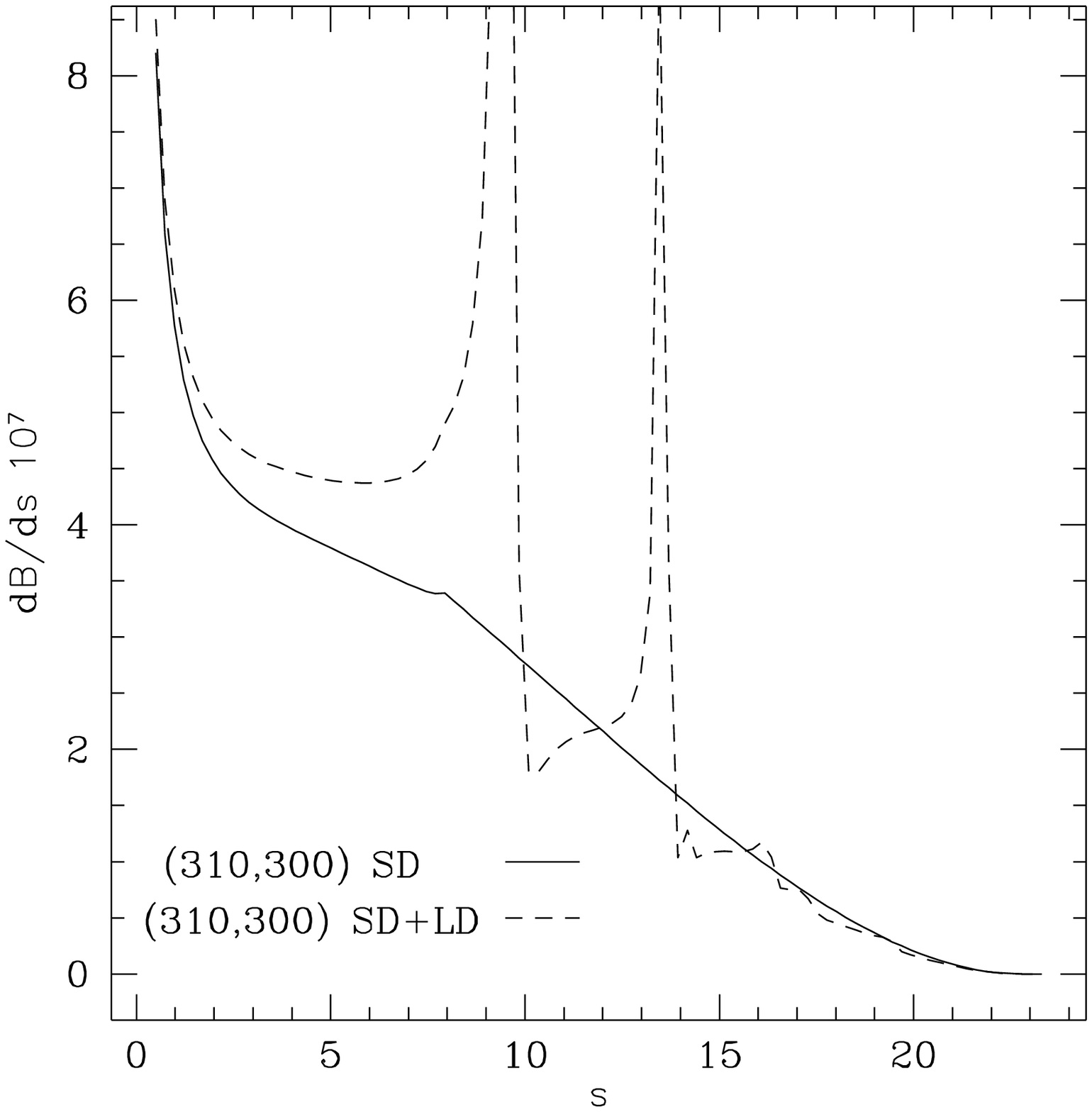,width=7.6cm}
     \end{minipage}
     \hspace{-0.4cm}
     \begin{minipage}[t]{6.3cm}
     \mbox{ }\hfill\hspace{3.1cm}(b)\hfill\mbox{ }
     \epsfig{file=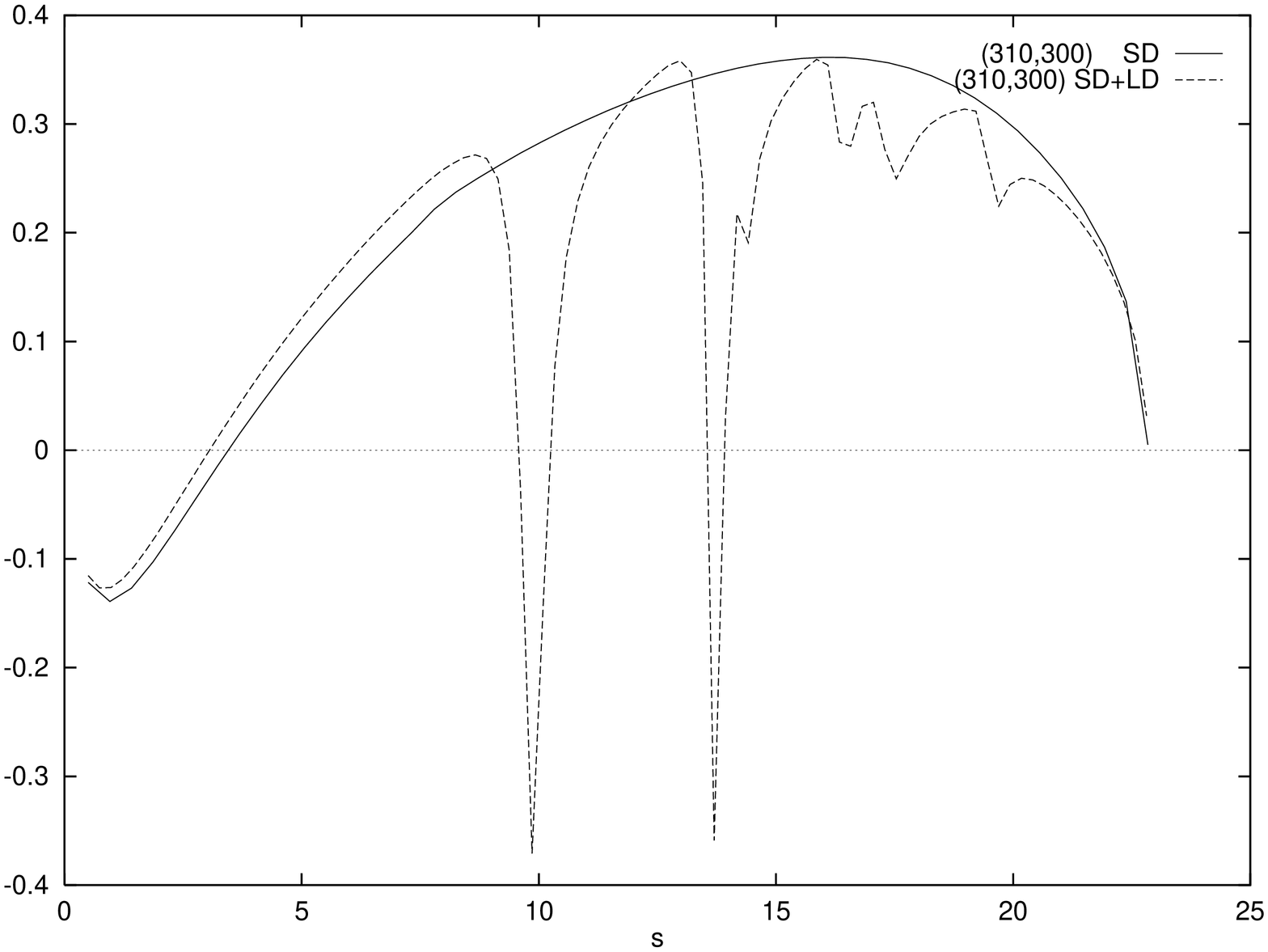,width=6.3cm,angle=270}
     \end{minipage}  
     \caption{\it 
Differential branching ratio $d{\cal B}/ds$ for $B \to X_s \ell^+ 
\ell^- $ (a) and normalized differential FB asymmetry 
$d\overline{{\cal A}}(s)/ds$ (b) calculated in the SM using the
next-to-leading order QCD corrections and Fermi motion effect (solid 
curve), and including the LD-contributions (dashed curve).
The Fermi motion model parameters $(p_F,m_q)$ in MeV 
are displayed in the figure.} 
\label{fig:lddb}
\end{figure}
\begin{figure}[t]
     \mbox{ }\hspace{-0.7cm}
     \begin{minipage}[t]{7.6cm}
     \mbox{ }\hfill\hspace{1cm}(a)\hfill\mbox{ }
     \epsfig{file=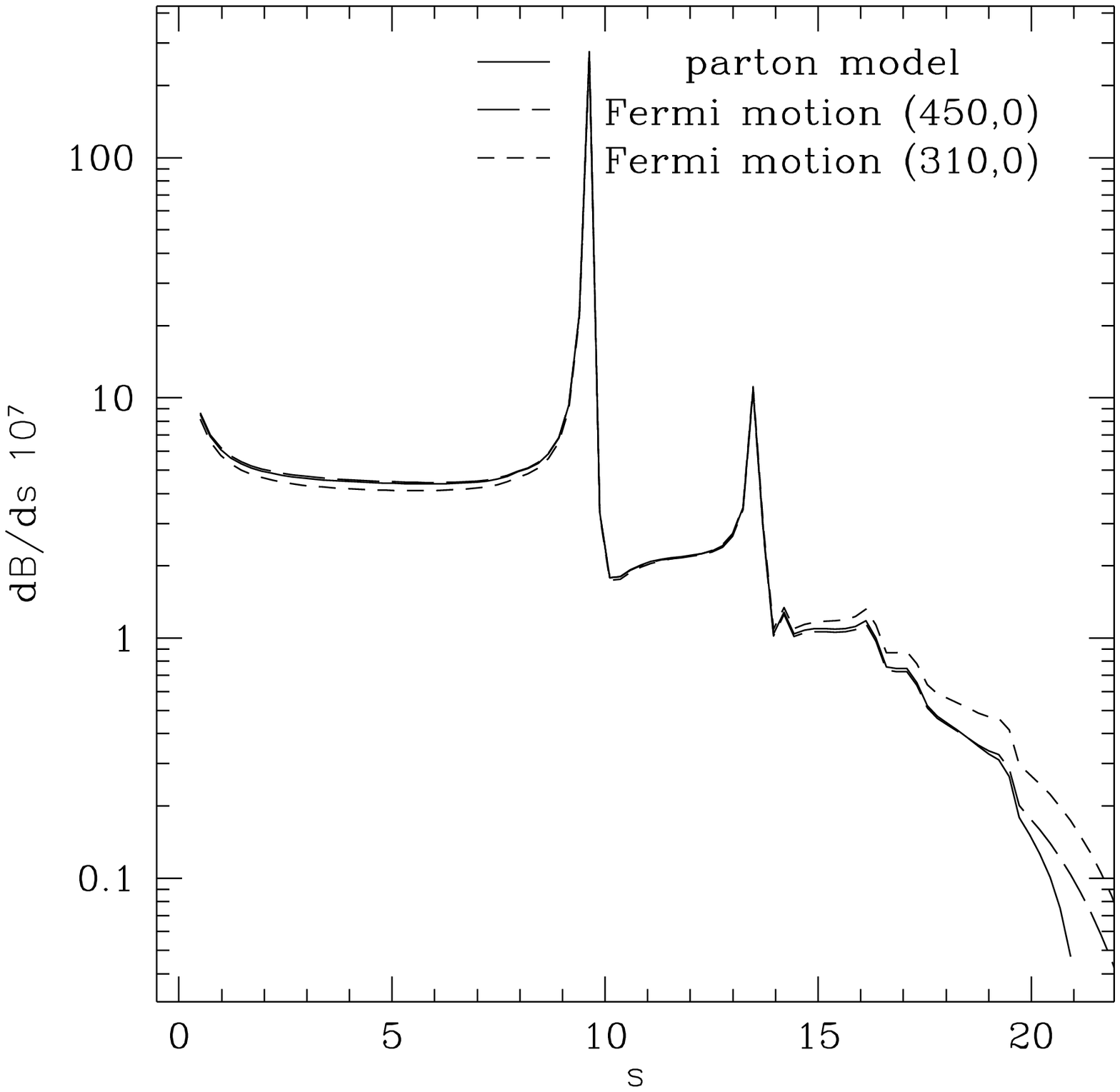,width=7.6cm}
     \end{minipage}
     \hspace{-0.4cm}
     \begin{minipage}[t]{6.3cm}
     \mbox{ }\hfill\hspace{3.1cm}(b)\hfill\mbox{ }
     \epsfig{file=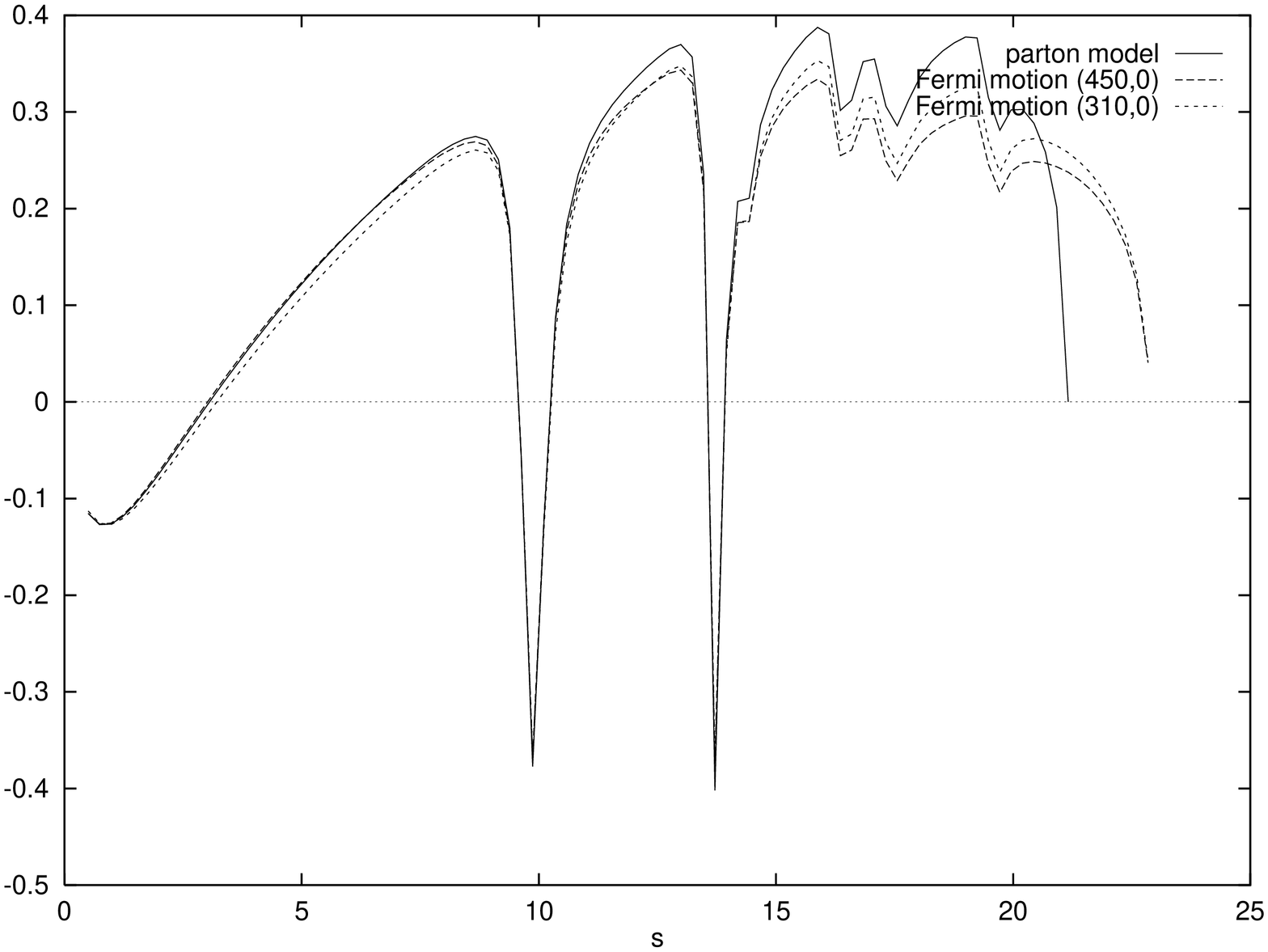,width=6.3cm,angle=270}
     \end{minipage}  
     \caption{\it 
Dilepton invariant mass distribution in $B \to X_s \ell^+ \ell^-$ (a)
and normalized differential FB asymmetry 
$d\overline{{\cal A}}(s)/ds$ (b) in the SM including
next-to-leading order QCD correction and LD effects. The solid curve
corresponds to the parton model and the short-dashed and long-dashed
curves correspond to including the Fermi motion effects. The
values of the FM parameters in MeV are indicated in the figure.}
\label{fig:dbrnsm}
\end{figure}
We show the resulting dilepton invariant mass distribution in 
Fig.~\ref{fig:fermi1} (a) and the FB-asymmetry in Fig.~\ref{fig:fermi1} (b), 
where for the sake of illustration we have
used  the values $(p_F, m_q)=(252,300)$ in (MeV,MeV), which  
correspond to an allowed set of 
parameters obtained from the analysis of the measured photon energy spectrum
in \bxsg, using the same model \cite{ag2}. We see that the dilepton mass 
distribution is stable against Fermi motion effects over most part
of this spectrum, as expected. 
We emphazise here that the end-point spectrum 
extends to the physical kinematic limit in \bxsll 
$s^{max}= (m_B-m_X)^2$ with $m_X=max(m_K,m_s+m_q)$ ($m_q$ is the spectator 
mass), which has to be imposed on the FM program.
It corresponds to the invariant hadronic mass of the lowest physical state 
with total strangeness number $s=1$, $m(X_s) = m_K$, 
as opposed to the parton model, in which $s^{max}=(m_b-m_s)^2$.  
The two thresholds can be made
to coincide for only unrealistically values of $m_b$ and $m_s$.
The FB-asymmetry shows a more marked dependence on the model parameters,
which becomes significant in the high dilepton mass region.

%\cite{BI98}

As the parameters of the Fermi motion model are not presently very 
well-determined from the fits of the existing data \cite{ag2,CLEOslfm},
one has to vary these parameters and estimate the resulting dispersion on
the distributions in \bxsll. We show in Figs.~\ref{fig:fermi5} 
the dilepton mass distribution (a) and the FB asymmetry (b), respectively, 
indicating also the ranges
of the parameters $(p_F,m_q)$. The resulting theoretical 
uncertainty in the distributions is found to be modest.

\section{LD Contributions in \bxsll (I) \label{sec:LD1}}

Next, we implement the effects of LD contributions in the
processes \bxsll. The issues involved here have been discussed in 
\cite{LW96,KS96,Ahmady96}.
The LD contributions due to the vector 
mesons $J/\psi$ and $\psi^\prime$ and higher resonances, as well as the
$(c\bar{c})$ continuum contribution, which we have already included in
the coefficient $C_9^{\mbox{eff}}$, 
appear in the $(\bar{s}_L \gamma_\mu b_L)(\bar{e} \gamma^\mu e)$
interaction term only, i.e., in the coefficient of the operator $O_9$.
 This implies that such LD-contributions should change
$C_9$ effectively, but keep $C_7^{\mbox{eff}}$ and $C_{10}$ unchanged. 
In principle,
one has also a LD contribution in the effective coefficient $C_7^{\mbox{eff}}$;
this, however, has been discussed extensively in the context of the \bxsg 
decay and  estimated to be small \cite{bsgamld,pakvasa,deshpande,eilam95}. 
The LD-contribution is negligible in 
$C_{10}$. Hence, the three-coefficient fit of the data on \bxsll 
and \bxsg, proposed in ref.~\cite{agm94} on the basis of the SD-contributions,
can be carried out also including the LD-effects.

In accordance with this, to incorporate the LD-effects in \bxsll, the 
function $Y(\s)$ introduced earlier is replaced by, \begin{equation}
        Y(\s) \rightarrow Y^\prime(\s) \equiv Y(\s) + 
                Y_{res}(\s) \; ,
\label{eq:yprime}
\end{equation}
where $Y_{res}(\s)$ accounts for the charmonium resonance contribution
via $B \to X_s (J/\Psi, \Psi^\prime, \dots) \to X_s \ell^{+} \ell^{-}$.
Its origin lies in the diagram displayed in Fig.~\ref{fig:ycharm}, where the
internal charm loop hadronizes before decaying into a photon.
We take the representation \cite{amm91},
\begin{equation}
        Y_{res}(\s) = \frac{3}{\alpha^2} \kappa \, C^{(0)}
                \sum_{V_i = \psi(1s),..., \psi(6s)}
                \frac{\pi \, \Gamma(V_i \rightarrow \ell^+ \ell^-)\, m_{V_i}}{
                {m_{V_i}}^2 - \s \, {m_b}^2 - i m_{V_i} \Gamma_{V_i}} ,
\label{LDeq}
\end{equation}
where
$C^{(0)} \equiv 3 C_1 + C_2 + 3 C_3 + C_4 + 3 C_5 + C_6$.
We adopt $\kappa = 2.3 $ for the numerical calculations \cite{LW96}.
This is a fair representation of present data in the factorization approach
\cite{neubert}; also the phase of $\kappa$, which is fixed
in eq.~(\ref{LDeq}), is now supported by data which finds it
close to its perturbative value \cite{BHP96}.
Of course, the data determines only the combination 
$\kappa \, C^{(0)} = 0.88$. The relevant parameters of the charmonium 
resonances $(1S,\dots,6S)$ are given in the Particle Data Group
\cite{PDG}, and we have averaged the leptonic widths for the decay modes
$V \to \ell^+ \ell^-$ for $\ell=e$ and $\ell=\mu$. Note that in
extrapolating the dilepton masses away from the
resonance region, no extra $q^2$-dependence
is included in the $\gamma^{*}(q^2)$-$V_i$ junction.
(The $q^2$-dependence written explicitly in eq.~(\ref{LDeq})
is due to the Breit-Wigner shape of the resonances.)
This is an assumption and may lead to an underestimate of the LD-effects in 
the low-$s$ region. 
However, as the present phenomenology is not equivocal on this
issue, any other choice at this stage would have been on a similar footing.

The resulting dilepton mass spectrum and
the FB asymmetry are shown in Fig.~\ref{fig:lddb} (a) and
Fig.~\ref{fig:lddb} (b), respectively. 
We recall that the two curves labeled SD and SD$+$LD 
include explicit $O(\alpha_s)$-improvement, calculated in the parton model
\cite{misiakE,burasmuenz} and non-perturbative effects related with the bound 
state nature of the
$B$-hadrons and the physical threshold in the final state in \bxsll,
using the Fermi motion model.
In addition, the SD$+$LD case also includes the LD-effects due to the 
vector resonances, contributing to $C_9^{\mbox{eff}}$ as discussed earlier.
The parametric dependence due to the FM 
is shown in Figs.~\ref{fig:dbrnsm} for the dilepton
mass spectrum (a) and the FB asymmetry (b), respectively, and compared with the
case of the parton model in which case no wave function effects are 
included. 
These figures give a fair estimate of the kind of uncertainties present in 
these distributions from non-perturbative
effects. In particular, we draw attention to the marked dependence of the FB 
asymmetry to both the LD-(resonances) and wave function effects,
which is particularly noticeable in the region $s > m_{\psi ^\prime}^2$. 
The dilepton invariant mass spectrum, on the other hand, is 
very stable except at the very end of the spectrum, which is clearly
different in all three cases shown.
This closes the first part of this chapter.

\subsubsection{Bridge}

Concerning our approach to include resonant charm effects 
eq.~(\ref{eq:yprime}), we compare it in section \ref{sec:LD2} with 
two other LD prescriptions, given in ref.~\cite{KS96} and \cite{LSW97} and 
estimate the resulting uncertainties
in the dilepton mass spectrum and the FB asymmetry. 
Further, we discuss a possible double counting, 
inherent in our procedure adding SD and LD amplitudes.
The determination of these uncertainties is very important, 
as a measurement of the partly integrated spectra
$\triangle \cb,\triangle \ca$  in \bxsll decay
will be used to extract the SD coefficients, testing the SM.

%%%%%%%%%%%%%%%second part %%%%%%%%%%%%%%%%%%%%%%%%%%
\section{Introduction to Hadron Spectra 
and Spectral Moments in the Decay \bxsll \label{sec:hadronintro}}

In this second part of this chapter we present spectra in inclusive
\bxsll decay in kinematical variables different from the dilepton invariant 
mass $q^2$,  
the hadronic energy and the hadronic invariant mass.
Further, we calculate lowest moments in these hadronic variables.
We include perturbative 
${\cal{O}}(\alpha_s)$ corrections, $1/m_b$ power corrections by means of
the heavy quark expansion technique (HQET) and studies in the Fermi motion 
model (FM).
This part is based on refs.~\cite{AH98-1,AH98-2,AH98-3}.

A similar program of investigations  
%\cite{FLS,greubrey,FLW,bargerkim90,DU97}
\cite{FLS,greubrey}, \citer{FLW,DU97} 
has been run for the charged current induced semileptonic 
$B \to X_{u,c} \ell \nu_\ell$ decay.
Here the  main interest is focused on testing HQET and on the determination  
of the CKM matrix elements $V_{cb}$ and $V_{ub}$.
To be more specific, the HQE parameters $\lambda_1$
and $\bar{\Lambda}$ have been extracted from moments of the hadronic 
invariant mass spectrum
in $B \to X_{u,c} \ell \nu_\ell$ decay \cite{FLSphenom}.
We recall that these non-perturbative parameters appear in the relation
$m_B=m_b+\bar{\Lambda}-(\lambda_1+3 \lambda_2)/2 m_b$ between 
the mass of the $B$-meson to the $b$-quark mass (see section \ref{sec:hqet}).
Explicit calculation \cite{AH98-1,AH98-2} shows that also in \bxsll decay
the hadronic invariant mass moments are sensitive to the HQET parameters
$\lambda_1$ and $\bar{\Lambda}$.
This provides potentially an independent determination of these quantities.
We think that the hadron spectra in \bxsll and $B \to X_u \ell \nu_\ell$
can be related to each other over limited phase space and this could help
in improving the present precision on  $V_{ub}$
\cite{PDG} and the parameters
$\lambda_1$ and $\bar{\Lambda}$ \cite{gremm,neubert98}.
Of course, $B \to X_{u,c} \ell \nu_\ell$ decays involve much less problems 
than FCNC \bxsll decay, as the charged current mode has simpler short-distance
(SD) couplings and no $c \bar{c}$ resonances present in the spectra.
Besides these obvious differences, we will point out 
in the following sections similarities between  
rare \bxsll and the charged current $B \to X_{u,c} \ell \nu_\ell$ decays.

What can we learn from the study of hadron spectra and moments in \bxsll ?
Our motivation is manifold:
\begin{itemize}
%\begin{enumerate}
\item Hadron spectra have an interest on their own, they complete the 
profile of \bxsll decay which has been given in the previous sections,
i. e. the dilepton invariant mass distribution and the FB asymmetry.
\item In their search for \bxsll the CLEO collaboration \cite{cleobsll97} 
imposed a cut on the 
hadronic invariant mass $S_H$ to suppress the $B \bar{B}$ background in 
measuring the dilepton invariant mass distribution.
The hadronic invariant mass spectrum is absolutely necessary to acquire 
control over the signal after a cut in $S_H$.
\item A possible determination of non-perturbative HQE parameters 
$\lambda_1,\bar{\Lambda}$ from the first two moments of the hadronic 
invariant mass in  
\bxsll decay, complementing the constraint from the 
charged current $B \to X_c \ell \nu_\ell$ decay. The constraints from these
decays can be used to reduce the present dispersion on $\lambda_1$ and
$\bar{\Lambda}$.
\item Test of the Fermi motion model in \bxsll decay.
%\end{enumerate}
\end{itemize}
The power corrections
presented here in the hadron spectrum and
hadronic spectral moments in \bxsll are the first results in this decay.

\subsection{Hadron kinematics}

Besides the parton level kinematics already introduced in 
section \ref{sec:kin}, the corresponding kinematics at hadron level can 
be written as:
\begin{equation}
B (p_B) \to X_s (p_H)+\ell^+ (p_{+})+\ell^{-} (p_{-})~.
\end{equation}
The hadronic invariant mass is denoted by  $S_H \equiv p_H^2$ and 
$E_H$ denotes the hadron energy in the final state. 
The corresponding quantities at parton level are the 
invariant mass $s_0$ and the scaled parton energy $x_0\equiv \frac{E_0}{m_b}$.
In parton model without gluon bremsstrahlung, this simplifies to
$s_0=m_s^2$ and $x_0$ becomes directly related to the dilepton invariant mass 
$x_0=1/2(1-\s +\ms^2)$.
{}From momentum conservation the following equalities hold in the $b$-quark,
equivalently $B$-meson, rest frame ($v=(1,0,0,0)$):
\begin{eqnarray}
x_0  &=& 1- v \cdot \q \, \, ,
~~~\s_0 = 1 -2 v \cdot \q + \s \, \, ,\label{eq:kin} \\
E_H &=& m_B-v \cdot q \, \, ,
~~~S_H = m_B^2 -2 m_B v \cdot q  + s \, \, .
\end{eqnarray}
The relations between the kinematic variables of the parton model and the 
hadronic states , using the
HQET mass relation, can be written as
\begin{eqnarray}
  E_H&=&\bar\Lambda-{\lambda_1+3\lambda_2\over2m_B}+\left(m_B-\bar
  \Lambda+
  {\lambda_1+3\lambda_2\over2m_B}\right) x_0+\dots\,, \nonumber\\
  S_H&=&m_s^2+\bar\Lambda^2+(m_B^2-2\bar\Lambda m_B+\bar\Lambda^2
  +\lambda_1+3\lambda_2)\,(\hat s_0-\hat m_s^2) \nonumber \\
  &&\qquad\qquad\mbox{}+(2\bar\Lambda
m_B-2 \bar\Lambda^2-\lambda_1-3\lambda_2) x_0
  +\dots\,,
\label{ehe0shs0}
\end{eqnarray}
where the ellipses denote terms higher order in
$1/m_b$.

\section{Perturbative $O(\alpha_s)$ Corrected Hadron Spectra in \bxsll Decay
\label{sec:pertqcd}}

In this section the $O(\alpha_s)$ corrections to the hadron spectra 
are investigated. 
Following the argument given in section \ref{sec:NLO},
only $O_9$ is subject to $\alpha_s$ corrections and the 
corresponding Feynman diagrams
can be seen in Fig.~\ref{fig:o9}.
The effect of a finite $s$-quark mass on the 
${\cal O}(\alpha_s)$ correction function is found to be very 
small. After showing this, we 
have neglected the $s$-quark mass in the numerical calculations of the
${\cal{O}}(\alpha_s)$ terms.

\subsection{Hadron energy spectrum}

\begin{figure}[htb]
\vskip -0.2truein
\centerline{\epsfysize=3.5in
{\epsffile{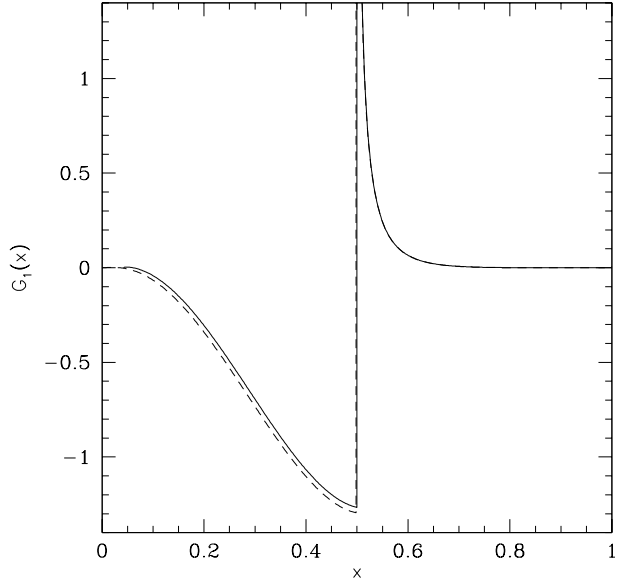}}}
\vskip -0.4truein
\caption[]{ \it The function $G_1(x)$ is shown for 
$m_s=0.2 \, {\mbox{\rm GeV}}$ (solid line)
and for the massless case corresponding to eq.~(\ref{eq:g1}) (dashed line).}
\label{fig:g1}
\end{figure}

The explicit order $\alpha_s$ correction to $O_9$ can be obtained
by using the existing results in the literature as follows:
The vector current $O_9$ can be decomposed as
$V=(V-A)/2 + (V+A)/2$.
We recall that the $(V-A)$ and $(V+A)$ currents yield the same  
hadron energy spectrum \cite{aliold}
and there is no interference term  present in this spectrum for massless 
leptons. So, the correction for the vector current case in \bxsll  
can be taken from the corresponding result for the charged $(V-A)$ case
\cite{aliqcd,jezkuhn}, yielding
\begin{eqnarray}
C_9^{{\mbox{eff}}}(x_0)=C_9 \rho(x_0) + Y(x_0)
\label{c9eff}
\end{eqnarray}
with
\begin{eqnarray}
\rho(x)&=& 1+ \frac{\alpha_s}{\pi} \sigma(x) \; , \\
\sigma(x)&=&
\frac{1}{(3 x -4 x^2-2 \ms^2+3 \ms^2 x)} \frac{G_1(x)}{3 \sqrt{x^2-\ms^2}} \; ,
%&=&\frac{1}{( 3 - 4 x) x^2} 
%\frac{G_1(x)}{3} \; ,
\end{eqnarray}
where
$Y(x_0)\equiv Y(\hat{s})$ with $\hat{s}=1-2 x_0 +\ms^2$.
The expression for $G_1(x)$ with  $m_s\neq 0$ has been calculated in 
\cite{jezkuhn}.
The effect of a finite $m_s$ is negligible in $G_1(x)$,
as can be seen in Fig.~\ref{fig:g1}, where this function
is plotted both with a 
finite $s$-quark mass, $m_s=0.2$ GeV, and for the  massless case, $m_s=0$.
A numerical difference occurs at the lowest order
end-point $x_0^{max}=1/2 (1+\ms^2)$ (for $m_l=0$), where the function 
develops a 
singularity from above ($x_0 >x_0^{max}$) and the position of which 
depends on the value of $m_s$.        
The function $G_1(x)$ for a massless $s$-quark  is given and 
discussed below \cite{jezkuhn}.
 \begin{eqnarray}
G_1(x)&=& x^2 \{ \frac{1}{90} (16 x^4 -84 x^3 +585 x^2-1860 x+1215) +
(8 x-9) \ln(2 x) \nonumber \\
&+& 2 (4 x-3) \left[ \frac{\pi^2}{2} +Li_2(1-2 x) \right] \} \, \, \, \, 
{\mbox{for}} \, \,  0 \leq x \leq 1/2 \nonumber \, \, , \\
G_1(x)&=&\frac{1}{180} (1-x)(32 x^5-136 x^4+1034 x^3-2946 x^2+1899 x+312) 
\nonumber \\
&-&\frac{1}{24} \ln(2 x-1) ( 64 x^3-48 x^2-24 x-5) \nonumber \\
&+& x^2 (3 -4 x) \left[ \frac{\pi^2}{3}-4 Li_2(\frac{1}{2 x}) +\ln^2(2 x-1)  -2
\ln^2(2 x) \right] \, \, \, {\mbox{for}} \, \, 1/2 < x \leq 1 \; . 
\label{eq:g1}
\end{eqnarray}

The ${\cal{O}}(\alpha_s)$ correction has a double logarithmic 
(integrable) singularity for $x_0 \to 1/2$ from above ($x_0 >1/2$).
Further, the value of the order $\alpha_s$ corrected Wilson coefficient
$C_9^{{\mbox{eff}}}(x_0)$ is reduced compared to its value 
with  $\alpha_s=0$,
therefore also the hadron energy spectrum is 
reduced after including the explicit order $\alpha_s$ QCD correction
for $0 < x_0 <1/2$.
Note that the hadron energy spectrum for \bxsll  
receives contributions for $1 \geq x > 1/2 $ only from the order 
$\alpha_s$ bremsstrahlung corrections.

\subsection{Hadronic invariant mass spectrum\label{sec:qcdmass}}

\begin{figure}[htb]
\vskip -0.0truein
\centerline{\epsfysize=3.5in
%{\epsffile{pmsudmass0.ps}}}
{\epsffile{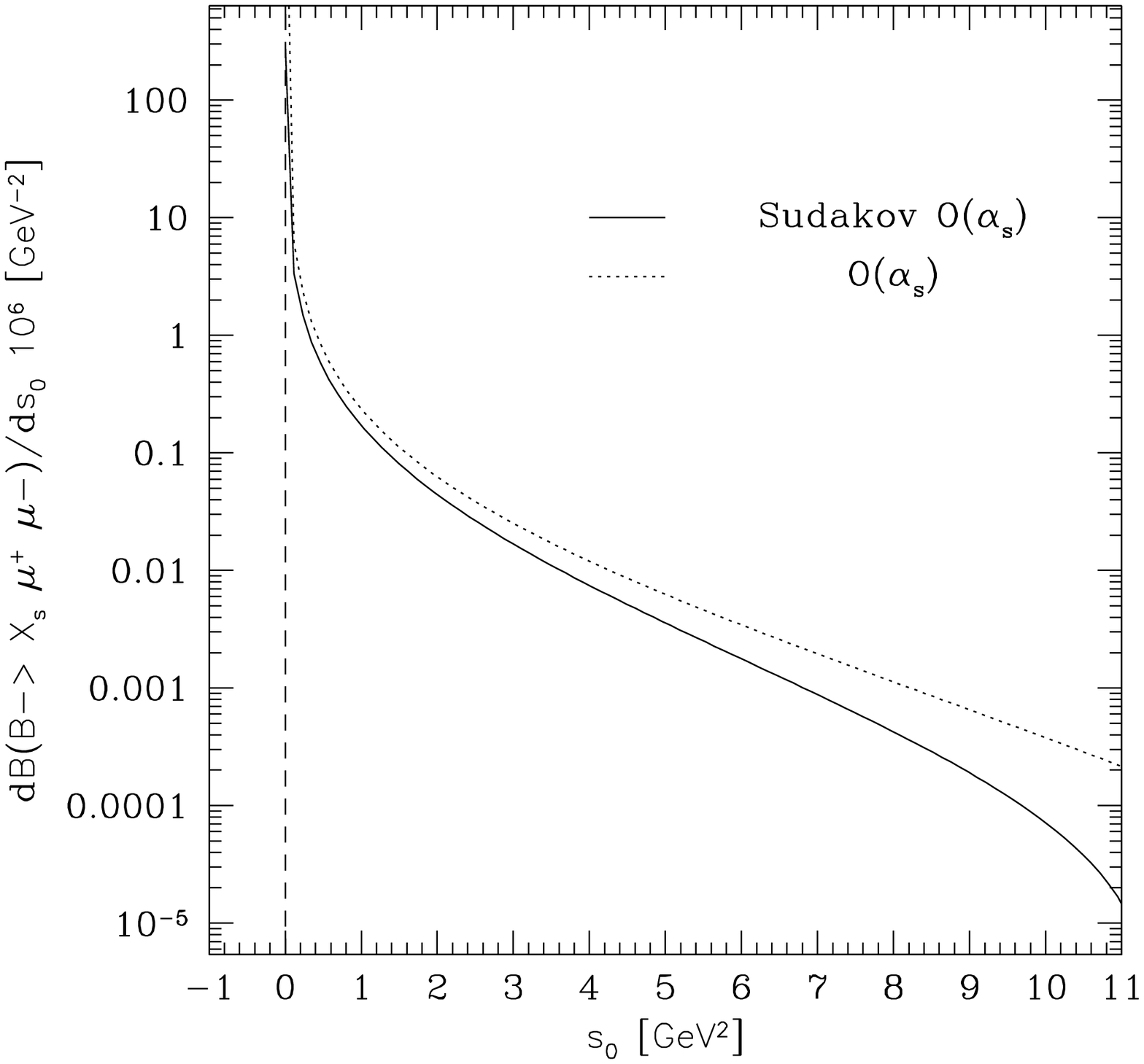}}}
\vskip -0.0truein
\caption[]{ \it The differential branching ratio 
$ \frac{{\rm d}{\cal B}(B \to X_s \ell^+ \ell^-)}{{\rm d} s_0}$ 
in the parton model 
is shown in the ${\cal{O}}(\alpha_s)$ bremsstrahlung region.
The dotted (solid) line corresponds to eq.~(\ref{dbds0}), (eq.~(\ref{eq:sud})).
The vertical line denotes the one particle pole from $b \to s \ell^+ 
\ell^-$. 
We do not show the full spectra in the range $0 \leq s_0 \leq m_b^2$ as 
they tend to zero for larger values of $s_0$.}
\label{fig:sh0}
\end{figure}
\begin{figure}[htb]
\vskip -0.0truein
\centerline{\epsfysize=3.5in
{\epsffile{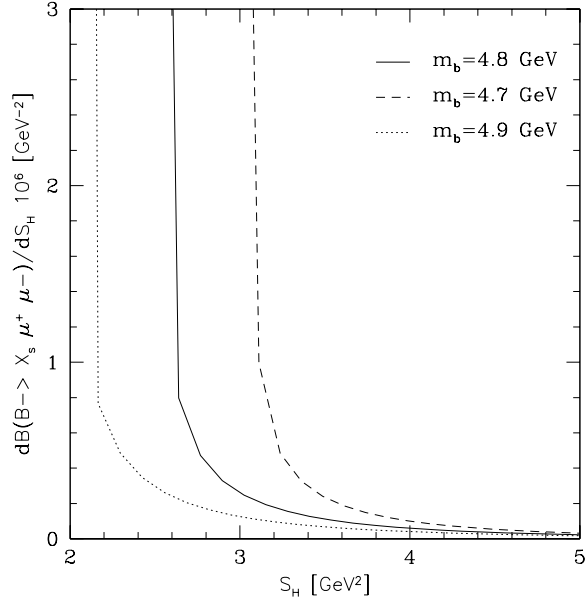}}}
\vskip -0.0truein
\caption[]{ \it The differential branching ratio 
$ \frac{{\rm d}{\cal B}(B \to X_s \ell^+ \ell^-)}{{\rm d} S_H}$  
in the hadronic invariant mass, $S_H$, shown for different values of 
$m_b$ in the range where only bremsstrahlung diagrams 
contribute. We do not show the result in the full kinematic range 
as the spectra tend monotonically to zero for larger values of $S_H 
\leq m_B^2$.} \label{fig:sH}
\end{figure}
We have calculated the  order $\alpha_s$  perturbative QCD correction
for the hadronic invariant mass in the range
$\ms^2 < \s_0 \leq 1 $. Since the decay $b \to s + \ell^+ + \ell^-$
contributes in the parton model only at $\s_0 =\ms^2$,  only 
the bremsstrahlung graphs $b \to s + g + \ell^+ + \ell^-$ contribute in this 
range.
This makes the calculation much simpler than in the full $\s_0$ range 
including virtual gluon diagrams. 
%After integrating $\sqrt{\s_0} \leq x_0 \leq \frac{1+\s_0}{2}$
We find 
\begin{eqnarray}
\frac{{\rm d}{\cal{B}}}{{\rm d } \s_0}=\frac{2}{3} {\cal{B}}_0
{\alpha_s\over \pi}{1\over\s_0}
  \{ \frac{(\s_0-1)}{27} (93-41\hat s_0-95 \hat s_0^2+55 \s_0^3)
+ {4\over 9}\ln\hat s_0 (-3-5\hat s_0+9 \hat s_0^2-2 \hat s_0^4) 
\} C_9^2 \, \, .
\label{dbds0}
\end{eqnarray}
Our result for the spectrum in \bxsll is in agreement with the corresponding 
result for the $(V-A)$ 
current obtained for the decay $B \to X_q \ell \nu_{\ell}$ in the $m_q=0$ 
limit in \cite{FLS} (their eq.~(3.8)), once one takes into account 
the difference in the  normalizations.
We display the hadronic invariant mass distribution  in Fig.~\ref{fig:sh0} 
as a function of $s_0$ (with 
$s_0=m_b^2 \s_0$), where we also show the Sudakov improved spectrum,
obtained from the ${\cal{O}}(\alpha_s)$ spectrum 
in which the double logarithms have been resummed. For the decay 
$B \to X_u \ell \nu_{\ell}$, this has been derived in \cite{greubrey}, 
where all further details can be seen.
We confirm eq.~(17) of \cite{greubrey} for the Sudakov exponentiated 
double differential 
decay rate $\frac{{\rm d^2}{\Gamma}}{{\rm d } x {\rm d } y}$  
and use it after changing the normalization 
$\Gamma_0 \to {\cal{B}}_0 \frac{2}{3} C_9^2$ for the decay
\bxsll . The constant ${\cal{B}}_0$ is given in eq.~(\ref{eqn:seminorm}). 
Defining the kinematic variables $(x,y)$ as
\begin{eqnarray}
q^2 &=& x^2 m_b^2 \, , \nonumber \\
v \cdot q &=& (x+\frac{1}{2} (1-x)^2 y) m_b \, ,
\end{eqnarray}
the Sudakov-improved Dalitz distribution  is given by
\begin{eqnarray}
\label{doubleexpon}
{d^2 {\cal{B}} \over d x d y}(B \to X_s \ell^+ \ell^-) &=& - {\cal{B}}_0 
 \frac{8}{3} x (1 - x^2)^2 (1 + 2 x^2) \,
\exp\Big( - {2 \alpha_s \over 3 \pi} \ln^2 (1 - y) \Big)
  \\
&\times &  \left\{
{4 \alpha_s \over 3 \pi} {\ln(1-y) \over (1-y)}
\Big[ 1 - {2 \alpha_s \over 3 \pi} \big( G(x) + H(y) \big) \Big]
-{2 \alpha_s \over 3 \pi} {d H \over d y}(y) \right\}  C_9^2
 \, ,\nonumber
\end{eqnarray}
where \cite{greubrey}
\begin{eqnarray}
G(x) &=&
\frac{[8x^2(1-x^2 -2x^4) \ln x
+ 2(1-x^2)^2 (5 + 4 x^2) \ln(1-x^2) -
(1-x^2)(5 + 9x^2 - 6 x^4) ]}{2 (1-x^2)^2 (1 + 2x^2)}
\nonumber \\
&& + \pi^2 + 2 Li_2 (x^2) - 2 Li_2 (1-x^2) \quad \, , \\
H(y)& = & \int_0^{y} dz \Big( {4 \over 1 - z}
\ln{2 - z(1-x) + \kappa
\over 2}          \nonumber \\
&&- {(1-x)(3 + x + xz - z) \over (1+x)^2}
\Big[\ln(1-z) - 2 \ln{2 - z(1-x) + \kappa \over 2} \Big]
\nonumber \\
&&- { \kappa \over 2 (1 + x)^2 (1 + 2x^2) }
\Big[{7 (1+x) (1 + 2 x^2) \over 1 - z} + (1-x)(3 - 2 x^2) \Big]\Big) \, .
\label{h0}
\end{eqnarray}
The quantity $\kappa$ in eq. (\ref{h0}) is defined as
$\kappa \equiv \sqrt {z^2 (1-x)^2 + 4 xz}$.

To get the hadronic invariant mass spectrum for a 
$b$-quark decaying at rest we change variables from $(x,y)$
to $(q^2,s_0)$ 
% using $s_0=m_b^2-2 m_b v \cdot q +q^2$ 
followed by an integration over $q^2$,
\begin{eqnarray}
\frac{{\rm d}{\cal{B}}}{{\rm d } s_0}=\int_{4 m_l^2}^{(m_b-\sqrt{s_0})^2}
{\rm d} q^2 \frac{{\rm d^2}{\cal{B}}}{{\rm d } x {\rm d } y} 
\frac{1}{2 m_b^4 x (1-x)^2} \;  .
\label{eq:sud}
\end{eqnarray}

The most significant effect of the bound state is the difference between
$m_B$ and $m_b$,  
which is dominated by $\bar{\Lambda}$. 
Neglecting $\lambda_1, \, \lambda_2$, i.e., using
$\bar{\Lambda}= m_B-m_b$, 
the spectrum $\frac{{\rm d}{\cal{B}}}{{\rm d } S_H}$
is obtained along the lines as given above for $\frac{d{\cal{B}}}{ds_0}$, 
after changing variables
from $(x,y)$ to $(q^2,S_H)$ and performing an integration over $q^2$.
It is valid in the region
$m_B \frac{m_B\bar{\Lambda}-\bar{\Lambda}^2+m_s^2}{m_B-\bar{\Lambda}}
< S_H \leq m_B^2 $ (or $m_B \bar{\Lambda} \leq S_H \leq m_B^2$, neglecting
$m_s$) which excludes the zeroth order and virtual gluon
 kinematics ($s_0=m_s^2$), yielding
\begin{eqnarray}
\frac{{\rm d}{\cal{B}}}{{\rm d } S_H}=\int_{4 m_l^2}^{(m_B-\sqrt{S_H})^2}
{\rm d} q^2 \frac{{\rm d^2}{\cal{B}}}{{\rm d } x {\rm d } y} 
\frac{1}{2 m_b^3 m_B x (1-x)^2} \; .
\label{eq:sudSH}
\end{eqnarray}
The hadronic invariant mass spectrum thus found depends rather
sensitively on $m_b$ (or equivalently $\bar{\Lambda}$),
as can be seen from Fig.~\ref{fig:sH}.
An analogous analysis for the charged current semileptonic $B$ decays
$B \to X_u \ell \nu_\ell$  has been performed in \cite{FLW},
with similar conclusions.

\section{Power Corrected Hadron Spectra in \bxsll Decay \label{sectionhqet}}

We start directly with the structure functions $T_i$ of the hadronic tensor, 
which are calculated up to $O(m_B/m_b^3)$ in \cite{AHHM97}.
They have been decomposed into a sum of
various terms,
$T_{i}(v.\hat{q},\s) = \sum_{j=0,1,2,s,g,\delta} T_{i}^{(j)}(v.\hat{q}, \s)$
where the individual parts $T_{i}^{(j)}$ can be seen in appendix \ref{app:Ti}.
After contracting the hadronic and leptonic tensors, one arrives at
eq.~(\ref{eq:tlr}). 
Now we are interested in a different set of kinematical variables.
We transform $(\z, \s) \to (x_0,\s_0)$
with the help of the kinematic identities given in eq.~(\ref{eq:kin}),
and make the dependence on $x_0$ and $\s_0$ explicit,
\begin{equation}
        {T^{L/R}}_{\mu \nu} \, {L^{L/R}}^{\mu \nu}\! = \! 
                {m_b}^2 \left\{ 2  (1-2 x_0+\s_0)  {T_1}^{L/R} 
                + \left[ x_0^2 - \frac{1}{4} \u^2 - \s_0 \right] {T_2}^{L/R} 
                \mp  (1-2 x_0+\s_0) \u \, {T_3}^{L/R} \right\} 
        \label{eqn:tlr2} \, .
\end{equation}
With this we are able to derive the double differential power corrected 
spectrum $\frac{{\rm d} {\cal{B}}}{{\rm d} x_0 \, {\rm d}\s_0}$ for
the decay \bxsll. 
Integrating eq.~(\ref{eqn:dgds}) over $\u$ first, where 
the variable $\u$ is bounded by 
\begin{eqnarray}
-2 \sqrt{x_0^2-\s_0} \leq \u \leq +2\sqrt{x_0^2-\s_0} \, \, ,
\end{eqnarray}
we arrive at the following expression \cite{AH98-2}
\begin{eqnarray}
\frac{{\rm d}^2 {\cal{B}}}{{\rm d} x_0 \, {\rm d}\s_0} = \! -\frac{8}{\pi} 
{\cal B}_{0}
{\mbox{Im}}\sqrt{x_0^2-\s_0}
\left\{ (1-2 x_0+\s_0)T_1(\s_0,x_0)+\frac{x_0^2-\s_0}{3}T_2(\s_0,x_0) \right\}
+ {\cal{O}}(\lambda_i \alpha_s)
\label{doublediff} \, ,
\end{eqnarray}
where 
\begin{eqnarray}
T_1(\s_0,x_0) \! \! &=& \! \! \frac{1}{x} \left\{ 
\left( 8 x_0-4  (\frac{\lo}{3}+\lt) \right) \left( |C_9^{\mbox{eff}}(\s)|^2 
+ |C_{10}|^2 \right)
\right. \nonumber \\
&+& \! \!
 \left(
32 (-2 \ms^2-2 \s_0-4 \ms^2 \s_0 +x_0+5 \ms^2 x_0+\s_0 x_0+\ms^2 \s_0 x_0) 
+ 16 (\frac{\lo}{3}+\lt) \right. \nonumber \\
&\times& \left. (-5-11 \ms^2+5 \s_0-\ms^2 \s_0+10 x_0 + 22 \ms^2 x_0-10 
x_0^2-10 \ms^2 x_0^2) \right) \frac{|C_7^{\mbox{eff}}|^2}{(\s_0-2 x_0 +1)^2} 
\nonumber \\
&+&  \! \! \left.
 \left( 
\frac{-32}{\s_0-2 x_0+1} (\ms^2+\s_0-x_0-\ms^2 x_0)-48(\frac{\lo}{3}+\lt) \right) 
Re(C_9^{\mbox{eff}}(\s)) \, C_7^{\mbox{eff}}
\right\} \nonumber \\
&+&  \! \! 
\frac{1}{x^2}  \left\{ 
 \left( \frac{8 \lo}{3} (-2 \s_0-3 x_0+5 x_0^2)+8 \lt (-2 \s_0+x_0+5 x_0^2)
\right)
\left( |C_9^{\mbox{eff}}(\s)|^2 + |C_{10}|^2 \right) \right.
 \nonumber \\
&+& \! \! \! 
\left( 
\frac{32 \lo}{3} (6 \ms^2+ 12 \s_0+ 18 \ms^2 \s_0-2 \s_0^2-2 \ms^2 \s_0^2-3 x_0-21 \ms^2 x_0-13 \s_0 x_0-19 \ms^2 \s_0 x_0 
\right.
\nonumber \\
&-&  \! \!
3 x_0^2+9 \ms^2 x_0^2+ 5 \s_0 x_0^2+ 5 \ms^2 \s_0 x_0^2+ 4 x_0^3+4 \ms^2 x_0^3)
\nonumber \\
&+&  \! \! 
32 \lt(-2 \ms^2-2 \ms^2 \s_0-2 \s_0^2-2 \ms^2 \s_0^2+x_0-\ms^2 x_0-5 \s_0 x_0-11 \ms^2 \s_0 x_0 + x_0^2 
\nonumber \\
&+&  \! \! \left. 13 \ms^2 x_0^2 + 5 \s_0 x_0^2+ 5 \ms^2 \s_0 x_0^2)\right)
\frac{|C_7^{\mbox{eff}}|^2}{(\s_0-2 x_0 +1)^2} 
 \nonumber \\
&+&  \! \!
\left( 
\frac{-32 \lo}{3} (-3 \ms^2-5 \s_0+2 \ms^2 \s_0+ 3 x_0+ 6 \ms^2 x_0+ 3 \s_0 x_0-x_0^2-5 \ms^2 x_0^2) \right.
\nonumber \\
&-& \! \! \left.\left.
32 \lt(\ms^2+\s_0+ 2 \ms^2 \s_0-x_0+2 \ms^2 x_0 + 3 \s_0 x_0 -3 x_0^2-5 \ms^2 x_0^2) \right)
\frac{Re(C_9^{\mbox{eff}}(\s)) \, C_7^{\mbox{eff}}}{\s_0-2 x_0+1}
\right\} \nonumber \\
&+&  \! \! \frac{1}{x^3}  \lo (\s_0-x_0^2)\left\{ 
 \frac{32 x_0}{3} \left( |C_9^{\mbox{eff}}(\s)|^2 + |C_{10}|^2 \right) \right.
\nonumber \\
&+&  \! \!
 \frac{128}{3} (-2 \ms^2 -2 \s_0-4 \ms^2 \s_0+x_0+5 \ms^2 x_0+ \s_0 x_0+\ms^2 \s_0 x_0) \frac{|C_7^{\mbox{eff}}|^2}{(\s_0-2 x_0 +1)^2}  \nonumber \\
&+&  \! \! \left.
\frac{-128}{3} (\ms^2+\s_0-x_0-\ms^2 x_0)
 \frac{Re(C_9^{\mbox{eff}}(\s)) \, C_7^{\mbox{eff}}}{\s_0-2 x_0+1}
\right\}  \, \, ,\nonumber \\
T_2(\s_0,x_0) \! \! &=& \! \! \frac{1}{x} \left\{ 
 \left( 16-40 (\frac{\lo}{3}+\lt) \right)
\left( |C_9^{\mbox{eff}}(\s)|^2 + |C_{10}|^2 \right) \right. \nonumber \\
&+& \!  \!\left.
 \left( -64+160(\frac{\lo}{3}+\lt) \right) (1+\ms^2) 
\frac{|C_7^{\mbox{eff}}|^2 }{\s_0-2 x_0+1}
\right\}\nonumber \\
&+& \! \! \frac{1}{x^2} \left\{ 
 \left( \frac{112 \lo}{3} (-1+x_0)+ 16 \lt (-3+5 x_0) \right)
\left( |C_9^{\mbox{eff}}(\s)|^2 + |C_{10}|^2 \right) \right.
 \nonumber \\
&+& \!  \! \! \left.
 \left( \frac{448 \lo}{3} (1-x_0)+ 64 \lt (5 x_0-1)\right) (1+\ms^2) 
\frac{|C_7^{\mbox{eff}}|^2 }{\s_0-2 x_0+1} 
 -64 \lt Re(C_9^{\mbox{eff}}(\s)) \, C_7^{\mbox{eff}}
\right\} \nonumber \\
&+& \!  \! \frac{1}{x^3} \lo (\s_0-x_0^2)\left\{ 
 \frac{64}{3} \left( |C_9^{\mbox{eff}}(\s)|^2 + |C_{10}|^2 \right)+
 \frac{-256}{3} (1+\ms^2) \frac{|C_7^{\mbox{eff}}|^2}{\s_0-2 x_0+1}
\right\} \; .
\end{eqnarray}
Here, $x=\s_0 -\ms^2 +i \epsilon$,
$\lo=\lambda_{1}/m_{b}^2$ and $\lt=\lambda_{2}/m_{b}^2$.
As the structure function $T_3$ does not contribute to the branching ratio, 
we did not consider it in the calculation of the hadron spectra.
The Wilson coefficient 
$C_9^{\mbox{eff}}(\s)$ depends both on the variables
$x_0$ and $\s_0$ arising from the matrix element of the four-Fermi-operators.
Here the normalization constant ${\cal B}_0$, defined in 
eq.~(\ref{eqn:seminorm}), expresses the branching ratio for \bxsll as usual 
in terms of the semileptonic decays $B \to X_c \ell \nu_\ell$.
The double differential ratio given in eq.~(\ref{doublediff}) agrees in the 
$(V-A)$ limit 
%(see appendix \ref{app:vminusa}) 
given in eqs.~(\ref{eq:c9CC}) - (\ref{eq:CKMCC})
with the corresponding expression derived for 
$B \to X_c \ell \nu_{\ell}$ decay in \cite{FLS} (their eq.~(3.2)). 

The hadron energy spectrum can now be obtained by integrating over $\s_0$.
Using eq.~(\ref{eq:imx}), 
the following replacements are equivalent to taking the imaginary part
\begin{eqnarray}
\frac{1}{x}   & \to & \delta (\s_0 -\ms^2) \, \, , \nonumber \\
\frac{1}{x^2} & \to & -\delta' (\s_0 -\ms^2) \, \, , \nonumber \\
\frac{1}{x^3} & \to & \frac{1}{2} \delta'' (\s_0 -\ms^2) \, \, .
\end{eqnarray}
The kinematic boundaries are given as:
\begin{eqnarray}
max(\ms^2,-1+2 x_0 +4 \ml^2) \leq &\s_0& \leq x_0^2 \, \, ,\nonumber \\
\ms \leq  & x_0 &  \leq \frac{1}{2} (1+\ms^2-4 \ml^2) \, \, .
\end{eqnarray}
Here we keep $\ml$ as a regulator wherever it is necessary and abbreviate
$C_9^{\mbox{eff}}\equiv C_9^{\mbox{eff}}(\s=1-2 x_0+\ms^2)$.
Including the leading power corrections, the  
hadron energy spectrum in the decay  \bxsll is given below:
\begin{eqnarray}
        \frac{{\rm d}{\cal B}}{{\rm d} x_0} & = & \; {\cal B}_0
 \left\{ 
     \left[
g_0^{(9,10)} + \lo g_1^{(9,10)} +\lt g_2^{(9,10)}
\right]
 \left( |C_9^{\mbox{eff}}|^2 + |C_{10}|^2 \right) \right.  \nonumber \\
&+&\left[
g_0^{(7)} + \lo g_1^{(7)} +\lt g_2^{(7)}
\right]
 \frac{|C_7^{\mbox{eff}}|^2}{x_0-\frac{1}{2}(1+\ms^2)} 
+  \left[
g_0^{(7,9)} + \lo g_1^{(7,9)} +\lt g_2^{(7,9)}
\right]
 Re(C_9^{\mbox{eff}}) \, C_7^{\mbox{eff}}  \nonumber \\
&+&  (\lo h_1^{(9)}+  \lt h_2^{(9)})
\frac{d |C_9^{\mbox{eff}}|^2}{d\s_0}
+  \lo k_1^{(9)}
\frac{d^2 |C_9^{\mbox{eff}}|^2}{d\s_0^2} \nonumber \\
&+&  ( \lo h_1^{(7,9)} + \lt  h_2^{(7,9)} )
\frac{d Re(C_9^{\mbox{eff}})}{d\s_0}  \, C_7^{\mbox{eff}}  
+  \left.  \lo  k_1^{(7,9)}
\frac{d^2 Re(C_9^{\mbox{eff}})}{d\s_0^2}  \, C_7^{\mbox{eff}} 
 \right\} 
 \nonumber \\
&+& \delta(x_0-\frac{1}{2}(1+\ms^2-4 \ml^2)) f_{\delta}(\lo,\lt)
+\delta'(x_0-\frac{1}{2}(1+\ms^2-4 \ml^2)) f_{\delta'}(\lo,\lt) \, \, .
\label{singlediff}
\end{eqnarray}
The functions 
$g_i^{(9,10)},g_i^{(7)},g_i^{(7,9)},h_i^{(9)},h_i^{(7,9)},
k_1^{(9)},k_1^{(7,9)}$  in the above expression are the coefficients of the 
$1/m_b^2$ 
power expansion for different combinations of Wilson coefficients, with
$g_0^{(j,k)}$ being the lowest order (parton model) functions.
They are functions of the variables $x_0$ and $\ms$ and are given 
in appendix \ref{app:auxfunc1}.
The singular functions $ \delta, \delta'$ have support only at the 
lowest order
end-point of the spectrum, i.e., at $x_0^{max} \equiv \frac{1}{2}(1+\ms^2-4 
\ml^2)$.
The auxiliary functions $f_{\delta}(\lo,\lt)$ and $f_{\delta'}(\lo,\lt)$ 
vanish in the limit $\lo=\lt=0$.
They are given in appendix {\ref{app:auxfunc}}.
The derivatives of $C_9^{\mbox{eff}}$ are defined as 
$ \frac{d^{n} C_9^{\mbox{eff}}}{d\s_0^n} \equiv 
\frac{d^{n} C_9^{\mbox{eff}}}{d\s^n}(\s=1-2 x_0+\s_0; \s_0=\ms^2)$ $(n=1,2)$.
In the $(V-A)$ limit our eq.~(\ref{singlediff}) for the
hadron energy spectrum in \bxsll agrees with the
corresponding spectrum in $B \to X \ell \nu_\ell$ given in ref.~\cite{FLS}
(their eq.~(A1)). Integrating also over $x_0$ the resulting total width
for \bxsll agrees again in the $(V-A)$
limit with the well known result \cite{georgi}.

\begin{figure}[htb]
\vskip -0.0truein
\centerline{\epsfysize=3.5in
{\epsffile{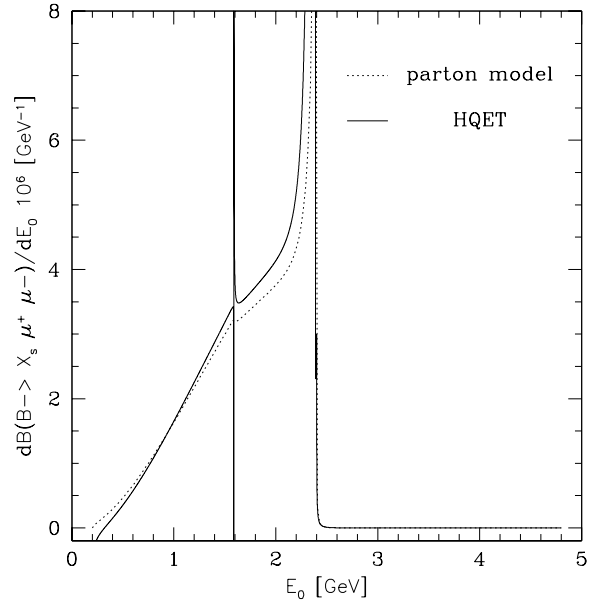}}}
\vskip -0.0truein
\caption[]{ \it Hadron energy spectrum
$ \frac{{\rm d}{\cal B}(B \to X_s \ell^+ \ell^-)}{{\rm d} E_0}$
in the parton model (dotted line) and including leading power corrections 
(solid line). For $m_b/2 < E_0 \leq m_b$ the distributions coincide. 
The parameters used for this plot are the central values given in 
Table~\ref{parameters} and the default values of the HQET parameters
specified in text. }
\label{fig:dbdx0}
\end{figure}
The power-corrected hadron energy spectrum 
$ \frac{{\rm d}{\cal B}(B \to X_s \ell^+ \ell^-)}{{\rm d} E_0}$ 
(with $E_0=m_b x_0$)
is displayed in Fig.~{\ref{fig:dbdx0}} through the solid curve, however, 
without the singular $\delta, \delta^{\prime}$ terms.
Note that before reaching the kinematic lower end-point, the 
power-corrected spectrum  becomes negative, as a result of the $\lt$ 
term. This behavior is analogous to what has already been reported for the 
dilepton mass spectrum 
$ \frac{{\rm d}{\cal B}(B \to X_s \ell^+ \ell^-)}{{\rm d} q^2}$
in the high $q^2$ region \cite{AHHM97},
%and for the lepton energy spectrum in semileptonic B decays \cite{MW}
signaling a breakdown of the $\frac{1}{m_b}$ expansion in this region.
The terms with the derivatives of $C_9^{\mbox{eff}}$ in 
eq.~(\ref{singlediff}) give rise to a 
singularity in the hadron energy spectrum at the charm threshold due to 
the cusp in the function $Y(\s)$, when approached from either 
side. The hadron energy spectrum for the parton model is also 
shown in Fig.~{\ref{fig:dbdx0}}, which is finite for all ranges of $E_0$.

What is the region of validity of the hadron energy spectrum derived
in HQET?
It is known that in \bxsll decay there are resonances present,
from which the known six \cite{PDG} populate the $x_0$ (or $E_0$) range 
between the lower end-point and the charm threshold. Taking this into
account and what has been remarked earlier, 
one concludes that the HQET spectrum cannot be used near the resonances, 
near the charm threshold and around the lower endpoint. Excluding these
regions, the spectrum calculated in HQET is close to the partonic 
perturbative spectrum as the power corrections are shown to be small.  
The authors of \cite{buchallaisidorirey},
who have performed an $1/m_c$ expansion for the dilepton mass spectrum 
$\frac{{\rm d}{\cal B}(B \to X_s \ell^+ \ell^-)}{{\rm d} q^2}$ 
and who 
also found a charm-threshold singularity, expect a reliable prediction of 
the spectrum for $q^2 \leq 3 m_c^2$ corresponding to
$E_0 \geq \frac{m_b}{2} (1+\ms^2-3 \hat m_c^2) \approx 1.8$ GeV. In this
region, the effect of the $1/m_b$ power corrections on the energy spectrum
is small and various spectra in \bxsll calculated here and in 
\cite{AHHM97} can be compared with data. 

The leading power corrections to the invariant mass spectrum is found by 
integrating eq.~(\ref{doublediff}) with respect to $x_0$.
We have already discussed the non-trivial hadronic invariant mass spectrum
which results  from the ${\cal O}(\alpha_s)$ bremsstrahlung   
and its Sudakov-improved version.
Since we have consistently dropped everywhere terms of
${\cal O}(\lambda_i \alpha_s)$ (see eq.~(\ref{doublediff})), this is the
only contribution to the invariant mass spectrum also in HQET away from 
$\s_0=\ms^2$, as the result of 
integrating the terms involving power corrections in eq.~(\ref{doublediff}) 
over $x_0$ is a singular function with support only at $\s_0=\ms^2$.
Of course, these corrections contribute to the normalization (i.e., branching
ratio) but leave the perturbative spectrum intact for  $\s_0 \neq \ms^2$.

\section{Hadronic Moments in \bxsll~in HQET}

We start with the derivation of the lowest spectral moments
in the decay \bxsll  at the parton 
level. These moments are worked out by taking into account the leading 
power $1/m_b$ and the perturbative ${\cal{O}}(\alpha_s)$ corrections.
To that end, we define:
\begin{equation}
{\cal M}^{(n,m)}_{\ell^{+} \ell^{-}} \equiv 
   {1\over {\cal B}_0}\int (\hat s_0-\hat m_s^2)^n  x_0^m\,
   {{\rm d}^2 {\cal B}\over{\rm d}\hat s_0{\rm d} x_0}
   \,{\rm d}\hat s_0{\rm d} x_0\,,
\end{equation}
for integers $n$ and $m$.  These moments are related to the 
corresponding moments $\langle x_0^m(\hat s_0-\hat m_s^2)^n\rangle$  
obtained at the parton level 
by a scaling factor which yields the corrected branching
ratio ${\cal B}={\cal B}_0 {\cal M}_{\ell^+ \ell^-}^{(n,m)}$.
Thus, 
\begin{equation}
   \langle x_0^m(\hat s_0-\hat m_s^2)^n\rangle =
{{\cal B}_0\over {\cal B}}\,
   {\cal M}^{(n,m)}_{\ell^{+} \ell^{-}}\,.
\label{momdef}
\end{equation}
The correction factor ${\cal B}_0/{\cal B}$ is given below 
in eq.~(\ref{eq:bb0}).
We remind that one has to Taylor expand it in terms of the 
${\cal O}(\alpha_s)$ and power corrections. 
The moments can be expressed as double expansion in ${\cal{O}}(\alpha_s)$
and $1/m_b$ and to the accuracy of our calculations can be 
represented in the following form:
 \begin{eqnarray}
 {\cal M}^{(n,m)}_{\ell^{+} \ell^{-}}=D_0^{(n,m)}+
\frac{\alpha_s}{\pi} {C_9}^2 A^{(n,m)}+
\lo D_1^{(n,m)} + \lt D_2^{(n,m)} \,\, ,
\end{eqnarray}
with a further decomposition into pieces from different Wilson 
coefficients for $i=0,1,2$:
\begin{eqnarray}
\label{momentexp}
D_i^{(n,m)}=\alpha_i^{(n,m)} {C_7^{{\mbox{eff}}}}^2+
\beta_i^{(n,m)} C_{10}^2+
\gamma_i^{(n,m)} C_7^{{\mbox{eff}}} +\delta_i^{(n,m)}.
\end{eqnarray}
The terms $\gamma_i^{(n,m)}$ and $\delta_i^{(n,m)}$ in              
eq.~(\ref{momentexp}) result from the terms proportional 
to ${\it{Re}}(C_9^{{\mbox{eff}}})C_7^{{\mbox{eff}}}$ and
$|C_9^{{\mbox{eff}}}|^2$  in  eq.~(\ref{doublediff}), respectively.   
The results for   
$\alpha_i^{(n,m)},\beta_i^{(n,m)},  \gamma_i^{(n,m)}, \delta_i^{(n,m)}$
are presented in appendix {\ref{app:moments}}. Out of these, the 
functions $\alpha_i^{(n,m)}$ and $\beta_i^{(n,m)}$
are given analytically, but the other two  
$\gamma_i^{(n,m)}$ and $\delta_i^{(n,m)}$ are given in terms of a 
one-dimensional integral over $x_0$, as these latter functions
involve the coefficient $C_9^{\mbox{eff}}$, which is a complicated
 function of $x_0$.

The leading perturbative contributions for the hadronic invariant mass and 
hadron energy 
moments can be obtained analytically by integrating eq.~(\ref{dbds0}) and 
eq.~(\ref{eq:g1}), respectively, yielding
\begin{eqnarray}
A^{(0,0)}&=&\frac{25-4 \pi^2}{9} \, ,
~~A^{(1,0)}=\frac{91}{675} \, ,
~~A^{(2,0)}=\frac{5}{486} \; , \nonumber \\
A^{(0,1)}&=&\frac{1381-210 \pi^2}{1350} \, ,
A^{(0,2)}=\frac{2257-320 \pi^2}{5400} \; .
\label{eq:A10}
\end{eqnarray}
The zeroth moment $n=m=0$ is needed for the normalization and we recall 
that the result for $A^{(0,0)}$ was derived some time ago \cite{CM78}.
Likewise, the first mixed moment $A^{(1,1)}$ can be extracted from 
the results given  
in \cite{FLS} for the decay $B \to X \ell \nu_{\ell}$ after changing the 
normalization, 
\begin{eqnarray}
\label{eq:A11}
A^{(1,1)}&=&
%\frac{2}{3} {9\over100} =
\frac{3}{50} \, \, .
\end{eqnarray}
For the lowest order parton model contribution 
$D_0^{(n,m)}$, we find, in agreement 
with \cite{FLS}, that the first two hadronic invariant mass moments 
$\langle \s_0-\ms^2 \rangle, \, \langle(\s_0-\ms^2)^2 \rangle$ and the first 
mixed moment $\langle x_0 (\s_0-\ms^2) \rangle$ vanish:
\begin{eqnarray}
D_0^{(n,0)}=0 \, \, \,  \mbox{for} \, \, n=1,2 \, \, \, \mbox{and}\, \, \,
D_0^{(1,1)}=0 \, .
\end{eqnarray}
We remark that we have included the $s$-quark mass dependence  in the 
leading term 
and in the power corrections, but omitted it throughout our work in the 
calculation of the explicit $\alpha_s$ term. 
All the expressions derived here for the moments agree in the
$V-A$ limit (and with $\ms=0$ in the perturbative $\alpha_s$ correction 
term) with the corresponding expressions given in ref.~\cite{FLS}.
{}From here the full ${\cal{O}}(\alpha_s m_s)$ expressions can be inferred 
after
adjusting the normalization $\Gamma_0 \to {\cal{B}}_0 \frac{2}{3} C_9^2$.
We have checked that a finite $s$-quark mass effects the 
values of the $A^{(n,m)}$  given in eqs.~(\ref{eq:A10}-\ref{eq:A11})
by less than $8 \%$ for $m_s=0.2$ GeV.

We can eliminate the hidden dependence on the non-perturbative parameters 
resulting from the $b$-quark mass in the moments 
${\cal M}^{(n,m)}_{\ell^{+} \ell^{-}}$ with the help of the HQET mass relation.
As $m_s$ is of order $\Lambda_{QCD}$, to be consistent we keep only 
terms up to order $m_s^2/m_b^2$ \cite{FLSphenom}. An additional 
$m_b$-dependence is in the mass ratios $\ml=\frac{m_l}{m_b}$.
Substituting $m_b$ by the $B$-meson mass using the HQET relation introduces 
additional ${\cal{O}}(1/m_B,1/m_B^2)$ 
terms in the Taylor expansion of eq.~(\ref{momdef}).
We get for the following normalization factor for ${\cal B}/{\cal B}_0 =
{\cal M}_{\ell^+ \ell^-}^{(0,0)}$:
\begin{eqnarray}
{{\cal B}\over {\cal B}_0}&=& \frac{32}{9 m_B^2}(-4 m_B^2-13 m_s^2-
3 (m_B^2-2 m_s^2) \ln(4 \frac{m_l^2}{m_B^2})){C_7^{\mbox{eff}}}^2+
\frac{2}{3 m_B^2} (m_B^2-8 m_s^2)C_{10}^2 \nonumber \\
&+&
\int_{m_s/m_B}^{\frac{1}{2}(1+m_s^2/m_B^2)} d x_0 \frac{64}{m_B^2}
(-m_s^2-4 m_s^2 x_0+2 m_B^2 x_0^2+2 m_s^2 x_0^2)
{\it{Re}}(C_9^{{\mbox{eff}}})C_7^{{\mbox{eff}}} \nonumber \\
&+&
\int_{m_s/m_B}^{\frac{1}{2}(1+m_s^2/m_B^2)} d x_0
\frac{16}{3 m_B^2}(-3 m_s^2+6 m_B^2 x_0^2+6 m_s^2 x_0^2-8 m_B^2 x_0^3)
|C_9^{{\mbox{eff}}}|^2\nonumber \\
&+&\frac{\alpha_s}{\pi} A^{(0,0)} C_9^2+ 
\frac{-64}{3} {C_7^{\mbox{eff}}}^2 \frac{\bar{\Lambda}}{m_B}+
\frac{-32}{3}{C_7^{\mbox{eff}}}^2\frac{\bar{\Lambda}^2}{m_B^2}
+ \left[ \frac{16}{9} (2-3\ln(4 \frac{m_l^2}{m_B^2})){C_7^{\mbox{eff}}}^2+
 \frac{C_{10}^2}{3}  \right. \nonumber \\
&+& \left.
\int_{0}^{\frac{1}{2}} d x_0
(64 x_0^2{\it{Re}}(C_9^{{\mbox{eff}}})C_7^{{\mbox{eff}}}
+\frac{16}{3} (3-4 x_0)x_0^2 |C_9^{{\mbox{eff}}}|^2 ) \right] 
\frac{\lambda_1}{m_B^2}
\nonumber \\
&+& \label{eq:bb0}
 \left[ \frac{16}{3} (4+9\ln(4 \frac{m_l^2}{m_B^2})){C_7^{\mbox{eff}}}^2-
 3 C_{10}^2   \right.\\
&+& \left.
\int_{0}^{\frac{1}{2}} d x_0
(64 (-1-4 x_0+7 x_0^2){\it{Re}}(C_9^{{\mbox{eff}}})C_7^{{\mbox{eff}}}
+16 (-1+15 x_0^2-20 x_0^3)|C_9^{{\mbox{eff}}}|^2 ) \right] 
\frac{\lambda_2}{m_B^2}
\nonumber  \, .
\end{eqnarray}
Here, the $\frac{\bar{\Lambda}}{m_B}$ and $ \frac{\bar{\Lambda}^2}{m_B^2}$ 
terms proportional to ${C_7^{\mbox{eff}}}^2$ result from the expansion of $\ml$
\begin{eqnarray}
\ln(\frac{4 m_l^2}{m_b^2})=\ln(\frac{4 m_l^2}{m_B^2})+
2 \frac{\bar{\Lambda}}{m_B}+\frac{\bar{\Lambda}^2}{m_B^2} -\frac{\lambda_1+3 \lambda_2}{m_B^2}+\dots \; .
\end{eqnarray}
The first two moments and the first mixed moment,
$\langle x_0 \rangle {\cal B}/{\cal B}_0$, $\langle x_0^2 \rangle {\cal 
B}/{\cal B}_0$, $\langle \hat{s}_0 - \hat{m}_s^2 \rangle {\cal B}/{\cal 
B}_0$, $\langle (\hat{s}_0 - \hat{m}_s^2)^2 \rangle {\cal B}/{\cal B}_0$
and $\langle x_0 (\hat{s}_0 - \hat{m}_s^2) \rangle {\cal B}/{\cal
B}_0$  are presented in appendix {\ref{app:lowmoments}}. 

With this we obtain the moments for the 
physical quantities valid up to ${\cal{O}}(\alpha_s/m_B^2,1/m_B^3)$,
where the second equation corresponds to a further use of
$m_s={\cal{O}}(\Lambda_{QCD})$.
We get for the first two hadronic invariant mass moments
{\footnote{Our first expression for 
$\langle S_H^2\rangle$, eq.~(\ref{sHmoments}), 
does not agree in the coefficient of 
$\langle\hat s_0-\hat m_s^2\rangle $ with the one given in ref.~\cite{FLS} 
(their eq.~(4.1)). We point out that $m_B^2$ should have been replaced by 
$m_b^2$ in this expression.
This has been confirmed by Adam Falk (private communication).
Dropping the higher order terms given in their expressions, 
the hadronic moments in HQET derived here and in
 \cite{FLS} agree.}}
\begin{eqnarray}\label{sHmoments}
   \langle S_H\rangle&=&m_s^2+\bar\Lambda^2+(m_B^2-2\bar\Lambda m_B
)\,\langle\hat s_0-\hat m_s^2\rangle
+(2\bar\Lambda m_B-2\bar\Lambda^2-\lambda_1-3\lambda_2)
  \langle  x_0\rangle\,, \nonumber\\
  \langle S_H^2\rangle&=&
m_s^4+2\bar\Lambda^2 m_s^2+
  2 m_s^2 (m_B^2-2\bar\Lambda m_B)
  \langle\hat s_0-\hat m_s^2\rangle
+2m_s^2 (2\bar\Lambda m_B-2\bar\Lambda^2-\lambda_1-3\lambda_2)
\langle  x_0\rangle
  \nonumber\\
  &&\quad\mbox{}+ 
(m_B^4-4\bar\Lambda m_B^3
)\langle (\hat s_0-\hat m_s^2)^2\rangle
+ 4\bar\Lambda^2 m_B^2 \langle x_0^2\rangle+
  4\bar\Lambda m_B^3\langle x_0(\hat s_0-\hat m_s^2)\rangle\,,  \\
&=&
(m_B^4-4\bar\Lambda m_B^3)\langle (\hat s_0-\hat m_s^2)^2\rangle
+ 4\bar\Lambda^2 m_B^2 \langle x_0^2\rangle+
  4\bar\Lambda m_B^3\langle x_0(\hat s_0-\hat m_s^2)\rangle \,,
\nonumber
\end{eqnarray}
and for the hadron energy moments:
\begin{eqnarray}\label{EHmoments}
  \langle E_H\rangle &=& \bar\Lambda-{\lambda_1+3\lambda_2\over2m_B}
  +\left(m_B-\bar\Lambda+{\lambda_1+3\lambda_2\over2m_B}\right)\langle
   x_0\rangle\,,\nonumber\\
  \langle E_H^2\rangle &=& \bar\Lambda^2 + (2\bar\Lambda m_B -
2\bar\Lambda^2
  -\lambda_1-3\lambda_2)\langle  x_0\rangle\\
  &&\quad +(m_B^2-2\bar\Lambda m_B+\bar\Lambda^2+\lambda_1+3\lambda_2)
  \langle x_0^2\rangle\,.\nonumber
\end{eqnarray}

\subsection{Numerical estimates of the hadronic moments in HQET   
\label{numerics:hqet}}

Using the expressions for the HQET moments given in 
appendix \ref{app:lowmoments},
we  present the numerical results for the hadronic moments in \bxsll,
valid up to ${\cal{O}}(\alpha_s/m_B^2,1/m_B^3)$.
We find: 
\begin{eqnarray}
\langle x_0\rangle \!  \!  \! & \!=\! & \! \!  \! 0.367 \, (1+0.148 \frac{\alpha_s}{\pi} 
-0.204 \frac{\bar{\Lambda}}{m_B} \frac{\alpha_s}{\pi}
-0.030 \frac{\bar{\Lambda}}{m_B}-0.017\frac{\bar{\Lambda}^2}{m_B^2}
+ 0.884 \frac{\lambda_1}{m_B^2}+3.652\frac{\lambda_2}{m_B^2}) \nonumber \, ,\\
\langle x_0^2\rangle \!  \!  \! & \! = \! & \! \!  \! 0.147 \, (1+0.324 \frac{\alpha_s}{\pi}
-0.221\frac{\bar{\Lambda}}{m_B}  \frac{\alpha_s}{\pi}
-0.058\frac{\bar{\Lambda}}{m_B}-0.034 \frac{\bar{\Lambda}^2}{m_B^2}
+ 1.206 \frac{\lambda_1}{m_B^2}+4.680\frac{\lambda_2}{m_B^2})\nonumber \, ,\\
\langle x_0(\s_0-\ms^2)\rangle \!  \!  \! & \! = \! & \! \!  \! 0.041 \frac{\alpha_s}{\pi} 
(1 + 0.083\frac{\bar{\Lambda}}{m_B})
+ 0.124 \frac{\lambda_1}{m_B^2}+0.172 \frac{\lambda_2}{m_B^2}\nonumber \, ,\\
\langle \s_0-\ms^2 \rangle \!  \!  \! & \! = \! & \! \!  \! 0.093 \frac{\alpha_s}{\pi}
(1+0.083\frac{\bar{\Lambda}}{m_B})
+ 0.641 \frac{\lambda_1}{m_B^2}+0.589\frac{\lambda_2}{m_B^2}\nonumber \, ,\\
\langle (\s_0-\ms^2)^2\rangle \!  \!  \! & \! = \! & \! \!  \! 0.0071 \frac{\alpha_s}{\pi} 
(1+0.083\frac{\bar{\Lambda}}{m_B})
-0.196 \frac{\lambda_1}{m_B^2} \, . 
\end{eqnarray}
As already discussed earlier, the normalizing factor
${\cal B}/ {\cal B}_0$ is also expanded in a Taylor series. 
Thus, in deriving the above results, we have used
 \begin{eqnarray}
{{\cal B}\over {\cal B}_0}&=&25.277 \,\, (1-1.108 \frac{\alpha_s}{\pi} 
-0.083 \frac{\bar{\Lambda}}{m_B}-0.041 \frac{\bar{\Lambda}^2}{m_B^2}
+ 0.546 \frac{\lambda_1}{m_B^2}-3.439\frac{\lambda_2}{m_B^2}) \nonumber \, .
\end{eqnarray}
%Note that we have omitted terms of order $\alpha_s \ms$.
The parameters used in arriving at the numerical coefficients are given in 
Table~\ref{parameters} and Table~\ref{wilson}.

Inserting the expressions for the moments calculated at the partonic
level into 
eq.~(\ref{sHmoments}) and eq.~(\ref{EHmoments}), 
we find the following expressions for the short-distance hadronic moments, 
valid up to ${\cal{O}}(\alpha_s/m_B^2,1/m_B^3)$:
\begin{eqnarray}
\langle S_H\rangle \!  \!  \! & \! = \! & \! \!  \! m_B^2 (\frac{m_s^2}{m_B^2}
+0.093 \frac{\alpha_s}{\pi} 
-0.069 \frac{\bar{\Lambda}}{m_B} \frac{\alpha_s}{\pi}
+0.735 \frac{\bar{\Lambda}}{m_B}+0.243 \frac{\bar{\Lambda}^2}{m_B^2}
+ 0.273 \frac{\lambda_1}{m_B^2}-0.513\frac{\lambda_2}{m_B^2}) \nonumber \, ,\\
\label{eq:hadmoments}
\langle S_H^2\rangle \!  \!  \! & \! = \! & \! \!  \! m_B^4 (0.0071 \frac{\alpha_s}{\pi} 
+0.138 \frac{\bar{\Lambda}}{m_B} \frac{\alpha_s}{\pi}
+0.587\frac{\bar{\Lambda}^2}{m_B^2}
-0.196 \frac{\lambda_1}{m_B^2}) \, ,\\
\langle E_H\rangle \!  \!  \! & \! = \! & \! \!  \!  0.367 m_B  (1+0.148 \frac{\alpha_s}{\pi} 
-0.352 \frac{\bar{\Lambda}}{m_B} \frac{\alpha_s}{\pi}
+1.691 \frac{\bar{\Lambda}}{m_B}+0.012\frac{\bar{\Lambda}^2}{m_B^2}
+ 0.024 \frac{\lambda_1}{m_B^2}+1.070\frac{\lambda_2}{m_B^2}) \, ,\nonumber \\
\langle E_H^2\rangle \!  \!  \! & \! = \! & \! \!  \! 0.147 m_B^2 (1+0.324 \frac{\alpha_s}{\pi} 
-0.128 \frac{\bar{\Lambda}}{m_B} \frac{\alpha_s}{\pi}
+2.954 \frac{\bar{\Lambda}}{m_B}+2.740\frac{\bar{\Lambda}^2}{m_B^2}
-0.299 \frac{\lambda_1}{m_B^2}+0.162\frac{\lambda_2}{m_B^2}) \, .\nonumber
\end{eqnarray}
One sees that
there are linear power corrections, ${\cal O}(\bar{\Lambda}/m_B)$,
present in all these hadronic quantities except 
$\langle S_H^2 \rangle$ which starts in
$\frac{\alpha_s}{\pi} \frac{\bar{\Lambda}}{m_B}$.

\begin{figure}[htb]
\vskip -0.0truein
\centerline{\epsfysize=3.5in
{\epsffile{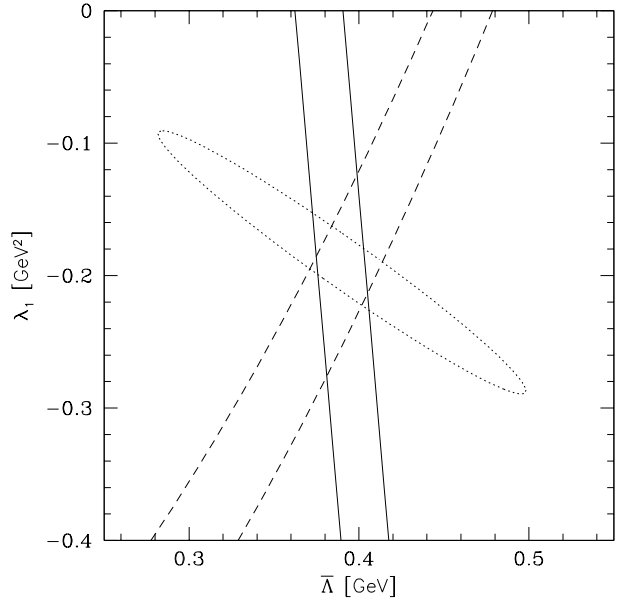}}}
\vskip -0.0truein
\caption[]{ \it $\langle S_H \rangle$ (solid bands) and  $\langle S_H^2 
\rangle$ (dashed bands)
correlation in ($\lambda_1$-$\bar{\Lambda}$) space for fixed values 
$\langle S_H \rangle =1.64$ GeV$^2$ and $\langle S_H^2 \rangle =4.48$ 
GeV$^4$, corresponding to the central values in Table \ref{tab:emoments}.
The curves are forced to meet at the point 
$\lambda_1=-0.2$ GeV$^2$ and $\bar{\Lambda}=0.39$ GeV.
The correlation
from $B \to X \ell \nu_\ell$ \cite{gremm} is also shown here (ellipse).}
\label{fig:laml1}
\end{figure}

Setting $m_s=0$ changes the numerical value of the coefficients in the 
expansion given above (in which we already neglected $\alpha_s m_s$) 
by at most $1 \% $.
With the help of the expressions given above,
we have calculated numerically the hadronic moments in HQET for the decay 
$B \to X_s \ell^{+} \ell^{-}$, $\ell=\mu,e$ and have estimated the errors
by varying the parameters within their $\pm 1 \sigma$ ranges given in 
Table~\ref{parameters}. They are presented in Table {\ref{tab:emoments}}
where we have used $\bar{\Lambda}=0.39 \, {\mbox{GeV}}$,  
$\lambda_1=-0.2 \, {\mbox{GeV}}^2$ and $\lambda_2=0.12 \, {\mbox{GeV}}^2$.
Further, using $\alpha_s(m_b)=0.21$,
the explicit dependence of the hadronic moments given in 
eq.~(\ref{eq:hadmoments}) on the HQET parameters
$\lambda_1$ and $\bar{\Lambda}$ can be worked out \cite{AH98-1}
\begin{eqnarray}
   \langle S_H\rangle&=&0.0055 m_B^2(1+
132.61 \frac{\bar{\Lambda}}{m_B}+44.14 \frac{\bar{\Lambda}^2}{m_B^2}
+ 49.66 \frac{\lambda_1}{m_B^2}) \nonumber \, ,\\
 \langle S_H^2\rangle&=& 0.00048 m_B^4(1+
19.41 \frac{\bar{\Lambda}}{m_B} 
+1223.41\frac{\bar{\Lambda}^2}{m_B^2}
-408.39 \frac{\lambda_1}{m_B^2}) \, ,\\
   \langle E_H\rangle&=& 0.372 m_B  (1+
1.64 \frac{\bar{\Lambda}}{m_B}+0.01 \frac{\bar{\Lambda}^2}{m_B^2}
+ 0.02 \frac{\lambda_1}{m_B^2}) \, ,\nonumber \\
 \langle E_H^2\rangle&=&0.150 m_B^2 (1+
2.88 \frac{\bar{\Lambda}}{m_B}+2.68\frac{\bar{\Lambda}^2}{m_B^2}
-0.29 \frac{\lambda_1}{m_B^2}) \, .\nonumber
\end{eqnarray}
While interpreting these numbers, one should bear in mind that there are two
comparable expansion parameters $\bar{\Lambda}/m_B$ and $\alpha_s/\pi$ 
and we have fixed the latter in showing the numbers.
As expected, the
dependence of the energy moments $\langle E_H^n\rangle$ on $\bar{\Lambda}$
and $\lambda_1$ is very weak.
The correlations on the HQET parameters $\lambda_1$ and $\bar{\Lambda}$
which follow from (assumed) fixed
values of the hadronic invariant mass moments  $\langle S_H \rangle$
 and  $\langle S_H^2 \rangle$ are shown in Fig.~\ref{fig:laml1}. We 
have taken the values for the decay $B \to X_s \mu^+ \mu^-$ from Table
\ref{tab:emoments}
for the sake of illustration and have also shown the presently 
irreducible
theoretical errors on these moments following from the input parameters
$m_t$, $\alpha_s$
and the scale $\mu$, given in Table \ref{parameters}. The errors were
calculated
by varying these parameters in the indicated range, one at a time,
and adding the individual errors in quadrature. 
Further the correlation following from the analysis of data on semileptonic 
$B \to X \ell \nu_{\ell}$ decays \cite{gremm} is shown
in Fig.~\ref{fig:laml1} (ellipse). As can be seen, it gives a 
complementary constraint to the one from \bxsll decay \cite{AH98-1},
which allows in principle a precise determination of 
$\bar{\Lambda}, \lambda_1$ from data on the latter.

The theoretical stability of the moments has to be checked against
higher order corrections and the
error estimates presented here will have to be improved.
 The ``BLM-enhanced" two-loop corrections
\cite{BLM} proportional to $\alpha_s^2\beta_0$, where $\beta_0 = 11 -2
n_f/3$ is the first coefficient in the QCD beta function, can be included
at the parton level as has been done in other decays \cite{FLS,gremmstewart}, 
but not being crucial to our point we have not
done this. More importantly, higher order corrections in
$\alpha_s$ and $1/m_b^3$ are not included here.
 While we do not think that
the higher orders in $\alpha_s$ will have a significant influence, the  
second moment $\langle S_H^2 \rangle$ is susceptible to the presence of
$1/m_b^3$ corrections as shown for the decay $B \to X
\ell \nu_\ell$ \cite{FL98}. This will considerably enlarge the theoretical
error represented by the dashed band for $\langle S_H^2 \rangle$ in
Fig.~\ref{fig:laml1}. Fortunately, the coefficient of the
$\bar{\Lambda}/m_B$ term in $\langle S_H \rangle$ is large. 
Hence, a measurement of this moment alone constrains
$\bar{\Lambda}$ effectively.
Of course, the utility
of the hadronic moments calculated above is only in conjunction
with the experimental cuts. Since
the optimal experimental cuts in \bxsll remain to be defined, we hope to
return to this and related issue of doing an improved
theoretical error estimate in a future publication.
We remark here that care has to be taken in a general HQE calculation with 
cuts. For an extraction of meaningful observables the calculation must be 
smeared by integration.
If the remaining phase space gets too restricted the OPE, which is based on 
parton-hadron duality, breaks down.
This happens for example near the high-$q^2$ end-point of the invariant 
dilepton mass spectrum in \bxsll decay \cite{AHHM97}.

Related issues in other decays have been studied in literature. The 
classification of the operators 
contributing in ${\cal O}(1/m_b^3)$, estimates of their matrix elements, 
and effects on the decay rates and spectra in the decays 
$B \to X \ell \nu_\ell$ and $B \to (D,D^*) \ell \nu_\ell$ 
have been studied in \cite{Shifmanetal94,Mannel94,GK96-2}.
Spectral moments of the photon energy in the decay $B \to X_s \gamma$
have been studied in \cite{KL95}.
For studies of ${\cal O}(1/m_b^3)$ contributions in this decay and the 
effects of the experimental cut (on the photon energy) on the photon energy 
moments, see \cite{Bauer97}.
An HQE analysis of the first two hadronic invariant mass moments with a 
lepton energy cut has been worked out in ref.~\cite{FL98}.

\begin{table}[h]
        \begin{center}
        \begin{tabular}{|c|l|l|l|l|}
        \hline
        \multicolumn{1}{|c|}{HQET}      &
                \multicolumn{1}{|c|}{$\langle S_H\rangle$  } &
\multicolumn{1}{|c|}{$\langle S_H^2\rangle$ } &
                \multicolumn{1}{|c|}{$\langle E_H\rangle$  } &
\multicolumn{1}{|c|}{$\langle E_H^2\rangle$ } \\
 \hline
\multicolumn{1}{|c|}{{\mbox{}}} &
\multicolumn{1}{|c|}{$({\mbox{GeV}}^2)$ } &
\multicolumn{1}{|c|}{$({\mbox{GeV}}^4)$ } &
\multicolumn{1}{|c|}{$({\mbox{GeV}})$ } &  
\multicolumn{1}{|c|}{$({\mbox{GeV}}^2)$ } \\
 \hline
$\mu^+ \mu^-$&$1.64 \pm 0.06$ &$4.48 \pm 0.29$ & $2.21 \pm 0.04 $&
$5.14 \pm 0.16$ \\
$e^+ e^-$   &$1.79 \pm 0.07 $ &$4.98 \pm 0.29$ & $2.41 \pm 0.06 $&
$6.09 \pm 0.29$\\
        \hline   
        \end{tabular}
        \end{center} 
\caption{\it Hadronic spectral moments for $B \to X_s \mu^{+} \mu^{-}$
and $B \to X_s e^{+} e^{-}$
in HQET with $\bar{\Lambda}=0.39 \, GeV$, $\lambda_1=-0.2 \, GeV^2$,
and $\lambda_2=0.12 \, GeV^2$.
The quoted errors result from varying $\mu, \alpha_s$ and the
top mass within the ranges given in Table \ref{parameters}.}   
\label{tab:emoments}
\end{table}
Finally, concerning the power corrections related to the
$c\bar{c}$ loop in \bxsll, it has been suggested
in \cite{buchallaisidorirey} that an
${\cal{O}}(\Lambda^2_{QCD}/m_c^2)$ expansion in the context of HQET
can be carried out to
take into account such effects  in
the invariant mass spectrum away from the resonances.
Using the expressions (obtained with $m_s=0$)
for the $1/m_c^2$ amplitude, we have calculated the
partonic energy moments
$\triangle \langle x_0^n \rangle$,
which correct the short-distance result at order $\lambda_2/m_c^2$:
\begin{eqnarray}
\triangle \langle x_0^n \rangle {{\cal B}\over {\cal B}_0}
\! \! \! & \! =\! & \! \! -\frac{256 C_2 \lambda_2}{27 m_c^2}
\int_0^{1/2(1-4 \ml^2)} \! dx_0 x_0^{n+2} {\rm Re} \left[
F(r) \left( C_9^{\mbox{eff}} (3-2 x_0)+2 C_7^{{\mbox{eff}}}
\frac{-3+4 x_0+2 x_0^2}{2 x_0-1} \right) \right]  , \nonumber \\
r \! \!  \! &\! = \! & \! \! \frac{1-2 x_0}{4 \mc^2} \; ,
\end{eqnarray}
\begin{equation}\label{frl1}
F(r)=\frac{3}{2r}\left\{ \begin{array}{ll}
\dis\frac{1}{\sqrt{r(1-r)}}\arctan\sqrt{\frac{r}{1-r}}
   -1 &  \qquad\qquad 0< r < 1~, \\
 \dis\frac{1}{2\sqrt{r(r-1)}}\left(
\ln\frac{1-\sqrt{1-1/r}}{1+\sqrt{1-1/r}}+i\pi\right)-1 &
\qquad\qquad r > 1~. \end{array} \right.
\end{equation}
The invariant mass and mixed moments give zero contribution in the order we
are working, with $m_s=0$.
Thus, the correction to the hadronic mass moments are vanishing, if we
further neglect terms proportional to
$\frac{\lambda_2}{m_c^2} \bar{\Lambda}$ and $ \frac{\lambda_2}{m_c^2}
\lambda_i$, with $i=1,2$.
For the hadron energy moments we obtain numerically
\begin{eqnarray}
\triangle \langle E_H \rangle_{1/m_c^2}&=& 
m_B \triangle \langle x_0 \rangle= -0.007 \, {\mbox{GeV}} \; ,
\nonumber \\
\triangle \langle E_H^2 \rangle_{1/m_c^2}&=& 
m_B^2 \triangle \langle x_0^2 \rangle= -0.013 \, {\mbox{GeV}}^2 \; ,
\end{eqnarray}
leading to a correction of order $-0.3 \%$
 to the short-distance values presented in
Table \ref{tab:moments}.

\section{Hadron Spectra in the 
Fermi Motion Model \label{hadronspectra}}

In this section, we study the non-perturbative effects associated with the 
bound state nature of the $B$-hadron on the hadronic invariant mass and 
hadron energy distributions in the decay \bxsll. These effects are studied 
in the Fermi motion model (FM) \cite{aliqcd} introduced in section 
\ref{subsec:FM}.
In the context of rare $B$ decays, this model has 
been employed to calculate the
energy spectra in the decay $B \to X_s+\gamma$ in \cite{effhamali,ag1}, which
was used subsequently by the CLEO collaboration in their successful search of
this decay \cite{CLEO95inkl}.
It has also been used in calculating the dilepton
invariant mass spectrum and FB asymmetry in \bxsll \cite{AHHM97}, see section
\ref{wavefunction}.

\subsubsection{Comparison with HQET}

The FM has received a lot of phenomenological
attention in $B$ decays, partly boosted by studies in the context of
HQET showing that this model can be made to mimic the effects associated with 
the HQET parameters $\bar{\Lambda}$ and $\lambda_1$ \cite{Bigietal94,MW}. 
We can further quantify this correspondence. 
The HQET parameters are calculable in terms of the FM parameters $p_F$ and 
$m_q$ with 
\begin{eqnarray}
\label{fmtohqet}
\bar{\Lambda} &=& \int_0^\infty dp \, p^2 \phi(p) \sqrt{m_q^2+p^2}, 
\nonumber\\
\lambda_1 &=& -  \int_0^\infty dp \, p^4 \phi(p) = - \frac{3}{2} p_F^2~.
\end{eqnarray}
In addition, for $m_q=0$, one can show that $\bar{\Lambda}=2p_F/\sqrt{\pi}$.
There is, however, no parameter in the FM model analogous to $\lambda_2$ in 
HQET. Curiously, much of the 
HQET {\it malaise} in describing the spectra in the end-point regions is 
related to $\lambda_2$, as also shown in \cite{MW,AHHM97}. 

The relation between $m_B$, $m_b$, $\bar{\Lambda}$, $\lambda_1$ and $\lambda_2$
in HQET has already been stated (eq.~(\ref{hqetmass})). With the 
quantity $m_b^{\mbox{eff}}$ defined in eq.~(\ref{effbmass}) and  
the relations in eqs.~(\ref{fmtohqet}) for $\lambda_1$ and $\bar{\Lambda}$, 
the relation
\begin{equation}
\label{mbfm}
m_B=m_b^{\mbox{eff}}+ \bar{\Lambda} -\lambda_1/(2m_b^{\mbox{eff}})~,
\end{equation}
is found to be satisfied in the FM model to a high accuracy
(better than $0.7 \%$), which is shown in 
Table \ref{tab:FMhqet} for some representative values of the 
HQET parameters and their FM model equivalents. We shall use the 
HQET parameters $\bar{\Lambda}$ and $\lambda_1$ to characterize also the
FM model parameters, with the relations given in eqs.~(\ref{fmtohqet})
and (\ref{effbmass}) and in Table \ref{tab:FMhqet}.  
\begin{table}[h]
        \begin{center}
        \begin{tabular}{|l|l|l|l|}
        \hline
        \multicolumn{1}{|c|}{$p_F,m_q$ (MeV,MeV)}       & 
                \multicolumn{1}{|c|}{$m_b^{{\mbox{eff}}}$ (GeV)} & 
 \multicolumn{1}{|c|}{$\lambda_1$ $(\mbox{GeV}^2)$} &  
\multicolumn{1}{|c|}{$\bar{\Lambda}$ (GeV) } \\
        \hline \hline
        $(450,0)    $     & $4.76 $  & -0.304 & 0.507 \\
        $(252,300)  $    & $4.85 $  & -0.095 & 0.422  \\
        $(310,0)    $      & $4.92 $  & -0.144 & 0.350  \\
        $(450,150)  $     & $4.73 $  & -0.304 & 0.534 \\
        $(500,150)  $     & $4.68 $  & -0.375 & 0.588  \\
        $(570,150)  $     & $4.60 $  & -0.487 & 0.664  \\
        \hline
        \end{tabular}
        \end{center}
\caption{\it Values of non perturbative parameters
$m_b^{{\mbox{eff}}}$, $\lambda_1$ and $\bar{\Lambda}$   
for different sets of the FM model parameters $(p_F,m_q)$
taken from various fits of the data on $B \to X_s + (J/\psi,\gamma)$
decays discussed in \cite{AH98-3}.}
\label{tab:FMhqet}
\end{table}

\subsubsection{Calculation of the hadron spectra}

\begin{figure}[htb]
\vskip 0.0truein
\centerline{\epsfysize=3.5in
{\epsffile{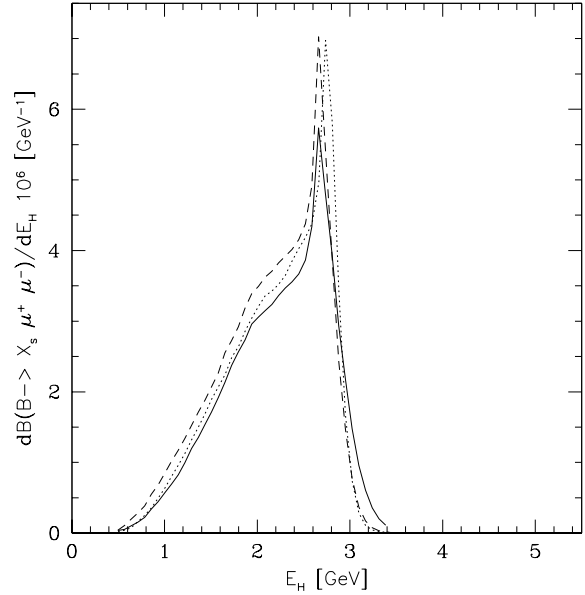}}}
\vskip 0.0truein
\caption[]{ \it Hadron energy spectrum in \bxsll in the Fermi motion 
model based on the perturbative contribution only. The solid, dotted, dashed 
curve corresponds to the parameters
$ (\lambda_1, \bar{\Lambda})=(-0.3,0.5),(-0.1,0.4),(-0.15,0.35)$ in 
(GeV$^2$, GeV), respectively.}
\label{fig:SDEh}   
\end{figure}

We turn to discuss the hadron energy spectrum in the 
decay \bxsll in the FM model 
including the ${\cal{O}}(\alpha_s)$ QCD corrections.
The spectrum $\frac{{\rm d}{\cal{B}}}{{\rm d } E_H}(B \to X_s \ell^+ \ell^-)$ 
is composed of a Sudakov improved piece from
$C_9^2$ and the remaining lowest order contribution.
The latter is based on the parton model distribution, which is well 
known and given below for the sake of completeness:
\begin{eqnarray}
        \frac{{\rm d}{\cal B}}{{\rm d} s} & = & 
{\cal{B}}_0 
\frac{\hat{u}}{m_b^6}
    \left\{ 
        \frac{4}{3} (m_b^4-2 m_s^2 m_b^2+m_s^4+m_b^2 s+m_s^2 s-2 s^2)
                \left( |C_9^{\mbox{eff}}(s)|^2 + |C_{10}|^2 \right)
             \right.   \nonumber \\
        & + &
   \frac{16}{3} (2 m_b^6 -2 m_b^4 m_s^2-2 m_b^2 m_s^4+2 m_s^6-m_b^4 s-
14 m_b^2 m_s^2 s-m_s^4 s-m_b^2 s^2-m_s^2 s^2)
                 \frac{|C_7^{\mbox{eff}}|^2}{s}
                \nonumber \\
        &+&     
             \left.  
 16  (m_b^4-2 m_s^2 m_b^2+m_s^4-m_b^2 s -m_s^2 s)
                         Re(C_9^{\mbox{eff}}(s)) \, C_7^{\mbox{eff}} 
                \right\}
\; , \\
{\cal{B}}_0&=&\frac{{\cal B}_{sl}}{\Gamma_{sl}} 
\frac{G_F^2 |V_{ts}^{\ast} V_{tb}|^2}{192 \pi^3} 
\frac{3 \alpha^2}{16 \pi^2} m_b^5\, , \nonumber \\
\Gamma_{sl}&=&\frac{G_F^2 V_{cb}^2 m_b^5}{192 \pi^3} f(\mc) \kappa(\mc) \; ,
\label{eq:Gsl}
\end{eqnarray}  
where $\u$ is given in eq.~(\ref{eq:boundaries}).
Note that in the lowest order expression just given, we have    
$|C_9^{\mbox{eff}}(s)|^2=|Y(s)|^2+2 C_9 Re (Y(s))$
with the rest of $C_9^{\mbox{eff}}(s)$  now included in the 
Sudakov-improved piece as can be seen in eq.~(\ref{doubleexpon}).
To be consistent, the total semileptonic width $\Gamma_{sl}$, which enters
via the normalization constant ${\cal{B}}_0$, has also to be calculated
in the FM model with the same set of the model parameters.
We implement the correction in the decay width 
by replacing the $b$-quark mass in $\Gamma_{sl}$ given in 
eq.~(\ref{eq:Gsl}) by $m_b^{\mbox{eff}}$ \cite{AHHM97}.
The hadronic invariant mass spectrum in the decay \bxsll in this model is 
calculated very much along the same lines.
 The kinematically allowed ranges for the distributions are
$m_X \leq E_H \leq m_B$ and $m_X^2 \leq S_H \leq m_B^2$,
and we recall here that the physical threshold has been implemented by 
demanding that the lowest hadronic invariant mass produced in the decay
\bxsll satisfies $m_X = max(m_K,m_q+m_s)$.
The results for the hadron energy and the hadronic invariant mass spectra
are presented in Figs.~\ref{fig:SDEh} and ~\ref{fig:SDSh}, respectively.  
We do not show the $S_H$ distribution in the entire range, as it tends
monotonically to zero for larger values of $S_H$.

%
%
%  SD figures, no cut
%
\begin{figure}[htb]
\vskip -1.0truein
\centerline{\epsfysize=8.5in
{\epsffile{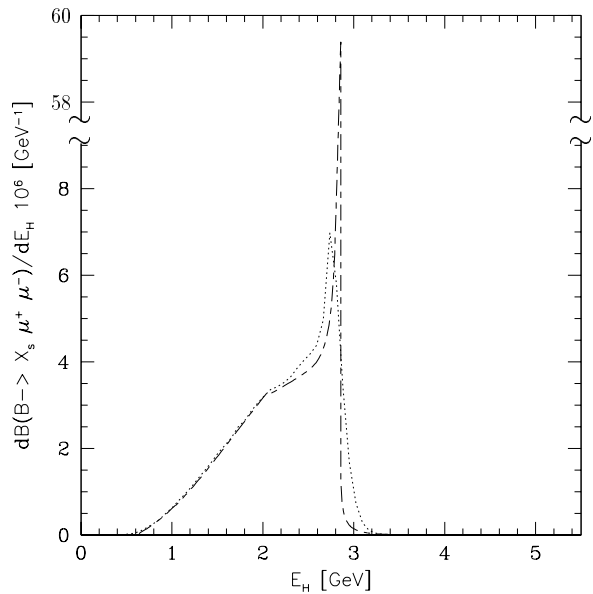}}}
\vskip -4.0truein
\caption[]{ \it Hadron energy spectrum in \bxsll
based on the perturbative contribution only, in the Fermi motion 
model (dotted curve) for $ (p_F, m_q)=(252,300)$ $(MeV,MeV)$, yielding
$m_b^{{\mbox{eff}}}=4.85$ GeV,  and in the parton 
model (long-short dashed curve) for $m_b=4.85$ GeV.}
\label{fig:SDEh485}   
\end{figure}
\begin{figure}[htb]
\vskip -0.0truein
\centerline{\epsfysize=3.5in   
{\epsffile{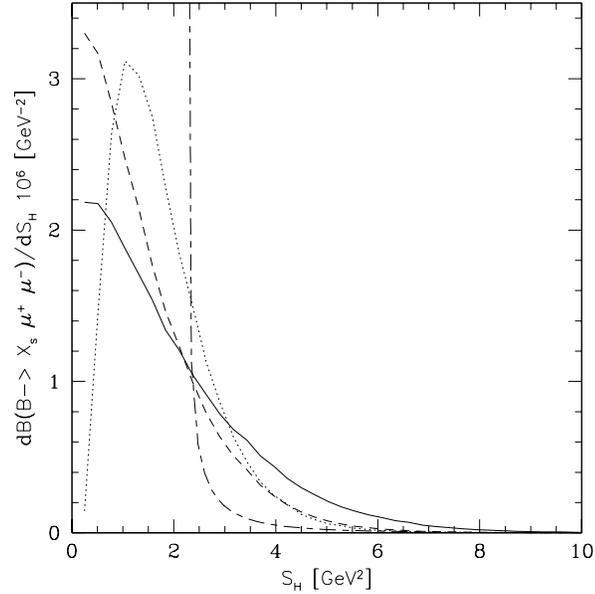}}}
\vskip -0.0truein
\caption[]{ \it Hadronic invariant mass spectrum in the Fermi motion 
model and parton model, based on the perturbative contribution only.
The solid, dotted, dashed
curve corresponds to the parameters
$ (\lambda_1, \bar{\Lambda})=(-0.3,0.5),(-0.1,0.4),(-0.15,0.35)$ in
(GeV$^2$, GeV), respectively.
The parton model (long-short dashed) curve is drawn for $m_b=4.85$ 
GeV.} \label{fig:SDSh}
\end{figure}

A number of remarks is in order:
\begin{itemize}
\item The hadron energy spectrum in \bxsll is rather insensitive to the 
model parameters. Also, the difference between the spectra in the FM and
the parton model is rather small as can be seen in Fig.~\ref{fig:SDEh485}. 
Since, away from the lower end-point and
the $c\bar{c}$ threshold, the parton model and HQET have very 
similar spectra (see Fig.~\ref{fig:dbdx0}), the estimates presented in  
Fig.~\ref{fig:SDEh} provide a good phenomenological profile of this
spectrum for the short-distance contribution. Very 
similar conclusions were drawn in
\cite{greubrey} for the corresponding spectrum in the decay $B \to X_u 
\ell \nu_\ell$, where, of course, the added complication of the $c\bar{c}$
threshold is not present.
\item In contrast to the hadron energy spectrum, the hadronic invariant
mass spectrum in \bxsll ~is sensitive to the model parameters, as can be 
seen in Fig.~\ref{fig:SDSh}. Again, one sees a close parallel in the
hadronic invariant mass spectra in \bxsll and   $B 
\to X_u \ell \nu_\ell$, with the latter worked out in \cite{FLW}.
 We think that the present 
theoretical dispersion on the hadron spectra in the decay \bxsll can be 
considerably reduced by the analysis of data in $B \to X_u \ell \nu_\ell$. 
\item The hadronic invariant mass distribution obtained by the 
$O(\alpha_s)$-corrected partonic spectrum and the HQET mass relation 
can only 
be calculated over a limited range of $S_H$, $S_H > m_B \bar{\Lambda}$, 
as shown in Fig.~\ref{fig:sh0}.
The larger is the value of $\bar{\Lambda}$, the smaller is this region. 
Also, in the range where it can be calculated, it depends on the
non-perturbative parameter $m_b$ (or $\bar{\Lambda}$). A comparison of this 
distribution 
and the one in the FM model may be made for the same values of $m_b$ and 
$m_b^{\mbox{eff}}$. This is shown for $m_b=4.85$ GeV in
Fig.~\ref{fig:SDSh} for HQET (long-short dashed curve) to be 
compared 
with the dotted curve in the FM model, which corresponds to
$m_b^{\mbox{eff}}=4.85$ GeV. We see that the two distributions differ though
they are qualitatively similar for larger values of 
$S_H > 3 \, {\mbox{GeV}}^2$.

\end{itemize}

%%%%%%%%%%%%%%%%%%%%%%%%%%%%%%%%%%%%%%%%%%%%%%%%%%%%%%%%%%

\section{LD Contributions in \bxsll (II) \label{sec:LD2}}
\begin{figure}[htb]
\vskip 0.0truein
\centerline{\epsfysize=3.5in
{\epsffile{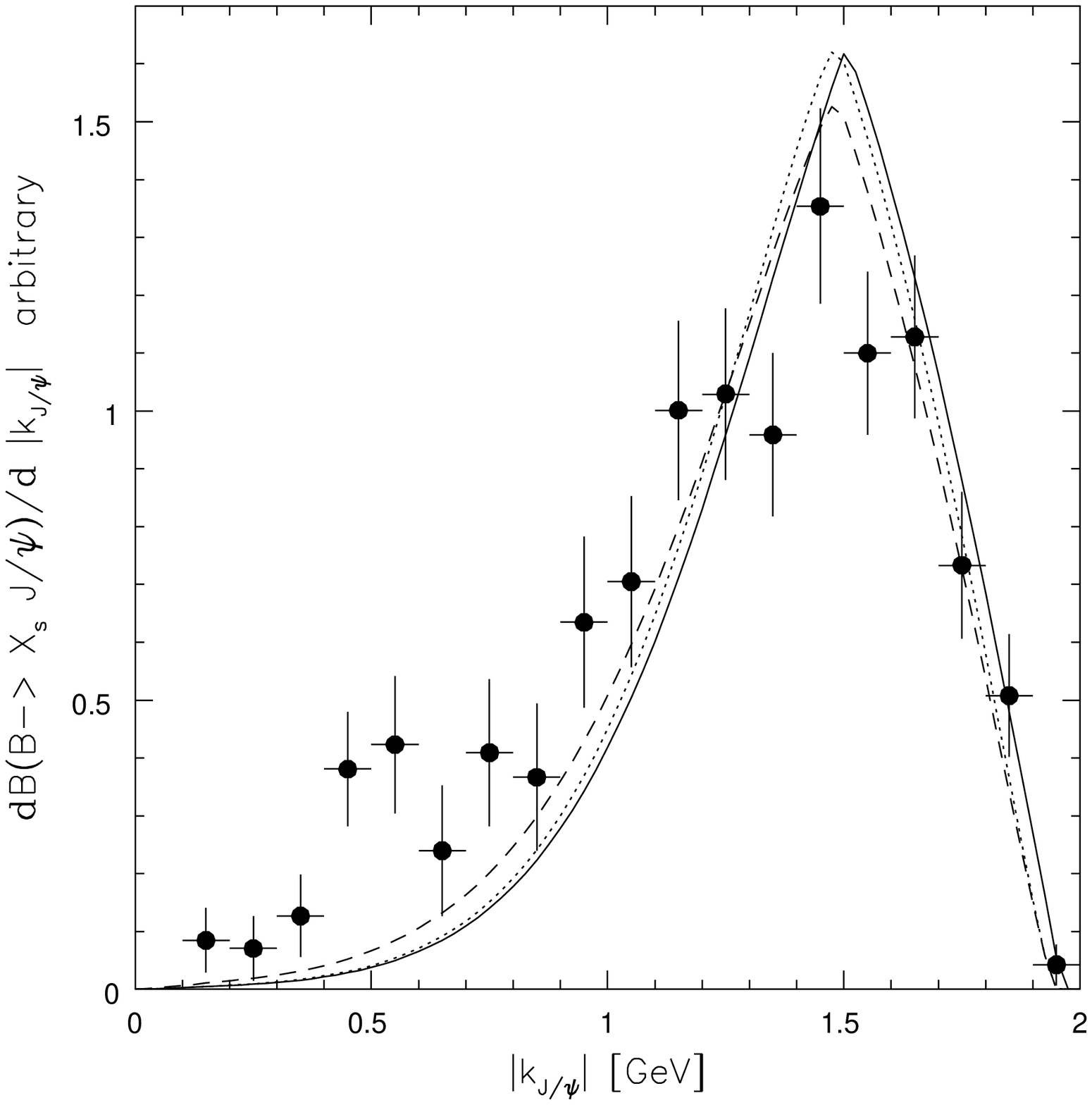}}}
\vskip 0.0truein
\caption[]{ \it Momentum distribution of $J/\psi$ in the decay
$B \to X_s J/\psi$ in the FM model.
The
solid, dotted, dashed curve corresponds to the parameters
$ (\lambda_1, \bar{\Lambda})=(-0.3,0.5),(-0.3,0.53),(-0.38,0.59)$ in
(GeV$^2$, GeV), respectively. The data points are from the CLEO measurements
\cite{CLEOjpsi94}.}
\label{fig:jpsi}
\end{figure}
\begin{figure}[htb]
\vskip 0.0truein
\centerline{\epsfysize=3.5in   
{\epsffile{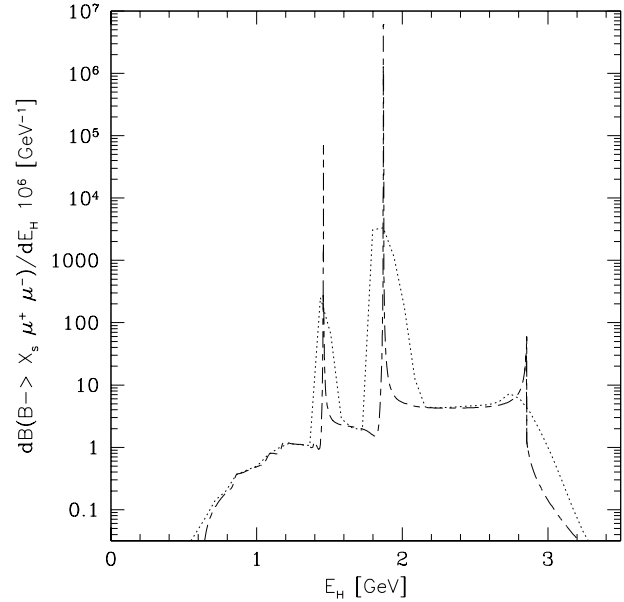}}}
\vskip 0.0truein
\caption[]{ \it Hadron energy spectrum in \bxsll
including the resonance and perturbative contributions in the Fermi motion
model (dotted curve) for $(\lambda_1,
\bar{\Lambda})=(-0.1~\mbox{GeV}^2,0.4~\mbox{GeV})$, and in the parton
model (long-short dashed curve) for $m_b=4.85$ GeV.}
\label{fig:LDEh485}
\end{figure}
\begin{figure}[htb]
\vskip -0.0truein
\centerline{\epsfysize=3.5in  
{\epsffile{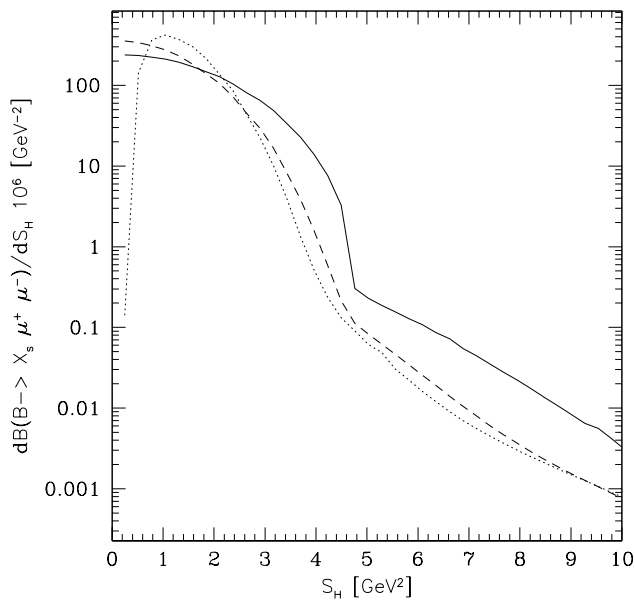}}}
\vskip -0.0truein
\caption[]{ \it Hadronic invariant mass spectrum in \bxsll including the
perturbative and resonance contributions in the Fermi motion model. The
solid, dotted, dashed curve corresponds to the parameters
$ (\lambda_1, \bar{\Lambda})=(-0.3,0.5),(-0.1,0.4),(-0.15,0.35)$ in
(GeV$^2$, GeV), respectively.}
\label{fig:LDSh} 
\end{figure}
\begin{figure}[t]
     \mbox{ }\hspace{-0.7cm}
     \begin{minipage}[t]{8.2cm}
     \mbox{ }\hfill\hspace{1cm}(a)\hfill\mbox{ }
     \epsfig{file=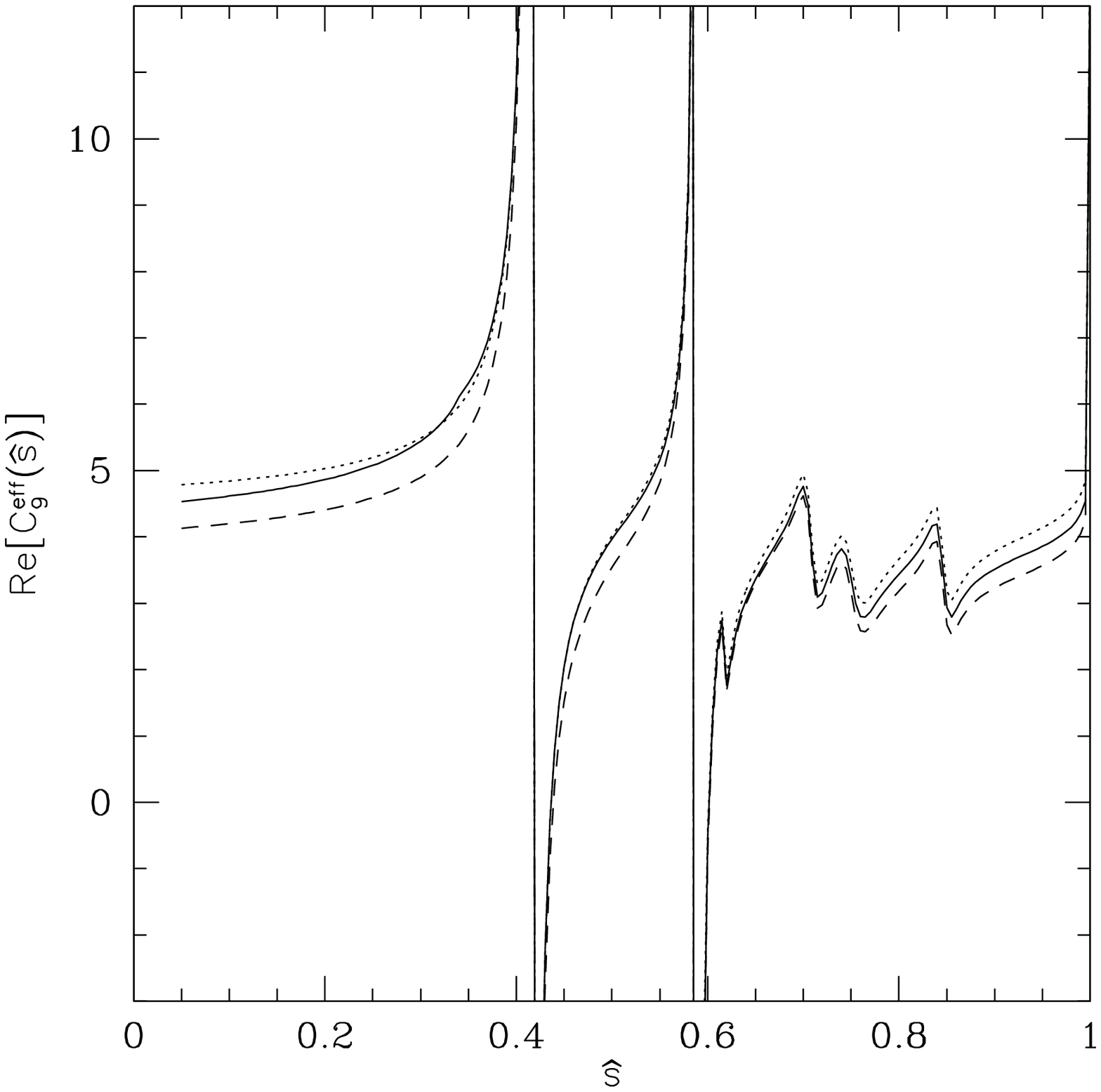,width=8.2cm}
     \end{minipage}
     \hspace{-0.4cm}
     \begin{minipage}[t]{8.2cm}
     \mbox{ }\hfill\hspace{1cm}(b)\hfill\mbox{ }
     \epsfig{file=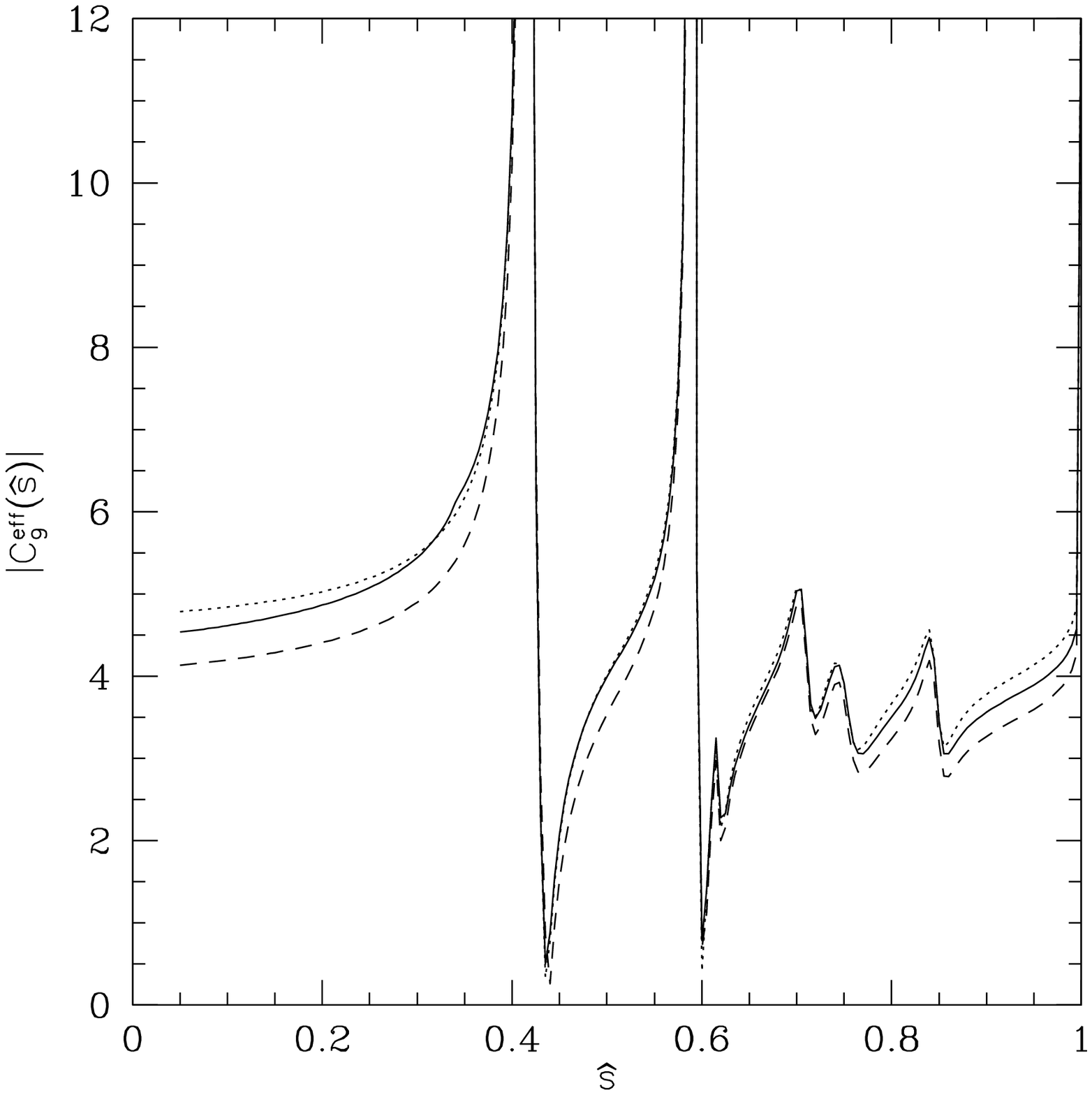,width=8.2cm}
     \end{minipage}  
     \caption{\it 
The real part (a) and the absolute value (b) of $C_9^{{\mbox{\rm eff}}}(\s)$ 
are shown as a function of $\s$, where $C_9^{\mbox{\rm eff}}(\s)=C_9 
\eta(\s) + Y(\s)  +Y_{res}(\s)$. 
The solid line corresponds to $Y(\s)$ calculated using the
perturbative $c\bar{c}$ contribution $g(\mc,\s)$ given in
eq.~(\ref{gpert}), and the dotted curve corresponds to using 
$\tilde{g}(\mc, \s)$ in eq.~(\ref{eq:gtilde}). The dashed one corresponds to
the approach by \cite{KS96}.}
\label{fig:c9real}
\end{figure}
%
%
% LD figure, no cut
%
%
\begin{figure}[t]
     \mbox{ }\hspace{-0.7cm}
     \begin{minipage}[t]{8.2cm}
     \mbox{ }\hfill\hspace{1cm}(a)\hfill\mbox{ }
     \epsfig{file=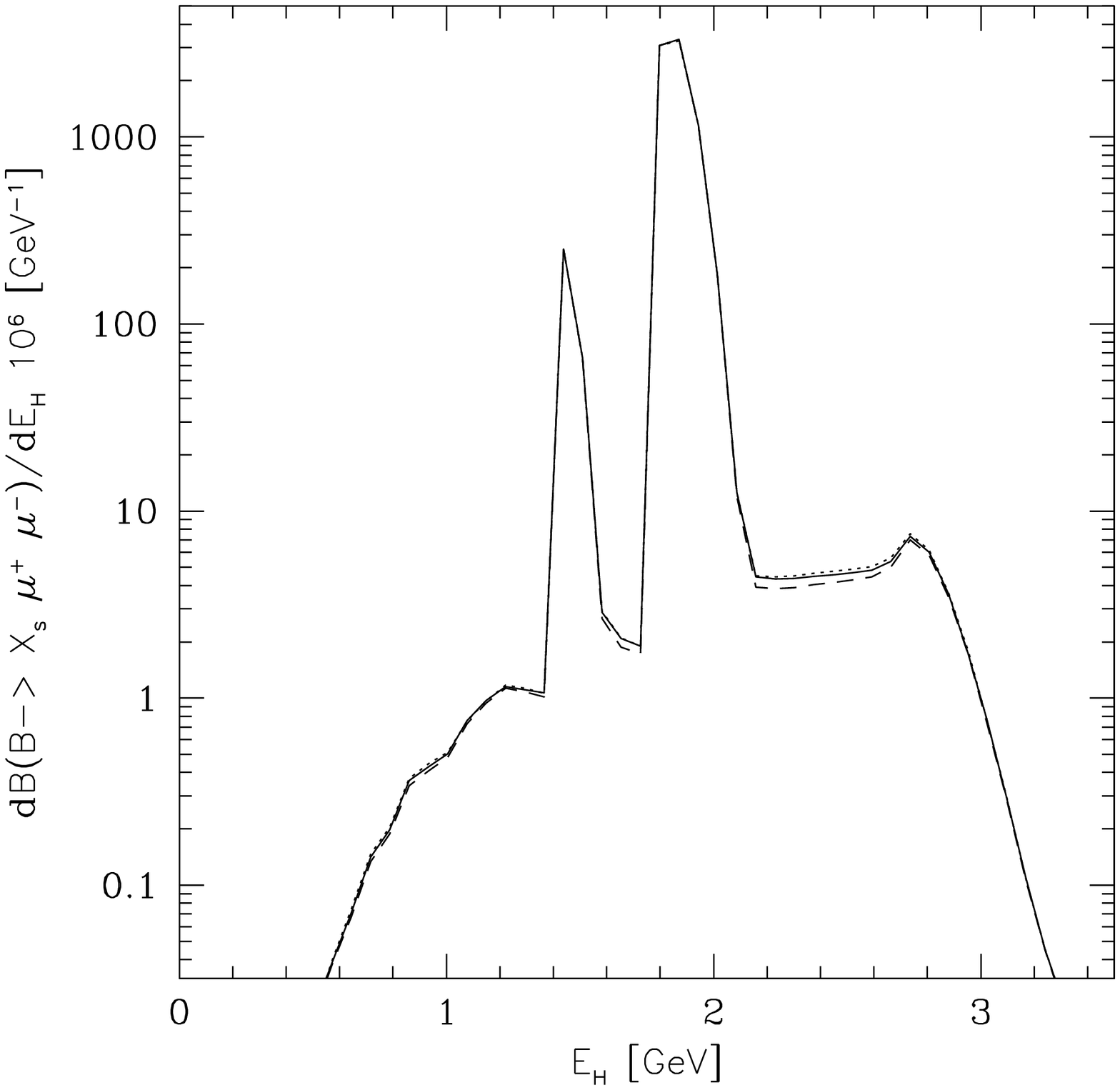,width=8.2cm}
     \end{minipage}
     \hspace{-0.4cm}
     \begin{minipage}[t]{8.2cm}
     \mbox{ }\hfill\hspace{1cm}(b)\hfill\mbox{ }
     \epsfig{file=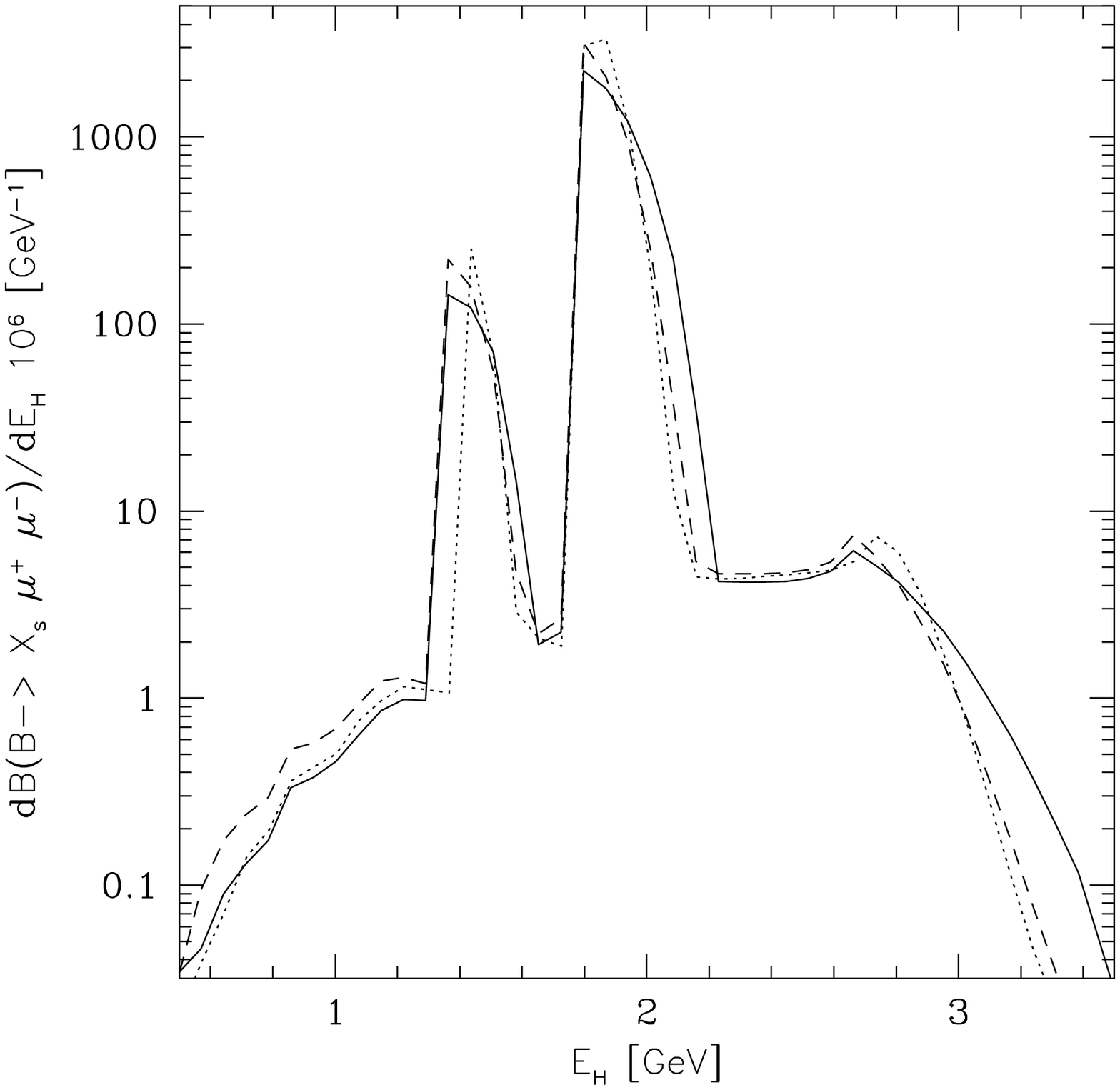,width=8.2cm}
     \end{minipage}  
     \caption{\it 
Hadron energy spectrum in \bxsll including the resonance and perturbative 
contributions in the Fermi motion model.
In (a), the FM model parameters are fixed at $(\lambda_1, 
\bar{\Lambda})=(-0.1~\mbox{GeV}^2,0.4~\mbox{GeV})$.  
 The almost overlapping curves differ in the perturbative
$c\bar{c}$ contribution with the solid curve obtained using 
eq.~(\ref{gpert})
for $g(\mc,\s)$, the dotted curve using $\tilde{g}(\mc,\s)$
given in eq.~(\ref{eq:gtilde}). The dashed curve corresponds to the 
approach by \cite{KS96}. 
In (b), the solid, dotted, dashed
curve corresponds to the parameters
$ (\lambda_1, \bar{\Lambda})=(-0.3,0.5),(-0.1,0.4),(-0.15,0.35)$ in
(GeV$^2$, GeV), respectively.}
\label{fig:LDEh}
\end{figure}
This section is devoted to various aspects of the $c\bar{c}$ resonance effects.
Following the procedure adopted in \cite{AHHM97}, we include the 
long-distance 
(LD) resonance effects in the decay \bxsll  
and {\it simply add} the $c\bar{c}$ resonant contribution with the 
perturbative $c\bar{c}$ contribution expressed through the function
 $g(\hat{m_c},\hat{s})$ in section \ref{sec:NLO} 
(see, eq.~(\ref{gpert})). Thus, in our method,
\begin{equation}
\label{simpleadd}
 C_9^{\mbox{eff}}(\s)=C_9 \eta(\s) + Y(\s)  +Y_{res}(\s)~.
 \end{equation}
The function $Y_{res}(\s)$ accounts for the $c \bar{c}$ resonance
contribution
via $B \to X_s (J/\Psi, \Psi^\prime, \dots) \to X_s \ell^{+} \ell^{-}$
and can be seen in eq.~(\ref{LDeq}). 
Note that in this 
approach, the effective coefficient $C_9^{\mbox{eff}}(\hat{s})$ has a
$\hat{s}$-dependence, which is not entirely due to the propagators
in the function $Y_{res}(\hat{s})$ as also  the perturbative
$c\bar{c}$ contribution $g(\hat{m_c}, \hat{s})$ is a function of 
$\hat{s}$. 
In the resonant region, the perturbative part is not noticeable  due to the 
fact that the resonant part in $C_9^{\mbox{eff}}(\hat{s})$ completely 
dominates. However, when 
the $c\bar{c}$ pair is sufficiently off-shell, the
$\hat{s}$-dependence of the function $C_9^{\mbox{eff}}(\hat{s})$
is not (and should not be) entirely determined by the $c\bar{c}$ resonant
contribution. This is the motivation of the representation in 
eq.~({\ref{simpleadd}).

We start with an analysis of the constraints from existing data on the 
FM model parameters.
Especially the question if the FM reproduces the measured $J/\psi$ momentum 
distribution  in $B \to X_s J/\psi$ will be investigated.
Then we turn to the effect of the Lorentz boost on the hadron spectra including
LD effects according to eq.~({\ref{simpleadd}) and present
$E_H,S_H$ distributions for \bxsll decay in the FM.
Further, we study the uncertainties in the \bxsll spectra
resulting from the ambiguities in the parametrization of the LD effects. 
Differences in the distributions from different approaches to treat the 
$c\bar{c}$ resonances 
are shown as well for hadron spectra as for the $q^2$ spectra discussed 
before, namely, the dilepton invariant mass distribution and the FB asymmetry.
Finally, the hadronic moments are calculated in the FM and compared with the 
ones in the HQE approach for identical values of equivalent parameters. 
\begin{figure}[t]
     \mbox{ }\hspace{-0.7cm}
     \begin{minipage}[t]{8.2cm}
     \mbox{ }\hfill\hspace{1cm}(a)\hfill\mbox{ }
     \epsfig{file=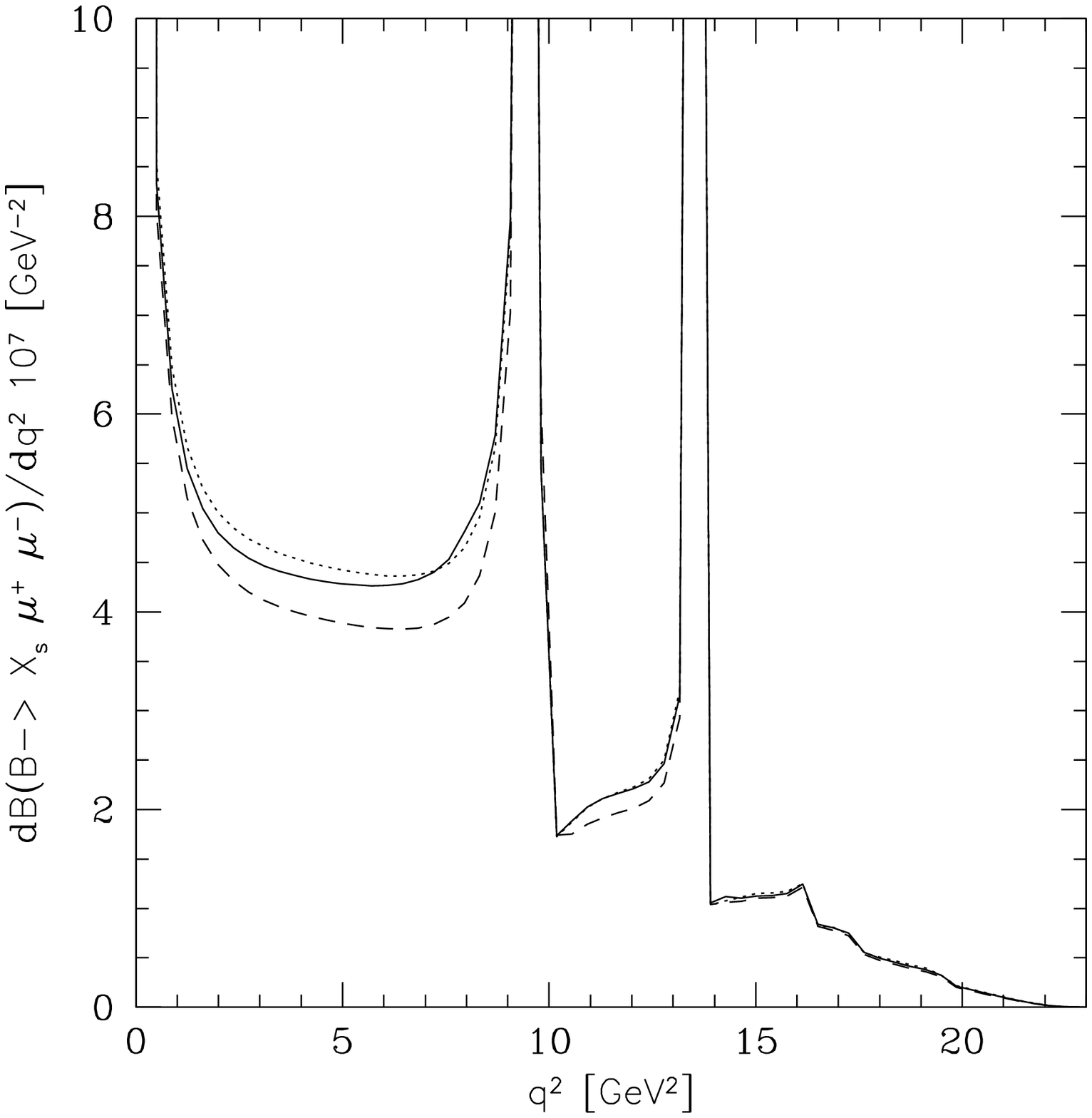,width=8.2cm}
     \end{minipage}
     \hspace{-0.4cm}
     \begin{minipage}[t]{8.2cm}
     \mbox{ }\hfill\hspace{1cm}(b)\hfill\mbox{ }
     \epsfig{file=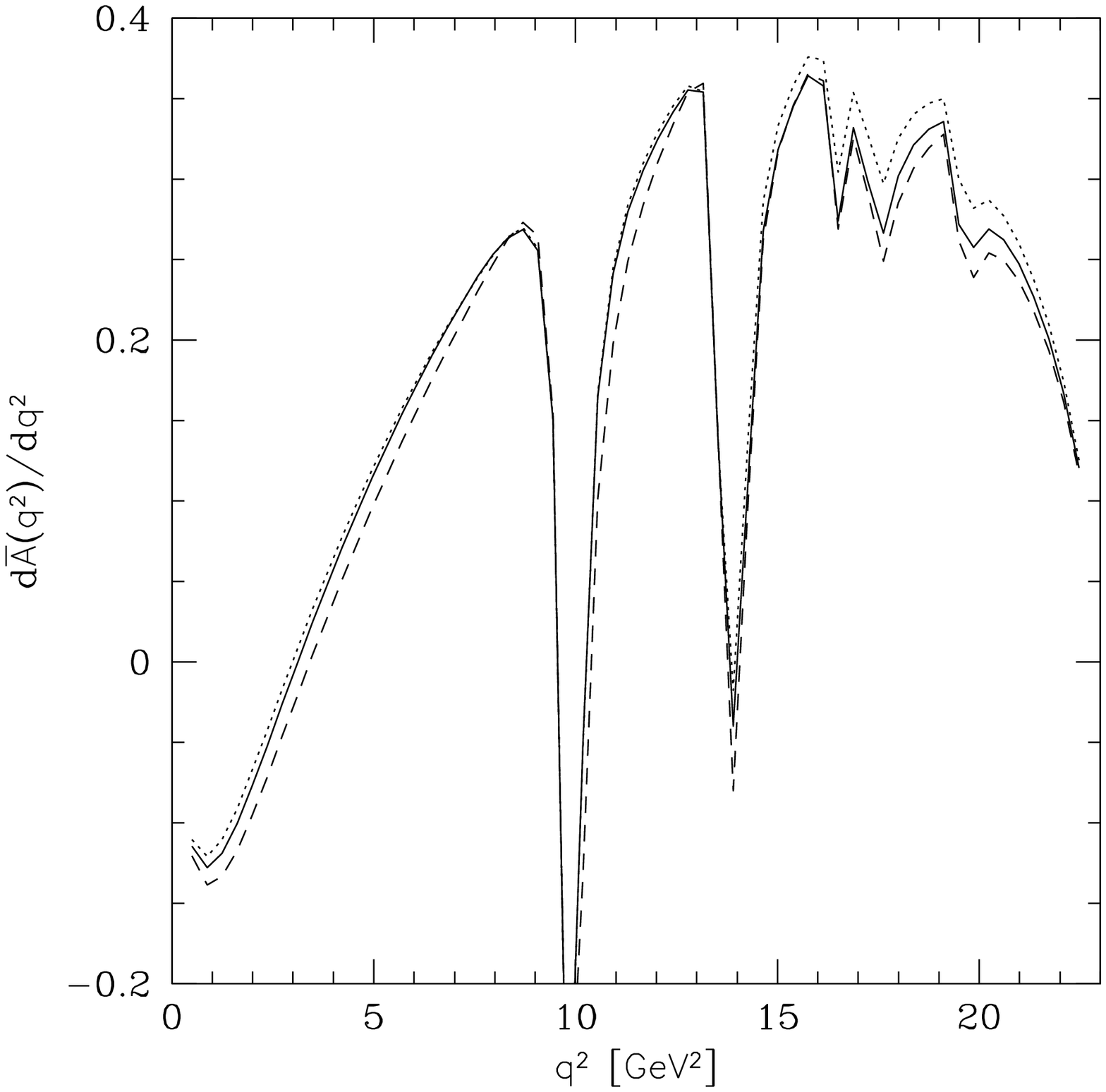,width=8.2cm}
     \end{minipage}  
     \caption{\it 
Dilepton invariant spectrum (a) and the (normalized)
Forward-Backward asymmetry (b) 
in \bxsll including the resonance and perturbative 
contributions in the Fermi motion model.
The FM model parameters are fixed at $(\lambda_1, 
\bar{\Lambda})=(-0.1~\mbox{GeV}^2,0.4~\mbox{GeV})$.  
The curves differ in the perturbative
$c\bar{c}$ contribution with the solid curve obtained using 
eq.~(\ref{gpert})
for $g(\mc,\s)$, and the dotted curve using $\tilde{g}(\mc,\s)$
given in eq.~(\ref{eq:gtilde}). The dashed curve corresponds to 
the approach given in ref.~\cite{KS96}. }
\label{fig:LDshat}
\end{figure}

\subsection{Constraints on the FM model parameters from existing data}
The FM model parameters $p_F$ and $m_q$ (equivalently $\lambda_1$ and
$\bar{\Lambda}$) can, in principle, be determined from an analysis of the
energy spectra in the decays $B \to X_u \ell \nu_\ell$
and   $B \to X_s + \gamma$,
as all of them involve the decay of a $b$-quark into (an almost)
massless ($u$ or $s$) quark. Assuming that the parameters of the FM models
are universal, these parameters can also
be constrained from the lepton energy spectrum in the decay $B \to X_c \ell
\nu_\ell$ and from the shape of the $J/\psi$- and $\psi^\prime$-
momentum distributions in the decays $B \to X_s (J/\psi,\psi^\prime)$. 
We review the presently available analyses of the photon- and lepton-energy 
spectra in $B$ decays in the FM model (and HQET, as the two are very 
similar) and also present an analysis
of the $J/\psi$-momentum spectrum in $B \to X_s J/\psi$.

\begin{itemize}
\item Analysis of the photon energy spectrum in  $B \to X_s + \gamma$ 
\end{itemize}
The photon energy- and invariant hadronic mass distributions
in $B \to X_s \gamma$ were calculated in the FM model  
using the leading order (in $\alpha_s)$ corrections in \cite{ag1,effhamali}.
These spectra were used in the analysis
of the CLEO data on $B \to X_s + \gamma$ \cite{CLEO95inkl}, in which the
values $p_F=270 \pm 40$ MeV suggested by the analysis of the CLEO data
on $B \to X \ell \nu_\ell$  were used, together with the effective $b$-quark
mass $m_b^{\mbox{eff}}=4.87 \pm 0.10$ GeV, which gave reasonable fits of the
data. We translate these parameters in terms of  $\lambda_1$
and $\bar{\Lambda}$ using the relations given in eqs. ~(\ref{fmtohqet})
and (\ref{mbfm}), yielding
\begin{equation}
\label{cleofmpar}
\lambda_1=-0.11 ^{-0.035}_{+0.030}~\mbox{GeV}^2, ~~~
\bar{\Lambda} = 0.40 \pm 0.1 ~\mbox{GeV}~.
\end{equation} 
The same data was fitted in \cite{ag2} in the FM model, yielding 
$(p_F,m_q) =(0.45~\mbox{GeV},0~\mbox{GeV})$ as the best-fit
solution, with $(p_F,m_q) =(0.310~\mbox{GeV},0.3~\mbox{GeV})$ differing 
from the best-fit solution by one unit in $\chi^2$. The quality of the
CLEO data \cite{CLEO95inkl} is not good enough to draw very quantitative
conclusions. The best-fit values translate into
\begin{equation} 
\label{agfmpar}
\lambda_1=-0.3 ~\mbox{GeV}^2~, ~~~\bar{\Lambda}=0.5 ~\mbox{GeV}~.
\end{equation}
\begin{itemize}
\item Analysis of the lepton energy spectrum in  $B \to X \ell \nu_\ell$ 
\end{itemize}
 A fit of the
lepton energy spectrum in the semileptonic decay $B \to X\ell \nu_\ell$
in the context of HQET has been performed in \cite{gremm}. Using the
CLEO data \cite{CLEObsl96}, the authors of \cite{gremm} find:
\begin{equation}
\label{hqetgremm}
\lambda_1= -0.19 \pm 0.10 ~\mbox{GeV}^2~,
~~~\bar{\Lambda}=0.39 \pm 0.11 ~\mbox{GeV} ~.
\end{equation}
Since the FM model and HQET yield very similar lepton energy spectra apart
from the end-point, one can take the analysis of \cite{gremm} also
holding approximately for the FM model.
\begin{itemize}
\item Analysis of the $J/\psi$-momentum spectrum in  $B \to X_s J/\psi$
\end{itemize}
An analysis
of the $J/\psi$-momentum spectrum in $B \to X_s (J/\psi,\psi^\prime)$
measured by the CLEO collaboration \cite{CLEOjpsi94} in the FM model
has been reported in \cite{PPS97}. The authors 
of \cite{PPS97} addressed both the shape and
normalization of the $J/\psi$-data, using the NRQCD formalism for
the inclusive color singlet and color octet charmonium 
production in $B \to X_sJ/\psi$ and the FM model. The
preferred FM parameters from this analysis are: $(p_F,m_q) 
=(0.57~\mbox{GeV},0.15~\mbox{GeV})$, where $m_q$ only plays a role in 
determining
the position  of the peak but otherwise does not influence the small momentum
tail of the $J/\psi$ momentum distribution. 
 This  yields values of the
parameter $p_F$ which are consistent with the ones obtained in 
\cite{Hwangetal} $p_F=0.54 ^{+0.16}_{-0.15}$, GeV based on an
analysis of the CLEO data on $B \to X \ell \nu_\ell$ \cite{CLEObsl96}.
The central values of $p_F$ in \cite{Hwangetal} as well as in
\cite{PPS97} correspond to  
$m_b^{\mbox{eff}} \simeq 4.6$ GeV, which is on the lower side of the present 
theoretical estimate of $m_b$ pole mass, namely $m_b=4.8 \pm 0.2$ GeV
\cite{neubertsachr}. 

  We have redone an analysis of the $J/\psi$-momentum distribution which
is shown in Fig.~\ref{fig:jpsi}. As shown in this figure, and also 
discussed in \cite{PPS97}, the low-momentum 
$J/\psi$, in particular in the region $|k_{J/\psi}| \leq 0.6$ GeV, are 
problematic for inclusive decay models, including also the FM model.
The $|k_{J/\psi}|$-spectrum appears to have a secondary bump;
an inclusive spectrum  behaving as a Gaussian tail or 
having a power-like behavior $\propto |k_{J/\psi}|^{-\delta}$ in this 
region is hard put to explain this data. There are also suggestions in
literature \cite{BN97} that the spectrum in this region is dominated by
the three-body decay $B \to J/\psi \Lambda \bar{p}$ and hence the bump
reflects the underlying dynamics of this exclusive decay. In view of this,
we have taken out the first six points in the low-$|k_{J/\psi}|$ spectrum
and fitted the FM model parameters in the rest of the 
$|k_{J/\psi}|$-spectrum. The three curves shown correspond to the FM model
parameters $(p_F,m_q) =(0.45~\mbox{GeV}, 0~\mbox{GeV})$ (solid curve),
$(p_F,m_q) =(0.45~\mbox{GeV}, 0.15~\mbox{GeV})$ (dotted curve) and
$(p_F,m_q) =(0.50~\mbox{GeV}, 0.15~\mbox{GeV})$ (dashed curve). They 
all have reasonable $\chi^2$, with $\chi^2/dof=1.6, 1.6$ and 1.1, 
respectively. Excluding also the seventh lowest point, the $\chi^2$
improves marginally, with the resulting $\chi^2$ being $\chi^2/dof=1.4, 
1.4$ and 0.94. Including the sixth point, the fits become slightly
worse. However, they are all acceptable fits. It is interesting that the
best-fit solution of the photon energy spectrum in $B \to X_s + \gamma$,
$(p_F,m_q) =(0.45~\mbox{GeV}, 0~\mbox{GeV})$, is also an acceptable
fit of the $|k_{J/\psi}|$-data. The corresponding $\lambda_1$, 
$\bar{\Lambda}$ and $m_b$ values from these two analyses are compatible
within $\pm 1 \sigma$ with the HQET-based constraints from the semileptonic 
$B$ decays
\cite{gremm}, quoted above. Thus, the values in eq.~(\ref{agfmpar})
appear to be  a reasonable guess of
the FM model parameters. But, more importantly for the present study, the
phenomenological profile of the LD contribution $B \to X_s 
(J/\psi,\psi^\prime,...) \to X_s \ell^+ \ell^-$ presented here is certainly 
consistent with present data and theoretical constraints.

\subsection{Effects of the Lorentz boost on the hadron spectra in \bxsll}
We now discuss the $B$-meson
wave function effects in the FM model on the hadron spectra in \bxsll.  
Since the resonances in \bxsll are in the dilepton invariant mass 
variable  $s$
and not in $S_H$, and noting that neither $E_0$ (partonic energy) nor $E_H$
are Lorentz-invariant quantities, it is expected on general grounds that
the effect of the Lorentz boost in the FM model on $E_H$-and
$S_H$-distributions will be more marked
than what was found on the invariant dilepton mass spectrum in
\cite{AHHM97}. We recall that for the dilepton invariant mass, the Lorentz
boost
involved in the FM model leaves the spectrum invariant
and there is only a subleading effect due to the momentum dependent $b$-quark
mass. Not so in the hadron spectra. In the hadron energy spectrum,
the $c\bar{c}$-resonances, which are narrowly peaked in the
parton model, are broadened by the Lorentz boost of the FM model.
To show this, the hadron energy spectrum in the FM model is compared with the 
spectrum in the parton model in Fig.~\ref{fig:LDEh485} for identical values 
of $m_b$ and $m_b^{\mbox{eff}}$, taken as $4.85$ GeV.
In terms of the hadronic invariant mass,
the resonance structure is greatly smeared.
The reason is that each $q^2$-bin
contributes to a  range of $E_H$ and $S_H$.
The different-$q^2$ regions overlap in  $S_H$ resulting in a smearing of
the resonances over a wide range. This can be seen in   
Fig.~\ref{fig:LDSh} for the hadronic invariant mass. Various curves
illustrate the sensitivity of this spectrum on the FM model parameters.

\subsection{Ambiguities in adding LD and SD contributions in \bxsll
\label{sec:ambiguities}}
Since we are simply adding the short-distance (SD) and resonant charmonium 
amplitudes, it can not be ruled out that possibly some double 
counting has crept in in the coefficient $C_9^{\mbox{eff}}(\hat{s})$ ,
once as a continuum $c\bar{c}$ contribution and then
again as $J/\psi,\psi^\prime,...$ resonances. In the absence of a clear 
separation of the LD and SD physics in the spectra, we 
can not plead the case one way or the other.  
In the meanwhile, 
the question is whether the addition of the $c\bar{c}$-continuum 
and resonating pieces as being done here and in \cite{AHHM97} compromises 
the 
resulting theoretical precision significantly. This can only be studied by
comparing the theoretical scenario in question with other trial 
constructions which have no $c\bar{c}$ double counting.
% \footnote{It, however,
%does not imply that the perturbative contribution has been uniquely fixed.}.
For example, one could retain in the perturbative function $g(\mc, \s)$  just 
the constant part in $\s$ by replacing $g(\mc, \s)$ by $\tilde{g}(\mc, \s)$, 
where
\begin{eqnarray}    
\label{eq:gtilde}   
\tilde{g}(\mc, \s) =
-\frac{8}{9} \ln(\frac{m_b}{\mu})-\frac{8}{9} \ln \mc +\frac{8}{27} \; .
\end{eqnarray}
This function (with $\mu=m_b$) has been proposed in \cite{LSW97,KS96}
as an alternative representation of the $c\bar{c}$ perturbative
contribution and represents the (minimal) short-distance contribution.
We denote this ansatz for $C_9^{\mbox{eff}}$ defined as
$C_9^{\mbox{eff}}(\s)=C_9 \eta(\s)+Y(\s)(\mbox{``}g(\mc, \s) \to 
\tilde{g}(\mc, \s) \mbox{"})+Y_{res}(\s)$ by LSW.
Another approach is based on a dispersion relation, as proposed by 
\cite{KS96}, here and in the following denoted by KS.
The KS parametrization of the $c \bar{c}$ resonant part 
differs from ours (eq.~(\ref{eq:yprime})), and
% in the $\s$ behaviour.
%, but its imaginary part is identical on the resonances. 
the non-resonant part has been extracted from data, see \cite{KS96} for 
details.
The advantage of the KS procedure is that there is certainly no double 
counting.
To study the difference numerically, we plot both the real 
part $Re C_9^{\mbox{eff}}(\s)$ and the absolute value  
$|C_9^{\mbox{eff}}(\s)|$ as functions of $\hat{s}$ in Fig.~\ref{fig:c9real} 
by using the complete perturbative expression for $g(\mc,\s)$ in 
eq.~(\ref{eqn:y}) (our approach) and $\tilde{g}(\mc, \s)$
given in eq.~(\ref{eq:gtilde}) (LSW) and the KS parametrization.
Both figures (a) and (b) show that the KS curve is always below our, which is 
again always below the LSW one.
One sees from Figs.~\ref{fig:c9real} (a) and \ref{fig:c9real} (b) that the 
difference in these functions in the variable $\s$ is visible. 
However, the three
parametrizations of the perturbative $c\bar{c}$ part give almost identical
hadron spectra, with the resulting uncertainty
in the hadron energy and the hadronic invariant mass spectra
being at most $12.1 (4.5) \%$ and $4.1 (2.5) \%$, respectively.
The difference between our approach and the KS one (first numbers)
is larger than the one between ours and the LSW one, given in parentheses.
These differences are already difficult to see in the hadron energy spectrum in
Fig.~\ref{fig:LDEh} (a); 
the effect on the hadronic invariant mass is even less noticeable and hence 
is not shown. Since, other uncertainties on
the  hadronic distribution are much larger, see, for example, 
Fig.~\ref{fig:LDEh} (b) showing the sensitivity of the hadron energy spectra
on the $B$-meson wave function parameters,
the much talked about $c\bar{c}$-continuum related ambiguity
in literature is numerically small.  
Fig.~\ref{fig:LDEh} shows that it is not the dominant uncertainty in
predicting the theoretical profiles of hadron spectra in \bxsll.

\subsubsection{LD uncertainties in the dilepton invariant mass and the 
FB asymmetry}

We further analyse the uncertainties resulting from different parametrizations of the short and long-distance amplitudes in $q^2$-spectra which we have 
investigated in the previous section \ref{sec:LD1}. 
We show in Fig.~\ref{fig:LDshat} the 
dilepton invariant mass (a) and the Forward-Backward (FB) asymmetry (b) with 
all three ``SD$+$LD" approaches discussed before.
Unfortunately, the difference between our and the KS one \cite{KS96} in the 
dilepton spectrum is maximal in the low-$q^2$ region below the $J/\psi$-peak, 
$q^2 < 9 \mbox{GeV}^2$. It amounts up to $15 \%$. 
In the other distribution, the FB asymmetry, the difference is found to be 
moderate over the full $q^2$-range . It does not exceed $5\%$ in the range
$5 \mbox{GeV}^2 \leq q^2 \leq 9 \mbox{GeV}^2$.
However, we remark here that the position of the first zero of the FB asymmetry
is affected by the parametrization of the $c \bar{c}$ states.

\subsection{Numerical estimates of the hadronic moments in FM model and
HQET   \label{numerics}}

To underline the similarity of the HQET and FM descriptions in \bxsll,
and also to make comparison with data,
we have calculated the hadronic moments in the FM model using
the spectra which we have presented in the previous sections.
The  moments based on the SD-contribution are defined as:
\begin{eqnarray}
\langle X_H^n\rangle=(\int X_H^n\frac{d{\cal{B}}}{dX_H} dX_H)/{\cal{B}}
\hspace{1cm} {\mbox{for}} \, \, \,\,  X=S,E ~.
\end{eqnarray}
The moments $\langle X_H^n\rangle_{\bar{c} c}$ are defined by taking
into account in addition to the SD-contribution also the contributions 
from the $c\bar{c}$ resonances. The values of the moments in both the
HQET approach and the FM for $n=1,2$ are shown in
Table~\ref{tab:moments}, with the numbers in the parentheses 
corresponding
to the former. They are again based on using the central values of the
parameters given in Table~\ref{parameters}, and are
calculated for the same
values of the HQET parameters $\bar{\Lambda}$ and $\lambda_1$, using 
the
transcriptions given in eqs.~(\ref{fmtohqet}).
Both the HQET and the FM model lead to strikingly similar results 
for the
hadronic moments shown in this table. However, the moments
$\langle X_H^n\rangle_{\bar{c} c}$ with $X=S,E$ are significantly 
lower than their SD-counterparts $\langle X_H^n\rangle$ calculated for the
same values of the FM model parameters. This shows, at least in this
model study, that the $c\bar{c}$ resonances are important also in
moments. The hadronic invariant mass spectra in \bxsll for both the 
SD and inclusive contributions are expected to be dominated by multi-body
states, with $\langle S_H\rangle \simeq (1.5 - 2.1) \, \mbox{GeV}^2$ and
$\langle S_H\rangle_{\bar{c} c} \simeq (1.2 - 1.5) \, \mbox{GeV}^2$ .
Note that the difference in the numerical values of
the hadronic mass moments $\langle S_H\rangle_{\bar{c} c}$ and
$\langle S_H^2\rangle_{\bar{c} c}$ shown in
Table~\ref{tab:moments} caused by different LD parametrizations is less than
$0.22 \%,0.42 \%$, respectively, as can be see in Table \ref{SHcutmoments}.

\begin{table}[h]
        \begin{center}   
        \begin{tabular}{|c|l|l|l|l|}
        \hline
        \multicolumn{1}{|c|}{{\mbox{}}}      &
                \multicolumn{1}{|c|}{$\langle S_H\rangle$  } &
\multicolumn{1}{|c|}{$\langle S_H\rangle_{\bar{c} c}$ } &
                \multicolumn{1}{|c|}{$\langle S_H^2\rangle$  } &
\multicolumn{1}{|c|}{$\langle S_H^2\rangle_{\bar{c} c}$ } \\
 \hline
\multicolumn{1}{|c|}{($\lambda_1,\bar{\Lambda})$ in (GeV$^2$, GeV)} &
\multicolumn{2}{|c|}{$({\mbox{GeV}}^2)$ } &
\multicolumn{2}{|c|}{$({\mbox{GeV}}^4)$ } \\
        \hline
    $(-0.3,0.5)$  & 2.03 (2.09)&1.51 &6.43 (6.93)&3.10   \\
    $(-0.1,0.4)$  & 1.75 (1.80)&1.36 &4.04 (4.38)&2.17   \\
    $(-0.14,0.35)$  & 1.54 (1.49)&1.19 &3.65 (3.64)&1.92   \\
        \hline
\hline
        \multicolumn{1}{|c|}{{\mbox{}}} &
               \multicolumn{1}{|c|}{$\langle E_H\rangle$  } &
\multicolumn{1}{|c|}{$\langle E_H\rangle_{\bar{c} c} $} &
                \multicolumn{1}{|c|}{$\langle E_H^2\rangle$  } &
\multicolumn{1}{|c|}{$\langle E_H^2\rangle_{\bar{c} c}$ } \\
 \hline
\multicolumn{1}{|c|}{$(\lambda_1,\bar{\Lambda})$ in (GeV$^2$, GeV)} & 
\multicolumn{2}{|c|}{$({\mbox{GeV)}} $ } &
\multicolumn{2}{|c|}{$({\mbox{GeV}}^2)$ }
\\
        \hline
$(-0.3,0.5)$   &2.23 (2.28)&1.87 &5.27 (5.46)& 3.52   \\
$(-0.1,0.4)$  &2.21 (2.22)&1.85 &5.19 (5.23)& 3.43   \\
$(-0.14,0.35)$  &2.15 (2.18)&1.84 &4.94 (5.04)& 3.39   \\
        \hline
        \end{tabular}
        \end{center}
\caption{\it Hadronic spectral moments for $B \to X_s \mu^{+} \mu^{-}$
in the Fermi motion model (HQET) for the indicated
values of the parameters $(\lambda_1,\bar{\Lambda})$.
 }
\label{tab:moments}
\end{table}

\section{Branching Ratios and Hadron Spectra in \bxsll with Cuts on 
Invariant Masses}

In experimental searches for the decay \bxsll, the 
short-distance contribution is expected to be visible away from the resonances.
So, cuts on the invariant dilepton mass are imposed to get rid of 
the dilepton mass range where the charmonium resonances $J/\psi$ and
$\psi^{\prime}$ are dominant. 
 For example, the cuts imposed in the recent CLEO analysis
\cite{cleobsll97} given below are typical: 
\begin{eqnarray}
{\mbox{cut}}\, A&:& 
q^2 \leq  (m_{J/\psi}-0.1 \, {\mbox{GeV}})^2 = 8.98 \, {\mbox{GeV}}^2 \, ,
\nonumber \\
{\mbox{cut}} \, B&:& 
q^2 \leq  (m_{J/\psi}-0.3 \, {\mbox{GeV}})^2 = 7.82 \, {\mbox{GeV}}^2 \, ,
\nonumber \\
{\mbox{cut}} \, C&:& 
q^2 \geq  (m_{\psi^{\prime}}+0.1 \, {\mbox{GeV}})^2 = 14.33 \, {\mbox{GeV}}^2
\, . \label{eq:cuts}
\end{eqnarray}
The cuts $A$ and $B$ have been chosen to take into account the
QED radiative corrections as these effects are different in the
$e^+ e^-$ and $\mu^+ \mu^{-}$ modes. In the following, we compare the
hadron spectra with and without the resonances after imposing these
experimental cuts.
For the low-$q^2$ cut for muons (cut $A$), the hadron energy spectra and the
hadronic invariant mass spectra are shown  
in Fig.~\ref{fig:EhLO} (a), (b) and Fig.~\ref{fig:ShLO} (a), (b), respectively.
The results for the low-$q^2$ cut for electrons (cut $B$), are shown in
Fig.~\ref{fig:EhLO} (c), (d) and Fig.~\ref{fig:ShLO} (c), (d), respectively. 
Finally, the hadronic spectra for the high-$q^2$ cut (cut $C$) for 
$e^{+} e^{-}$ and $\mu^{+} \mu^{-}$ can be seen in 
Fig.~\ref{fig:EhLO} (e), (f) for the hadronic energy and in 
Fig.~\ref{fig:ShLO} (e), (f)
for the hadronic invariant mass.
We see that the above  
cuts in $q^2$ greatly reduce the resonance contributions.
Hence, the resulting distributions essentially test the non-resonant 
$c\bar{c}$ and short-distance contributions. These figures will be used
later to quantify the model dependence of the integrated branching ratios
in \bxsll.

 As mentioned in \cite{cleobsll97}, the dominant $B\bar{B}$ background 
to the decay \bxsll comes from two semileptonic decays of $B$ or $D$
mesons, which produce the lepton pair with two undetected neutrinos.
 To suppress this $B\bar{B}$ background,
it is required that the invariant mass of the final hadronic state is 
less than $t=1.8 \, {\mbox{GeV}}$, which approximately equals $m_D$. 
We define the survival probability of the \bxsll signal 
after the hadronic invariant mass cut:
\begin{eqnarray}
S(t)\equiv (\int_{m_{X}^2}^{t^2} \frac{d{\cal{B}}}{dS_H} dS_H)/{\cal{B}}~,
\label{eq:eff}
\end{eqnarray}
and present $S(t=1.8 ~\mbox{GeV}))$ as the
fraction of the branching ratio for \bxsll surviving these cuts  
in Table~\ref{SHoutcome}. To estimate the model dependence
of this probability, we vary the FM 
model parameters. Concentrating on the SD piece, we note that the effect
of this cut alone is that between $83\%$ to $92\%$ of the signal for
$B \to X_s \mu^+ \mu^-$ and between $79\%$ to $90\%$ of the signal
in $B \to X_s e^+ e^-$ survives, depending on the FM model parameters.
The corresponding numbers for the inclusive spectrum including the
SD and LD contribution, here and in the following abbreviated as
$tot=$SD$+$LD, is $96\%$ to $99.7\%$ for both the dimuon and 
dielectron case. This shows that while this cut removes a good fraction of
the $B\bar{B}$ background, it allows a very large fraction of the \bxsll
signal to survive. However, this cut does not discriminate between the
SD and (SD+LD) contributions, for which the cuts $A$ - $C$ are effective.
The numbers for the survival probability $S(t=1.8 ~\mbox{GeV})$ reflect 
that the hadronic 
invariant mass distribution of the LD-contribution is more steep than
the one from the SD contribution.

\begin{table}[h]
        \begin{center}
        \begin{tabular}{|l|l|l|l|l|l|l|l|l|}
        \hline
        \multicolumn{1}{|c|}{$(\lambda_1,\bar{\Lambda})$ }       & 
\multicolumn{1}{|c|}{ $ {\cal{B}}\cdot 10^{-6}$} & 
\multicolumn{1}{|c|}{${\cal{B}}\cdot 10^{-6}$} &  
\multicolumn{1}{|c|}{No $s$-cut} & 
\multicolumn{1}{|c|}{No $s$-cut } &  
\multicolumn{1}{|c|}{cut A}  &
\multicolumn{1}{|c|}{cut B}  &
\multicolumn{1}{|c|}{cut C}  &
\multicolumn{1}{|c|}{cut C}  \\
 \multicolumn{1}{|c|}{GeV$^2$, GeV}           & 
 \multicolumn{1}{|c|}{$\mu^{+} \mu^{-}$}   & 
 \multicolumn{1}{|c|}{$e^+ e^-$}           &  
 \multicolumn{1}{|c|}{$\mu^{+} \mu^{-}$}   & 
 \multicolumn{1}{|c|}{$e^+ e^-$}           &  
 \multicolumn{1}{|c|}{$\mu^{+} \mu^{-}$}   &
 \multicolumn{1}{|c|}{$e^+ e^-$}           &
 \multicolumn{1}{|c|}{$\mu^{+} \mu^{-}$}   &
 \multicolumn{1}{|c|}{$e^+ e^-$}           \\
        \hline \hline
   $(-0.3,0.5)  $  &5.8 &8.6 & $83 \% $  & 79 \%&$57 \% $  & $57 \% $ 
&$6.4\% $&$4.5\%$\\
   $(-0.1,0.4) $ &5.7 &8.4 & $93 \% $  & 91 \%&$63 \% $  & $68 \% $
&$8.3\%$&$5.8\%$\\
   $(-0.14,0.35)  $  &5.6 &8.3 & $92 \% $  & 90 \%&$65 \% $  & $67 \% $ 
&$7.9\%$&$5.5\%$\\
\hline
   $(-0.3,0.5)_{tot} $ &562.5&563.9 & $96 \% $  & 96 \%&$0.8 \% $ & $1.0 
\% $ &$0.06\%$&$0.06\%$\\
   $(-0.1,0.4)_{tot} $&564.0&565.6 & $99.7 \% $&99.7\%&$0.8 \% $ & $1.1 
\% $ &$0.08\%$&$0.08\%$\\
   $(-0.14,0.35)_{tot} $ &566.5&568.2 & $99 \% $  & 99 \%&$0.9 \% $ & $1.2 
\% $ &$0.08\%$&$0.08\%$\\
        \hline
        \end{tabular}
        \end{center}
\caption{\it Branching ratios for \bxsll, $\ell=\mu,e$ for different FM model 
parameters are given in the second and third columns. The values 
given in percentage in the fourth to ninth columns represent  the  
survival probability $S(t=1.8 {\mbox{ {\rm {GeV}}}})$  
defined in eq.~(\ref{eq:eff}) 
for different FM model parameters and cuts on the dilepton invariant mass
as defined in eq.~(\ref{eq:cuts}).
The subscript $tot=SD+LD$ denotes that both the short and the long-distance 
contributions are included in the branching ratios and $S(t)$. } 
\label{SHoutcome}
\end{table}

With the additional cut $A$ ($B$) imposed on the dimuon (dielectron) invariant 
mass, between $57\%$ to $65\%$ ($57\%$ to $68\%$) of the \bxsll signal 
survives the additional cut
on the hadronic invariant mass for the SD contribution. However, as
expected, the cuts $A$ and $B$ result in drastic reduction of the inclusive
branching ratio for the decay \bxsll, as they effectively remove the 
dominant $c\bar{c}$-resonant part. In this case only $0.8\%$ to $0.9\%$
($1.0\%$ to $1.2\%$ of the inclusive signal survives for the cut $A$ ($B$).
The theoretical branching ratios for both the dielectron and dimuon
cases, calculated using the central values in Table ~\ref{parameters} are 
also given in Table~\ref{SHoutcome}. As estimated in eq.~(\ref{eq:Brnumbers}),
the uncertainty on the branching ratios resulting from the errors on the
parameters in Table~\ref{parameters} is about $\pm 23\%$ (for the dielectron 
mode) and
$\pm 16 \%$ (for the dimuon case). The wave function-related uncertainty in
the branching ratios is smaller, as can be seen in Table ~\ref{SHoutcome}.
 With the help of the theoretical branching ratio and the survival
probability $S(t=1.8)$ GeV, calculated for three sets of the FM parameters,
 the cross section can be calculated for all six 
cases:\\
 (i) no cut on the dimuon invariant mass [(SD) and (SD + LD)],
(ii) no cut on the dielectron invariant mass [(SD) and (SD + LD)],
(iii) cut $A$ on the dimuon invariant mass [(SD) and (SD + LD)],
(iv) cut $B$ on the dielectron invariant mass [(SD) and (SD + LD)],
(v)  cut $C$ on the dimuon invariant mass [(SD) and (SD + LD)],
(vi)  cut $C$ on the dielectron invariant mass [(SD) and (SD + LD)].
This gives a fair estimate of the theoretical uncertainties on
the partially integrated branching ratios from the $B$-meson wave 
function and $c\bar{c}$ resonances. This table shows that with
$10^7$ $B\bar{B}$ events, ${\cal 
O}(70)$ dimuon and (${\cal O}(100)$ dielectron) signal events from 
\bxsll should survive the CLEO cuts $A$ ($B$) with $m(X_s) <1.8$ GeV.
With cut $C$, one expects an order of magnitude less events, making 
this region interesting for the LHC experiments. 
Given enough data, one can compare
the experimental distributions in \bxsll directly with the ones presented 
here. 
%The phenomenological success of the FM model in describing spectra in
%$B$ decays and its close proximity to HQET make us confident that
%the hadron spectra in \bxsll will be no exception either!    

\subsection{Hadronic spectral moments with cuts in the FM}

We have calculated the first two moments of the hadronic invariant mass in 
the FM model by imposing a cut $S_H < t^2$ with $t=1.8 \, \mbox{GeV}$ 
and an optional cut on $q^2$.
\begin{eqnarray}
\label{eq:SHcut}
\langle S_H^n\rangle=
(\int_{m_{X}^2}^{t^2} S_H^n\frac{d^2{\cal{B}}_{cut X}}{dS_Hdq^2} dS_Hdq^2)
/(\int_{m_{X}^2}^{t^2} \frac{d^2{\cal{B}}_{cut X}}{dS_Hdq^2} dS_Hdq^2)
\hspace{1cm} {\mbox{for}} \, \, \,\,  n=1,2 ~.
\end{eqnarray}  
Here the subscript $cut X$ indicates whether we evaluated 
$\langle S_H\rangle$ and $\langle S_H^2 \rangle$ with the cuts 
on the invariant dilepton mass as defined in 
eq.~(\ref{eq:cuts}), or without any cut on the dilepton mass. 
The results are collected in Table \ref{SHcutmoments}.
The moments given in Table \ref{SHcutmoments} can be
compared directly with the data to extract the FM model parameters.
The entries in this table  give a fairly
good idea of what the effects of the experimental cuts on the corresponding
moments in HQET will be, as the FM and HQET yield very similar moments
for equivalent values of the parameters. The 
functional dependence of 
the hadronic moments on the HQET parameters taking into account the 
experimental cuts still remains to be worked out.

\begin{table}[h]
        \begin{center}
        \begin{tabular}{|c|l|l|l|l|l|l|l|l|l|l|}
        \hline
        \multicolumn{1}{|c|}{FM}       & 
\multicolumn{2}{|c|}{No $s$-cut} & 
\multicolumn{2}{|c|}{No $s$-cut } &  
\multicolumn{2}{|c|}{cut A}  &
\multicolumn{2}{|c|}{cut B}  &
\multicolumn{2}{|c|}{cut C}  \\
 \multicolumn{1}{|c|}{parameters}           & 
 \multicolumn{2}{|c|}{$\mu^{+} \mu^{-}$}   & 
 \multicolumn{2}{|c|}{$e^+ e^-$}           &  
 \multicolumn{2}{|c|}{$\mu^{+} \mu^{-}$}   & 
 \multicolumn{2}{|c|}{$e^+ e^-$}           &  
 \multicolumn{2}{|c|}{$\ell^{+} \ell^{-}$}   \\
 \multicolumn{1}{|c|}{($\lambda_1,\bar{\Lambda})$}           & 
 \multicolumn{1}{|c|}{$\langle S_H \rangle$ } &
\multicolumn{1}{|c|}{$\langle S_H^2\rangle$} &
 \multicolumn{1}{|c|}{$\langle S_H\rangle$  } &
\multicolumn{1}{|c|}{$\langle S_H^2\rangle$ } &
 \multicolumn{1}{|c|}{$\langle S_H\rangle$  } &
\multicolumn{1}{|c|}{$\langle S_H^2\rangle$ } &
 \multicolumn{1}{|c|}{$\langle S_H\rangle$  } &
\multicolumn{1}{|c|}{$\langle S_H^2\rangle$ } &
 \multicolumn{1}{|c|}{$\langle S_H\rangle$  } &
\multicolumn{1}{|c|}{$\langle S_H^2\rangle$ } \\
 \multicolumn{1}{|c|}{GeV$^2$, GeV}           & 
\multicolumn{1}{|c|}{${\mbox{GeV}}^2$ } &
\multicolumn{1}{|c|}{${\mbox{GeV}}^4$ } &
\multicolumn{1}{|c|}{${\mbox{GeV}}^2$ } &
\multicolumn{1}{|c|}{${\mbox{GeV}}^4$ } &
\multicolumn{1}{|c|}{${\mbox{GeV}}^2$ } &
\multicolumn{1}{|c|}{${\mbox{GeV}}^4$ } &
\multicolumn{1}{|c|}{${\mbox{GeV}}^2$ } &
\multicolumn{1}{|c|}{${\mbox{GeV}}^4$ } &
\multicolumn{1}{|c|}{${\mbox{GeV}}^2$ } &
\multicolumn{1}{|c|}{${\mbox{GeV}}^4$ } \\
        \hline \hline
$(-0.3,0.5)  $ &1.47&2.87&1.52&3.05&1.62&3.37&1.66&3.48&0.74&0.69\\
$(-0.1,0.4)  $ &1.57&2.98&1.69&3.37&1.80&3.71&1.88&3.99&0.74&0.63 \\
$(-0.14,0.35)$ &1.31&2.34&1.38&2.55&1.47&2.83&1.52&2.97&0.66&0.54\\
        \hline
$(-0.3,0.5)_{tot}  $ &1.41&2.61&1.41&2.62&1.61&3.32&1.66&3.47&0.74&0.68\\
$(-0.1,0.4)_{tot}  $ &1.35&2.14&1.36&2.15&1.77&3.60&1.87&3.94&0.74&0.62 \\
$(-0.14,0.35)_{tot}$ &1.17&1.84&1.18&1.85&1.45&2.76&1.51&2.95&0.66&0.54\\
        \hline
$\triangle \, (\%)$        &$0.15 $&$0.19 $&$0.22 $&$0.42 $&$0.90 $&$1.56 $&$0.32
$&$0.58 $&$0.01 $&$ 0.32$\\
\hline
        \end{tabular}
        \end{center}
\caption{\it $\langle S_H\rangle$ and $\langle S_H^2\rangle$ for 
\bxsll, $\ell=\mu,e$ for different FM model 
parameters and a hadronic invariant mass cut  $S_H <3.24 \, \mbox{GeV}^2$ 
are given with and without additional cuts on the 
dilepton invariant mass as defined in eq.~(\ref{eq:cuts}).
The $S_H$-moments with cuts are defined in eq.~(\ref{eq:SHcut}).
The subscript $tot=SD+LD$ denotes that both the short and the long-distance 
contributions are included in these moments. 
The value of $\triangle$ estimates the uncertainty from different 
approaches to take into account the
effect of the $c \bar{c}$ continuum and resonances, see text.} 
\label{SHcutmoments}
\end{table}

Further, we have calculated $\langle S_H\rangle$ 
and $\langle S_H^2 \rangle$ with a cut $S_H < 3.24 \, \mbox{GeV}^2$
and optional ones on $q^2$ (cut $A$-$C$ according to eq.~(\ref{eq:cuts}))
with the  approaches KS \cite{KS96} and LSW \cite{LSW97} 
for $(\lambda_1,\bar{\Lambda})=(-0.1,0.4)$ in $\mbox{GeV}^2,\mbox{GeV}$.
They differ from ours (eq.~(\ref{simpleadd})) in the parametrization of 
the resonant and non-resonant $c \bar{c}$ contributions, as discussed in
section \ref{sec:ambiguities}. 
We compare the values of the moments for the same set of FM parameters.
Denoting our approach by $y$, we 
define by $\triangle$ the maximal deviation in $\%$ between ours and KS and 
LSW, generically written as:
$\triangle=max(|y-{\mbox{LSW}}|/|y|,|y-{\mbox{KS}}|/|y|)$ and present it 
in the last row of Table \ref{SHcutmoments}. 
We see that the uncertainties in the hadronic mass moments from different 
``SD$+$LD" parametrizations are small, namely below $1.6\%$ in the worst case.
%

%%%%%%%%%%%%%%%%%%%%%%%%%%%%%%%%%%%%%%%%%%%%%%%%%%%%%%%%%%%%%%%%%%%%%%%%
%
%    Figures with the experimental cuts
%
%
%
% 
%  SD and SD+LD for energy
% with cuts 
%
\begin{figure}[t]
\mbox{}\vspace{-2cm}\\
\begin{center}
     \mbox{ }\hspace{-0.7cm}
     \begin{minipage}[t]{7.0cm}
     \mbox{ }\hfill\hspace{1cm}(a)\hfill\mbox{ }
     \epsfig{file=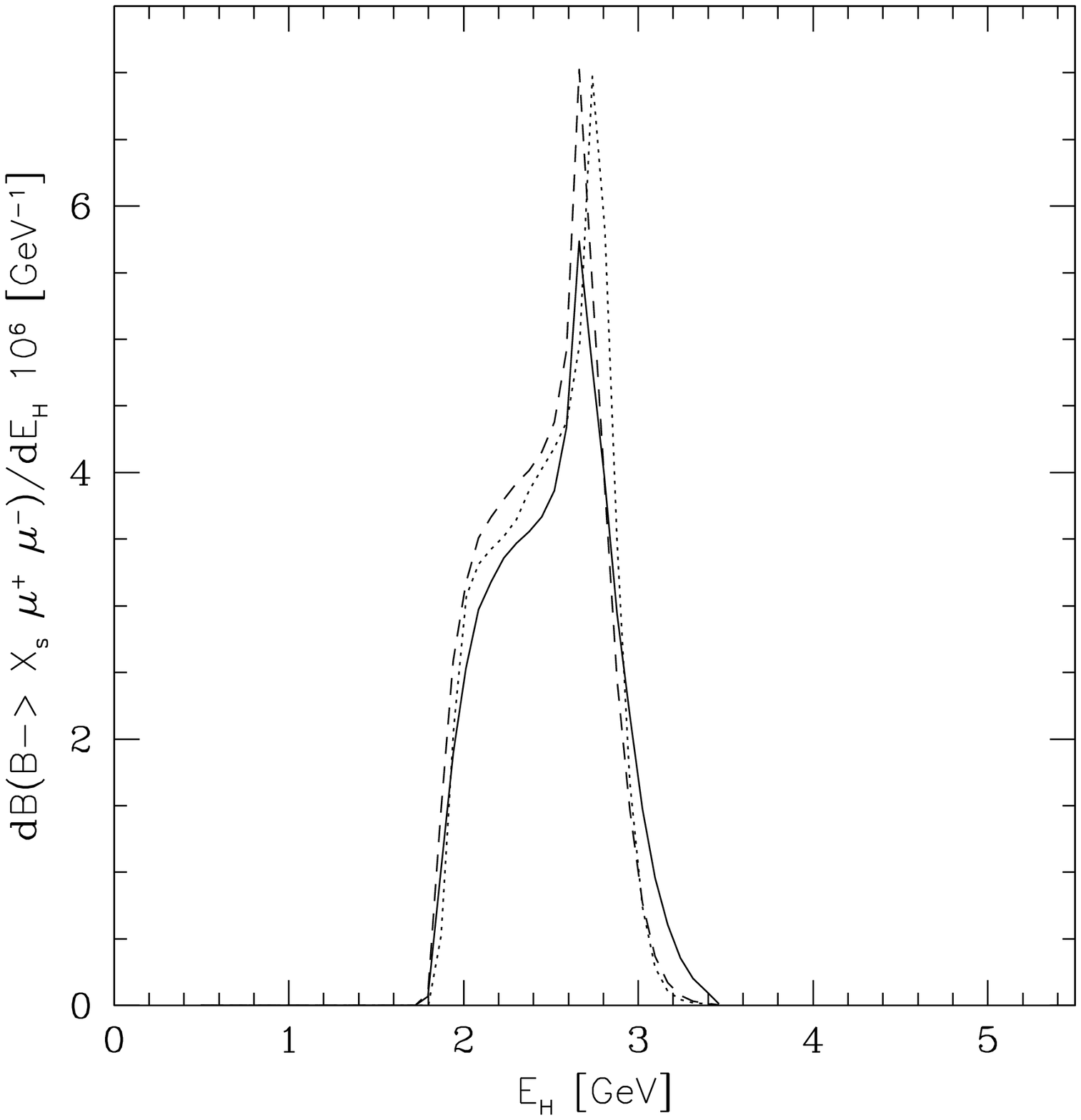,width=7.0cm}
     \end{minipage}
     \hspace{-0.4cm}
     \begin{minipage}[t]{7.0cm}
     \mbox{ }\hfill\hspace{1cm}(b)\hfill\mbox{ }
     \epsfig{file=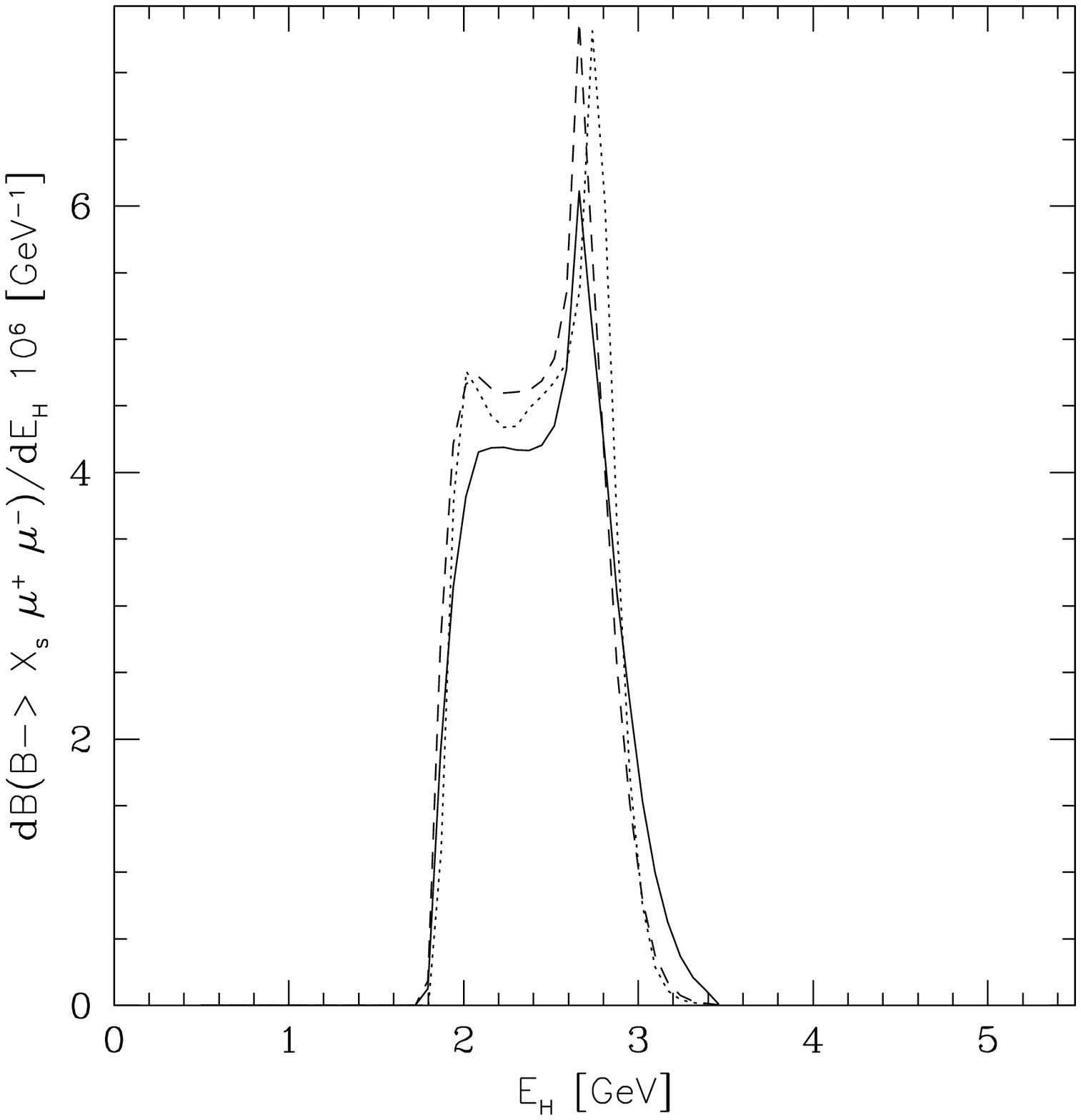,width=7.0cm}
     \end{minipage} \\ 
 \mbox{ }\hspace{-0.7cm}
     \begin{minipage}[t]{7.0cm}
     \mbox{ }\hfill\hspace{1cm}(c)\hfill\mbox{ }
     \epsfig{file=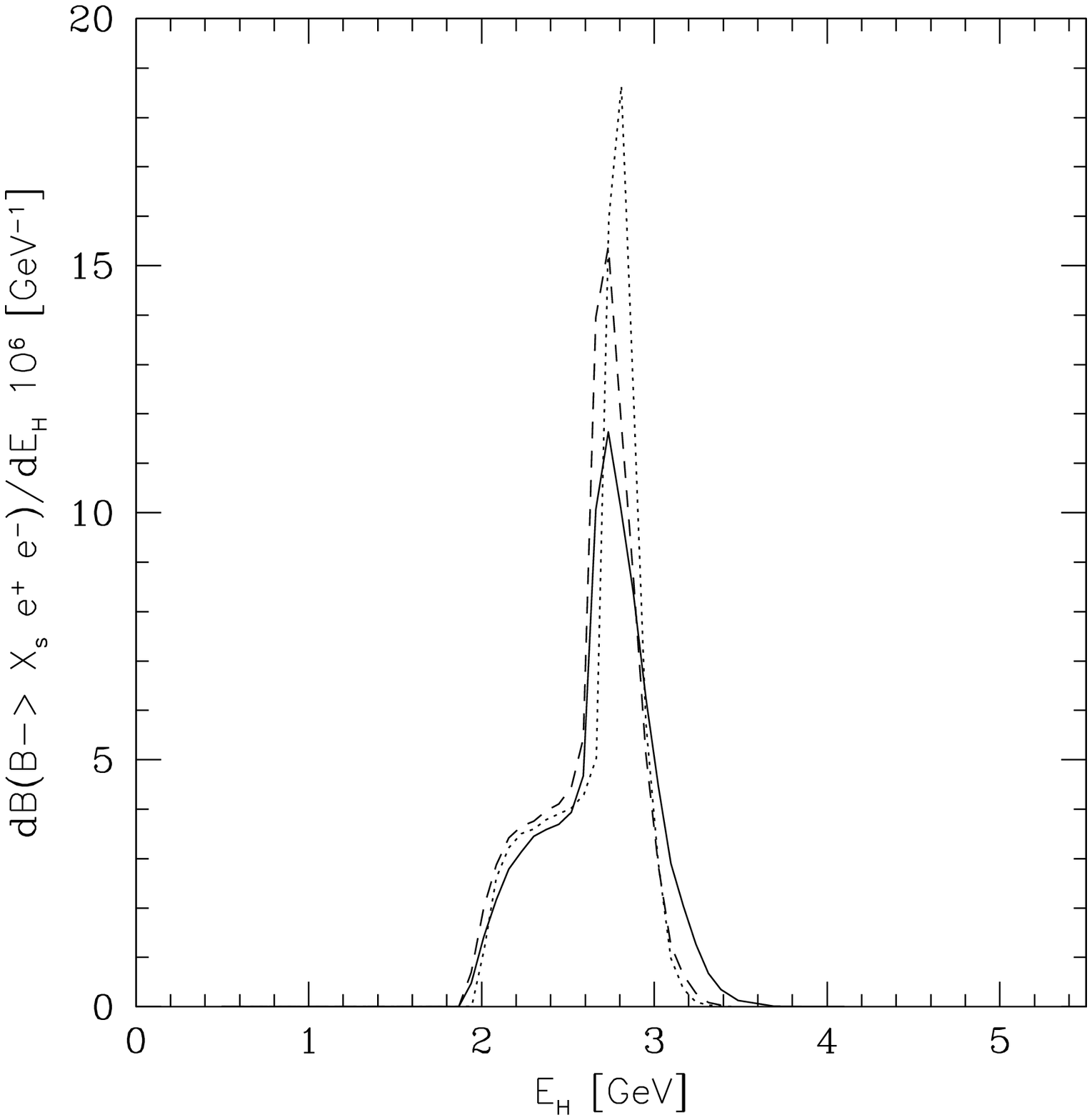,width=7.0cm}
     \end{minipage}
     \hspace{-0.4cm}
     \begin{minipage}[t]{7.0cm}
     \mbox{ }\hfill\hspace{1cm}(d)\hfill\mbox{ }
     \epsfig{file=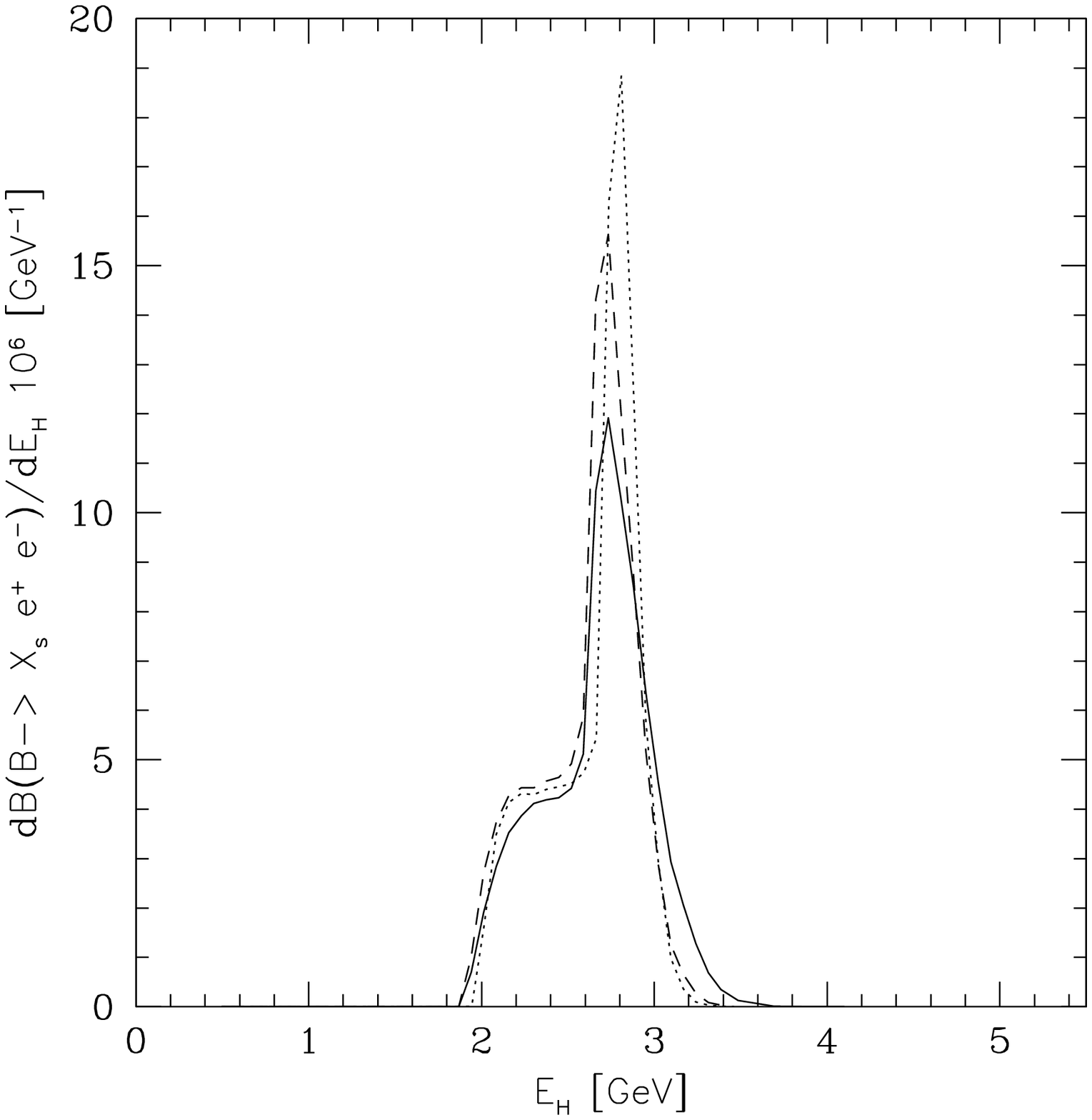,width=7.0cm}
     \end{minipage} \\ 
 \mbox{ }\hspace{-0.7cm}
     \begin{minipage}[t]{7.0cm}
     \mbox{ }\hfill\hspace{1cm}(e)\hfill\mbox{ }
     \epsfig{file=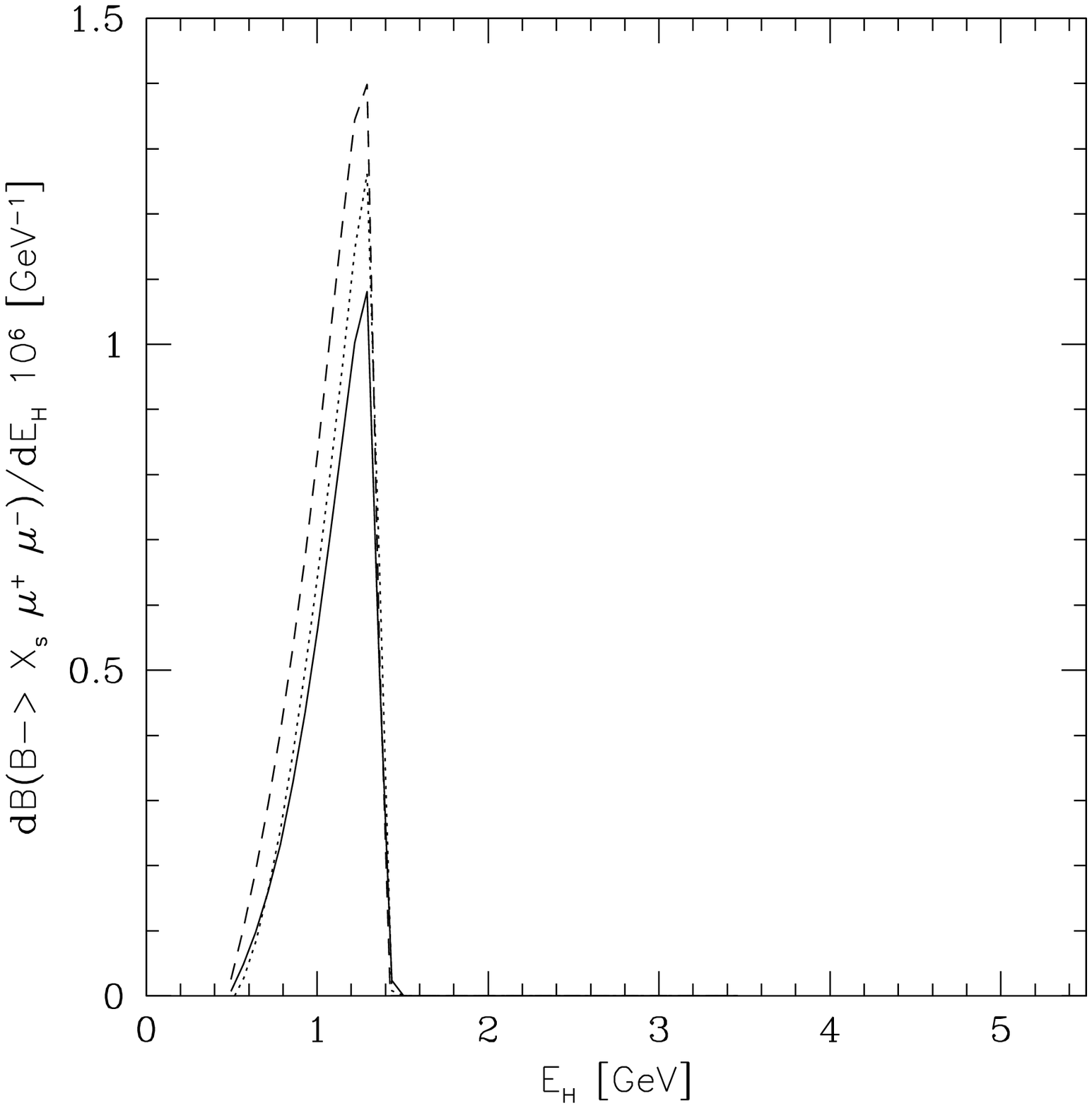,width=7.0cm}
     \end{minipage}
     \hspace{-0.4cm}
     \begin{minipage}[t]{7.0cm}
     \mbox{ }\hfill\hspace{1cm}(f)\hfill\mbox{ }
     \epsfig{file=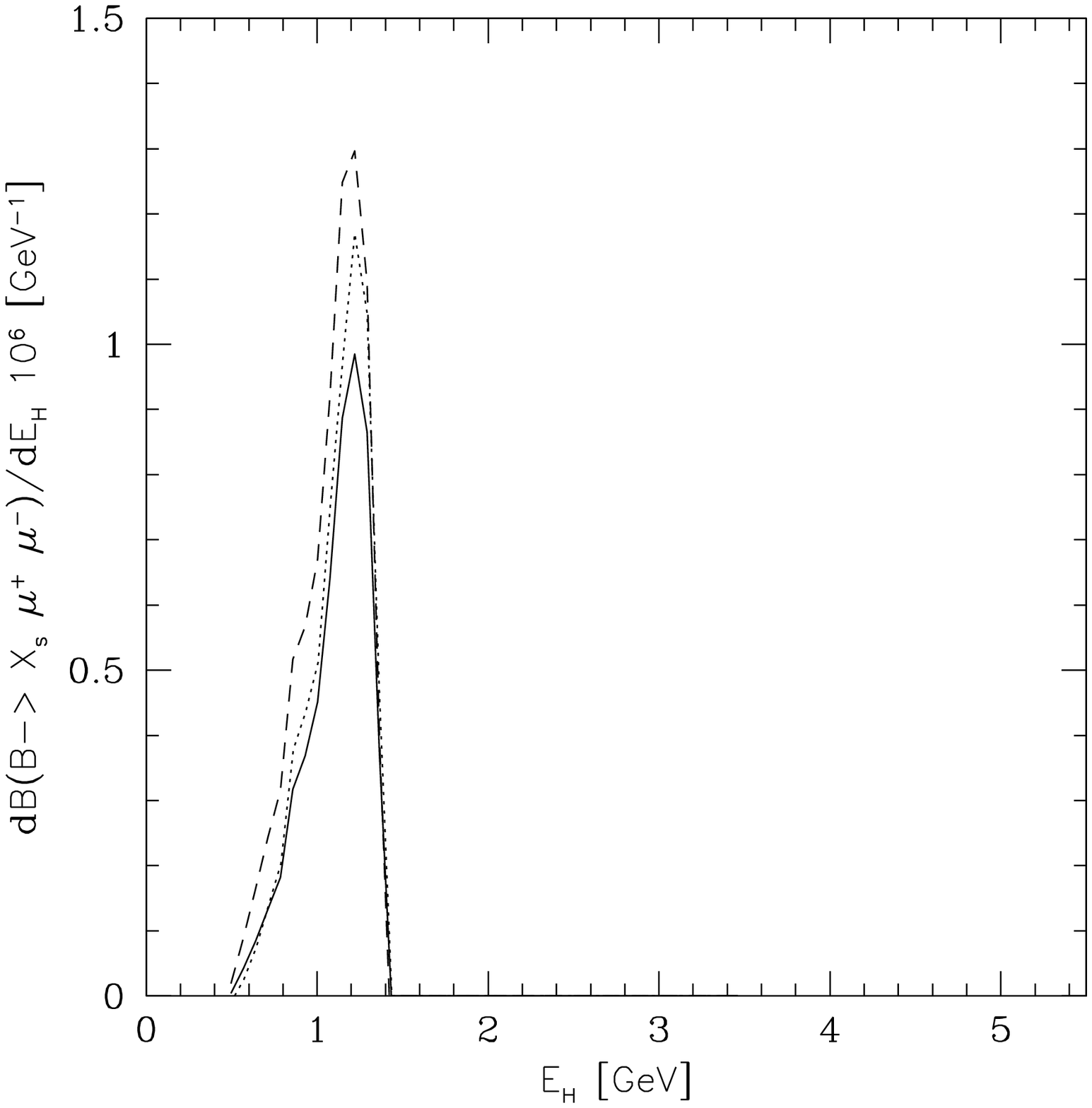,width=7.0cm}
     \end{minipage}
\end{center}  
     \caption{\it 
Hadron energy spectrum in \bxsll in the Fermi motion model with the
cuts on the dilepton mass defined in eq.~(\ref{eq:cuts}); (a),(c),(e) without 
and (b),(d),(f) with the $c\bar{c}$-resonance contribution corresponding to cut
A,B,C, respectively. The  
solid, dotted, dashed curves correspond to the parameters
$ (\lambda_1, \bar{\Lambda})=(-0.3,0.5),(-0.1,0.4),(-0.15,0.35)$ in
(GeV$^2$, GeV), respectively.
}\label{fig:EhLO}
\end{figure}
%
%
% 
%  SD and SD+LD for hadronic inv mass
% with cuts 
%
\begin{figure}[t]
\mbox{}\vspace{-2cm}\\
\begin{center}
     \mbox{ }\hspace{-0.7cm}
     \begin{minipage}[t]{7.0cm}
     \mbox{ }\hfill\hspace{1cm}(a)\hfill\mbox{ }
     \epsfig{file=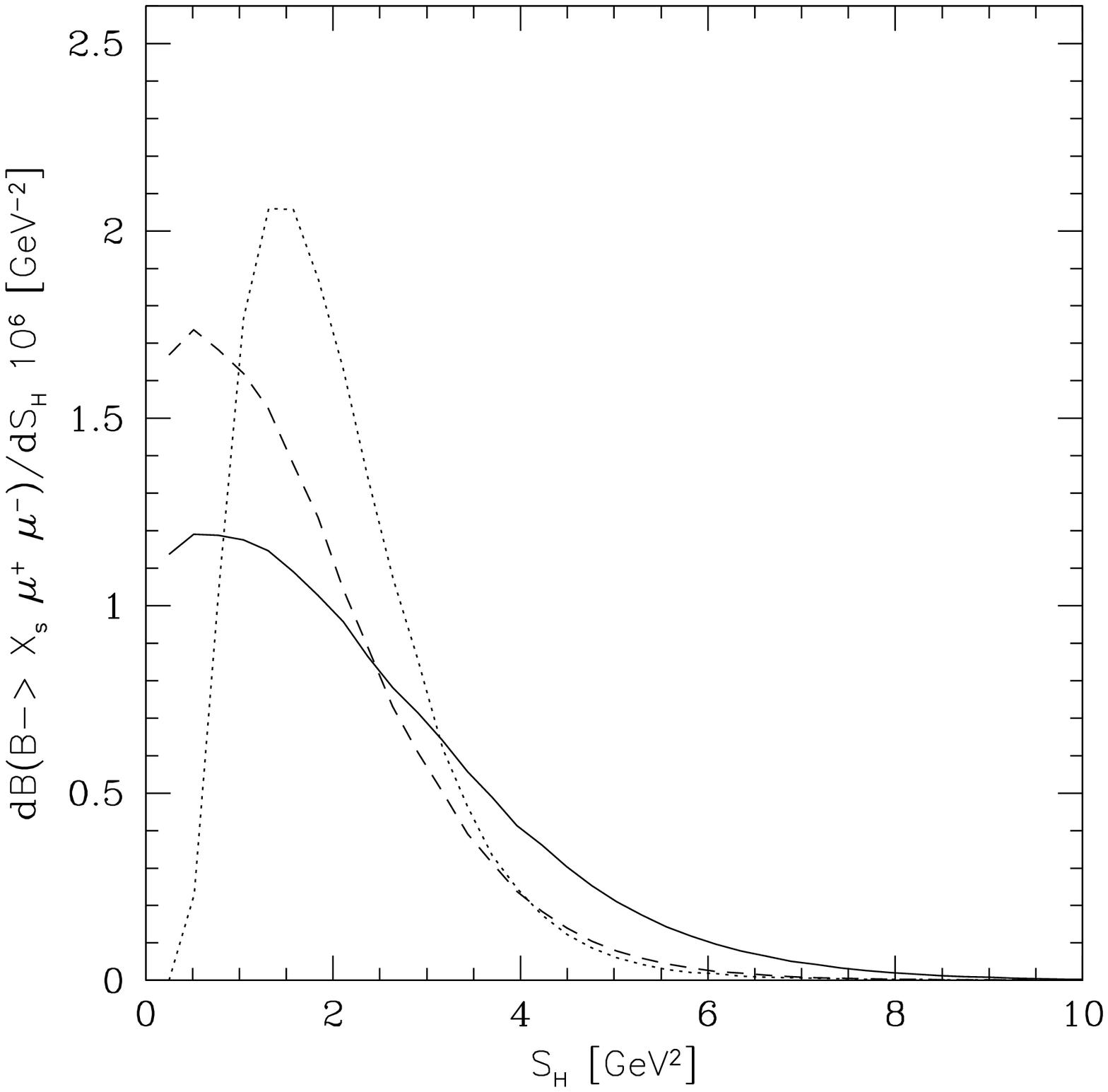,width=7.0cm}
     \end{minipage}
     \hspace{-0.4cm}
     \begin{minipage}[t]{7.0cm}
     \mbox{ }\hfill\hspace{1cm}(b)\hfill\mbox{ }
     \epsfig{file=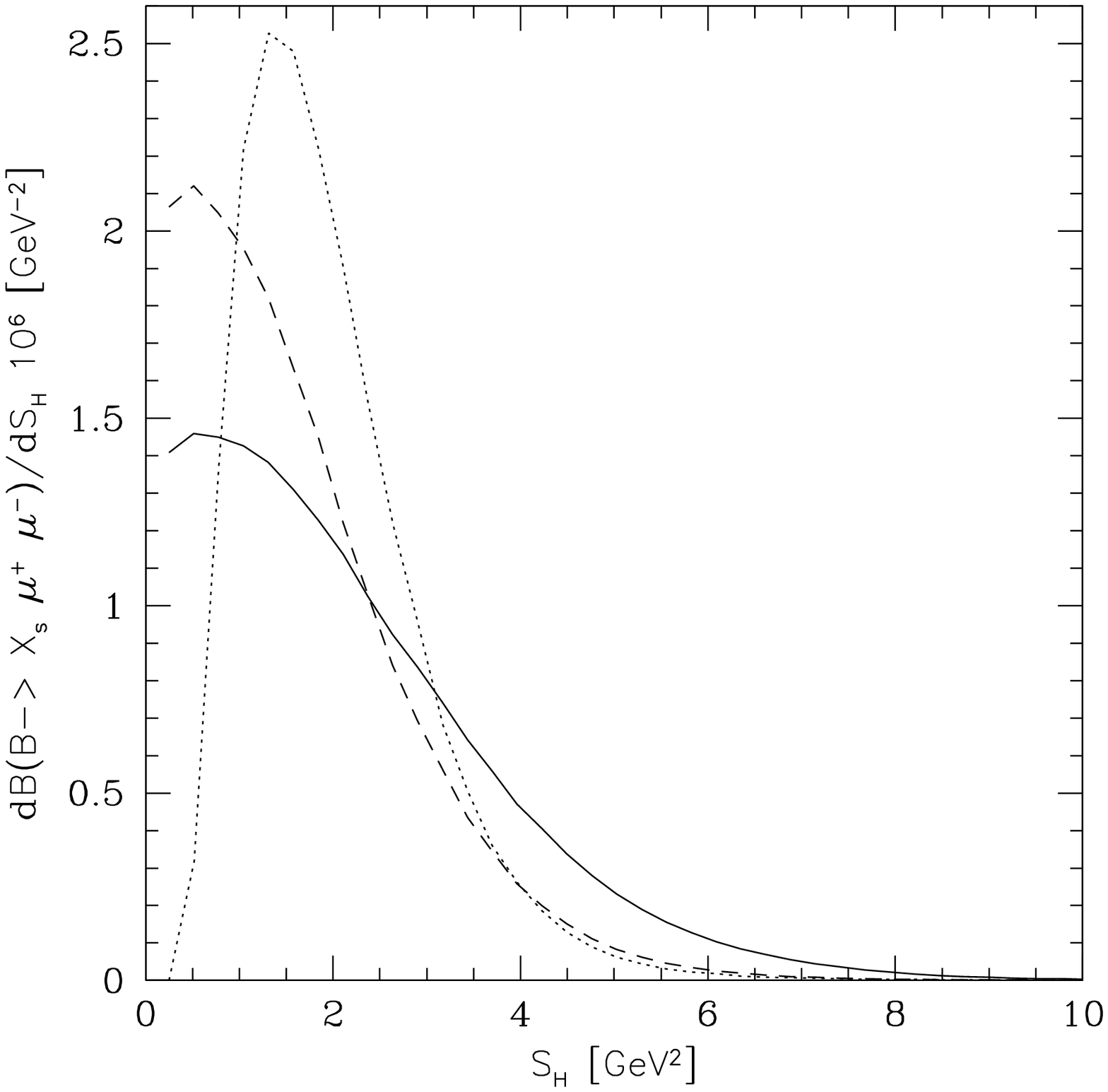,width=7.0cm}
     \end{minipage} \\ 
 \mbox{ }\hspace{-0.7cm}
     \begin{minipage}[t]{7.0cm}
     \mbox{ }\hfill\hspace{1cm}(c)\hfill\mbox{ }
     \epsfig{file=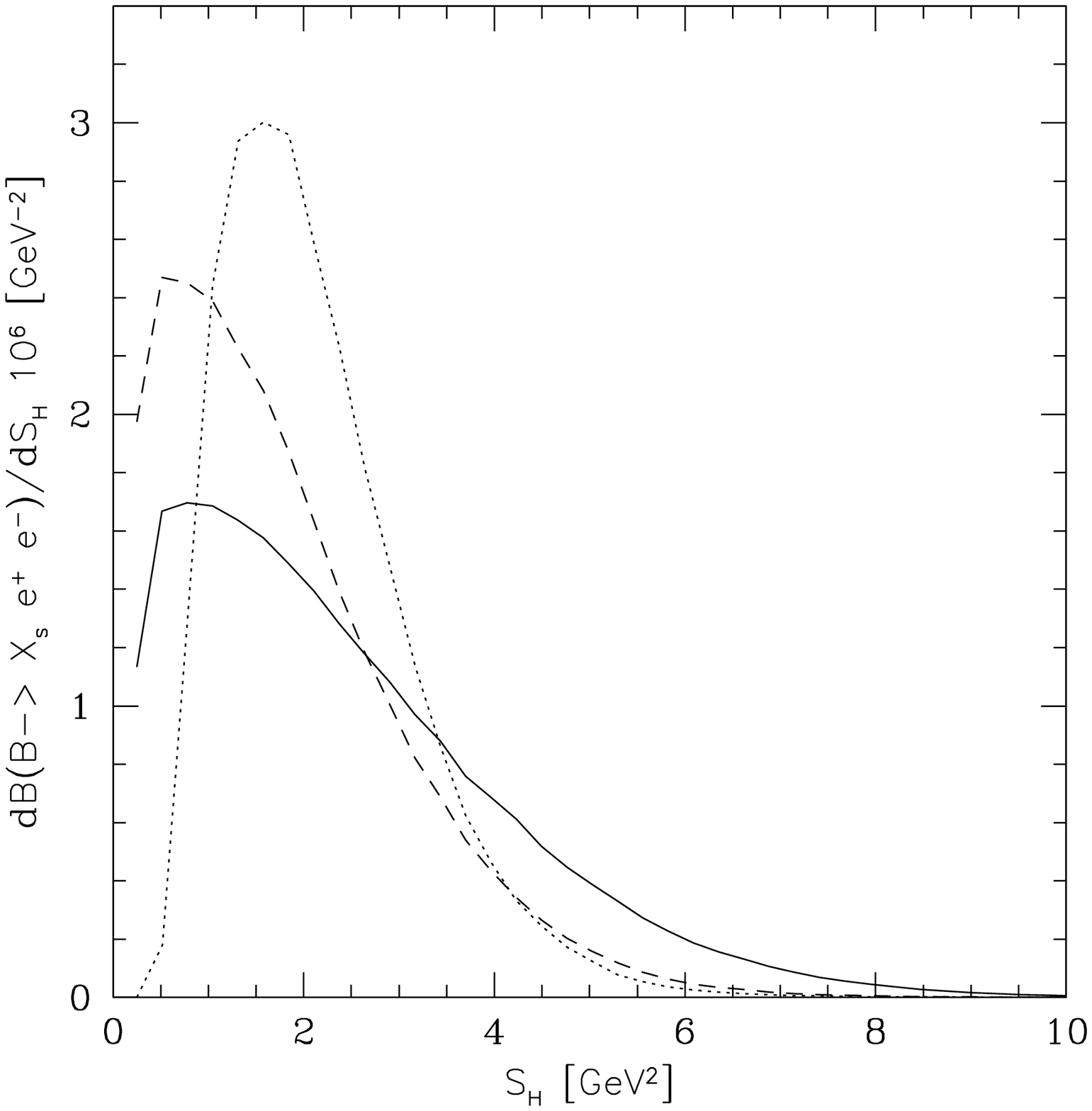,width=7.0cm}
     \end{minipage}
     \hspace{-0.4cm}
     \begin{minipage}[t]{7.0cm}
     \mbox{ }\hfill\hspace{1cm}(d)\hfill\mbox{ }
     \epsfig{file=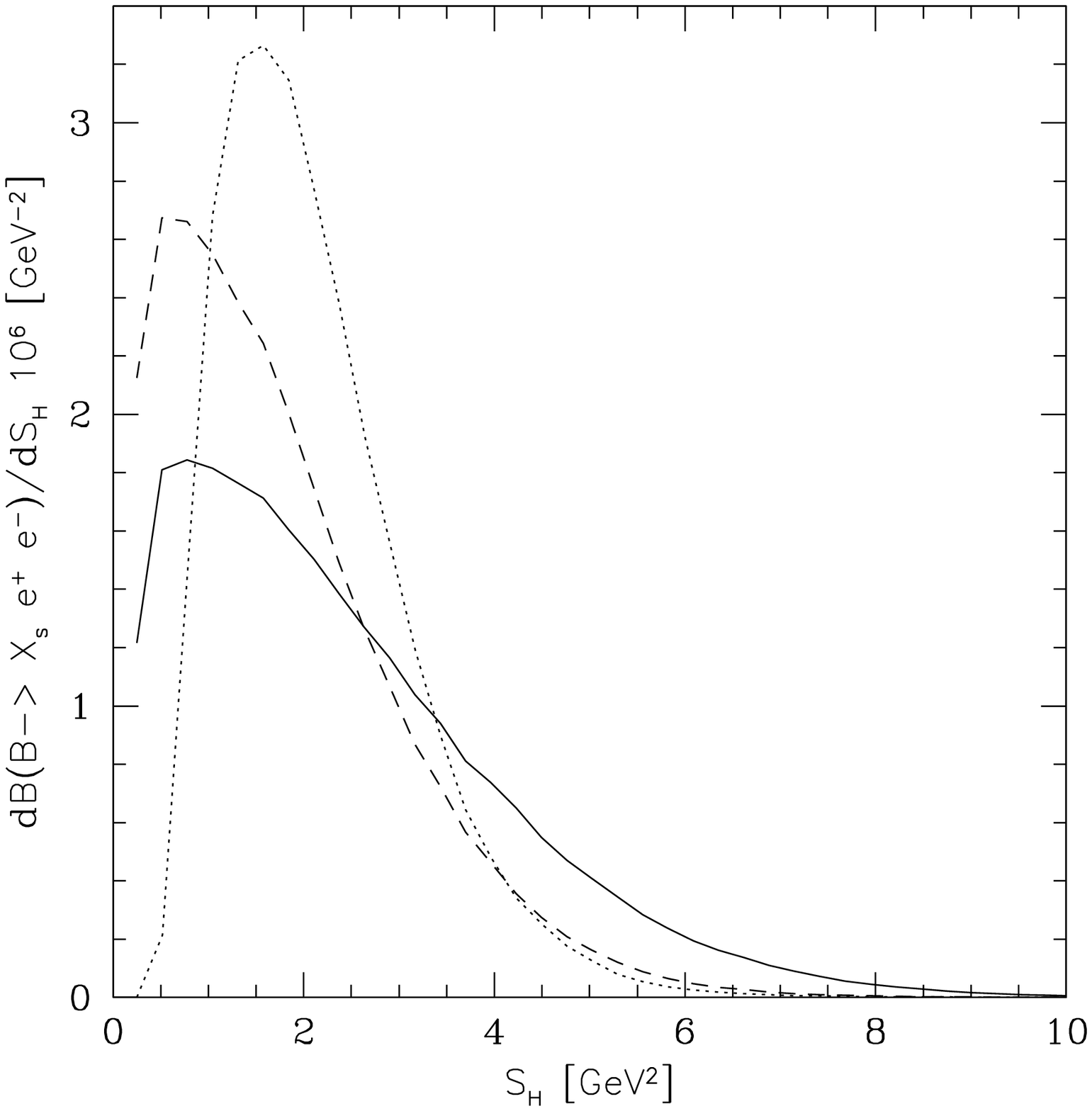,width=7.0cm}
     \end{minipage} \\ 
 \mbox{ }\hspace{-0.7cm}
     \begin{minipage}[t]{7.0cm}
     \mbox{ }\hfill\hspace{1cm}(e)\hfill\mbox{ }
     \epsfig{file=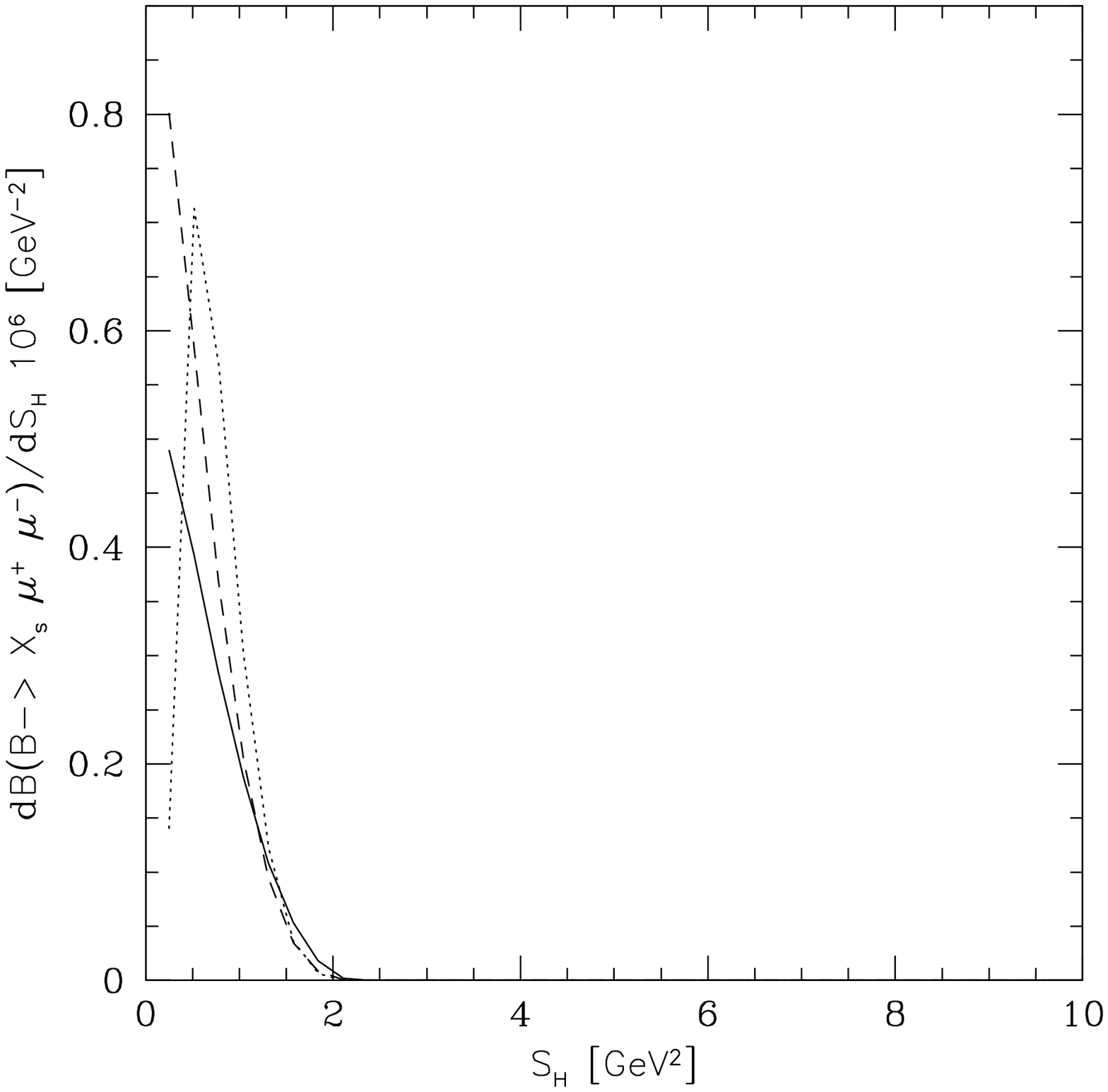,width=7.0cm}
     \end{minipage}
     \hspace{-0.4cm}
     \begin{minipage}[t]{7.0cm}
     \mbox{ }\hfill\hspace{1cm}(f)\hfill\mbox{ }
     \epsfig{file=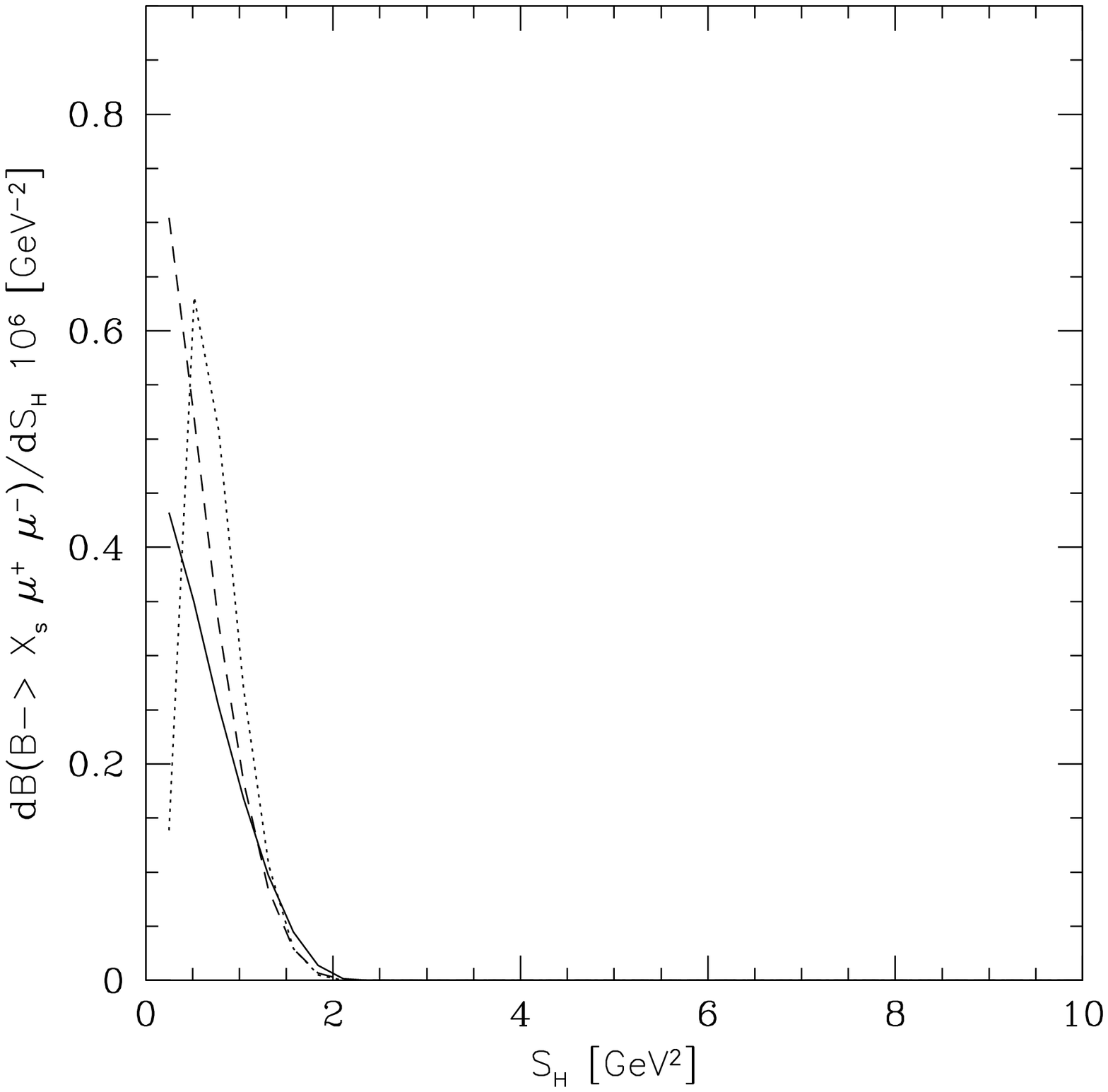,width=7.0cm}
     \end{minipage}
\end{center}  
     \caption{\it 
Hadronic invariant mass spectrum in \bxsll in the Fermi motion model with the
cuts on the dilepton mass defined in eq.~(\ref{eq:cuts}); (a),(c),(e) without 
and (b),(d),(f) with the $c\bar{c}$-resonance contribution corresponding to cut
A,B,C, respectively. The  
solid, dotted, dashed curves correspond to the parameters
$ (\lambda_1, \bar{\Lambda})=(-0.3,0.5),(-0.1,0.4),(-0.15,0.35)$ in
(GeV$^2$, GeV), respectively.
}\label{fig:ShLO}
\end{figure}
\section{Summary and Concluding Remarks on the Decay \bxsll}

In this chapter we have investigated distributions, decay rates and moments in
rare inclusive \bxsll decay in the standard model.
In the first part, we have concentrated mainly on the study of
distributions in the dilepton invariant mass $q^2$, the differential 
branching ratio in this variable and the FB asymmetry.
Our findings can be summarized as follows \cite{AHHM97}:
\begin{itemize}
\item We have calculated the leading $1/m_b$ power corrections with HQE
techniques in the dilepton invariant mass distribution in \bxsll decay
and have explicitly kept the $s$-quark dependence.
Our calculation is at variance 
with an earlier one \cite{FLSold} in the limit $m_s=0$ and has been confirmed 
recently by \cite{buchallaisidorirey} for the massless $s$-quark case.
\item 
We find that the $1/m_b^2$ corrections are stable over a good part of the 
dilepton mass spectrum. However, near the high $q^2$ end-point the
distribution becomes unphysical due to the HQE parameter $\lambda_2$, 
signaling a breakdown of the heavy quark expansion. 
\item The \bxsll decay rate in the HQE method 
decreases by about $4 \%$ and the branching ratio by about $1.5 \%$ from their
corresponding parton model values. 

\item Alternatively, we have implemented a Gaussian Fermi motion model in the 
decay \bxsll to model $B$-meson bound state effects. 
We have analysed the dilepton invariant mass distribution and the FB asymmetry 
within this framework, showing the dependence on the parameters of this model.
Non-perturbative effects are found to be perceptible in both distributions in 
the high $q^2$ region.

\item The theoretical uncertainties in the short-distance branching ratios in 
\bxsll decay are found to be $\pm 23 \%$ ($\pm 16 \%$) for the electron 
(muon) case in the FM.
\item We have modeled the long-distance contributions from intermediate 
charmonium resonances with a VMD ansatz and presented the dilepton invariant 
mass distribution and the FB asymmetry in the FM, including
next-to-leading order perturbative QCD corrections in figures.
\end{itemize}
We have completed the description of final states in  \bxsll decay in
the second part of this chapter, which is devoted to the study of hadron 
spectra and 
hadronic spectral moments. We summarize \cite{AH98-1,AH98-2,AH98-3}:
\begin{itemize}
\item We have calculated the ${\cal O}(\alpha_s)$ perturbative 
QCD and leading ${\cal O}(1/m_b)$ corrections to the hadron spectra in the 
decay \bxsll, including the Sudakov-improvements in the
perturbative part.
\item We find that the hadronic invariant mass spectrum is calculable
in HQET over a limited range $S_H > m_B \bar{\Lambda}$ and it depends 
sensitively on the parameter $\bar{\Lambda}$ (equivalently $m_b$). 
These features are qualitatively very
similar to the ones found for the hadronic invariant mass spectrum in 
the decay $B \to X_u \ell \nu_\ell$ \cite{FLW}.
\item The $1/m_b$-corrections to the parton model hadron energy spectrum in 
\bxsll are small over most part of this spectrum. However, heavy quark
expansion breaks down near the lower end-point of this spectrum and 
close to the
$c\bar{c}$ threshold. The behavior in the former case 
has a similar origin as the breakdown of HQET near the high end-point
in the dilepton invariant mass spectrum, which we have presented here and in
ref.~\cite{AHHM97}.
\item We have calculated the hadronic spectral moments $\langle S_H^n 
\rangle$ and $\langle E_H^n \rangle$ for $n=1,2$ using HQET.
The dependence of these moments on the HQET parameters is worked out 
numerically. In particular, the moments $\langle S_H^n
\rangle$ are sensitive to the parameters $\bar{\Lambda}$ and $\lambda_1$
and they provide complementary constraints on them than the ones
following from the analysis of the decay $B \to X \ell \nu_\ell$.
The simultaneous fit of the data in \bxsll and $B \to X \ell \nu_\ell$ could 
then be used to determine these 
parameters very precisely. This is illustrated in Fig.~\ref{fig:laml1}.
\item The corrections to the hadron energy
moments $\triangle \langle E_H \rangle_{1/m_c^2}$ and
$\triangle \langle E_H^2 \rangle_{1/m_c^2}$ from the leading
${\cal{O}}(\Lambda^2_{QCD}/m_c^2)$ power corrections have been worked out, 
using the results of \cite{buchallaisidorirey}. 
We find that these corrections are very
small. The corresponding corrections in $\triangle \langle S_H^n 
\rangle_{1/m_c^2}$ vanish in the theoretical accuracy we are working.
\item
We think that the quantitative knowledge of $\bar{\Lambda}$ and 
$\lambda_1$ from the moments can be used to remove much of the theoretical
uncertainties in the partially integrated decay rates 
in $B \to X_u \ell \nu_\ell$ and \bxsll. Relating the two decay rates would 
enable a precise determination of the CKM matrix element $V_{ub}$.

\item As a phenomenological alternative to HQET, we have worked out the 
hadron spectra and moments in \bxsll in the Fermi motion model.
We find that the hadron energy spectrum is stable
against the variation of the FM model parameters. However, the hadronic
invariant mass is sensitive to the input parameters. Present
theoretical dispersion on this spectrum can be reduced with the experimental
measurements of the corresponding spectrum in the decay $B \to X_u \ell 
\nu_\ell$, which will determine these parameters. Conversely, with good
measurements of the decay \bxsll, one could fix the input parameters in
the decay $B \to X_u \ell \nu_\ell$ and determine the CKM matrix element
$V_{ub}$ rather precisely. 

\item 
For equivalent values of the FM and HQET parameters, 
the hadronic spectral moments in \bxsll decay are found to be remarkably 
close to each other.

\item We have worked out the effect of the $c\bar{c}$ resonances
in the effective coefficient $C_9^{\mbox{eff}}(\hat{s})$ on the
hadron spectra in \bxsll, by parametrizing the present data on the
resonant part from the decays $B \to X_s (J/\psi,\psi^\prime,...)$.
The resonances are incorporated at the parton level and the broadening
of these resonances from the wave function effects in the FM model are then 
worked out. These spectra will provide an important test of the FM model
in \bxsll.

\item We find that the $c \bar{c}$ long-distance effects in \bxsll decay are 
also important in the hadronic moments.

\item We have quantitatively studied possible double counting effects 
which may have entered in simply adding 
the resonant contribution via Breit-Wigner functions and the 
complete perturbative  $c\bar{c}$ contribution in
the coefficient $C_9^{\mbox{eff}}(\hat{s})$ in \bxsll.
The numerical difference between this approach, followed here 
\cite{AHHM97,AH98-3}, and alternative ones 
\cite{KS96} and \cite{LSW97}, are found to be small in the dilepton 
invariant mass
spectrum and negligible in the hadron spectra and moments. Theoretical
spectra are found to be more sensitive to the wave function effects,
which dominate the uncertainty in the shape. 
\item We have worked out the hadron spectra by imposing the experimental
cuts designed to suppress the resonant $c\bar{c}$ contributions,
as well as the dominant $B\bar{B}$ background
leading to 
the final state $B\bar{B} \to X_s \ell^+ \ell^-$ (+ missing energy). The
parametric dependences of the resulting spectra are studied in the FM 
model. In particular, the survival probability of the \bxsll signal
by imposing a cut on the hadronic invariant mass $S_H < 3.24 
~\mbox{GeV}^2$, as used in the CLEO analysis, is estimated and its model
dependence studied.
This quantifies the statement that with the indicated cuts, these spectra 
essentially test the physics 
of the short-distance (and non-resonant $c\bar{c}$) contribution.

\end{itemize}
The CLEO collaboration has already been searched for inclusive \bsll decay 
with $\ell=e,\mu$. Their results are \cite{cleobsll97}:
\begin{eqnarray}
{\cal{B}}(b \to s e^{+} e^{-})^{\mbox{{\small{CLEO}}}}& < &
5.7 \cdot 10^{-5} \; ,  \nonumber \\
{\cal{B}}(b \to s \mu^{+} \mu^{-})^{\mbox{{\small{CLEO}}}}& < &
5.8 \cdot 10^{-5} \; .
\end{eqnarray}
Comparing this with our estimates for the SD branching ratios
in the decay \bxsll given in eq.~(\ref{eq:Brnumbers}), we see that the above 
CLEO upper bounds are approximately one order of magnitude away from the 
theoretical standard model prediction.

%We hope that the work presented here will contribute to precise 
%determinations of the HQET parameters and $V_{ub}$ using the inclusive decays
%\bxsll~and $B \to X_u \ell \nu_\ell$ in forthcoming B facilities.

%%%%%%%%%%%%%%%%%%%%%%%%%%%%%%%%%%%%%%%%%%%%%%%%%%%%%%%%%%%%%%%

\clearpage
\chapter{The Decay $B_s \to \gamma \gamma$ 
\label{chap:bsgg}}

%{\it "seems to me, I could live my life, a lot better than I think I am,
%I guess that's why they call me, they call me the working man! "} \\
%"working man", Rush

Besides the rare decays \bxsg and \bxsll, $B_{s}\rightarrow \gamma \gamma$ is 
another potential candidate to explore perturbative and non-perturbative 
aspects of QCD and test the standard model (SM). 
The L3 collaboration has already been searched 
for $B_{s}\rightarrow \gamma \gamma$ decay.
Their upper bound is the best present limit of this channel
\cite{L3} 
\begin{equation}
{\cal B}(B_s \rightarrow \gamma \gamma) < 1.48 \cdot 10^{-4} \, \, .
\label{eq:L3}
\end{equation}
The first theoretical analysis of a rare decay into 2 photons is contained
in the pioneering work by Gaillard and Lee \cite{gaillard}, who considered
$K_{S,L} \to \g \g$.
Exclusive $B_{s}\rightarrow \gamma \gamma$ decay has been 
investigated in the lowest order in 
%refs.~\cite{yao,simma,herrlichkalinowski,aliev}.
refs.~\citer{yao,aliev}.
The branching ratio found is 
$4.5 \cdot 10^{-7}$ in the SM context for $m_s=0.5$ GeV and 
other parameters given in Table~\ref{parameters}.
The large value of the $s$-quark mass here results from using the constituent 
quark mass $m_s \sim m_K$.
As learned from studies of \bxsg decay, the flavour changing neutral current
(FCNC) $b \to s \g$ vertex 
receives in leading logarithmic (LLog) QCD approximation a large enhancement 
about a factor of 2-3. We suggest similar effects for 
the $b \to s \g \g $ transition.
To see whether $B_s \rightarrow \g \g $ decay is worth more effort to be 
searched for at future experiments,
a more advanced analysis of its branching ratio is required.
A branching ratio of order $10^{-6}$ is a benchmark of a decay to be 
measured at present $B$ experiments like CLEO or Hera-B 
with reasonable statistics.
Upgrades and planned $B$-factories will be 
sensitive to branching ratios of order $10^{-8}$.
However, the $B_s \bar{B}_s$ pair is too heavy to be produced at the 
$\Upsilon(4s)$ resonance.

The $B_s \to \gamma \gamma$ final state consists of a CP-odd $T^-$ and a 
CP-even $T^+$ amplitude. It offers,
besides the branching ratio, another observable, the so called 
CP ratio $|T^+|^2/|T^-|^2$ \cite{herrlichkalinowski,aliev}. 
With it CP violating effects can be studied.

The work reported in this thesis, which has already been published
\cite{gudi}, \cite{hilleriltanpsi} differs from the previous 
ones \cite{aliev,simma,yao} with respect to three points:
a) We calculate and use the QCD-improved LLog amplitudes.
b) In contrast to previous works using the constituent quark model
we model the bound state effects of the $B_s$ meson through an 
heavy quark expansion technique (HQET) inspired approach following \cite{MW}.
This introduces an additional dispersion on the 
$B_{s}\rightarrow \gamma \gamma$ branching ratio and CP ratio, 
which serves as an estimate of the hadronic uncertainties.
c) We include long distance effects due to 
decay chains via intermediate vector mesons in our analysis. 
To be definite,
we estimate the additional contribution in the decay 
$B_{s}\rightarrow \gamma \gamma$ through $B_{s}\rightarrow \phi \gamma$
followed by $\phi\rightarrow \gamma$ using 
Vector Meson Dominance (VMD) \cite{sakurai}. 
Further, the $B_{s}\rightarrow \phi\psi$ decay is modeled by 
inclusive $b\rightarrow s\psi$ decay. Using the VMD model,
the amplitude for the chain process 
$B_{s}\rightarrow\phi\psi\rightarrow\phi\gamma\rightarrow\gamma\gamma$
is presented.

Leading logarithmic QCD corrections for the
short-distance part of the decay $B_s \rightarrow \gamma \gamma$ 
have also been calculated by Chang et al. \cite{yaonew}. 
They fix $\bar{\Lambda}=m_{B_s}-m_b$, which corresponds to $m_s$ in the naive 
constituent quark model, and the renormalization scale $\mu=m_b$. 
We emphasize here that the decay rate (and the CP ratio) is sensitive to 
both of these parameters and requires further theoretical investigation.
Soni et al. \cite{sonireina} calculated also LLog QCD calculations in
$B_{s}\rightarrow \gamma \gamma$ and in addition in the decay 
$B \rightarrow X_s \gamma \gamma$.
Analyses of $B_s \rightarrow \gamma \gamma$ decay in non standard models have 
been done in ref.~\cite{aliev} in the lowest order and in 
ref.~\cite{alievhillererhan}
including LLog QCD corrections
in the two Higgs doublet model (2HDM), in ref.~\cite{matias97} in the 
minimal supersymmetric model (MSSM) and in ref.~\cite{alieverhan98} in 
the 2HDM with flavour changing neutral currents allowed at tree level.

\section{Leading Logarithmic Improved Short-Distance Contributions in 
$B_{s}\rightarrow \gamma \gamma$ Decay \label{sec:llog}}
In this section we present the leading logarithmic QCD-improved rates for
exclusive $B_s \rightarrow \gamma \gamma$ decay. 
We use the free quark model and make the connection between quark and mesons 
states by means of the $B_s$ meson decay constant $f_{B_{s}}$. 
Dot products of kinematical variables, which are not fixed by this
are estimated in an approach inspired by HQET.

QCD-improved rates in $b$-quark decays
can in general be obtained through the following procedure:
Matching of the full theory with an effective theory at a scale $\mu=m_W$,
using an effective Hamiltonian
and performing an evolution of the Wilson coefficients from 
$m_W$ down to $\mu \sim {\cal O}(m_b)$, thus 
resumming all large logarithms of the form 
$\alpha_s^n(m_b) \ln^m(\frac{m_b}{m_W})$, where $m \leq n$ 
($n=0,1,2, \ldots$). In the leading logarithmic approximation, which we use
here, $m=n$.
In our case, which is $B_s \to \g \g$ now 
an enormous short cut is possible from observing that the effective 
Hamiltonian in eq.~(\ref{eq:operatorbasis}) for $b \rightarrow s \gamma$ 
is identical for $b \rightarrow s \gamma \gamma$ to this order 
of $\frac{1}{m_W^2}$:
\begin{eqnarray}
\label{eq:heff2gamma}
{\cal{H}}_{eff}(b\to s \gamma)={\cal{H}}_{eff}(b\to s \gamma \gamma) 
+{\cal{O}}(\frac{1}{m_W^4}) \; .
\end{eqnarray}

The proof goes as follows:
One can ask, if there are more operators needed for $b \to s \gamma \gamma$ 
than included in ${\cal{H}}_{eff}(b\to s \gamma)$ 
and try to find an operator $\bar{s} X b$ with $dim(X) \leq 3$ 
and containing two photons! Here
$X$ must be a gauge and Lorentz invariant structure made out of
quark and photon fields, masses and covariant derivatives 
$D_{\mu}=\partial_\mu+i e Q_q A_\mu$.
(For the moment we shall work in zeroth order of the strong interactions.)
When constructing the full set of physical operators, the equations of 
motion (EOM) can be used to reduce the operator basis:
\begin{eqnarray}
i \Slash{D} q=m_q q \, ,
~~D_{\mu} F^{\mu \nu}=e Q_q \bar{q} \gamma^{\nu} q \, ,
~~\partial_{\mu} \tilde{F}^{\mu \nu}=0  \; .
\end{eqnarray}
Here $F^{\mu \nu}, \tilde{F}^{\mu \nu}
=\frac{1}{2} \epsilon^{\mu \nu \alpha \beta} F_{\alpha \beta}$ 
denote the photon field strength tensor and its dual, respectively.
For chiral fermions we have the following EOM, which can be obtained with
the identities given in eq.~(\ref{eq:chiralfields})
\begin{eqnarray}
i \Slash{D} b_L=m_b b_R \; , 
~~\Bar{s_L} i \Slash{D}=m_s \Bar{s} L=m_s \Bar{s_R} \; . 
\end{eqnarray}
Other useful identities in this context are:
\begin{eqnarray}
~~D^2=\Slash{D}^2-\frac{1}{2} e Q_q  \sigma F  \; ,
~~D_{\mu}=\frac{1}{2}( \Slash{D} \gamma_{\mu}+\gamma_{\mu} \Slash{D} ) \; ,
~~[D_{\mu},D_{\nu}]=i e Q_q F_{\mu \nu} \; .
\end{eqnarray}
As a result, using the EOM gives either a mass, a current or
remains in the operator basis $O_{1 \dots 8}$ 
eq.~(\ref{eq:operatorbasis}) or gives contributions to the FCNC self energy. 
The latter will be absorbed in the on-shell renormalization
and does not give any contribution to $b\to s \gamma \gamma$ 
\begin{eqnarray}
\Sigma(p^2=m_b^2) b=\bar{s}\Sigma(p'^2=m_s^2)=0 \; . 
\end{eqnarray}
Here $p,p'$ is the 4-momentum of the incoming $b$-quark, outgoing $s$-quark,
respectively.
To display the foregoing we give some examples of EOM operator identities:
\begin{itemize}
\item $F_{\mu \nu} \Bar{s_L} \gamma^{\mu} D^{\nu} b_L =
\frac{1}{4} \bar{s} \sigma_{\mu \nu} (m_b R+m_s L) b F^{\mu \nu} \sim O_7$
\item $\bar{s} (i \Slash{D})^3 b=m_b^3 \bar{s} b $
\item $m_b \bar{s} D^2 b=-m_b^3 \bar{s} b 
-\frac{m_b}{2} e Q_q \bar{s} \sigma F b $
\item $\bar{s}\gamma_{\nu} (D_{\mu} F^{\mu \nu}) b= e Q_q
\bar{s}\gamma_{\nu} b \bar{q} \gamma^{\nu} q$
\end{itemize}
Since after applying the equations of motion there exists no gauge-invariant 
FCNC-2-photon operator with field dimension $\leq 6$,
the set of operators given in eq.~(\ref{eq:operatorbasis}) is a complete basis 
for both $b\to s \g$ and $b \to s \g \g$ decay \cite{grinstein,grinstein90}.
Hence all the results obtained for the former, a collection of which can be
seen in section \ref{sec:effham}, can be used for the latter, 
like the LLog evolution of the Wilson coefficients.

\subsection{$B_s \to \g \g$ decay in the effective Hamiltonian theory}
Having convinced us that the set-up eq.~(\ref{eq:heff2gamma}) is correct, 
we can now turn to an explicit calculation of LLog improved FCNC 2 photon
amplitudes.
The amplitude for the decay $B_{s}\rightarrow \gamma \gamma$ can be 
decomposed as
\cite{yao,simma,aliev}
\begin{equation}
{\cal A}(B_{s}\rightarrow \gamma \gamma)=
\epsilon_1^{\mu}(k_1) \epsilon_2^{\nu}(k_2)
(A^{+} g_{\mu \nu} +
i A^{-} \epsilon_{\mu \nu \alpha \beta} k_1^{\alpha}  k_2^{\beta}) \, \, ,
\label{am}
\end{equation}
where the $k_i$ and $\epsilon_i^{\nu}(k_i)$ denote the four-momenta and the 
polarization vectors of the outgoing photons, respectively 
\footnote{We adopt the convention 
$Tr(\gamma^{\mu} \gamma^{\nu} \gamma^{\alpha} \gamma^{\beta} \gamma_5)=
4i\epsilon^{\mu \nu \alpha \beta}$, with $\epsilon^{0 1 2 3}=+1$.}.
Alternatively, we can write the amplitude  in terms of photon field strength
tensors:
\begin{equation}
{\cal A}(B_{s}\rightarrow \gamma \gamma)=
T^{+} F^1_{\mu \nu} F^{\mu \nu}_2 +
i T^{-} F^1_{\mu \nu} \tilde{F}^{\mu \nu}_2 \; .
\label{eq:am2}
\end{equation}
We have for real photons $k_i^2=0, \epsilon_i . k_i=0$ with $i=1,2$, 
$k_1 . k_2=m_{B_s}^2/2$ and further for a $B_s$ decaying at rest the
additional conditions $ \epsilon_1 . k_2 =\epsilon_2 . k_1 =0$.
Then the following equations result:
\begin{eqnarray}
A^{+}=- m_{B_s}^2 T^+ \; ,
~~A^{-}=2 T^- \; .
\end{eqnarray}
Also, in the rest frame of the $B_s$ meson, the $CP=-1$ amplitude $A^{-}$
is proportional to the perpendicular spin polarization
$\vec{\epsilon_1}\times\vec{\epsilon_2}$, and the $CP=1$ amplitude $A^{+}$ is
proportional to the parallel spin polarization
$\vec{\epsilon_1}.\vec{\epsilon_2}$. 
The ratio 
\begin{eqnarray}
r_{CP} \equiv |T^{+}|^2/|T^{-}|^2= \frac{4 |A^{+}|^2}{ m_{B_s}^4|A^{-}|^2 }
\label{eq:rcp}
\end{eqnarray}
 can be used to study CP violating 
effects in $B_{s}\rightarrow\gamma\gamma$ decays. 
We will discuss it together with the $B_s \to \g \g $ branching ratio 
in section \ref{sec:LDo7estimate}.

Using the effective Hamiltonian in eq.~(\ref{eq:operatorbasis}),
the CP-even ($A^{+}$) and CP-odd ($A^{-}$) parts in the SM can be written 
as (for diagrams see Fig.~\ref{fig:o2} and Fig.~\ref{fig:o7})
in a HQET inspired approach:
\begin{eqnarray}
A^{+}&=&-\frac{\alpha G_F}{\sqrt{2} \pi} f_{B_s} \lambda_t 
\left( \frac{1}{3}
\frac{m^4_{B_s} (m_b^{eff}-m_s^{eff})}{\bar{\Lambda}_s 
(m_{B_s}-\bar{\Lambda}_s) (m_b^{eff}+m_s^{eff})} 
C_7^{\mbox{eff}}(\mu)
\nonumber
\right.\\
&-&
\left.
\frac{4}{9} \frac{m_{B_{s}^2}}{m_b^{eff}+m_s^{eff}}
(-m_b J(m_b)+ m_s J(m_s) ) D(\mu) 
\right)  ,\nonumber\\
A^{-}&=&-\frac{\alpha G_F}{\sqrt{2} \pi} 2 f_{B_s} \lambda_t 
\left( \frac{1}{3}
\frac{1}{m_{B_s} \bar{\Lambda}_s (m_{B_s}-\bar{\Lambda}_s)} g_{-}
C_7^{\mbox{eff}}(\mu) \nonumber
\right.\\
&-&
\left.
\sum_q Q_q^2 I(m_q) C_q(\mu) +
\frac{1}{9 (m_b^{eff}+m_s^{eff})} 
(m_b \triangle(m_b)+m_s \triangle(m_s)) D(\mu)
\right) \, \, ,
\label{amplitudes}
\end{eqnarray}
where we have used the unitarity of the CKM-matrix 
$\sum_{i=u,c,t} V_{is}^{*} V_{ib}=0 $ 
and have neglected the contribution due to 
$V_{us}^{*} V_{ub} \ll V_{ts}^{*} V_{tb}\equiv \lambda_t$. 
In eq.~(\ref{amplitudes}) $N_{c}$ is the colour factor ($N_c=3$ for QCD)
and $Q_q=\frac{2}{3}$ for $q=u,c$ and $Q_q=-\frac{1}{3}$ for $q=d,s,b$.
The QCD-corrected Wilson coefficients 
in leading logarithmic approximation \cite{effhamali,effhamburas},
which are discussed in section \ref{sec:effham}, 
$C_{1 \dots 6}(\mu)$ and $C_7^{\mbox{eff}}(\mu)$,
enter the amplitudes in the combinations
\begin{eqnarray}
\label{eq:combinations}
C_u(\mu)&=&C_d(\mu)=(C_3(\mu)-C_5(\mu)) N_c +C_4(\mu)-C_6(\mu) \, \, , 
\nonumber \\
C_c(\mu)&=&
(C_1(\mu)+C_3(\mu)-C_5(\mu)) N_c +C_2(\mu)+C_4(\mu)-C_6(\mu) \, \, ,
\nonumber \\
C_s(\mu)&=&C_b(\mu)=
(C_3(\mu)+C_4(\mu))(N_c+1)-N_c C_5(\mu)-C_6(\mu) \, \, , \nonumber \\
D(\mu)&=&C_5(\mu)+C_6(\mu) N_c \, \, .
\end{eqnarray} 
The Feynman rules used are given in appendix \ref{app:feynrules}.
Note that the chromomagnetic operator $O_8$ does not contribute here in this 
order of $\alpha_s$.
The functions $I(m_q), \, J(m_q)$ and $\triangle(m_q)$ come from the 
irreducible diagrams with an internal 
$q$-type quark propagating, see Fig.~\ref{fig:o2}, and are defined as
\begin{eqnarray}
I(m_q)&=&1+\frac{m_q^2}{m_{B_s}^2} \triangle (m_q) \, \, , \nonumber \\
J(m_q)&=&1-\frac{m_{B_s}^2-4 m_q^2}{4 m_{B_s}^2} \triangle(m_q)  \, \, ,
\nonumber \\
\triangle(m_q)&=&\left(
\ln(\frac{m_{B_s}+\sqrt{m_{B_s}^2-4 m_q^2}}
         {m_{B_s}-\sqrt{m_{B_s}^2-4 m_q^2}})-i \pi \right)^2 
\, \,{\mbox{for}}\, \, \frac{m^2_{B_s}}{4 m_q^2} \geq 1 , \nonumber \\
\triangle(m_q)&=&-\left(
2 \arctan(\frac{\sqrt{4 m_q^2-m_{B_s}^2}}
         {m_{B_s}})- \pi \right)^2 
\, \,{\mbox{for}}\, \, \frac{m^2_{B_s}}{4 m_q^2} <\ 1 .
\label{eq:IJdelta}
\end{eqnarray}
Detailed analysis shows that the diagram Fig.~\ref{fig:o2}
requires the calculation of three different types of insertions:
The current-current operators $O_{1,2}$ only give a contribution to a charm 
loop. They have the structure
$\gamma_{\mu} L \otimes \gamma_{\mu} L$, which leads after integration over 
the internal quark momentum to the function $I(m_q)$ given above. 
In contrast, each of the penguin operators
$O_{3 \dots 6}$ has two possible insertions, a ``direct" one and
one after Fierz ordering of the fields. 
In the former just internal $b$- and $s$-quarks appear and the operators have 
to be ``turned" by $\pm 90^o$ to generate a diagram consisting of one 
continuous $b \to s$ fermion line.  
In the latter the four-Fermi 
operators, which contribute to all 5 active flavours $q=u,d,s,c,b$, 
are rearranged with the help of the Fierz transformation given in 
appendix \ref{app:feynrules}. This simplifies the calculation as it 
circumvents a trace over $\gamma$-matrices. 
The procedure is legitimate since the resulting amplitude is 
(IR and UV) finite.
The operators $O_{3,4}$ have the same Dirac structure as $O_{1,2}$
$\sim \gamma_{\mu} L \otimes \gamma_{\mu} L$
which is reproduced after Fierz transformation, see eq.~(\ref{eq:fierzLL}).
Therefore, here no new integrals appear.
The operators $O_{5,6}$ are of $\gamma_{\mu} L \otimes \gamma_{\mu} R$
type. ``Direct" insertion leads to the functions $\triangle(m_q),J(m_q)$ 
given above. They contribute to the $B_s \to \g \g $ amplitude only via an
internal $s$- and $b$-quark.
Here care must be taken of the left-right structure, which is different for 
the $s$- and the $b$-quark and results in the sign difference in the
corresponding term proportional to $D(\mu)$ in the CP-even amplitude 
$A^+$ given in eq.~(\ref{amplitudes}).
The Fierz transformation of $O_{5,6}$ results in a scalar/pseudoscalar 
coupling $\sim R \otimes L $, see eq.~(\ref{eq:fierzLR}). 
The analytical expression 
for such an insertion is minus the one for 
$\gamma_{\mu} L \otimes \gamma_{\mu} L$, which can be checked after explicit 
calculation. From here the minus signs in the functions 
eq.~(\ref{eq:combinations}) can be understood.

The parameter $\bar{\Lambda}_s$ enters eq.~(\ref{amplitudes}) through the 
bound state kinematics. 
At the quark and meson level, the decay kinematics are given  
\begin{eqnarray}
b(p) &\to& s(p') \gamma (k_1,\epsilon_1) \gamma(k_2 ,\epsilon_2) \; , \\
B_s(P)& \to& \gamma (k_1,\epsilon_1) \gamma(k_2 ,\epsilon_2) \; ,
\end{eqnarray}
respectively.
A problem lies now in the intermediate propagators of the reducible diagrams, 
see Fig.~\ref{fig:o7}, where
we need to evaluate $p.k_i$ and $p'.k_i$, $i=1,2$.
The answer cannot be given just by using kinematics, energy/momentum 
conservation in a chosen frame and a model is necessary here.
For definiteness, we consider the decay 
$B_s\equiv (\bar{b} s) \rightarrow \gamma \gamma$.
We write the momentum of the $\bar{b}$-quark inside the meson as
$p=m_b v+k$, where $k$ is a small residual momentum, $v$ is the 4-velocity,
which connects the quark with the meson kinematics through $P=m_{B_s} v$
and $P$ is the momentum of the meson. In the $B_s$ rest frame, $v=(1,0,0,0)$.
Now following \cite{MW}, we average the residual momentum of the 
$\bar{b}$-quark through 
\begin{eqnarray}
<k_{\alpha}>&=&
-\frac{1}{2 m_b} (\lambda_1+3 \lambda_2) v_{\alpha}\, \, ,\nonumber\\
<k_{\alpha} k_{\beta}>&=&\frac{\lambda_1}{3} 
(g_{\alpha \beta } -v_{\alpha} v_{\beta}) \, \, ,
\end{eqnarray}
where $\lambda_1, \lambda_2$ are matrix elements from the heavy quark 
expansion. Using $P=p-p'$, $P.k_i=\frac{m^2_{B_s}}{2}$,  
$v.k_i=\frac{m_{B_s}}{2}$ and the HQET relation \cite{MW}
\begin{equation}
m_{B_s}=m_b +\bar{\Lambda}_s-\frac{1}{2 m_b} (\lambda_1+3 \lambda_2)
\label{hqe}
\end{equation}
one gets:
\begin{eqnarray}
p.k_i &=&\frac{m_{B_s}}{2} (m_{B_s}-\bar{\Lambda}_s) \, \, ,\nonumber \\
p'.k_i&=&-\frac{m_{B_s}}{2} \bar{\Lambda}_s \, \, ,\nonumber \\
(m_b^{eff})^2\equiv p^2&=&m_b^2-3 \lambda_2\, \, ,\nonumber \\
(m_s^{eff})^2\equiv p'^2&=&(m_b^{eff})^2
-m^2_{B_s}+2 m_{B_s} \bar{\Lambda}_s  \, \, .
\label{kin}
\end{eqnarray}
The non-perturbative parameter $\bar{\Lambda}_s$ can be related to 
$\bar{\Lambda}$, which 
has been extracted (together with $\lambda_1$) from data on semileptonic 
$B^{\pm},B^0$ decays in ref.~\cite{gremm}, and the measured mass
difference $\triangle m=m_{B_s}-m_B=90$ MeV \cite{PDG}, defining
$\bar{\Lambda}_s=\bar{\Lambda}+\triangle m$.
The matrix element $\lambda_2$ is well determined from the 
$B^{\ast}_{(s)}-B_{(s)}$
mass splitting, $\lambda_2=0.12 \, {\mbox{GeV}}^2$.
With the help of eq.~(\ref{hqe}), the correlated values of $\bar{\Lambda}$  
and $\lambda_1$ can be transcribed into a 
correlation between $\bar{\Lambda}_{(s)}$ and $m_b$.
We select 3 representative values
\footnote{We choose 
$(\lambda_1, \bar{\Lambda})=(-0.09,280),(-0.19,390),(-0.29,500)$ in 
$( {\mbox{GeV}}^2, {\mbox{MeV}} )$ from Fig.~1 in \cite{gremm}. }
$(m_b,\bar{\Lambda}_s)=(5.03,370),(4.91,480),(4.79,590)$ in 
$( {\mbox{GeV}}, {\mbox{MeV}} )$
to study the hadronic
uncertainties of our approach. 
Note that we assume here that $\lambda_1,\lambda_2$ are flavour independent.
Furthermore, we have used the definition 
\begin{eqnarray}
<0|\bar{s} \gamma_{\mu} \gamma_5 b|B_{s}(P)>&=&i f_{B_s} P_{\mu} \, \, ,
\end{eqnarray}
which leads together with
the off-shellness of the quarks inside the meson to the
matrix element of the pseudoscalar current 
\begin{eqnarray}
<0|\bar{s} \gamma_5 b|B_{s}(P)>&=&
-i f_{B_s} \frac{m^2_{B_s}}{m_b^{eff}+m_s^{eff}} \, \, .
\end{eqnarray}
The auxiliary function
$g_{-}=g_{-}(m_b^{eff},\bar{\Lambda}_s)$ is defined as
\begin{eqnarray}
g_{-}=m_{B_s}(m_b^{eff}+m_s^{eff})^2+
\bar{\Lambda}_s (m^2_{B_s}-(m_b^{eff}+m_s^{eff})^2) 
\,\, .
\end{eqnarray}
Note that in the limit $\bar{\Lambda}_s \to m_s$, 
$m_{b,s}^{eff} \to m_{b,s}$ and using $m_{B_s}=m_b+m_s$ we recover the 
result obtained by the constituent quark model \cite{yao,simma,aliev},
ignoring QCD corrections.
Using the above expressions, the partial decay width is then given by :
\begin{equation}
\Gamma(B_{s}\rightarrow \gamma \gamma)=\frac{1}{32 \pi m_{B_s}} 
(4 |A^{+}|^2+\frac{1}{2} m_{B_s}^4|A^{-}|^2) \, \, .
\label{br}
\end{equation}

Now, there are 2 new observations to be made:\\
First, the Wilson coefficients in eq.~(\ref{amplitudes}) depend on the scale 
$\mu$. Therefore, since the behaviour of these short-distance (SD) 
coefficients under renormalization is known from the studies of
$B \rightarrow X_s \gamma $ \cite{nlogreub,nlomisiak,effhamali,effhamburas}, 
one can give an improved width 
for $B_s \rightarrow \gamma \gamma$
by including the leading logarithmic QCD corrections by 
renormalizing the coefficients $C_{1\dots 6}$ and $C_7^{\mbox{eff}}$ from 
$\mu=m_W$ down to the relevant scale $\mu \approx {\cal O}(m_b)$.
The explicit ${\cal O}(\alpha_s)$ improvement in the decay width
$\Gamma(B_s \rightarrow \gamma \gamma)$ requires the calculation of a 
large number of virtual corrections, which we have not taken into account.
Varying the scale $\mu$ in the range
$\frac{m_b}{2} \leq \mu \leq 2 m_b$,
one introduces an uncertainty, which can be reduced only when the complete 
next-to-leading order (NLO)-analysis is available, similar to the recently 
completed calculation for 
the $B \rightarrow X_s \gamma$ decay \cite{nlogreub,nlomisiak}. 
%We identify the $b$-quark mass with its pole mass throughout this paper.
%
%
\begin{figure}[htb]
\vskip -0.6truein
\centerline{\epsfysize=7in
{\epsffile{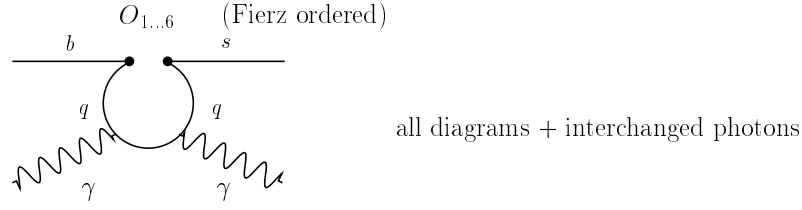}}}
\vskip -5.0truein
\caption[]{ \it The generic diagram contributing to $b \rightarrow s \gamma 
\gamma$ in the effective theory due to the (Fierz ordered) four-quark 
operators. 
The diagram with interchanged photons is not shown.}
\label{fig:o2}
\end{figure}
\begin{figure}[htb]
\vskip -0.6truein
\centerline{\epsfysize=7in
{\epsffile{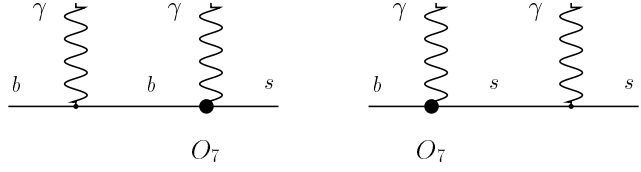}}}
\vskip -5.0truein 
\caption[]{ \it The reducible diagrams contributing to $b \rightarrow s \gamma 
\gamma$. The blob denotes the FCNC operator $O_7$. The diagrams with interchanged photons are not shown.}
\label{fig:o7}
\end{figure}

The second point concerns the strong dependence of the decay width
$\Gamma(B_s \rightarrow \gamma \gamma)$
on $\bar{\Lambda}_s$,
$\Gamma \sim {\cal{O}}(\frac{1}{\bar{\Lambda}_s^2})$ in eq.~(\ref{br}). 
It originates in the $s$-quark propagator in the diagram with an intermediate 
$s$-quark in Fig.~\ref{fig:o7}.
In the earlier 
work the authors of e.g. \cite{yao} evaluated the decay width with 
$m_s \approx m_K$, assuming that the constituent quarks are to be treated as 
static quarks in the meson.
This is a questionable assumption. 
In the HQET inspired approach, this gets replaced by $\bar{\Lambda}_s$, which 
is well-defined experimentally. This formalism implies, that the decay width
$\Gamma(B_d \rightarrow \gamma \gamma)$ will involve the parameter
$\bar{\Lambda}$, which avoids the unwanted uncertainty on $m_d$.

The branching ratio and the CP ratio as a function of 
the scale $\mu$ for different values of $(m_b,\bar{\Lambda}_s)$ are
discussed in section \ref{sec:LDo7estimate} including the
$O_7$-type long-distance (LD) estimate.

%\subsubsection{supplement, sidestep}
In the lowest order, which can be recovered at $\mu=m_W$, $C_7$ and $C_2$
are the only remaining non-zero Wilson coefficients.
The reducible diagram (1PR), which is proportional to $C_7$, contributes to
$A^\pm$, however, the irreducible one (1PI) ($\sim C_2(m_W)=1$), 
represented by the charm loop, enters only the CP-odd amplitude $A^{-}$.
In contrast, in neutral pion decays $\pi^0 \to \gamma \gamma$ the  
electromagnetic vector 
coupling results in an CP-odd amplitude and hence $A^+=0$ and the CP ratio
$\sim |A^+|^2/|A^{-}|^2$ vanishes.
The authors of \cite{yao}, analysed the $b \to s \gamma \gamma$ transition
in the lowest order in the full theory (SM). This amounts in the calculation
of in total $2 \times34$ diagrams (the factor 2 is due to the diagrams with interchanged photons).
They further interpreted the total 1PI amplitude as 
a (local) FCNC 2 photon operator with canonical field dimension 8:
\begin{equation}
O_{\g \g }=(Q_c e)^2 \Bar{s} \left( F_{\mu \nu}^1 \pa^{\nu} 
\Tilde{F}_2^{\mu \beta}+ 
F_{\nu \mu}^2 \pa^{\mu} \Tilde{F}_1^{\beta \nu}  \right) \gamma_{\beta} 
L b \; .
\end{equation}
After applying the EOM to the photonic part of $O_{\g \g}$ the following 
Lorentz structure is obtained in the $B_s$ rest frame
\begin{equation}
X=\left[ \epsilon_1.\epsilon_2 ( \slash{k_1}-\slash{k_2})-
\slash{\epsilon_1} \slash{\epsilon_2} ( \slash{k_1}-\slash{k_2}) \right] L \; ,
\end{equation}
which leads to the relation $O_{\g \g }=(Q_c e)^2 k_1.k_2  \Bar{s} X b$
and the 1PI amplitude:
\begin{equation}\label{heff}
{\cal A}(B_s \to  \gamma \gamma)_{1PI}(\mu=m_W)
  =  -\frac{4 G_F}{\sqrt{2}} \lambda_t C_{\g \g }(m_W) 
\langle O_{\g \g } \rangle  \; .
\end{equation}
Note that we used on-shell conditions for the photons.
The initial value of the ``Wilson'' coefficient is given as
\begin{equation}
C_{\g \g}(m_W)=-\frac{1}{k_1.k_2} (\frac{1}{8 \pi^2}) I(m_c) \; .
\end{equation}
Note that  $O_{\g \g}$ is the {\bf only} operator
with dimension $\leq 8$ after applying the EOM containing 2 photons and
2 fermions \cite{simma}.
One can try to renormalize this dim 8 operator as a point.
As a result, the leading order anomalous dimension of $O_{\g \g}$
vanishes.

%Axial anomaly

\section{QCD Sum Rule for 
        the $B_{s}\rightarrow\phi\gamma$ Form Factor \label{sec:qcdsumrule}}

In the description of exclusive $B$ decays 
hadronic matrix elements $<X|O_{i}|B>$ are involved.
Here $X$ is any hadron (with mass $m_X$) and the amplitudes just 
mentioned are  purely non-perturbative objects.
At present there are two methods to calculate them in a model independent way
and, depending on the mass value of X, one can be chosen. 
For so called ``heavy-to-heavy" transitions, where $X$ contains 
one heavy quark, the heavy quark expansion technique (HQET) is most 
appropriate.
Heavy quark symmetry implies that the form factors, 
which appear in the Lorentz decomposition of a decay into a hadron with 
certain spin, are all related to {\bf one} single function, 
the Isgur-Wise function $\xi(v.v')$ \cite{isgurwise}. 
Here $v,v'$ denote the velocities of the $B,X$, respectively.
Moreover, it can be shown that at the {\it zero recoil} point, 
that is where the final hadron is at rest in the rest frame of the 
decaying $B$, we have the normalization $\xi(1)=1$.
The behaviour of $\xi(v.v')$ for general values of the argument $v.v'\not=1$ 
cannot be calculated. It is of non-perturbative nature. The advantage is 
that there is just one universal function describing all transitions in the 
heavy quark limit.
From kinematical considerations one can get the possible range of the dot 
product as 
$v.v' =(m_B^2+m_X^2-q^2)/2 m_B m_X$, where $q^2=(m_B v-m_X v')^2$.
The maximal momentum transfer
$q^2_{max}=(m_B-m_X)^2$ corresponds to the minimal value of $v.v' =1$.
The decay under consideration $B_{s}\rightarrow \phi \gamma$ requires
$q^2=0$ for an on-shell photon, and we have $v.v'=2.73$. 
For such ``heavy-to-light" decay an extrapolation too far from the
zero recoil point is needed, and the QCD sum rule method is more useful.

The decay $B_{s}\rightarrow \phi \gamma$ is CKM allowed 
and, in the 
language of the operator basis given in eq.~(\ref{eq:operatorbasis}), 
involves the operator $O_7$, (see eq.~(\ref{eq:o7}) below).
It has been studied in the literature 
in the framework of Light-cone QCD sum rules \cite{alibraunsimma}, which is 
based on the approximate conformal invariance of QCD.
Here the sum rule is evaluated in terms of meson wave functions on the 
light-cone.
These universal functions with increasing twist replace the expansion
in ``classical" QCD sum rules into many vacuum expectation values of 
operators with increasing dimension. 
We show the calculation of the form factor $F_1$ in the decay
$B_{s}\rightarrow\phi\gamma$ in the framework of the ordinary QCD sum rules
\cite{shifman}, 
including the contribution from the gluon condensate \cite{gudi}.

\subsection{Calculation of the sum rule}
The amplitude for the  $B_{s}\rightarrow\phi\gamma$  transition 
${\cal{A}}(B_{s}\rightarrow \phi\gamma)= <\phi\gamma|{\cal{H}}_{eff}|B_{s}>$
reduces to
\begin{eqnarray}
{\cal{A}}(B_{s}\rightarrow \phi\gamma)
=\epsilon^{\mu} C m_b <\phi(p')|\bar{s}\sigma_{\mu\nu} R q^{\nu}b|B_{s}(p)>
\label{amphigam}
\end{eqnarray}
with the constant C 
\begin{eqnarray}
C=\frac{G_{F}}{\sqrt{2}}\frac{e}{2 \pi^2} 
V_{ts}^{*} V_{tb} C^{\mbox{eff}}_{7}(\mu)\, \, ,
\label{const}
\end{eqnarray}
where
we just take the contribution due to the electromagnetic penguin 
operator $O_7$
\begin{eqnarray}
 O_7 = \frac{e}{16 \pi^2}
          \bar{s}_{\alpha} \sigma_{\mu \nu} (m_b R + m_s L) b_{\alpha}
                F^{\mu \nu} \; ,
\label{eq:o7}
\end{eqnarray}
into account and put $m_s=0$, justified by
 $m_s \ll m_b$.
Here $\epsilon$ and q are the photon polarization and the (outgoing) photon 
momentum, respectively.
Lorentz decomposition gives further:
\begin{eqnarray}
<\phi(p')|\bar{s}\sigma_{\mu\nu} R q^{\nu}b|B_{s}(p)>&=&i \epsilon_{\mu\nu\rho
\sigma} \epsilon^{\phi \nu} p^{\rho} p'^{\sigma} F_{1}(q^{2})\nonumber\\
&+& (\epsilon^{\phi}_{\mu} p.q-p_{\mu} q.\epsilon^{\phi})G(q^{2}) \, \, ,
\label{f1g}
\end{eqnarray}
where $p, \, p'$ denote the four-momenta of the initial $B_s$-meson and the 
outgoing $\phi$, respectively and $\epsilon_{\mu}^{\phi}$ is the polarization 
vector of the $\phi$-meson.
At $q^{2}=0$ both form factors coincide \cite{alibraun} and it is sufficient 
to calculate $F_{1}(0)$.
Note, that the form factors introduced above are in general functions of two 
variables $q^2$ and $p'^2$. Since the $\phi$ is on-shell, we abbreviate here 
and in the following unless otherwise stated
$F_1(q^2) \equiv F_1(q^2, p'^2=m_{\phi}^2)$.

The starting point for the sum rule is the three-point 
function \cite{colangelo}
\begin{eqnarray}
T_{\alpha \mu}&=& -\int d^{4}x e^{ipx-ip'y}
<0|T[J_{\alpha} (x)T_{\mu}(0)J_{5}(y)]|0> \, \, ,
\label{talfabeta}
\end{eqnarray}
where $J_{\alpha}=\bar{s}\gamma_{\alpha}s$, $J_{5}=\bar{s}i\gamma_{5}b$ and
$T_{\mu}=\bar{s} \frac{1}{2} \sigma_{\mu\nu} q^{\nu} b$ correspond to the electromagnetic,
pseudoscalar currents and the penguin operator, respectively. 
Performing now an operator product expansion (OPE) of $T_{\alpha \mu}$, we 
obtain a perturbative term, the so-called bare loop, and non-perturbative power corrections, diagrammatically shown in Fig.~\ref{fig:diagrams}.
The bare loop diagram can be obtained using a double dispersion relation in
$p^{2}$ and $p'^2$,
\begin{eqnarray}
T_{bare}=\frac{1}{\pi^2} \int_{m_b^2}^{\infty} ds \int_0^{\infty} ds' 
\frac{\rho(s,s')}{(s-p^2)(s'-p'^2)}
+ {\mbox{subtractions}} \, \, .
\label{bare}
\end{eqnarray}
Technically, the spectral density $\rho(s,s')$ can be calculated by
using the Cutkosky rule, namely, by replacing the usual propagator 
denominator by a delta function: \\
$\frac{1}{k^{2}-m^{2}} \rightarrow -2\pi i \delta(k^{2}-m^2) \theta(k_0)$. 
As a result we get 
\begin{eqnarray}
\rho(s,s')=\frac{N_{c}}{8} m_{b}^{4}\frac{s'}{(s-s')^3} \, \, .
\end{eqnarray}

\begin{figure}[htb]
\vskip -0.5truein
\centerline{\epsfysize=12cm
{\epsffile{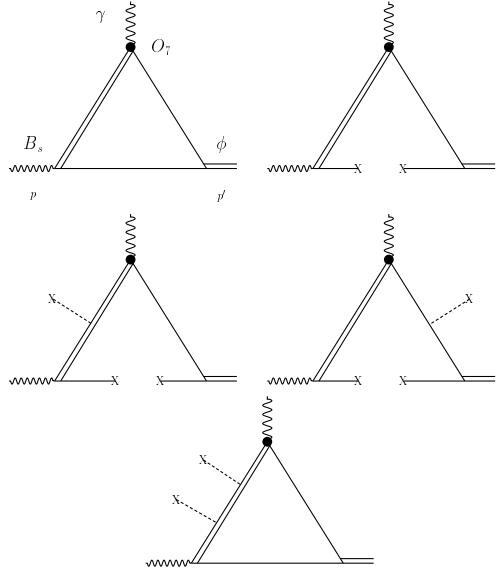}}}
\vskip -0.9truein
\caption[]{ \it
Contributions of perturbation theory and of vacuum condensates to the 
$B_{s}\rightarrow\phi\gamma$ decay. The dashed lines denote soft gluons.}
\label{fig:diagrams}
\end{figure}
OPE enables us further to parametrize the non-perturbative effects in 
terms of vacuum expectation values of gauge-invariant operators up to a 
certain dimension,  the so-called condensates.
We consider up to dimension-5 operators; i.e. the quark 
condensate, gluon condensate and the quark-gluon (mixed) condensate 
contributions 
(Fig.~\ref{fig:diagrams}). 
This calculation is carried out in the fixed point gauge, 
i.e. $A_{\mu} x^{\mu}=0$.
We get
\begin{eqnarray}
T_{dim-3}&=& \frac{-m_{b}}{2}
<\bar{s}s>\frac{1}{(p^{2}-m_{b}^{2}) p'^{2}}  \, \, ,\nonumber\\ 
T_{dim-4}&=&\frac{\alpha_s}{144 \pi} <G^2> \int_{0}^{1} dx \int_{0}^{1-x} dy
\int_{0}^{\infty} d \alpha \alpha^3 \nonumber \\ 
&\cdot& ( c_1 + c_2 P^2 + c_3 P'^{2}) e^{-\alpha (d_1 + d_2 P^2 +d_3 P'^2)}
\, \, , \nonumber \\
T_{dim-5}&=&\frac{m_{b}}{2}g<\bar{s}\sigma G s> [ \frac{m_{b}^{2}}
{2 (p^{2}-m_{b}^{2})^{3} p'^{2}}+\frac{m_{b}^{2}}{3
(p^{2}-m_{b}^{2})^{2} p'^{4}}
\nonumber \\ &+&\frac{1}{2 (p^{2}-m_{b}^2)^{2} p'^{2}} ] \, \, ,
\label{Ti}
\end{eqnarray}
where
%
% full ms
%\begin{eqnarray}
%c_1&=&(m_b+m_s) x^2 (3 m_s^3+ 4 m_b m_s^2 x - 3 m_s^3 x + m_b^3 x^2-
%m_b m_s^2 x^2)  \nonumber \\
%c_2&=&(m_b+m_s) x^3 (3 m_s + m_b x) (1 - x- y) \nonumber \\
%c_3&=&(m_b+m_s) x^2 (3 m_b x + 3 m_s y +m_b x y) (1 - x - y) \nonumber \\
%d_1&=&m_b^{2} x+ m_s^2 (1 - x) \nonumber \\
%d_2&=&x (1 - x - y ) \nonumber \\
%d_3&=&y (1 - x - y ) \, \, .
%\end{eqnarray}
%
\begin{eqnarray}
c_1&=&m_b^4 x^4  \, \, , \nonumber \\
c_2&=&m_b^2 x^4 (1 - x -y )  \, \, ,  \nonumber \\
c_3&=&m_b^2 x^3 (3+y) (1 - x - y)  \, \, ,  \nonumber \\
d_1&=&m_b^{2} x  \, \, ,  \nonumber \\
d_2&=&x (1 - x - y)  \, \, ,  \nonumber \\
d_3&=&y (1 - x - y ) \, \, .
\label{cidi}
\end{eqnarray}
Here we used the exponential representation for the gluon condensate
contribution:
\begin{equation}
\frac{1}{D^n}=\frac{1}{(n-1)!} \int_{0}^{\infty} d\alpha \, \, \alpha^{n-1}
e^{-\alpha D} \, \, .
\end{equation}
The momenta $P, \, P' $ in eq.~(\ref{Ti}) are Euclidean.

For the calculation of the physical part of the sum rules we insert a complete set of on-shell states with the same quantum numbers as  $B_{s}$ and $\phi$ in
eq.~(\ref{talfabeta}) and get a double dispersion relation
\begin{eqnarray}
T_{phys}=\frac{m_{B_s}^2 f_{B_s}} {m_{b}} f_{\phi} m_{\phi}
\frac{1}{(p^{2}-m_{B_s}^2)(p'^{2}-m_{\phi}^2)} 
F_{1}(0)+ {\mbox{continuum}} \, \, ,
\end{eqnarray}
where $f_{\phi}$ and $f_{B_s}$ are the leptonic decay constants of the $\phi$ 
and $B_{s}$ mesons respectively, defined as usual by
\begin{eqnarray}
<0|J_{\alpha}|\phi>&=&m_{\phi} f_{\phi}
\epsilon_{\alpha}^{\phi}  \, \, , \nonumber \\ 
<0|J_{5}|B_{s}(p)>&=&f_{B_s} m_{B_s}^{2}/m_{b} \, \, .
\end{eqnarray}
We have absorbed all higher order states and resonances in the continuum.

Now, we equate the hadron-world with 
the quark-world by $T_{phys}=T_{bare}+T_3+T_4+T_5$. 
Using quark-hadron duality, we model the continuum contribution by purely 
perturbative QCD. To be definite, it is the part in eq.~(\ref{bare})
above the so-called continuum thresholds $s_0$ and $s'_{0}$.
To get rid of subtractions and to 
suppress the contribution of higher order states, we apply a double Borel 
transformation $\hat{B}$ \cite{shifman} with respect to $p^2$ and $p'^{2}$. 
We make use of the following properties of the Borel transform:
\begin{eqnarray}
\hat{B}(\frac{1}{(p^2-m^2)^n})&=&\frac{(-1)^n}{(n-1)!} 
\frac{e^{-m^2/M^2}}{(M^2)^n}  \, \, , \\
\hat{B}(e^{-\alpha p^2})&=& \delta(1-\alpha M^2) \, \, .
\end{eqnarray}
Finally, this yields the sum rule:
\begin{eqnarray}
F_{1}(0)&=&
\exp(\frac{m_{B_s}^{2}}{M^{2}}+\frac{m_{\phi}^{2}}{M'^{2}}) \frac{m_{b}}
{f_{B_s} f_{\phi} m_{\phi} m_{B_s}^{2}} \{ \frac{1}{ \pi^{2}}
\int_{m_{b}^{2}}^{s_{0}} ds \int_{0}^{\bar{s}} ds' \rho (s,s')
e^{-s/M^{2}-s'/M'^{2}} \nonumber\\
&-&  \frac{m_{b}}{2} <\bar{s} s> e^{(-m_{b}^{2}/M^{2})} [1-
m_{0}^2(\frac{m_{b}^{2}}{4 M^{4}}+\frac{m_{b}^{2}}{3 M^{2} M'^{2}}-
\frac{1}{2 M^{2}} ) ] 
\nonumber \\
&+& \frac{\alpha_s}{\pi} <G^2> \int_{0}^{x_{max}} N(x) dx 
\} \, \, ,
\label{sumrule}
\end{eqnarray}
where $\bar{s}=min(s-m_{b}^{2},s'_{0})$
and $x_{max}=\frac{M'^{2}}{M^2+M'^{2}}$. 
Here we used the parametrization
\begin{eqnarray}
g<\bar{s} \sigma G s>&=m_{0}^{2} <\bar{s} s> \, \, .
\end{eqnarray}
\begin{figure}[htb]
\vskip -0.6truein
\centerline{\epsfysize=12.0cm
{\epsffile{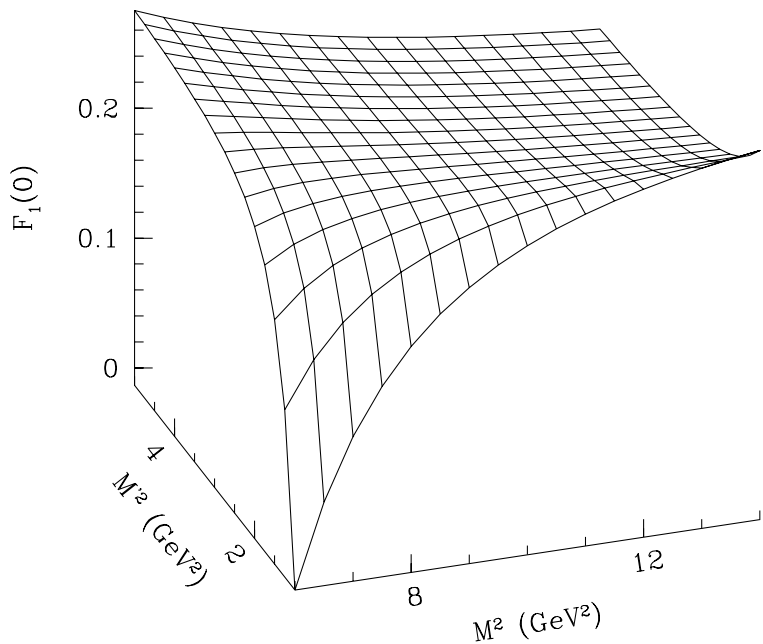}}}
\vskip -0.6truein
\caption[]{ \it The dependence of the decay constant $F_{1}(0)$
on the Borel parameters $M^{2}$ and $M'^{2}$ for 
$s_{0}=33\,\, {\mbox{GeV}}^{2}$.}
\label{fig:3d}
\end{figure}
\begin{figure}[htb]
\vskip -2.0truein
\centerline{\epsfysize=12.0cm
{\epsffile{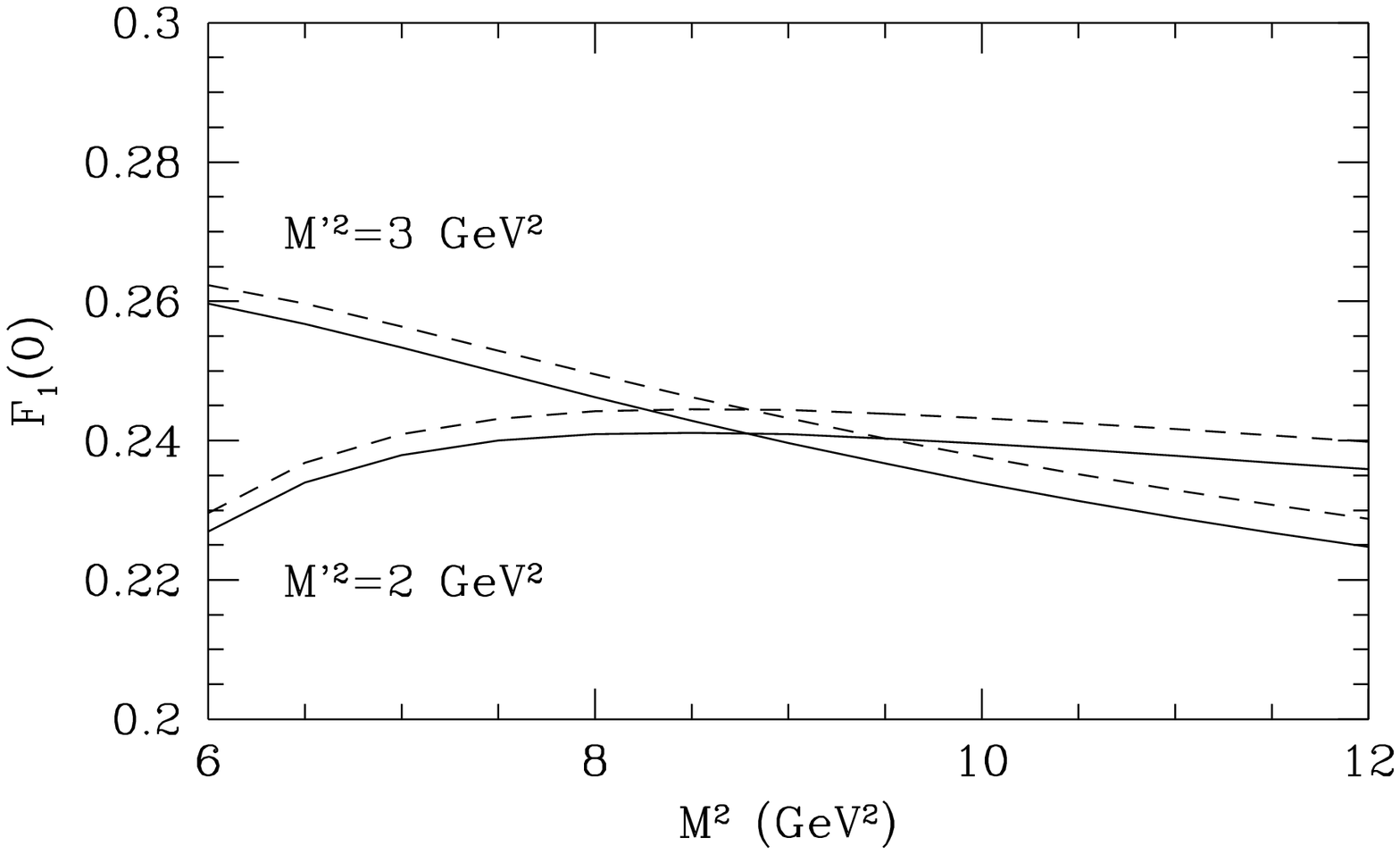}}}
\vskip -0.3truein
\caption[]{ \it The dependence of the decay constant $F_{1}(0)$
on the Borel parameter $M^{2}$ for fixed $M'^{2}$ at 
$s_{0}=33\,\,{\mbox{GeV}}^{2}$ (solid) and 
$s_{0}=35\,\,{\mbox{GeV}}^{2}$ (dashed). }
\label{fig:2d}
\end{figure}
The last term in eq.~(\ref{sumrule}) is
due to the gluon condensate contribution and the function $N(x)$ is defined
by:
\begin{eqnarray}
N(x)&=&\frac{1}{48} \exp(-\frac{m_b^2}
{M^2 (1 - x- x M^2/M'^{2})}) m^2_{b} M'^{6} x
(m^2_b M'^{4} - 4 M^2 M'^{4}+5 M^2 M'^{4} x \nonumber \\
&+&5 M^4 M'^{2} x- M^2 M'^{4} x^2 -2 M^4 M'^{2} x^2 - M^6 x^2)
/ (M^{4}(-M'^{2} + M'^{2} x + M^{2} x)^5) \, .
\end{eqnarray}

\subsection{Analysis of the sum rule \label{sec:sumruleanalysis}}
First we list the values of the input parameters entering the sum rules
(eq.~(\ref{sumrule})), which are not included in Table \ref{parameters}: 
$m_{0}^{2}=0.8\,\,{\mbox{GeV}}^{2} $ \cite{mzero},
$<\bar{s}s>=-0.011 \, \, {\mbox{GeV}}^{3}$ \cite{sbars}, 
$\frac{\alpha_{s}}{\pi}<G^{2}>=0.012 \,\, {\mbox{GeV}}^{4}$ \cite{shifman}, 
$m_{\phi}=1.019 \,\,{\mbox{GeV}}$ and 
$f_{\phi}=0.23 \, \, {\mbox{GeV}}$ \cite{chernyak}.

We do the calculations for two different continuum threshold values
$s_{0}=33 \,\,{\mbox{GeV}}^2$ and $s_{0}=35\,\,{\mbox{GeV}}^2$ and take 
$s'_{0}=1.8 \,\, {\mbox{GeV}}^{2}$.
In Fig.~\ref{fig:3d} we present the dependence of $F_{1}(0)$
on $M^{2}$ and $M'^{2}$ for $s_{0}=33 \,\, {\mbox{GeV}}^{2}$.
According to the QCD sum rules method,
it is necessary to find a range of $M^{2}$ and $M'^{2}$, where the
dependence of $F_{1}(0)$ on these parameters is very weak and, at the
same time, the power corrections and the continuum contribution remain
under control. 
{}From Fig.~\ref{fig:3d} and Fig.~\ref{fig:2d} follows that the best
stability region for $F_{1}(0)$ is 
$7\,\,{\mbox{GeV}}^{2}\leq M^{2}\leq 9\,\,{\mbox{GeV}}^{2}$,
$2\,\,{\mbox{GeV}}^{2}\leq M'^{2}\leq 3\,\,{\mbox{GeV}}^{2}$ for 
$s_{0}=33,35\,\,{\mbox{GeV}}^{2}$. We get:
\begin{eqnarray}
F_{1}(0)= 0.24 \pm 0.02 \, \, .
\label{eq:F1}
\end{eqnarray}
This agrees for our value of $m_b$ within errors with the result
given in the literature, 
based on Light-cone QCD sum rule calculations \cite{alibraunsimma}.

Numerical analysis shows, as also mentioned in \cite{colangelo}, that the 
natural hierarchy of the bare loop, the power corrections and 
continuum contributions does not hold due to the smallness of the integration 
region, and the power corrections exceed the bare loop contribution.
The gluon condensate contribution is $\leq 1 \% $ of
the dim-3 $+$ dim-5 condensate contributions 
and can therefore be safely neglected in numerical calculations.

\section{The $B_{s} \rightarrow \phi\gamma\rightarrow \gamma \gamma$ Amplitude
using VMD  \label{sec:bsggVMD}}

Starting with the amplitude for the decay
$B_{s}\rightarrow\phi\gamma$ as input, we calculate the CP-odd and 
CP-even amplitudes in $B_{s}\rightarrow \gamma \gamma$  by using a 
$\phi\rightarrow \gamma$ conversion factor supplied by the VMD model.
Here an extrapolation of the
$B_{s}\rightarrow\phi\gamma$ decay amplitude from $p'^{2}=m_{\phi}^{2}$ 
(needed for $B_s \rightarrow \phi \gamma$)
to $p'^{2}=0$ (required for $B_s \rightarrow \gamma \gamma$) is necessary,
such that the $\phi$ meson propagates as a massless virtual
particle before converting into a photon.
Note that we suppressed in our notation the dependence of the form factor
$F_1(q^2)=F_1(q^2,p'^2= m_{\phi}^2)$ on the second argument $p'^2$. 
We define here
$\bar{F_1}(Q^2) \equiv F_1(q^2=0,Q^2)$ for virtual momenta $Q^{2}=-p'^{2}$.
We assume, that the form factor $\bar{F_1}(Q^2)$ is dominated by a single pole,
which is a good approximation for light mesons and write:
\begin{eqnarray}
\bar{F_1}(Q^2)=\frac{\bar{F_1}(0)}{1-Q^{2}/m_{pole}^{2}} \; .
\end{eqnarray}
Using an $m_{pole}$ of order 
$1.7-1.9$ GeV, which corresponds to the mass of the higher resonances of 
the $\phi$ meson, 
we estimate $\bar{F_1}(0)= 0.16 \pm 0.02$.

With the help of VMD \cite{sakurai,pakvasa,deshpande} and factorization we 
can now present the 
amplitude for $B_{s}\rightarrow\gamma\gamma$.
Using the intermediate propagator 
$\frac{-1}{Q^{2}+m_{\phi}^{2}}$ at 
$Q^{2}=0$, the $\phi\rightarrow \gamma$ conversion vertex from the VMD 
mechanism 
\begin{eqnarray}
<0|J_{\mu\,\, em}|\phi(p',\epsilon)>=e Q_{s} f_{\phi}(0) m_{\phi} \epsilon_{\mu} \, \, ,
\end{eqnarray}
and the ${\cal A}(B_{s}\rightarrow\phi\gamma)$ 
amplitude, see eq.~({\ref{amphigam}}),
we get:
\begin{eqnarray}
{\cal A}(B_{s}\rightarrow\phi\gamma\rightarrow\gamma\gamma)
&=&\epsilon_1^{\mu}(k_1) \epsilon_2^{\nu}(k_2)
(A^{+}_{LD_{O_7}} g_{\mu \nu} +
i A^{-}_{LD_{O_7}} \epsilon_{\mu \nu \alpha \beta} k_1^{\alpha}  k_2^{\beta})
\, \, ,
\label{amo7}
\end{eqnarray}
with the CP-even ($A^{+}_{LD_{O_7}}$) and CP-odd ($A^{-}_{LD_{O_7}}$) parts:
\begin{eqnarray}
A^{+}_{LD_{O_7}}&=&2 \chi C m_b \frac{m_{B_s}^2-m_{\phi}^2}{2} \bar{F_1}(0)
\nonumber\\
&=&\sqrt{2} \frac{\alpha G_F}{\pi} \bar{F_1}(0) f_{\phi}(0) \lambda_t 
\frac{m_b (m_{B_s}^2-m_{\phi}^2)}{3 m_{\phi}} C_{7}^{\mbox{eff}}(\mu)  
 \, \, , \nonumber\\
A^{-}_{LD_{O_7}}&=&2 \chi C m_b \bar{F_1}(0)
\nonumber\\
&=&2 \sqrt{2} \frac{\alpha G_F}{\pi} \bar{F_1}(0) f_{\phi}(0) 
\lambda_t \frac{m_b}{3 m_{\phi}} C_{7}^{\mbox{eff}}(\mu) \, \, ,
\label{eq:o7amplitudes}
\end{eqnarray}
where $f_{\phi}(0)=0.18$ GeV \cite{terasaki}, $Q_{s}=-1/3$ and
$C$ is defined in eq.~(\ref{const}).
The factor 2 stems from the addition of the diagrams with interchanged photons.
While for the analysis of the sum rule for 
$B_s \rightarrow \phi \gamma$ we have used 
$f_{\phi}\equiv f_{\phi}(m_{\phi}^2)$, here we take 
into account the suppression in $f_{\phi}(Q^2)$ going from
$Q^2=m_{\phi}^2$ to $Q^2=0$.
We treated the polarization vector $\epsilon^{\phi}$ as transversal and 
replaced $\epsilon \to \epsilon_1, \,\epsilon^{\phi} \to \epsilon_2, 
\, q \to k_1, \, p' \to k_2 $.
The conversion factor $\chi$ is defined as 
$\chi=-e Q_{s} \frac{f_{\phi}(0)}{m_{\phi}}$.

\begin{figure}[htb]
\vskip 0.2truein
\centerline{\epsfysize=9.0cm
{\epsffile{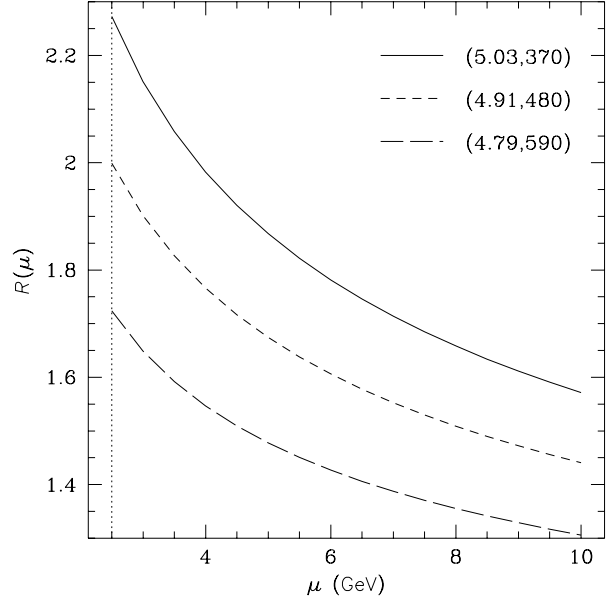}}}
\vskip -0.2truein
\caption[]{ \it Scale dependence of the ratio ${\it R}(\mu)$ defined in 
eq.~(\ref{rat}). The solid, short-dashed and long-dashed lines correspond to
the values $(m_b,\bar{\Lambda}_s)$ in $( {\mbox{GeV}}, {\mbox{MeV}} )$ 
as indicated in the figure.
The dotted line depicts the suggested choice of the scale $\mu$ from
$B \rightarrow X_s \gamma$ studies in NLO 
\cite{nlogreub,nlomisiak,effhamali,effhamburas}. 
The parameters used are given in Table \ref{parameters}.}
\label{fig:dependence}
\end{figure}

\section{Numerical Estimates of the $\mu, (m_b,\bar{\Lambda}_s)$ Uncertainties 
and the $O_7$ Mediated LD Effects  \label{sec:LDo7estimate}}

We combine in this section the results obtained in the previous sections 
\ref{sec:llog}-\ref{sec:bsggVMD}, i.~e., the LLog QCD corrections, 
the HQET inspired bound state model and the $O_7$-type LD effects in 
$B_s \to \g \g$ decay.
We give numbers for the $B_s \to \g \g$ branching ratio and CP 
ratio and discuss the dependences on and the uncertainties due to 
the renormalization scale $\mu$ and the bound state parameters 
$(m_b,\bar{\Lambda}_s)$.

Adding the $O_7$-type LD amplitudes (eq.~(\ref{eq:o7amplitudes})) 
to the short-distance ones (eq.~(\ref{amplitudes})), 
we obtain the $B_{s}\rightarrow\gamma\gamma$ width 
including the $B_s \to \phi \gamma \to \gamma \gamma$ contribution:
\begin{equation}
\Gamma(B_{s}\rightarrow \gamma \gamma)_{SD+LD_{O_7}}=\frac{1}{32 \pi m_{B_s}} 
(4 |A^{+}+A^{+}_{LD_{O_7}}|^2+\frac{1}{2} m_{B_s}^4|A^{-}+A^{-}_{LD_{O_7}}|^2)
\, \, .
\label{brld}
\end{equation}
Here a comment about double counting is in order.
The ``LD" amplitudes considered here, which involve the Wilson coefficient 
$C_7$, also contain a piece from perturbation theory. 
It originates in the bare loop diagram in the calculation of the 
sum rule and enters the value of the form factor $F_1$. Without this 
perturbative part, it is not possible to perform an operator product 
expansion; it corresponds to the leading term in the sum rule and
hence there is no way to avoid it.
The ``SD" amplitudes on the other hand include also contributions from small momenta. As a consequence,
by adding the perturbative and the non-perturbative parts
in eq.~(\ref{brld}) there is certainly some double counting present.
However, as usual, it is assumed that the SD parts are small in regions 
where the LD ones are large and hence the effect of this is small.

First we study the leading logarithmic $\mu$-dependence of the ratio
\begin{eqnarray}
{\it R}(\mu)=\frac{\Gamma(B_{s}\rightarrow \gamma \gamma)(\mu)_{SD+LD_{O_7}}}
{\Gamma(B_{s}\rightarrow \gamma \gamma)(m_W)_{SD+LD_{O_7}}} \, \, .
\label{rat}
\end{eqnarray}
In the numerical analysis we neglect the masses of the light quarks,
i. e. we use $I(m_q)=1$ for $q=u,d,s$ and $m_s \triangle(m_s)=m_s J(m_s)=0$
in eq.~(\ref{eq:IJdelta}).
{}From Fig.~\ref{fig:dependence} we find an 
enhancement factor of $1.3-2.3$ relative to the 
lowest  order result obtained by setting $\mu=m_W$, depending on the model 
parameter $(m_b,\bar{\Lambda}_s)$. Varying $\mu$ in the range
$2.5 \, \,{\mbox{GeV}} \leq \mu \leq 10.0 \, \,{\mbox{GeV}}$, gives an 
uncertainty
$\triangle {\it R}/{\it R}(\mu=5\,\,{\mbox{GeV}})\approx \pm (17,19,22) \%$
for $\bar{\Lambda}_s=(590,480,370)$ MeV, respectively.
Here one can argue, that the choice $\mu=\frac{m_b}{2}$ takes into account
effectively the bulk of the NLO correction as suggested by the
NLO calculation for $B \rightarrow X_s \gamma$ \cite{nlogreub,nlomisiak}.

Table~\ref{mums} shows the combined $\mu$ and model parameter dependence of 
the branching ratio 
\begin{equation}
{\cal B}(B_{s} \rightarrow \gamma \gamma)_{SD+LD_{O_7}}=
\frac{\Gamma(B_{s}\rightarrow \gamma \gamma)_{SD+LD_{O_7}}}
 {\Gamma_{tot}(B_s)}\,\,.
\end{equation}
The dependence of the form factor $\bar{F_1}(m_{\phi}^2)$ 
on the $b$-quark mass has been extrapolated from Fig.~3 \cite{alibraunsimma}. 
Here $\bar{F_1}(0)= 0.14, 0.15, 0.16$ has been used for 
$m_b=(5.03,4.91,4.79)$ GeV, respectively.
\begin{table}[h]
        \begin{center}
        \begin{tabular}{|c|l|l|l|}
        \hline
        \multicolumn{1}{|c|}{$\mu$}     & 
                \multicolumn{1}{|c|}{$\bar{\Lambda}_s=370$ MeV} 
& 
                \multicolumn{1}{|c|}{$\bar{\Lambda}_s=480$ MeV}
& 
                \multicolumn{1}{|c|}{$\bar{\Lambda}_s=590$ MeV}\\
\multicolumn{1}{|c|}{(GeV)}  
& $m_b=5.03$ GeV & $m_b=4.91$ GeV & $m_b=4.79$ GeV \\
        \hline \hline
$2.5$   & $1.43\cdot 10^{-6}$ & $8.1 \cdot 10^{-7}$ & $5.0 \cdot 10^{-7}$\\
$5.0$   & $1.18\cdot 10^{-6}$ & $6.8 \cdot 10^{-7}$ & $4.3 \cdot 10^{-7}$\\
$10.0$  & $0.99\cdot 10^{-6}$ & $5.9 \cdot 10^{-7}$ & $3.8 \cdot 10^{-7}$\\
\hline 
        \end{tabular}
        \end{center}
\caption{\it Branching ratio 
${\cal B}(B_{s}\rightarrow \gamma \gamma)_{SD+LD_{O_7}}$ 
for selected values $(m_b,\bar{\Lambda}_s)$ 
and the renormalization scale $\mu$.}
\label{mums}
\end{table}
Qualitatively, the influence of the LD contribution through
$B_{s}\rightarrow\phi\gamma \rightarrow \gamma \gamma$ reduces the width 
because of the destructive interference of the LD $+$ SD contributions.
To quantify this, we define 
\begin{equation}
\kappa \equiv \frac{{\cal B}(B_{s}\rightarrow\gamma\gamma)_{SD+LD_{O_7}}
 -{\cal B}(B_{s}\rightarrow\gamma\gamma)_{SD}}
{{\cal B}(B_{s}\rightarrow\gamma\gamma)_{SD}} \, \, ,
\end{equation}
with $\Gamma(B_{s}\rightarrow \gamma \gamma)_{SD}$ given in 
eq.~(\ref{br}).
We find, that $\kappa$ lies in the range:
\begin{equation}
-15 \% \leq \kappa \leq -27 \% \, \, ,
\end{equation}
depending mainly on $(m_b,\bar{\Lambda}_s)$.
To summarize, lowering the scale $\mu$ and $\bar{\Lambda}_s$
enhances the branching ratio ${\cal B}(B_{s}\rightarrow \gamma \gamma)$. 

\begin{table}[h]
        \begin{center}
        \begin{tabular}{|c|l|l|l|}
        \hline
        \multicolumn{1}{|c|}{$\mu$}     & 
                \multicolumn{1}{|c|}{$\bar{\Lambda}_s=370$ MeV} 
& 
                \multicolumn{1}{|c|}{$\bar{\Lambda}_s=480$ MeV}
& 
                \multicolumn{1}{|c|}{$\bar{\Lambda}_s=590$ MeV}\\
\multicolumn{1}{|c|}{(GeV)}  
& $m_b=5.03$ GeV & $m_b=4.91$ GeV & $m_b=4.79$ GeV \\
        \hline  \hline
$2.5$   & $0.79 \,(0.80)$ & $0.88 \,(0.88)$ & $0.89 \,(0.90)$\\
$5.0$   & $0.69 \,(0.71)$ & $0.73 \,(0.75)$ & $0.70 \,(0.73)$\\
$10.0$  & $0.61 \,(0.63)$ & $0.60 \,(0.63)$ & $0.55 \,(0.60)$\\
\hline 
$m_W$  & $0.38 \,(0.41)$ & $0.33 \,(0.36)$ & $0.26 \,(0.33)$\\
\hline 
        \end{tabular}
        \end{center}
\caption{\it The CP ratio $r_{CP \, SD+LD_{O7}}$ given in eq.~(\ref{eq:rcpo7})
for selected values $(m_b,\bar{\Lambda}_s)$ 
and the renormalization scale $\mu$. The values in parentheses correspond to
$r_{CP}$ as defined in
eq.~(\ref{eq:rcp}) without taking into account the $O_7$-type LD effects.}
\label{tab:cpratio}
\end{table}
The dependence of the CP ratio \cite{aliev}, \cite{alievhillererhan},
here including our LD  $O_7$-type estimate
\begin{eqnarray}
r_{CP \, SD+LD_{O7} } = \frac{4 |A^{+}+A^{+}_{LD_{O_7}}|^2}
{ m_{B_s}^4|A^{-}+A^{-}_{LD_{O_7}}|^2 } \; ,
\label{eq:rcpo7}
\end{eqnarray}
on the renormalization scale and on the bound state parameters can be inferred 
from Table \ref{tab:cpratio}. 
The values of $r_{CP}$ without taking into account the LD contribution
from the decay chain 
$B_s \to \phi \g \to \g \g$ are also shown in parentheses. 
As can be seen, the $O_7$-type LD effects 
reduce the ratio \cite{alievhillererhan}.
Further including the LLog QCD corrections enhance $r_{CP \, SD+LD_{O7} }$ 
by a factor of $1.6-3.4$ compared to the lowest order result ($\mu=m_W$), 
depending on the $(m_b,\bar{\Lambda}_s)$ parameter set.
As a rule, both lowering $\mu$ and increasing $\bar{\Lambda}_s$ enlarge the
value of the CP ratio.

\section{Estimate of the Long-Distance Contribution through 
$b\rightarrow s \psi$ in $B_{s}\rightarrow\gamma\gamma$ Decay}

In this section we estimate the additional LD effect 
due to the dominant four-quark operators $O_1$ and $O_{2}$ 
(see eq.~(\ref{O12})) through the 
$B_{s}\rightarrow \phi\psi\rightarrow\phi\gamma\rightarrow\gamma \gamma$
chain decay. 
We use at quark level $b\rightarrow s \psi$ followed by  
$b\rightarrow s\gamma$ decay \cite{deshpande} and we pass to the 
hadronic level using the transition form factor $F_{1}(0)$ 
from the amplitude ${\cal{A}}(B_{s}\rightarrow\phi)$ 
\cite{gudi}, \cite{alibraun} given in eq.~(\ref{eq:F1}).
For both the conversions $\psi\rightarrow\gamma$ and 
$\phi\rightarrow\gamma$ we employ the Vector Meson Dominance (VMD) model 
\cite{deshpande}. The conversion $\psi\rightarrow\gamma$ needs further
manipulation because of the strong contribution from the longitudinal part
of the $\psi$ meson. We extract the transverse part using the Golowich-Pakvasa
procedure \cite{deshpande}, \cite{pakvasa}.
Further, we calculate the $O_{1,\, 2}$-type LD effect to the $B_{s}\rightarrow
\phi\gamma$ decay using the method given in ref.~\cite{Ruckl}, namely, 
by taking
into account the virtual c-quark loop instead of the hadronization of the
$\bar{c}c$ pair. This procedure was originally applied to estimate the 
LD effect in $B \to K^{\ast} \gamma$ decay and uses operator product 
expansion and QCD sum rule techniques. 
Finally we present amplitudes for the decay chain 
$B_{s}\rightarrow\phi\psi\rightarrow\phi\gamma\rightarrow\gamma\gamma$.
 
\subsection{The chain process
$B_{s}\rightarrow\phi\psi\rightarrow\phi\gamma\rightarrow\gamma\gamma$}

We first consider the additional contribution to $b\rightarrow s\gamma$ 
from $b\rightarrow s\psi_{i} \rightarrow s \gamma$, 
where $\psi_{i}$ are all $\bar{c}c$
$J=1$ bound states, see Fig.~\ref{fig:vmd}.
The relevant part of the effective Hamiltonian describing this process
is given as
\begin{eqnarray}
{\cal{H}}_{eff}=4 \frac{G_{F}}{\sqrt{2}} V^{*}_{cs} V_{cb}
(C_{1}(\mu) O_{1}(\mu)+C_{2}(\mu) O_{2}(\mu)) \; ,
\label{effH2}
\end{eqnarray}
with the dominant four-Fermi operators 
\begin{eqnarray}
O_{1}&=&\bar{s}_{\alpha}\gamma_{\mu}\frac{1-\gamma_{5}}{2}c_{\beta}\,\,\bar{c}_{\beta}
\gamma_{\mu}\frac{1-\gamma_{5}}{2}b_{\alpha}\nonumber \, \, ,\\
O_{2}&=&\bar{s}\gamma_{\mu}\frac{1-\gamma_{5}}{2}c\,\,\bar{c}\gamma_{\mu}
\frac{1-\gamma_{5}}{2}b \, \, .
\label{O12}
\end{eqnarray}
Here $\alpha,\beta$ are $SU(3)$ colour indices
and $ V^{(*)}_{ij}$
are the relevant elements of the quark mixing matrix.
The initial values of the corresponding Wilson coefficients are
$C_{1}(m_W)=0$ and $C_{2}(m_W)=1$.
To include leading logarithmic QCD corrections we evaluate
$C_{1,2}(\mu)$ at the relevant scale, $\mu \approx m_b$ for $B$-decays,
and this takes into account short-distance effects from single gluon exchange.
The analytical expressions can be found in \cite{effhamali,effhamburas}. 
\begin{figure}[htb]
\vskip -0.6truein
\centering
\epsfysize=7in
\leavevmode\epsffile{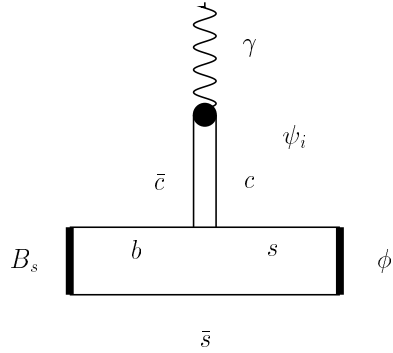}
\vskip -4.2truein
\caption[]{ \it The diagram contributing to 
$B_s \rightarrow \phi \psi \rightarrow \phi \gamma $. }
\label{fig:vmd}
\end{figure}
Further we have used the unitarity of the CKM matrix
$V_{cs}^{*}V_{cb}=-V_{ts}^{*}V_{tb}-V_{us}^{*}V_{ub}$ and have neglected 
the contribution due to an internal u-quark, since 
$V_{us}^{*}V_{ub} \ll V_{ts}^{*}V_{tb}=\lambda_t$.

Using factorization, we obtain the inclusive decay amplitude 
for the process $b\rightarrow s\psi$ \cite{deshpande} as
\begin{eqnarray}
{\cal{A}}(b\rightarrow s\,\psi(k_{1},\epsilon^{\psi}))=-i
\bar{C} f_{\psi}(m^{2}_{\psi}) m_{\psi} 
\bar{s}\gamma^{\mu}(1-\gamma_{5})b\, \epsilon_{\mu}^{\psi} \, \, .
\label{A1}
\end{eqnarray}
Here 
\begin{eqnarray}
\bar{C}=-\frac{G_{F}}{\sqrt{2}} \lambda_t a_2(\mu)
\label{const1}
\end{eqnarray}
with, assuming naive factorization,
\begin{eqnarray}
a_2(\mu)=C_{1}(\mu)+\frac{C_{2}(\mu)}{N_c} \, \, ,
\label{a2}
\end{eqnarray}
where $N_c=3$ in colour $SU(3)$ and
$k_1, \epsilon^{\psi}$ are the momentum and the polarization vector of the 
$\psi$, respectively.
In eq.~(\ref{A1}) we used the matrix element 
\begin{eqnarray}
<0|\bar{c}\gamma_{\mu} c|\psi(k_{1},\epsilon^{\psi})>=
f_{\psi} (m^{2}_{\psi}) m_{\psi}
\epsilon_{\mu}^{\psi} \, \, .
\label{psim}
\end{eqnarray}

At this stage there is a critical remark about factorization 
in order, concerning the value of $a_2(\mu)$ used.
The decay under consideration is a class II decay following the 
classification of \cite{BSW}. 
In general eq.~(\ref{a2}) is written as 
\begin{eqnarray}
a_2^{\mbox{eff}}=(C_{1}(\mu)+\frac{C_{2}(\mu)}{N_c}) 
\left[ 1+\epsilon_{1}(\mu) \right]+ 
C_{2}(\mu) \epsilon_{2}(\mu) \, \, ,
\label{a2eff}
\end{eqnarray}
where $\epsilon_{1}(\mu)$ and $\epsilon_{2}(\mu)$ parametrize the
non-factorizable
contributions to the hadronic matrix elements.
$a_{2}^{\mbox{eff}}$ takes into account all contributions of the matrix 
elements
in contrast to $a_{2}(\mu)$, which assumes naive factorization 
$\epsilon_1(\mu)=\epsilon_2(\mu)=0$.
Especially $\epsilon_{2}(\mu)$, which is the colour octet piece, has sizable
contributions to naive factorization in class II decays \cite{neubert}. 
Furthermore, the 
additional problem is not to know the correct factorization scale.
In order to include the non-factorizable corrections 
we use the effective coefficient 
$a_{2}^{\mbox{eff}}$, which is determined experimentally
from the world average branching ratio of
$\bar{B} \to \bar{K}^{(\ast)} \psi $ as \cite{neubert} 
\begin{eqnarray}
a_2^{\mbox{eff}}=0.21 \, \, .
\label{a2eff2}
\end{eqnarray}
This choice restores the correct scale and is $\mu$ independent.
Writing the ansatz
\begin{eqnarray}
a_{2}^{\mbox{eff}}=C_{1}(m_{b})+\xi C_{2}(m_{b}) \, \, ,
\label{a2xi}
\end{eqnarray}
it follows that $\xi\approx 0.41$ with 
$C_{1}(m_b)=-0.25$ and $C_{2}(m_b)=1.11$ for the input values given in 
Table~\ref{parameters}. For comparison, naive factorization would give 
$a_2(m_b)=0.12$. 

Our aim is to replace the $\psi$ meson with the photon $\gamma$ and
to construct a gauge invariant amplitude. 
We remove the longitudinal component of the meson $\psi$ and then 
$\epsilon_{\mu}^{\psi}$ can be converted into the
polarization vector $\epsilon_{\mu}^{\gamma}$ of the photon $\gamma$. 
We utilize the Golowich-Pakvasa \cite{deshpande},\cite{pakvasa} procedure 
making use of the Gordon identity, namely
$\gamma_{\mu}\gamma_{\alpha}=g_{\mu\alpha}-i\sigma_{\mu\alpha}$.
We start with the vertex $\bar{s}\gamma_{\mu}(1-\gamma_{5})b$ and using the 
equation of motion ${\slash{p}} b=m_b b$ and momentum conservation $p=p'+k_1$, 
we get
\begin{eqnarray}
\bar{s}\gamma_{\mu}(1-\gamma_{5})b=\frac{1}{m_{b}}\{\bar{s}\gamma_{\mu}\slash{p'}
(1+\gamma_{5})b + \bar{s}\gamma_{\mu}\slash{k_{1}} (1+\gamma_{5})b \} \, \, ,
\label{vert1}
\end{eqnarray}
where $p,p'$ are the momenta of the $b$- and $s$-quark, respectively.
We neglect the first term in eq.~(\ref{vert1}) since 
$\frac{m_{s}}{m_b} \ll 1$ and $p'^{\mu} \epsilon^{T}_{\mu}=0$, which follows 
from $\epsilon_{\mu}^{T} p^{\mu}=0$ 
in the rest frame of the $b$-quark and the transversality condition
$\epsilon_{\mu}^{T} k_{1}^{\mu}=0$, 
where $\epsilon_{\mu}^{T}$ is the transversal polarization 
vector of the $\psi$ meson \cite{deshpande}.
The second term can be written as
\begin{eqnarray}
\frac{1}{m_{b}}\bar{s}\gamma_{\mu}\slash{k}_{1} (1+\gamma_{5}) b=
\frac{1}{m_{b}} \{ \bar{s}(1+\gamma_{5}) k_{1 \mu} b 
-i \bar{s}\sigma_{\mu\alpha}k_{1}^{\alpha}(1+\gamma_{5})b \} \, \, .
\label{vert2}
\end{eqnarray}
Only the $\sigma_{\mu\alpha}$ term in eq.~(\ref{vert2}) couples to 
the transversal component of the $\psi$ and we obtain the corresponding 
amplitude as
\begin{eqnarray}
{\cal{A}}(b\rightarrow s \psi^{T})=
-2 \bar{C} f_{\psi}(m^{2}_{\psi}) \frac{m_{\psi}}{m_{b}}
\bar{s}\sigma_{\mu\alpha} k_{1}^{\alpha}\, R \, b \epsilon_{\mu}^{T} \; ,
\label{A2}
\end{eqnarray}
where $R=\frac{1+\gamma_{5}}{2}$ denotes the chiral right projection.
Note that the coupling structure is the same as due to a direct use of 
$O_7= \frac{e}{16 \pi^2} \bar{s} \sigma_{\mu \nu} m_b R b F^{\mu \nu} $ 
\cite{effhamali,effhamburas}
with the photon field strength tensor $ F^{\mu \nu}$ and $m_s=0$.
For the $\psi^{T}\rightarrow\gamma$ conversion following the VMD mechanism 
we have
\begin{eqnarray}
<0|J_{\mu, el}|\psi^{T}(k_{1}, \epsilon^{T})>=e Q_{c} f_{\psi}(0) m_{\psi}
\epsilon^{T}_{\mu} \, \, ,
\label{conv}
\end{eqnarray}
where $Q_{c}=2/3$ and $f_{\psi}(0)$ is the coupling 
at $k_1^2=0$, see eq.~(\ref{suppr}).
Using the intermediate propagator of the $\psi$ meson at $k_{1}^2=0$,
we get
\begin{eqnarray}
{\cal{A}}(b\rightarrow s \psi^{T}\rightarrow s\gamma )=
2 \bar{C} f^{2}_{\psi}(0) 
\frac{e Q_{c}}{m_{b}}
\bar{s}\sigma_{\mu\alpha} k_{1}^{\alpha} \,R \, b \, \epsilon_{\mu}^{T}
\, \, .
\label{A3}
\end{eqnarray}
The expression for the amplitude eq.~(\ref{A3}) can be completed by 
summing over all 
$\bar{c}c$ resonant states $\psi(1S)$,$\psi(2S)$,$\psi(3770)$,$\psi(4040)$,
$\psi(4160)$ and $\psi(4415)$
\begin{eqnarray}
{\cal{A}}(b\rightarrow s \psi_{i}^{T}\rightarrow s\gamma )=
2 \bar{C} \sum_{i} f^{2}_{\psi_{i}}(0) 
\frac{e Q_{c}}{m_{b}}
\bar{s}\sigma_{\mu\alpha} k_{1}^{\alpha} \,R \,b \epsilon_{\mu}^{T}
\label{A4} \, \, .
\end{eqnarray}
The various decay couplings $f_{\psi_{i}}=f_{\psi_{i}}(m^{2}_{\psi_{i}})$ are
calculated using 
\begin{eqnarray}
f^{2}_{\psi_{i}}=\Gamma(\psi_{i}\rightarrow e^{+}e^{-})\frac{3 m_{\psi_{i}}}
{Q_c^2 4 \pi\alpha^{2}} \; ,
\label{fpsi2}
\end{eqnarray}
and the measured widths from \cite{PDG}
and given in  Table~\ref{fpsi}. 
\begin{table}[h]
        \begin{center}
        \begin{tabular}{|l|l|}
        \hline
        \multicolumn{1}{|c|}{$\psi_i$} & 
                \multicolumn{1}{|c|}{$f_{\psi_{i}} [\mbox{GeV}]$}     \\
        \hline \hline
$f_{\psi(1S)}$& 0.405  \\
$f_{\psi(2S)}$& 0.282   \\
$f_{\psi(3770)}$& 0.099    \\
$f_{\psi(4040)}$& 0.175  \\
$f_{\psi(4160)}$& 0.180   \\
$f_{\psi(4415)}$& 0.145  \\
        \hline
        \end{tabular}
        \end{center}
\caption{\it Vector meson coupling constants used in the numerical
          calculations.}
\label{fpsi}
\end{table}

Now we have to extrapolate the couplings $f_{\psi_{i}}(k_1^{2})$ from 
$k_1^{2}=m_{\psi_i}^{2}$ to $k_1^{2}=0$. 
We use the suppression factor \cite{deshpande} 
\begin{eqnarray}
\kappa=f^{2}_{\psi(1S)}(0)/f^{2}_{\psi(1S)}(m_{\psi}^{2})=0.12
\label{suppr}
\end{eqnarray}
obtained from data on photo production of the $\psi$
and assume $\kappa$ to be  universal for the other (higher) resonances.
\footnote{This is consistent with $\kappa=0.11$ \cite{terasaki} based on a
dispersion relation calculation.}
We now use eq.~(\ref{A4}) to find the matrix element of
$B_{s}\rightarrow\phi\gamma$ through the 
$b\rightarrow s\psi^{T}\rightarrow s \gamma$
transition at quark level. The matrix element \cite{alibraun} 
is given as
\begin{eqnarray}
<\phi(p')|\bar{s}\sigma_{\mu\alpha}\, R \, k_{1}^{\alpha} b|B_{s}(p)>&=&i
\epsilon_{\mu\nu\rho\sigma}\epsilon^{\phi
\nu}p^{\rho}p'^{\sigma}F_{1}(k_1^{2})\nonumber \\
&+& (\epsilon_{\mu}^{\phi} p.k_1-p_{\mu} k_1.\epsilon^{\phi}) G(k_1^{2}) \; ,
\label{alibraun}
\end{eqnarray}
and we get the amplitude 
\begin{eqnarray}
{\cal{A}}(B_{s}\rightarrow\phi\gamma)&=&
2 \bar{C} \epsilon_{1}^{\mu} \epsilon^{\phi \nu} 
\sum_{i}\frac{f^{2}_{\psi_{i}}(0)}{m_{b}} e Q_{c} \{ i 
\epsilon_{\mu\nu\rho\sigma}k_{1}^{\rho}
p'^{\sigma}\nonumber \\&+& g_{\mu\nu}\frac{m_{B_{s}}^{2}-m_{\phi}^{2}}{2}  \}
F_{1}(0) \; ,
\label{A5}
\end{eqnarray}
where $\epsilon_{1 \mu}$, $\epsilon_{\nu}^{\phi}$ are the polarization vectors
and $k_{1}$, $p'$ are the momenta of the photon and
$\phi$ meson, respectively.
We used $G(k_1^{2}=0)=F_{1}(k_1^{2}=0)$ \cite{alibraun}.
Note, that the form factors introduced above are in general functions of two 
variables $k_1^2$ and $p'^2$. We abbreviated here 
$F_1(k_1^2) \equiv F_1(k_1^2, p'^2=m_{\phi}^2)$ and use the value 
$F_1(0)=0.24 \pm 0.02$ \cite{gudi} obtained in section 
\ref{sec:sumruleanalysis}.

\subsubsection{VMD vs soft gluon interaction}

Now we want to compare our result for ${\cal{A}}(B_{s}\rightarrow\phi\gamma)$
eq.~(\ref{A5})
with the same amplitude calculated by the method worked out in
\cite{Ruckl}. This method is based on the new effective quark-gluon operator
obtained by the interaction of the virtual charm quark loop with soft gluons,
in contrast to a phenomenological description in terms of $\psi$ resonances 
converting into a photon, as we used. In this approach, the operator $O_{1}$
does not give any contribution to the matrix element of
$B_{s}\rightarrow\phi\gamma$ for an on-shell photon. 
The Fierz transformation of the
operator $O_{2}$ reads (using eq.~(\ref{eq:TaTa}) and eq.~(\ref{eq:fierzLL}))
\begin{eqnarray}
O_{2}=1/N_{c} \,O_{1}+1/2 \,\,O_{octet} \, \, ,
\label{O2exp}
\end{eqnarray}
where  
\begin{eqnarray}
O_{octet}=4 (\bar{c}\,
\gamma_{\mu}\frac{1-\gamma_{5}}{2} T^a\, c)
(\bar{s}\,
\gamma_{\mu}\frac{1-\gamma_{5}}{2} T^a\, b) \, \, ,
\label{octet}
\end{eqnarray}
and $T^a=\lambda^{a}/2$ are the $SU(3)$ colour generators.
Then the only contribution comes from the colour octet part $O_{octet}$.
Using the operator $O_{octet}$ as a vertex of the virtual charm quark loop,
which emits a real photon, and taking into account the c-quark-soft gluon 
interaction, a new effective operator is obtained. The matrix element 
of this operator between $B_{s}$ and $\phi$ meson states gives the long
distance amplitude of $B_{s}\rightarrow \phi\gamma$ decay 
due to the $O_{1,\,2}$ operators and it is written as 
(see \cite{Ruckl} for details; there the amplitude for the decay 
$B \to K^{\ast} \gamma$ is given)
\begin{eqnarray}
{\cal{A'}}(B_{s}\rightarrow\phi\gamma)=
2 C' \epsilon_{1}^{\mu} \epsilon^{\phi\, \nu}
\{ i \epsilon_{\mu\nu\rho\sigma} k_{1}^{\rho}p'^{\sigma}\, L+
\frac{m_{B_{s}}^{2}-m_{\phi}^{2}}{2} g_{\mu\nu}\, \tilde{L} \}\,,
\label{Aruckl}
\end{eqnarray}
where $C'=\frac{e G_{F}\lambda_{t}}{8 \sqrt{2}\pi^{2}}\frac{C_{2}(\mu)}{9
m_{c}^{2}}$ . 
The form factors $L$ and $\tilde{L}$ are calculated 
using QCD sum rules \cite{Ruckl},
\begin{eqnarray}
L&=& \frac{m_{b}}{m_{\phi} m_{B_{s}}^{2} f_{B_{s}} f_{\phi}} 
\, \exp(\frac{m_{B_{s}}^{2}}{M^{2}}+\frac{m_{\phi}^{2}}{M'^{2}})
\nonumber \\  
&.& 
\{ \frac{m_{b}}{48}
\{ \frac{\alpha_{s}}{\pi} <G^2> \int_{m_{b}^{2}/M^{2}}^{\infty} ds\,\, 
e^{-s}\,
[ \frac{m_{b}^{2}}{s}-\frac{M^{4}}{M^{2}+M'^{2}}
(1-\frac{m_{b}^{2}}{s M^{2}})(1+\frac{M'^{2}}{s M^{2}})]\nonumber \\
&-& [\frac{m_{0}^{2} <\bar{s}s> m_{b}^{2}}{12}-\frac {4 \pi\alpha_{s}
<\bar{s}s>^{2} m_{b}}{27} (1+\frac{m_{b}^{2}}{M'^{2}})]
\exp(-\frac{m_{b}^{2}}{M^{2}}) \} \nonumber\,\, , \\
\tilde{L}&=& \frac{m_{b}}{m_{\phi} m_{B_{s}}^{2} f_{B_{s}} f_{\phi}} 
\exp(\frac{m_{B_{s}}^{2}}{M^{2}}+\frac{m_{\phi}^{2}}{M'^{2}}) 
\nonumber \\  
&.&
\{ \frac{m_{b}}{48}
\{ \frac{\alpha_{s}}{\pi}<G^2> \int_{m_{b}^{2}/M^{2}}^{\infty} ds\,\, 
e^{-s} \,
[ \frac{m_{b}^{2}}{s}+\frac{M^{4}}{M^{2}+M'^{2}}(1-\frac{m_{b}^{2}}{s
M^{2}})(1+\frac{M'^{2}}{s M^{2}})]\nonumber \\
&-& [\frac{m_{0}^{2} <\bar{s}s> m_{b}^{2}}{12}-\frac {16 \pi\alpha_{s}
<\bar{s}s>^{2} m_{b}}{27} (1+\frac{m_{b}^{2}}{M'^{2}})]
\exp(-\frac{m_{b}^{2}}{M^{2}}) \} \; . 
\label{LLt}
\end{eqnarray}
The Borel parameters $M$ and $M'$ are varied to find the
stability region for $L$ and $\tilde{L}$.
We use in the evaluation of the sum rules the input parameters given at the 
beginning of section \ref{sec:sumruleanalysis} and Table \ref{parameters}.
The stability region is reached for 
$6\, \mbox{GeV}^2 \leq M^{2}\leq 9\, \mbox{GeV}^2$ 
and $2\, \mbox{GeV}^2\leq M'^{2}\leq 4\, \mbox{GeV}^2$ and we get 
\begin{eqnarray}
L&=&(0.30 \pm 0.05)\,\, \mbox{GeV}^3 \nonumber \, \, , \\
\tilde{L}&=&(0.35\pm 0.05)\,\, \mbox{GeV}^3   \, \, .
\label{llt}
\end{eqnarray}
Writing the amplitude for $B_{s}\rightarrow\phi\gamma$ as
\begin{eqnarray}
{\cal{A^{(')}}}(B_{s}\rightarrow\phi\gamma)=
\epsilon_{1}^{\mu} \epsilon^{\phi\, \nu}
( i \epsilon_{\mu\nu\rho\sigma} k_{1}^{\rho} p'^{\sigma} \, A^{-(')}+
g_{\mu \nu} \, A^{+(')} ) \; ,
\label{ampgen}
\end{eqnarray}
and using eq.~(\ref{A5}),~(\ref{Aruckl}) and (\ref{llt}), we can compare the 
coefficients obtained by the two different methods and get
\begin{eqnarray}
\frac{|A^{-}-A'^{-}|}
{A^{-}}& \leq & 10 \% \, \, \nonumber ,\\
\frac{|A^{+}-A'^{+}|}
{A^{+}}& \leq & 5 \% \, \, .
\label{ratio}
\end{eqnarray}
This means, that the amplitudes agree within $10 \%$.

In our approach, the structure of the transition $b\rightarrow s
\psi^{T}\rightarrow s \gamma$ is proportional to
$\sigma_{\mu \alpha}\,\frac{1+\gamma_{5}}{2} k_1^{\alpha}$ 
(see eq.~\ref{A2}), since we removed the
longitudinal part of the $\psi$ meson from the amplitude. 
Further, the form factors $F_{1}(k_{1}^{2})$ 
and  $G(k_{1}^{2})$ in eq.~(\ref{alibraun}) are related for a real photon
($k_{1}^{2}=0)$,
$F_{1}(0)=G(0)$. Therefore, in the amplitude ${\cal{A}}(B_s \to \phi \gamma)$
only one form factor appears, which is $F_{1}(0)$ in eq.~(\ref{A5}).
However, the form factors $L$ and $\tilde{L}$ in 
${\cal{A}}'(B_s \to \phi \gamma)$ given in eq.~(\ref{Aruckl})
are not related. 
They are calculated separately 
using QCD sum rules and
this causes the difference between the ratios in eq.~(\ref{ratio}).
In spite of the fact that
the amplitudes $A^{\pm}$ and $A'^{\pm}$ are different from each other,
they coincide within the given approximation and theoretical uncertainties
lying in both methods.

\subsubsection{$O_{1,2}$-type LD amplitudes in $B_{s}\rightarrow\gamma\gamma$} 
We can now present the amplitude for $B_{s}\rightarrow\gamma\gamma$ due to the
chain reaction 
$B_{s}\rightarrow\phi\psi\rightarrow\phi\gamma\rightarrow\gamma\gamma$.
We use the intermediate propagator at zero momentum transfer
and the $\phi\rightarrow\gamma$ conversion vertex from the VMD model,
\begin{eqnarray}
<0|J_{\mu\,\, el}|\phi(p',\epsilon^{\phi})>=e Q_{s} f_{\phi}(0) m_{\phi}
\epsilon_{\mu}^{\phi} \; ,
\label{Jel}
\end{eqnarray}
where the polarization vector $\epsilon_{\mu}^{\phi}$ is treated as
transversal.
To apply the VMD mechanism to the amplitude eq.~(\ref{A5}), we have to 
know the form factor at $F_1(k_1^2=0,p'^2=0)$. We employ the extrapolated value
$\bar{F}_{1}(0)\equiv F_1(0,0)=0.16\pm 0.02 $
\cite{gudi} from section \ref{sec:bsggVMD}. 
Then the amplitude can be written with $p' \to k_2$, 
$ \epsilon^{\phi} \to \epsilon_2$ as
\begin{eqnarray}
{\cal{A}}
(B_{s}\rightarrow\phi\psi\rightarrow\phi\gamma\rightarrow\gamma\gamma)=
\epsilon_{1}^{\mu}(k_{1})\epsilon_{2}^{\nu}(k_{2}) ( g_{\mu\nu}
A^{+}_{\bar{LD}_{O_{2}}}+i \epsilon_{\mu\nu\alpha\beta}k_{1}^{\alpha} k_{2}^{\beta}
A^{-}_{\bar{LD}_{O_{2}}} ) \; ,
\label{A6}
\end{eqnarray}
with the CP-even $A^{+}$ and CP-odd $A^{-}$ parts
\begin{eqnarray}
A^{+}_{\bar{LD}_{O_{2}}}&=&4 \chi \frac{\bar{C}}{m_{b}} \bar{F}_{1}(0)
\sum_{i} f^{2}_{\psi_{i}}(0) e Q_{c}
\frac{m_{B_{s}}^{2}-m_{\phi}^{2}}{2} \, \, , \nonumber \\ 
A^{-}_{\bar{LD}_{O_{2}}}&=&4 \chi \frac{\bar{C}}{m_{b}} \bar{F}_{1}(0)
\sum_{i} f^{2}_{\psi_{i}}(0) e Q_{c} \, \, ,
\label{Apm}
\end{eqnarray}
where $\bar{C}$ is defined in eq.~(\ref{const1}) and 
the conversion factor $\chi$ is given as 
$\chi=-e Q_{s} \frac{f_{\phi}(0)}{m_{\phi}}$.
Here $f_{\phi}(0)=0.18 \, \, \mbox{GeV}$ 
\cite{terasaki} and $Q_{s}=-1/3$.
The extra factor 2 comes from the addition of the diagram with interchanged 
photons.

\section{Final Numbers and Conclusion on the Decay $B_s \to \g \g$}
In conclusion, we have reanalysed the decay rate 
$B_{s}\rightarrow\gamma\gamma$ in the SM and 
we included leading logarithmic QCD corrections.
Our model to incorporate the bound state effects in the $B_s$ meson is 
inspired by HQET, resulting in the parameters $(m_b,\bar{\Lambda}_s)$.
The strong parametric dependence
of the decay rate $\Gamma(B_s \rightarrow \gamma \gamma)$ and the
CP ratio $r_{CP}$ on 
$(m_b,\bar{\Lambda}_s)$ and on the renormalization scale $\mu$ has been 
studied by us.
Further we investigated the 
influence of the LD contributions due to the chain
$B_{s}\rightarrow\phi\gamma \rightarrow\gamma\gamma $. 
Depending on $\bar{\Lambda}_s$, the LD-contributions induced by the
operator $O_7$ become sizeable.

For typical values
$(m_b,\bar{\Lambda}_s)=(5 \,\mbox{GeV},370 \, \mbox{MeV})$ and $\mu$=5 GeV,
we get (including long-distance effects through $O_7$) the branching ratio
${\cal B}(B_{s}\rightarrow \gamma \gamma)_{SD+LD_{O_7}}=1.18 \cdot 10^{-6}$,
which is a factor $1.9$ larger compared to the lowest order estimate for the 
same values of the parameters.
However, varying $(m_b,\bar{\Lambda}_s)$ and $\mu$ in the allowed range 
results in significant variation on the branching ratio 
(see Table~\ref{mums}), yielding
\begin{equation}
0.38 \cdot 10^{-6} \leq
{\cal B}(B_s \rightarrow \gamma \gamma)_{SD+LD_{O_7}} 
\leq 1.43 \cdot 10^{-6} \, \, .
\label{eq:bsggBR}
\end{equation}
Improving this requires NLO calculation in the decay rate 
$B_s \rightarrow \gamma \gamma$ and further study of the bound state
effects.
The present best limit on the branching ratio in 
$B_s \rightarrow \gamma \gamma$ decay
\cite{L3} given in eq.~(\ref{eq:L3})
is still a factor $\approx 100-400$ away from the estimates given here.

Likewise the CP ratio $r_{CP \, SD+LD_{O7}}$ is rather uncertain.
Varying $m_b/2 \leq \mu \leq 2 m_b$ and $(m_b,\bar{\Lambda}_s)$ in the allowed
range, we get in the SM 
\begin{eqnarray}
0.55 \leq r_{CP \, SD+LD_{O7}} \leq 0.89 \; .
\end{eqnarray}

Further we presented a VMD model based calculation of the LD contribution
to CP-even $A^{+}$ and CP-odd $A^{-}$ decay amplitudes for
$B_{s}\rightarrow\gamma\gamma$ decay due to the inclusive process
$b\rightarrow s\psi$ via 
$B_{s}\rightarrow\phi\psi\rightarrow\phi\gamma \rightarrow\gamma\gamma$ decay.
The conversions to photons from both the $\psi_{i}$ 
resonances
and the $\phi$ meson lead to two suppressions and make the amplitudes
in eq.~(\ref{Apm}) smaller than the ones from the LD effect 
from the
$B_{s}\rightarrow\phi\gamma\rightarrow\gamma\gamma$ chain decay
%of the operator $O_{7}$ 
\cite{gudi}. To quantify this 
we estimated the ratio
\begin{eqnarray}
\rho=|\frac{A^{+(-)}_{\bar{LD}_{O_{2}}}(
B_{s}\rightarrow\phi\psi\rightarrow\phi\gamma\rightarrow\gamma\gamma)}
{A^{+(-)}_{LD_{O_{7}}}(B_{s}\rightarrow\phi\gamma\rightarrow\gamma\gamma)}|
=4 \pi^{2} Q_c \frac{a_2^{\mbox{eff}}}{|C_7^{\mbox{eff}}(\mu)|}
\sum_i \frac{f_{\psi_{i}}^2(0)}{m_b^2}
\label{eq:rho}
\end{eqnarray}
and found 
\begin{eqnarray}
2\% \leq \rho \leq 4 \% \; ,
\label{num}
\end{eqnarray}
while varying $\frac{m_b}{2} \leq \mu \leq 2 m_b$ and allowing
$a_2^{\mbox{eff}}$ to have a theoretical error of $25 \%$ as stated in 
\cite{neubert}. 
We compared the LD-contribution to 
$B_{s}\rightarrow\gamma\gamma$ decay resulting from intermediate
$\psi_{i}$ production
with the one obtained by the
interaction of the virtual charm loop with soft gluons \cite{Ruckl}.
We see that both amplitudes are in good agreement within the accuracy of the 
calculation. 
The new LD contribution resulting from the four-quark operators $O_1$ and
$O_2$ is smaller compared to the one 
of the operator $O_{7}$
\cite{gudi}
and affects our old estimate given in eq~(\ref{eq:bsggBR}) for the 
branching ratio
${\cal{B}}(B_{s}\rightarrow\gamma\gamma)_{SD+LD_{O_{7}}} $ \cite{gudi}
by less than $1\%$.

Another LD effect in $B_s \to \g \g$ decay is the one due to intermediate
$D_s,D_s^{*}$ mesons \cite{choudhuryellis98}.
At quark level this involves the four-Fermi operator transition 
$b \to c \bar{c} s$. The calculation cannot be done by first principles and 
hence is not straight forward, as the diagrammatic structure is the one of 
charmed mesons in a loop, which are no fundamental particles.
The $B_s D_s^{(*)}D_s^{(*)}$ vertex can be treated with a 
factorization approach, which is an approximation.
The next task is to give a prescription of the electromagnetic coupling of 
the (charged) $D_s^{(*)}$ mesons, which
can be solved by {\it minimal substitution} as a first approximation.
The authors of  \cite{choudhuryellis98} found a contribution to
the branching ratio $B_s \to \g \g$, which is even {\bf larger} than the SM 
short-distant one.
The channel $B_s \stackrel{D_s^{(*)}D_s^{(*)}}{\to} \g \g $ surely needs 
further investigation.

``New physics"-effects in $B_s \to \g \g$ decay are found to be small as 
in $b \to s \g \g$ they are mainly driven by the Wilson coefficient
$\cseff$, for which a strong constraint from data on $B \to X_s \gamma$
decay exists. This has accordingly been studied in 
ref.~\cite{alievhillererhan} in the 2HDM and in ref.~\cite{matias97} in the 
MSSM. However, a small enhancement 
of the branching ratio ${\cal{B}}(B_s \to \g \g)$ compared to the SM one is 
still possible in some regions of the parameter space.
In models with an extended operator basis the branching ratio and the CP ratio
can be much larger than the SM estimates \cite{alieverhan98}.

Once the necessary machines are running, 
$B_s \to   \g \g$ will certainly get 
the same attention as the single photon decay $b \to s \g$ has at present.
In particular, the branching ratio
${\cal{B}}(B_{s}\rightarrow\gamma\gamma) \sim {\cal{O}}(10^{-6})$ is large enough to be observed at the LHC.

\chapter{Summary \& Future \label{chap:out}}

Rare $B$ decays are one of the most active fields in recent particle physics. 
Its main theoretical principles and developments of the last 40 years, which 
are still used, are roughly given as:
Of course, the standard model (SM) (1961) \cite{GSW} and the quark mixing 
(CKM) matrix (1963) \cite{CKM};
further phenomenological approaches like vector meson dominance (VMD)(1969) 
\cite{sakurai} and the Fermi motion model (FM) (1979) \cite{aliqcd}.
The description of low energy weak processes (1981) \cite{inamilim}
made progress with the inclusion of QCD improved perturbative corrections by 
using renormalization group equation methods yielding the effective 
Hamiltonian theory (1991) 
%\cite{Ciuchini,grinstein,effhamali,effhamburas},
\citer{effhamali,Ciuchini}, \cite{grinstein},
together with the onset of the heavy quark expansion (HQE) (1990)\cite{georgi}.

We outlined these methods and applied them to the decays \bxsll with 
$\ell=e, \mu$ and (partly) to exclusive $B_s \to \g \g$ decay.
We presented quantitative SM based results in terms of distributions,
decay rates and moments which can be compared with experimental results.

Concerning \bxsll decays, the invariant dilepton mass spectrum and 
the Forward-Backward (FB) asymmetry can be used to extract the short-distance 
coefficients from data in conjunction with the branching ratio in \bxsg decay.
In this work we have analysed these spectra and their present uncertainties.

Further, apart from being a test of the SM, the decay \bxsll can help to 
improve our knowledge in certain aspects of long-distance effects:

a) HQE enables a description of $B$-meson bound state effects in terms of
higher dimensional operator matrix elements. 
We proposed in this thesis the
determination of the non-perturbative HQE parameters $\bar{\Lambda}$ and
$\lambda_1$ from moments of the hadronic invariant mass in \bxsll decays,
as it has been done
for the charged current induced decays $B \to X \ell \nu_{\ell}$ 
\cite{FLSphenom}.
Given enough data, these parameters can be extracted from \bxsll decays
and assuming universality, can be used in the analysis of,
e.g., the decay $B \to X_{u} \ell \nu_{\ell}$.
Likewise, \bxsll decay can be used to test the FM, which can be seen as a 
model dependent resummation of the 
theory into a so-called shape function \cite{Bigietal2,neubertbsg},
and/or to determine its parameters.
We remark here that some of the HQE and FM parameters are related.

b) Long-distance (LD) effects occur in \bxsll decays via the decay chain 
$B \to X_s (J/\Psi, \Psi^\prime, \dots) $ $\to X_s \ell^{+} \ell^{-}$,
which we have taken into account with a VMD ansatz.
Since in the literature there is no agreement 
about the implementation of this LD contribution together with the
short-distance one, 
we compared our approach \cite{AHHM97} 
with alternative ones \cite{KS96,LSW97} and 
estimated the resulting uncertainties in the observables.
We find that these uncertainties are not the dominant ones in \bxsll decay.
Further we have shown, that one can reduce the influence of the $c \bar{c}$ resonances by kinematical cuts.
At present, only an experimental analysis can identify the
correct procedure to implement the
charmonium resonances into an a parton model based calculation in the decays 
$b \to s \ell^+ \ell^{-}$.

Finally, by means of building appropriate ratios of (partly) integrated spectra
of rare \bxsll and semileptonic $B \to X_{u,c} \ell \nu_\ell$ decays,
the uncertainties resulting from bound state effects in the individual decays are expected to cancel out to a large extent.

The essential points reported in this thesis are:
\begin{itemize}
\item The calculation of leading power corrections in spectra and hadronic 
spectral moments in the decay \bxsll, 
including next-to-leading order perturbative 
${\cal{O}}(\alpha_s)$ corrections \cite{AHHM97,AH98-1,AH98-2,AH98-3}.

\item The presentation of leading logarithmic QCD corrections to 
$B_s \to \g \g$
decay and an estimate of the long-distance effects due to intermediate neutral 
vector mesons in $B_s \to \g \g $ decay \cite{gudi,hilleriltanpsi}.
\end{itemize}

Besides $B_s \to \g \g$, other exclusive decay modes relevant for future 
$B$-experiments are
$B_s \to \ell^+ \ell^{-}$, $B_s \to \nu \bar{\nu}$ and, of course,
$B \to (K,K^*) \ell^+ \ell^{-}$, $ B \to (K,K^*) \nu \bar{\nu}$.
The transitions $b \to s \nu \bar{\nu}$ are the cleanest 
theoretically among other $b \to s$ decays. 
The expected branching ratio is larger 
($\sim 4 \cdot 10^{-5}$ \cite{ali96}; 
note, that one has summed over all neutrino flavours) 
than the one from $b \to s \ell^+ \ell^{-}$. However, the decay 
$b \to s \nu \bar{\nu}$ is difficult to observe.
Moreover, the inclusive $B \to X_s \tau^+ \tau^{-}$ channel is interesting. 
In the SM, 
its branching ratio is smaller than the one involving light lepton species 
$(e,\mu)$, however, in non-standard multi-Higgs models it 
can be enhanced through large Higgs coupling of the $\tau$-lepton.

Upgrades of present experiments and planned $B$-facilities like Hera-B, CLEO, 
BaBar and Belle are about to start soon and they will be
sensitive to branching ratios of order $10^{-6}$ and below.
For an overview see Table \ref{tab:bfuture}.
\begin{table}[bth]
\begin{tabular}{|l|l|c|r|c|r|r|r|r|r|c|}
\hline
{Expt.} &  {Collider} &  
{Beams} & {$\sqrt{s}$}   &      {Year}   &   {${\cal L}$ ($10^{33}$} &
{$\sigma(b\bar b)$} &  {$b\bar b$ pairs}     & {$\beta\gamma c\tau$} 
              & {$\sigma(b\bar{b})$} \\ 
         &           & 
         & (GeV)   &    {online}   &   {$cm^{-1} s^{-1})$} & ($nb$)
           & {($10^7$/yr)} &  ($\mu m$)    &   {$/\sigma(q\bar{q})$}    \\   
\hline \hline
{CLEO III}   &  {CESR}   &  {$e^+e^-$}        &    {10} &  {1999}   &  {1.2}
      & {1}
& {1.2}     &  {30} & {$3\cdot10^{-1}$}  \\
            & {CESR-IV} &          &    {10} &    {?}      &    {30}    & {1}
& {30} & {30} & {$3\cdot10^{-1}$} \\
\hline
{BaBar}      &  {PEP-II} &  {$e^+e^-{\,^\dag}$}  &   {10} & {1999}   &  {3-10}
   & {1}  
& {3-10}    & {270} & {$3\cdot10^{-1}$}  \\
{Belle}      &  {KEK-B}  &  {$e^+e^-{\,^\dag}$}  &  {10} & {1999}   & {3-10}
   & {1}  
& {3-10}   & {200} &{$3\cdot10^{-1}$}  \\
\hline
{HERA-B}     &  {HERA}   &  {$pN$}              &    {40} &  {1998}   &  {---}
      & {6-12} 
& {50-100} & {9000}  & {$1\cdot10^{-6}$} \\
\hline
{CDF II}     & {Tevatron} & {$p\bar p$}      & {1800} &  {2000}   & {0.2-1.0}
  & $10^5$   
& {20000}   & {500}  & {$1\cdot10^{-3}$} \\
{D0}         &  & &  &  &  & & &    &     \\
{BTeV${\,^\ddag}$}       &  &  & & {2004} & {0.2} & & & {5000} & \\  
\hline
{LHC-B${\,^\ddag}$}      & {LHC}    &  {$pp$}            &
        { 14000} & {2005}   & {0.15}   & $5 \cdot 10^5$   
&{75000} & {7000} & {$5\cdot10^{-3}$} \\
{Atlas}      & & & &  &  &  & & {500} &  \\
{CMS}        & & & &  &  &  &       & &  \\
\hline
\end{tabular} 
\caption{ \it Future $B$ experiments. Parameters which do not
change between different experiments at the same collider are entered only
once. 
$^{\dag}$ asymmetric beam energies, $^{\ddag}$ forward detector.
\cite{skwarnicki97}}
\label{tab:bfuture}
\end{table}

This work will help the search for flavour changing neutral current \bxsll and 
$B_s \to \g \g$ decays and in particular, will contribute to precise 
determinations of the HQET parameters and $V_{ub}$ using the inclusive decays
\bxsll~and $B \to X_u \ell \nu_\ell$ in forthcoming $B$-facilities.

\newpage
{\Large \bf Acknowledgments}

I would like to thank Ahmed Ali, who suggested this work.
I have benefited from his strong interest in this work and advice in
numerous clarifying discussions.

Further the DESY theory group members and visitors are gratefully 
acknowledged for instructive discussions and practical help, 
especially Sven Moch, Ulrich Nierste, Tilman Plehn, 
Michael Pl\"umacher and Mathias Vogt.

Also I would like to thank Erhan Iltan for intense collaboration
on  $B_s \to \g \g$ decay.

%
%   This is Appendix A
\begin{appendix}
\chapter{Generalities \label{app:generalities}}
\setcounter{equation}{0}

\section{Input Parameters \label{app:input}}
\begin{table}[h]
        \begin{center}
        \begin{tabular}{|c|c||c|c|}
        \hline
        \multicolumn{1}{|c|}{Parameter} & 
                \multicolumn{1}{|c||}{Value} &
 \multicolumn{1}{|c|}{Parameter} & 
                \multicolumn{1}{|c|}{Value} 
    \\
        \hline \hline
$m_W$                 & $80.26$ GeV& $\alpha^{-1}$     & 129           \\
$m_Z$                 & $91.19$ GeV& $\alpha_s (m_Z) $ & $0.117 \pm 0.005$ \\
$\sin^2 \theta_W $    & $0.2325$   & $|V_{ts}^{*} V_{tb}|/|V_{cb}|$ & 1 \\
$m_s$                 & $0.2$ GeV  & ${\cal B}_{sl}$   & $(10.4 \pm 0.4)$ \% \\
$m_c$                 & $1.4$ GeV  & $\lambda_1$       & $-0.20$ GeV$^2$ \\
$m_b$                 & $4.8$ GeV  & $\lambda_2$       & $+0.12$ GeV$^2$ \\
$m_t$    & $175 \pm 5$ GeV& $\Gamma_{tot}(B_s)$& $4.09 \cdot 10^{-13}$ GeV \\
$\mu$                 & ${m_{b}}^{+m_{b}}_{-m_{b}/2}$&$f_{B_s}$& $0.2$ GeV \\
$\Lambda_{QCD}^{(5)}$ & $0.214^{+0.066}_{-0.054}$ GeV& $m_{B_s}$&$5.369$ GeV\\
        \hline
        \end{tabular}
        \end{center}
\caption{\it Values of the input parameters used in the numerical
          calculations, unless otherwise specified.}
\label{parameters}
\end{table}

\section{QCD \label{app:qcd}}

The QCD Lagrangian reads in covariant gauge \cite{tarrachbook}
($A^a_{\mu}$: gluon field) 
\begin{eqnarray}
\cl_{QCD}=\sum_{q=u,d,s,c,b,t} \bar{q} (i \Slash{D}-m_q) q-\frac{1}{4} 
G_{\mu \nu}^a G^{a \, \mu \nu} + \cl_{fix} +\cl_{ghosts} \; ,
\end{eqnarray}
where the gauge fixing term and the one for ghosts $c^a,\bar{c}^a$ are given as
\begin{eqnarray}
\cl_{fix} &= &-\frac{1}{2 \xi} (\partial \cdot A^a)^2 \; , \\
\cl_{ghosts} &= &\bar{c}^a \partial \cdot D^{ab} c^b \; .
\end{eqnarray}
The chromomagnetic field strength tensor and covariant derivative 
are written as
\begin{eqnarray}
G^a_{\mu \nu} &=& \partial_\mu A_\nu^a-\partial_\nu A_\mu^a+g f^{axy} A_\mu^x
A_\nu^y \; , \\
D_\mu&=& \partial_\mu -i g T^x A_\mu^x \; , 
\label{eq:textbookDmu} \\
D_\mu^{ab}&=&\delta^{ab} \partial_\mu-g f^{abx} A_\mu^x \; ,
\end{eqnarray}
where $f^{abx}$ are the structure constants of $SU(3)$, defined by
\begin{eqnarray}
[ T^a, T^b ]=i f^{abx} T^x \; .
\end{eqnarray}
We have the identities
\begin{eqnarray}
[D_\mu, D_\nu] &=&-i g T^x G^x_{\mu \nu} \; , \\
D^{ax}_\mu D^{xb}_\nu-D^{ax}_\nu D^{xb}_\mu & =& -g f^{abx} G^x_{\mu \nu} \; .
\end{eqnarray}
Often the abbreviation is used 
\begin{eqnarray}G_{\mu \nu}=G_{\mu \nu}^a T^a \; .
\label{eq:Gqcd}
\end{eqnarray}
$T^a, \, a=1, \dots ,8$ are the generators of QCD.
They are related to the Gell-Mann $(3\times3)$ matrices $\lambda^a$ through
$T^a=\frac{\lambda^a}{2}$.
The $T^a$ obey the following relations ($ i,j,k=1,2,3$)
\begin {eqnarray}
Sp(T_a)&=&0 \; , \\
Sp(T_a T_b)&=&\delta_{a b}/2 \; , \\
\label{eq:TaTa}
T^a_{i j} T^a_{k l}&=&-\frac{1}{2 N_c} \delta_{i j} \delta_{k l}+\frac{1}{2}
 \delta_{i k} \delta_{j l} \; , \\
T^a_{i l}   T^a_{l k}&=& \delta_{i k} C_F  \; ,
\end{eqnarray}
with the invariant $C_F$ is in an arbitrary $SU(N_c)$ given as
\begin{equation}
C_F=\frac{N_c^2-1}{2 N_c} \; \; (= \frac{4}{3} \; \; {\mbox{for}} \; N_c=3) 
\; .
\end{equation}
The coefficients of the QCD beta function (see eq.~(\ref{eq:qcdbeta}))
are written as:
\begin{eqnarray}
\beta_0&=&\frac{11 N_c-2 N_f}{3} \; , \\
\beta_1&=&\frac{34 N_c^2-10 N_c N_f-6 C_F N_f}{3} \; .
\end{eqnarray}
Here, $N_c$ denotes the number of colours ($N_c=3$ for QCD) and
$N_f$ denotes the number of {\it active} flavours ($N_f=5$ for the effective
Hamiltonian theory relevant for $b$ decays).

\section{Feynman Rules \label{app:feynrules}}

The covariant derivative consistent with our definition of the operator basis 
and the corresponding Wilson coefficients given in section \ref{sec:effham} is 
\cite{grinstein90}
\begin{eqnarray}
\label{eq:HeffDmu}
D_{\mu} \equiv \partial_{\mu}+i g T^a A_{\mu}^a+i e Q A_{\mu} \; ,
\end{eqnarray}
where $A^a, A$ denote the polarization four-vectors of the gluon, photon
respectively.
Note that the sign convention of the strong coupling here is opposite to the 
usual one appearing in QCD text books \cite{tarrachbook,Yndurainbook} given in
eq.~(\ref{eq:textbookDmu}), but can be made coincident with the substitution 
$g \to -g$.
The Feynman rules consistent with eq.~(\ref{eq:HeffDmu}) are given here with
boson propagators in Feynman gauge. In a general 
gauge with gauge parameter $\xi$ they are written as:
\begin{eqnarray}
-i \frac{g_{\mu \nu}+ (\xi-1) k_{\mu} k_{\nu}/(k^2+i \epsilon)}{k^2+i \epsilon}
\; ,
\end{eqnarray} 
with $\xi=1,0$ corresponding to Feynman, Landau gauge, respectively.
\begin{figure}[htb]
\vskip -1.0truein
\centerline{\epsfysize=10in
{\epsffile{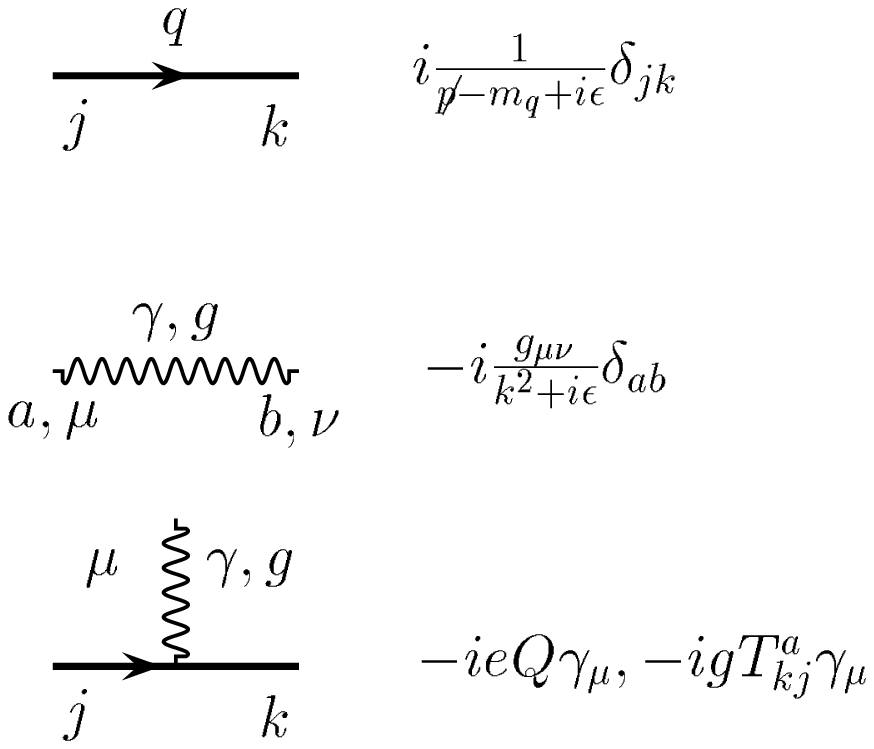}}}
\vskip -5.9truein
\label{fig:frules}
\end{figure}  
The rules should be complemented by
\begin{itemize}
\item evaluate fermion lines {\bf against} the momentum flow
\item add a $(-1)$ for a closed fermion loop and perform the trace over the 
string of $\gamma$ matrices
\end{itemize}
The rule for an $O_7$ operator insertion is, using $\partial_{\mu}=i q_{\mu}$
for an {\bf out going} photon and further $\epsilon \cdot q=0$ for a real
photon, and $F^{\mu \nu}=\partial^\mu A^\nu-\partial^\nu A^\mu$
\begin{eqnarray}
\sigma F =\sigma_{\mu \nu} F^{\mu \nu}=i [\slash{\partial},\Slash{A}]=
2 \gamma_{\mu} \slash{q} \epsilon^{\mu} \; .
\end{eqnarray}
The Fierz transformation in $d=4$ dimensions is defined as:
\begin{eqnarray}
\label{eq:fierzLL}
(\bar{q}_1 \gamma_{\mu} L q_2)(\bar{q}_3 \gamma_{\mu} L q_4)&=&
+(\bar{q}_1 \gamma_{\mu} L q_4)(\bar{q}_3 \gamma_{\mu} L q_2) \; , \\
(\bar{q}_1 \gamma_{\mu} L(R) q_2)(\bar{q}_3 \gamma_{\mu} R(L) q_4)&=&
(-2)(\bar{q}_1 R(L) q_4)(\bar{q}_3  L(R) q_2) \; .
\label{eq:fierzLR}
\end{eqnarray}

\subsection{Feynman Rules in the Heavy Quark Limit of QCD
\label{app:hqetrules} }

The effective Lagrangian in the limit of an infinitely heavy quark $h$
with mass
$m_Q \to \infty$ is given by ${\cal{L}}_{HQET}=\bar{h} i v . D h$.
%in eq.~(\ref{eq:Lhqet}). 
\begin{figure}[htb]
\vskip -1.0truein
\centerline{\epsfysize=10in
{\epsffile{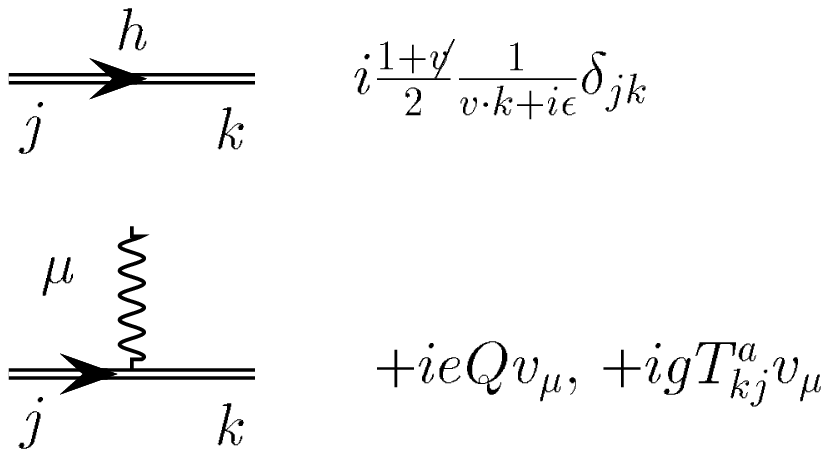}}}
\vskip -7.0truein
\label{fig:hqetrules}
\end{figure} 
Here $v$ denotes the velocity ($v^2=1$) of a heavy quark $h$ with momentum 
$p=m_Q v+k$ and small residual momentum $k$ of order $\Lambda_{QCD}$.
These rules are consistent with the definition of the covariant QCD derivative
in eq.~(\ref{eq:textbookDmu}), which causes a sign difference in the quark 
gluon/photon coupling compared to the  weak effective Hamiltonian rules above, 
based on the convention eq.~(\ref{eq:HeffDmu}).
A heavy quark $h$ is represented by a double line.

\section{Utilities}
A variety of tools for 1-loop calculations is collected in the 
appendix of ref.~\cite{Yndurainbook}. \\
Distributions
\begin{eqnarray}
\delta(x)&=&\frac{1}{2 \pi} \int_{\mbox{R}} dq e^{i q x} dq \; , \\
\theta(x)&=& \lim_{\epsilon \to 0} \frac{-i}{2 \pi} 
\int_{\mbox{R}} dq \frac{e^{i q x}}{q-i \epsilon} dq\; , \\
\frac{d \theta}{dx}(x)&=&\delta(x) \; . 
\end{eqnarray}
Geometrical series
\begin{eqnarray}
\frac{1}{1 \pm x}=1 \mp x + x^2  +\sum_{n=3}^{\infty}(\mp x)^n \; .
\end{eqnarray}
Fit quality $\chi^2$ (see also \cite{PDG} statistics section),
dof: degrees of freedom
\begin{eqnarray}
\chi^2&=&\sum_{\mbox{data points}} |\frac{\mbox{data-curve}}
{\mbox{error}}|^2 \; , \\
\frac{\chi^2}{dof}&=& \frac{\chi^2}{N_{\mbox{data points}}-
N_{\mbox{Fit parameters}}} \; . 
\label{eq:chidof}
\end{eqnarray}
$\frac{\chi^2}{dof}$ should be around 1, if it is much smaller, 
the errors are underestimated, if it is large, the model fails.

Special Functions useful for loops\\
Poly logarithms:
\begin{eqnarray}
Li_n(z)&=&\sum_{k=1}^{\infty} \frac{z^k}{k^n} ; \; |z| <1  \; , \\
Li_2(z)&=&- \int_0^z \frac{dt}{t} \ln(1-t)  \; .
\end{eqnarray}
Spence function 
\begin{eqnarray}
Sp(z) &\equiv& Li_2(z)=-\int_0^1 \frac{dt}{t} \ln(1-z t)  \; , \\
Sp(0) & = &0 \; \; , \; \;   Sp(1)=\frac{\pi^2}{6} \; \; , 
\; \;   Sp(-1)=\frac{\pi^2}{12}  \;  \; , \\
Sp(z)&=&-Sp(1-z) +\frac{\pi^2}{6}-\ln(z) \ln(1-z) \; , \\
Sp(z)&=&-Sp(\frac{1}{z}) -\frac{\pi^2}{6}-\frac{1}{2} \ln^2(-z) \; .
\end{eqnarray}
Useful identities for loops:
\begin{eqnarray}
\arctan(z) =\frac{1}{2i} \ln\frac{1+i z}{1-i z} \; , 
~~{\mbox{arctanh}}(z)=\frac{1}{2} \ln\frac{1+z}{1-z} \; .
\end{eqnarray}
Phase space element, 
$d^3 \vec{p}=|\vec{p}|^2 d|\vec{p}| d \cos{\theta} d\phi \; ,\cos{\theta} \in [-1,+1] \; ,
\phi \in [0,2 \pi[ $.
\begin{eqnarray}
\frac{d^3 \vec{p}}{2 E}=\int d^4 p \delta(p^2-m^2) \theta(E) \; \; ;~~
E=\sqrt{{\vec{p}}^2+m^2}\; .
\end{eqnarray}
Dirac algebra identities, for more see ref.~\cite{itzykson}, 
especially the appendix.
\begin{eqnarray}
\{ \gamma_{\mu}, \gamma_\nu \}=2 g_{\mu \nu} \; ,
~~\sigma_{\mu \nu}=\frac{i}{2}[\gamma_{\mu}, \gamma_\nu] \; ,
~~\gamma_{\mu} \gamma_\nu=g_{\mu \nu}-i \sigma_{\mu \nu} \; .
\end{eqnarray}
A useful tool within this context is the TRACER routine 
\cite{tracer} running under the symbolic algebra program  mathematica.

Chiral projectors $L(R)\equiv(1 \mp \gamma_5)/2$:
\begin{eqnarray}
\gamma_5^2&=&1 \; ,~~ \gamma_5^\dagger=\gamma_5 \; , \\
(L(R))^2&=&L(R)  \; ,~~ L R=R L=0 \; , ~~(L(R))^\dagger=L(R)  \; .
\end{eqnarray}
Further we have
\begin{eqnarray}
\gamma_0^2&=&1 \; ,~~ \gamma_0^\dagger=\gamma_0 \; .
\end{eqnarray}
Fermion fields 
\begin{eqnarray}
\Bar{\psi}& \equiv &\psi^{\dagger} \gamma_0=(\psi^{*})^T \gamma_0\; ,
~~ \psi_{L(R)} \equiv L(R) \psi \; , \nonumber \\
\Bar{\psi_{L(R)}}&=&\Bar{L(R) \psi}=(L(R) \psi)^\dagger \gamma_0=
\psi^\dagger (L(R))^\dagger \gamma_0= \psi^\dagger \gamma_0 R(L)
=\Bar{\psi} R(L) \; .
\label{eq:chiralfields}
\end{eqnarray}
%
%   This is Appendix A
%
\chapter[Dilepton Inv. Mass Distributions and FB Asymmetry]
{\bxsll Dilepton Invariant Mass Distributions and FB Asymmetry 
\label{app:dilepton}}

\setcounter{equation}{0}
\section{The Functions $T_{i}^{(j)}(v.\hat{q}, \hat{s})$ \label{app:Ti}}

In this appendix we list the functions $T_{i}^{(j)}(v.\hat{q}, \hat{s})$, 
$(i=1,2,3)$ with $j=0,1,2, s, g, \delta$, 
(defined in eq.~(\ref{eq:Tij})), representing the power
corrections in \bsll up to and including terms of order $m_B/m_b^3$.
The parton model contributions 
$T_{i}^{(0)}$ are given in eqs.~(\ref{eq:T01}) - (\ref{eq:T03}). 

\begin{eqnarray}
        {{T_1}^{(0)}}^{L/R} & = & - \frac{1}{x} \frac{m_B}{m_b} \left\{
                (1 - \z ) | \cnt|^2 
                \right.
                \nonumber \\
        & & \left. 
                \; \; \; \; \; \; \; \; \; \; 
                + \frac{4}{\s^2} \left[ 
                \left( 1 + \ms^2 \right)
                        \left( 2 \, (\z)^2 - \s (\z) - \s \right) 
                        - 2 \ms^2 \, \s \right] | C_7^{\mbox{eff}} |^2
                \right.
                \nonumber \\
        & & \left. 
                \; \; \; \; \; \; \; \; \; \; 
                + \frac{4}{\s} \left[ 
                \z - \s - \ms^2 \, (\z) 
                \right] {\rm Re} \left[ ( \cnt )^\ast C_7^{\mbox{eff}} \right]
                \right\}
                \, , 
\label{eq:T01} \\
        {{T_2}^{(0)}}^{L/R} & = & - \frac{2}{x} \frac{m_B}{m_b} \left\{
                |\cnt |^2
                - \frac{4}{\s} \left( 1 + \ms^2 \right) 
                        |C_7^{\mbox{eff}} |^2
                \right\}
                \, , \\
        {{T_3}^{(0)}}^{L/R} & = & \frac{1}{x} \frac{m_B}{m_b} \left\{
                - |\cnt |^2 
                - \frac{4}{\s^2} \left[ 2 \, (\z) - \s\right]
                        \left[ 1 - \ms^2 \right] |C_7^{\mbox{eff}}|^2
                \right.
                \nonumber \\
        & & \left. 
                \; \; \; \; \; \; \; \; \; \; 
                - \frac{4}{\s} \left( \ms^2 + 1 \right) 
                {\rm Re} \left[ (\cnt )^\ast C_7^{\mbox{eff}} \right] 
                \right\}
                \, , 
\label{eq:T03} \\
        {{T_1}^{(1)}}^{L/R} & = & - \frac{1}{3} \mb \, 
                ( \loo + 3 \lto ) \left\{
                \left[ \frac{1}{x} - \frac{2}{x^2} \left( \s - (\z)^2 
                        \right) \right] | \cnt |^2
                \right. \\
        & & 
                + \frac{4}{\s^2} \left[ 
                \frac{1}{x} \left( \s - 2 \, (\z)^2 \right) 
                - \frac{2}{x^2} \left( \s^2 - 2 \, \s (\z) - \s (\z)^2 
                        + 2 \, (\z)^3 \right) \right]\; (1+\ms^2) | C_7^{\mbox{eff}} |^2
\nonumber \\
& &   \left.
  - (\s-\z^2)\; \ms^2 \frac{8}{\s x^2}\; {\rm Re} (\cnt )^\ast C_7^{\mbox{eff}}
                \right\} \, , 
\nonumber \\
        {{T_2}^{(1)}}^{L/R} & = & - \frac{2}{3} \mb \, 
                ( \loo + 3 \lto ) \left[
                \frac{1}{x} + \frac{2}{x^2} \, \z \right]
                \left[ - | \cnt |^2 + \frac{4}{\s} \; (1+\ms^2) | C_7^{\mbox{eff}} |^2
                \right] \, , \\
        {{T_3}^{(1)}}^{L/R} & = & - \frac{2}{3} \mb \, 
                ( \loo + 3 \lto ) \left\{
                \frac{1}{x^2} ( 1 - \z ) | \cnt |^2    
                \right.
                \nonumber \\
        & &  
                - \frac{4}{\s^2} \left[ \frac{1}{x} \z 
                - \frac{1}{x^2} \left( \s + \s (\z) - 2 \, (\z)^2 \right)
                \right] \; (1-\ms^2) | C_7^{\mbox{eff}} |^2
\nonumber \\
& & \left. 
- \z \; \ms^2 \frac{4}{\s x^2}\; {\rm Re} (\cnt )^\ast C_7^{\mbox{eff}}
                \right\} \, , \\
        {{T_1}^{(2)}}^{L/R} & = & \frac{1}{3} \mb \, \loo 
                \left\{ 
                \left[ - \frac{4}{x^3} \left( \s - (\z)^2 \right)
                        + \frac{3}{x^2} \right] (1 - \z) 
                        | \cnt |^2 
                \right.
                \nonumber \\
        & & - \frac{4}{\s^2 x^3} \left[
-4 \s^2 -12 \ms^2 \s^2 + 3 \s x + 9 \ms^2 \s x -4 \s^2 \z -4 \ms^2 \s^2 \z +7 \s x \z
\right.
\nonumber \\
& &
 + 7 \ms^2 \s x \z + 12 \s \z^2+ 20 \ms^2 \s \z^2 -6 x \z^2 -6 \ms^2 x \z^2 + 4 \s \z^3
\nonumber \\
& & \left. + 4 \ms^2 \z^3 \s -4 x \z^3 -4 \ms^2 x \z^3 -8 \z^4 -8 \ms^2 \z^4
                \right] 
                 | C_7^{\mbox{eff}} |^2
                \nonumber \\
        & & 
                + \frac{4}{\s x^3} \left[
4 \s^2 - 5 \s x - 4 \s \z +4 \ms^2 \s \z +3 x  \z-3 \ms^2 x \z -4 \s \z^2+2 x \z^2 
\right.
\nonumber \\
& & \left. \left. +4 \z^3 -4 \ms^2 \z^3
                \right]
                {\rm Re} \left[ ( \cnt )^\ast \, C_7^{\mbox{eff}} \right]
                \right\} \, , \\
        {{T_2}^{(2)}}^{L/R} & = & - \frac{2}{3} \mb \, \loo 
                \left[ \frac{4}{x^3} \left( \s - (\z)^2 \right)
                        - \frac{3}{x^2} - \frac{2}{x^2} \, \z \right]
                \nonumber \\
        & &    \left( | \cnt |^2 - \frac{4}{\s} (1+\ms^2)\; |C_7^{\mbox{eff}} |^2
                \right) \, , \\
        {{T_3}^{(2)}}^{L/R} & = & - \frac{1}{3} \mb \, \loo 
                \left\{
                \left[ \frac{4}{x^3} \left( \s - (\z)^2 \right)
                        - \frac{5}{x^2} \right] | \cnt |^2 
                \right.
                \nonumber \\
        & & 
                + \frac{4}{\s^2 x^3} \left[
-4 \s^2 +5 \s x + 8 \s \z  -6 x \z + 4 \s \z^2-4 x  \z^2 -8 \z^3
                 \right]  (1-\ms^2) \, | C_7^{\mbox{eff}} |^2
                \nonumber \\
        & & \left. 
                + \frac{4}{\s x^3} \left[
(4 \s -3 x -4 \z^2)(1+\ms^2) -2 x \z 
                 \right]
                {\rm Re} \left[ ( \cnt )^\ast \, C_7^{\mbox{eff}} \right]
                \right\} \, , \\
        {{T_1}^{(s)}}^{L/R} & = & \frac{2}{\s \, x}
                \mb \, ( \loo + 3 \lto ) \left[ ( \s - \z ) 
                {\rm Re} \left[ ( \cnt )^\ast \, C_7^{\mbox{eff}} \right] +2 \ms^2 |C_7^{\mbox{eff}}|^2 \right] \, , \\
        {{T_2}^{(s)}}^{L/R} & = & 0 \, , \\
        {{T_3}^{(s)}}^{L/R} & = & - \frac{2}{\s \, x}
                \mb \, ( \loo + 3 \lto ) 
                {\rm Re} \left[ ( \cnt )^\ast \, C_7^{\mbox{eff}} \right] \, , \\
        {{T_1}^{(g)}}^{L/R} & = & \frac{1}{x^2} \mb \, \lto 
                \left\{ -( 1 - \z ) | \cnt |^2 
\right.
\nonumber \\
& & 
                + \frac{4}{\s^2} \left[ \s +3 \ms^2 \s + \s (\z) (1+\ms^2) - 2 \, (\z)^2 \; (1+\ms^2) \right] | C_7^{\mbox{eff}} |^2 
                \nonumber \\
        & & \left. 
                + \frac{4}{\s} ( \s - \z \; (1-\ms^2) ) 
                {\rm Re} \left[ ( \cnt )^\ast \, C_7^{\mbox{eff}} \right]
                \right\}
                \, , \\
        {{T_2}^{(g)}}^{L/R} & = & \frac{-2}{x^2} \mb \, \lto
                \left\{ - | \cnt |^2 - \frac{4}{\s} \; (1+\ms^2) \;| C_7^{\mbox{eff}} |^2 
                - 4  {\rm Re} \left[ ( \cnt )^\ast  C_7^{\mbox{eff}} \right]
                \right\} , \\
        {{T_3}^{(g)}}^{L/R} & = & \frac{-1}{x^2} \mb \, \lto
                \left\{ | \cnt |^2 + \frac{4}{\s^2} 
                \left[ 2 \, (\z) - \s \right] \; (1-\ms^2)\; | C_7^{\mbox{eff}} |^2 
\right.
\nonumber \\
& & \left.
                + \frac{4}{\s} \; (1+ \ms^2) \; {\rm Re} \left[ ( \cnt )^\ast \, C_7^{\mbox{eff}} \right]
                \right\} \, , \\
        {{T_1}^{(\delta)}}^{L/R} & = & 
                \frac{1}{2} \, \mb \, ( \loo + 3 \lto ) \left\{
                \left[ \frac{1}{x} - \frac{2}{x^2} 
                        \left( 1 - \z \right)^2 \right] 
                        | \cnt |^2
                \right.
                \nonumber \\
        & &  
                - \frac{4}{\s^2 x^2} \left[
-2 \s -6 \ms^2 \s + \s x + \ms^2 \s x +  4 \ms^2 \s \z + 4 \z^2 + 4 \ms^2 \z +2 \s \z^2
\right.
\nonumber \\
& & \left. + 2 \ms^2 \s \z^2-2 x \z^2 -2 \ms^2 x \z^2 -4 \z^3 -4 \ms^2 \z^3
                \right] | C_7^{\mbox{eff}} |^2 
                \nonumber \\
        & & 
                - \frac{4}{\s x^2} \left[ 
-2 \s + 2 \z -2 \ms^2 \z + 2 \s \z -x \z -2 \z^2 
\right.
\nonumber\\
& & \left. \left.
+ 2 \ms^2 \z^2
                \right] {\rm Re} \left[ ( \cnt )^\ast \, C_7^{\mbox{eff}} \right]
                \right\} \, , \\
        {{T_2}^{(\delta)}}^{L/R} & = &
                \mb \, ( \loo + 3 \lto ) \left[
                \frac{1}{x} - \frac{2}{x^2} \left( 1 - \z \right)
                \right] \left[ 
                | \cnt |^2 - \frac{4}{\s}\; (1+\ms^2)\; | C_7^{\mbox{eff}} |^2 \right] \, , \\
        {{T_3}^{(\delta)}}^{L/R} & = & 
                \mb \, ( \loo + 3 \lto ) \left\{
                - \frac{1}{x^2} \left( 1 - \z \right) | \cnt |^2 
                \right.
                \nonumber \\
        & & \left. 
                \; \; \; \; \; \; \; \; \; \; 
                + \frac{4}{\s^2} \left[
                \frac{1}{x} \z - 
                \frac{1}{x^2} \left( 1 - \z \right)
                        \left( 2 \, (\z) - \s \right) \right]\; (1-\ms^2)\; | C_7^{\mbox{eff}} |^2
                \right.
                \nonumber \\
        & & \left. 
                \; \; \; \; \; \; \; \; \; \; 
                - \frac{2}{\s x^2} \left[
2+2 \ms^2 -x -2 \z -2 \ms^2 \z
                \right] {\rm Re} \left[ ( \cnt )^\ast \, C_7^{\mbox{eff}} \right]
                \right\} \, . 
\end{eqnarray}
Here the variable $x$ is defined as
 $x \equiv 1 + \s - 2 \, (\z) - \ms^2 + i \, \epsilon $.

%
% This is Appendix B
%
\section{Auxiliary Functions $E_1(\hat{s}, \hat{u})$ and
 $E_2(\hat{s}, \hat{u})$   \label{app:Ei}}

 \setcounter{equation}{0}

In this appendix we give the auxiliary functions $E_1 (\s, \u)$
and $E_2 (\s, \u)$, multiplying the
delta-function $\delta [\u (\s,\hat{m}_s) -\u^2]$
and its first derivative 
$\delta^\prime [ \u (\s,\hat{m}_s) -\u^2 ]$, respectively,
appearing in the power corrected Dalitz distribution  $d^2{\cal 
B}/d\s d\u (b \to s \ell^+ \ell^-)$  in the HQE approach 
given in eq.~(\ref{eqn:dddw}).
 
\begin{eqnarray}
        E_1 (\s, \u) & = & 
                \frac{1}{3} \left\{ 
                2 \, \lo \left[ 1 -4 \ms^2+6 \ms^4-4 \ms^6 + \ms^8 -2 \ms^2 \s + 4 \ms^4 \s -2 \ms^6 \s + 2 \ms^2 \s^3  - \s^4
\right. \right.
\nonumber \\
& & \left. \left.
+ \u^2 \left( 1 -2 \ms^2 + \ms^4 -2 \ms^2 \s + 4 \, \s + \s^2 \right) \right]
        \right.
        \nonumber \\
& &      \left. 
                + 3 \, \lt \; (1-\ms^2+\s) \left[ -1 + 7 \ms^2 -11 \ms^4 +5 \ms^6 + 11 \, \s +10 \ms^2 \s -5 \ms^4 \s - 15 \, \s^2 \right. \right.
\nonumber \\
& &\left. \left.
-5 \ms^2 \s^2 +5 \s^3 
                + \u^2 \left( 1 -5 \ms^2 +5 \, \s  \right) \right]
                \right\}
        \nonumber \\
        & & 
        \times  \left( |C_9^{\mbox{eff}}|^2 + |C_{10}|^2 \right)
        \nonumber \\
        & &     + \frac{4}{3 \, \s} \left\{
                2 \, \lo \left[ 1 -3 \ms^2+2 \ms^4+2 \ms^6 -3 \ms^8+\ms^{10} -10 \ms^2 \s + 18 \ms^4 \s -6 \ms^6 \s -2 \ms^8 \s
\right. \right.
\nonumber \\
& &
 +16 \ms^4 \s^2 -6 \ms^2 \s^3 +2 \ms^4 \s^3 - \s^4 -\ms^2 \s^4 
\nonumber \\
& & \left. 
- \u^2 \left( 1 - \ms^2 - \ms^4 +\ms^6  + 4 \s + 2 \ms^2 \s - 2 \ms^4 \s + \s^2+ \ms^2 \s^2 \right)  \right]
        \nonumber \\
& &      
                + 3 \, \lt  \; (1-\ms^2+\s) \left[ 3 +2\ms^2-8 \ms^4 -2 \ms^6 +5 \ms^8+ 3 \s - 35 \ms^2 \s -27 \ms^4 \s -5 \ms^6 \s
\right.
\nonumber \\
& &
  -11 \s^2+8 \ms^2 \s^2 -5 \ms^4 \s^2  +5 \s^3 +5 \ms^2 \s^3
\nonumber \\
& & \left. \left.
 + \u^2 \left( 
                        3 +8 \ms^2+5 \ms^4 - 5 \s-5 \ms^2 \s  \right) \right]
                \right\} |C_7^{\mbox{eff}}|^2
        \nonumber \\
        & &     + 8 \left\{ \frac{2}{3} \lo (1 -4 \ms^2 +6 \ms^4 - 4 \ms^6+\ms^8 -\s-\ms^2 \s +5\ms^4 \s -3 \ms^6 \s  + \s^2 + 3\ms^4 \s^2
\right.
\nonumber \\
& &
 -\s^3- \ms^2 \s^3) 
\nonumber \\
 & & \left.
                + \lt (1 -\ms^2 + \s) \left[ 4-3 \ms^2 -6 \ms^4+5 \ms^6 -6 \s-4 \ms^2 \s -10  \ms^4 \s +2 \s^2+5 \ms^2 \s^2 + \u^2      \right]
                \right\} 
\nonumber \\
& &
        \, Re(C_9^{\mbox{eff}}) \, C_7^{\mbox{eff}}
        \nonumber \\
        & &     + 4 \, \s \, \u \left[ - \frac{4}{3} \lo \, \s 
                + \lt \left( 7 -2 \ms^2 -5 \ms^4 + 2 \, \s +10 \ms^2 \s - 5 \, \s^2 \right) \right] 
                \, Re(C_9^{\mbox{eff}}) \, C_{10}
        \nonumber \\
        & &     + \frac{8}{3} \u \left[ -4 \, \lo \, \s \; (1+\ms^2)
\right.
\nonumber \\
& &  \left. + 
        3  \lt \left( 5 + \ms^2 - \ms^4 -5 \ms^6+ 2  \s +4 \ms^2 \s+10 \ms^4 \s  - 3 \, \s^2-5 \ms^2 \s^2 \right) \right]
                 {C_{10}}^\ast \, C_7^{\mbox{eff}},
                \label{eqn:e1} 
\end{eqnarray}
% this is the simplified version of E2 -----------------------------------
\begin{eqnarray}
        E_2 (\s, \u) & = & 
                \frac{2}{3} \lo \left( 1 - \ms^2 + \s \right)^2 \u(\s,\ms)^2
\left[  
                        \left( 1 -2 \ms^2 +\ms^4 - \s^2 - \u^2 \right)
                        \left( |C_9^{\mbox{eff}}|^2 + |C_{10}|^2 \right)
\right.
        \nonumber \\
        & & 
  + 4   \left( 1 - \ms^2 -\ms^4+\ms^6-8 \ms^2 \s - \s^2-\ms^2 \s^2 + \u^2+\ms^2 \u^2 \right)
        \frac{|C_7^{\mbox{eff}}|^2}{\s}
        \nonumber \\
        & &  
                + 8  \left( 1 -2 \ms^2+\ms^4 - \s -\ms^2 \s \right) 
                        Re(C_9^{\mbox{eff}}) \,C_7^{\mbox{eff}}
\nonumber \\
 & & \left.
        + 4 \, \s \, \u \; 
                        Re(C_9^{\mbox{eff}}) \,C_{10}
                + 8 \, \u \; (1+\ms^2) \;
                                                Re(C_{10}) \, C_7^{\mbox{eff}} 
\right]         
\; .
                \label{eqn:e2} 
\end{eqnarray}

\section{Dalitz Distribution $d^2\Gamma(B \to X_s \ell^+ 
\ell^-)/ds du$ and FB Asymmetry in the Fermi Motion Model \label{app:dalitz}}

 \setcounter{equation}{0}
\begin{figure}[htb]
\vskip -1.4truein
\centerline{\epsfysize=5in
{\epsffile{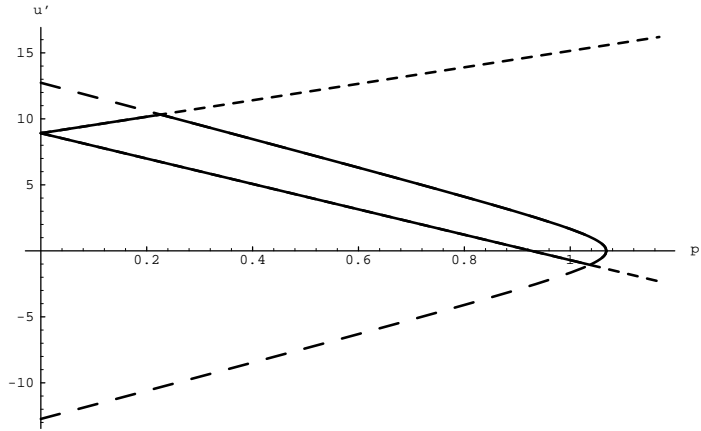}}}
\vskip -1.6truein
\caption[]{\it Phase space boundaries for the $u^\prime$ and $p$ integrations 
with fixed values of $s$ and $u$ drawn for $s=15$ GeV$^2$ and
$u= 8.9$ GeV$^2$. The integration region (solid 
curve) is given by the intersection of $u^\prime_\pm$ (short dashed)
and $\pm u(s,p)$ (long dashed curve). The Fermi motion parameters used are
$(p_F,m_q)=(450,0)$ in MeV.}
\label{fig:pspace}
\end{figure}

We start with the differential decay rate
 $d^3\Gamma_B/{\rm d}s \, {\rm d}u \, {\rm d}p$, describing the decay
\bsll of a moving $b$-quark having a mass $m_b \equiv m_b(p)$ and 
three momentum $\vert p \vert \equiv p$ 
with a distribution $\phi(p)$, which will be
taken as a Gaussian \cite{aliqcd},
\begin{equation}
        \frac{{\rm d}\Gamma_B}{{\rm d}s \, {\rm d}u \, {\rm d}p} = 
        \int^{u^\prime_{\rm max}}_{u^\prime_{\rm min}} \, {\rm d}u^\prime \, 
        \frac{{m_b}^2}{m_B} \, p \, \phi(p) \, 
        \frac{1}{\sqrt{{u^\prime}^2 + 4 \, {m_b}^2 \, s}} \, 
        \left[ \frac{{\rm d}^2 \Gamma_b}{{\rm d}s \, {\rm d}u^\prime} 
        \right] \; . 
\end{equation}
Here, ${\rm d}^2 \Gamma_b/{\rm d}s \, {\rm d}u^\prime$
is the  double differential decay rate of a b-quark at rest and
can be written in the case of \bsll as,  
\begin{equation}
        \frac{{\rm d}^2 \Gamma_b}{{\rm d}s \, {\rm d}u^\prime} =  
        \left | V_{ts}^\ast \, V_{tb} \right|^2 \, 
        \frac{{G_F}^2}{192 \, \pi^3} \, \frac{1}{{m_b}^3} \, 
        \frac{3 \, \alpha^2}{16 \, \pi^2} \left[ F_1(s,p) +  
        F_2(s,p) \, u^\prime + F_3(s,p) \, {u^\prime}^2 
        \right] \; , 
\end{equation}
and the three functions have the following expressions,
\begin{eqnarray}
    F_1(s,p) & = & \left[ \left( {m_b}^2 - {m_s}^2 \right)^2 - s^2 \right]
                \left( |{C_9^{\mbox{eff}}}|^2 + |{C_{10}}|^2 \right) 
                \nonumber \\
               & + &4 \left[ {m_b}^4 - {m_s}^2 \, {m_b}^2 - {m_s}^4 
                        + \frac{{m_s}^6}{{m_b}^2} - 8 \, s \, {m_s}^2 
                        - s^2 \, \left( 1 + \frac{{m_s}^2}{{m_b}^2} \right)
                     \right] \, \frac{{m_b}^2}{s} \, |{C_7^{\mbox{eff}}}|^2 
                \nonumber \\
             & -& 8 \left[ s \, \left( {m_b}^2 + {m_s}^2 \right) 
                        - \left( {m_b}^2 - {m_s}^2  \right)^2 \right] \, 
                        {\rm Re} (C_7^{\mbox{eff}} \, C_9^{\mbox{eff}})
                        \; , \\
        F_2(s,p) & = & 4 \, s \, {\rm Re}(C_9^{\mbox{eff}} \, C_{10}) + 
                8 \, \left( {m_b}^2 + {m_s}^2 \right) \, C_{10} \, C_7^{\mbox{eff}} 
                \; , \\
        F_3(s,p) & = & - \left(|{C_9^{\mbox{eff}}}|^2 + |{C_{10}}|^2 \right) 
                + 4 \left[ 1 + \left( \frac{{m_s}^2}{{m_b}^2} \right)^2 
              \right] \, \frac{{m_b}^2}{s} \, |{C_7^{\mbox{eff}}}|^2  \; , 
\end{eqnarray}
which can be read off directly from eq.~(\ref{eqn:dddw})
 in the limit $\lambda_{i} =0;~i=1,2$.
Note that the Wilson coefficient ${C_9}^{\mbox{eff}}$ also has an
implicit $m_b$ dependence, as can be seen in the text.
The integration limit for $u^\prime$ is determined through the equations 
\begin{eqnarray}
        u_{\rm max}^\prime & \equiv & {\rm Min} \, 
                \left[ u_+^\prime, u(s,p) \right] \; , \\
        u_{\rm min}^\prime & \equiv & {\rm Max} \, 
                \left[ u_-^\prime, -u(s,p) \right]      \; ,
\end{eqnarray}
where 
\begin{equation}
        u^\prime_{\pm} \equiv \frac{E_b}{m_B} \, u \pm 
        \frac{p}{m_B} \, \sqrt{4 \, s \, {m_B}^2 + u^2} \; , 
\end{equation}
\begin{equation}
E_b=\sqrt{m_b^2+p^2} \; ,
\end{equation}
and
\begin{equation}
        u(s,p) \equiv \sqrt{\left[ s- \left( m_b + {m_s} \right)^2  \right] 
                \left[ s -\left( m_b - {m_s} \right)^2 \right]} \; . 
\end{equation}

A typical situation in the phase space is displayed in Fig.~\ref{fig:pspace}.
 Integration over $p$ gives the double differential 
decay rate including the Fermi motion. The result is, 
\begin{eqnarray}
        \frac{{\rm d}^2 \Gamma_B}{{\rm d}s \, {\rm d}u} & = & 
          \left | V_{ts}^\ast \, V_{tb} \right|^2 \, 
        \frac{{G_F}^2}{192 \, \pi^3} \, 
        \frac{3 \, \alpha^2}{16 \, \pi^2} 
        \int_0^{p_{\rm max}} {\rm d}p  \, 
        \frac{1}{m_b 2 m_B} \, p \, \phi(p) \,  
         \nonumber \\ 
        & & \; \; \; \; \; \; 
        \left\{ F_1(s,p) \, \ln \left| \frac{u_{\rm max}^\prime + 
                \sqrt{{u_{\rm max}^\prime}^2 + 4 \, {m_b}^2 \, s}}{
                u_{\rm min}^\prime + 
                \sqrt{{u_{\rm min}^\prime}^2 + 4 \, {m_b}^2 \, s}} \right| 
        \right. \nonumber \\ 
        & & \; \; \; \; \; \; \left.
        + F_2(s,p) \, \left[ \sqrt{{u_{\rm max}^\prime}^2 + 4 \, {m_b}^2 \, s} 
                - \sqrt{{u_{\rm min}^\prime}^2 + 4 \, {m_b}^2 \, s} \right]
        \right. \nonumber \\ 
        & & \; \; \; \; \; \; \left.
        + F_3(s,p) \, \frac{1}{2} \, \left[ 
        u_{\rm max}^\prime \, \sqrt{{u_{\rm max}^\prime}^2 + 4 \, {m_b}^2 \, s} 
        - u_{\rm min}^\prime \, \sqrt{{u_{\rm min}^\prime}^2 + 4 \, {m_b}^2 \, s} 
        \right. \right. \nonumber \\ 
        & & \; \; \; \; \; \; \; \; \; \; \; \; \; \; \; \; \; \; \left. \left.
        - 4 \, {m_b}^2 \, s \, 
        \ln \left| \frac{u_{\rm max}^\prime + 
                \sqrt{{u_{\rm max}^\prime}^2 + 4 \, {m_b}^2 \, s}}{
                u_{\rm min}^\prime + 
                \sqrt{{u_{\rm min}^\prime}^2 + 4 \, {m_b}^2 \, s}} \right| 
        \right] \right\} \; .
\end{eqnarray}
Note that the upper limit in $p$ integration, $p_{\rm max}$ is 
determined such that $p$ satisfies, 
\begin{equation}
        u_{\rm max}^\prime(p_{\rm max},s,u) = 
                u_{\rm min}^\prime(p_{\rm max},s,u) \; .
\end{equation}
Lastly, the normalized differential FB asymmetry including the Fermi motion becomes, 
\begin{equation}
        \frac{{\rm d}\overline{\cal A}}{{\rm d}s} = 
        \frac{\int_{-u_{\rm ph}}^0 \frac{{\rm d}\Gamma_B}{
                {\rm d}s \, {\rm d}u} \, {\rm d}u 
        - \int^{u_{\rm ph}}_0 \frac{{\rm d}\Gamma_B}{
                {\rm d}s \, {\rm d}u} \, {\rm d}u 
        }{\int_{-u_{\rm ph}}^0 \frac{{\rm d}\Gamma_B}{
                {\rm d}s \, {\rm d}u} \, {\rm d}u 
        + \int^{u_{\rm ph}}_0 \frac{{\rm d}\Gamma_B}{
                {\rm d}s \, {\rm d}u} \, {\rm d}u } \; , 
\end{equation}
where 
\begin{equation}
        u_{\rm ph} \equiv \sqrt{
                \left[ s - \left( m_B + m_X \right)^2 \right]
                \left[ s - \left( m_B - m_X \right)^2 \right]} \; ,
\end{equation}
and 
\begin{equation}
        m_X \equiv {\rm Max} \left[ m_K, m_s + m_{\rm q} \right] \; , 
\end{equation}
with $m_{\rm q}$ the spectator quark mass and $m_K$ the 
kaon mass. Since the calculations are being done for an inclusive decay
\bxsll, we should have put this threshold higher, say starting from 
$m_K + m_\pi$, but as this effects the very end of a steeply falling 
dilepton mass spectrum, we have kept the threshold in \bxsll at 
$m(X_s) =m_K$.

\chapter{\bxsll Hadron Spectra and Moments \label{app:hadron}}
\setcounter{equation}{0}

\section{Coefficient Functions 
$g_i^{(9,10)},g_i^{(7)},g_i^{(7,9)},h_i^{(9)},h_i^{(7,9)},
k_1^{(9)},k_1^{(7,9)}$ \label{app:auxfunc1}}
These functions enter in the derivation of the leading $(1/m_b^2)$
corrections to the hadron energy spectrum in \bxsll, given in 
eq.~(\ref{singlediff}).
\begin{eqnarray}
g_0^{(9,10)}&=&
\sqrt{x_0^2-\ms^2} \frac{32}{3}(-2 \ms^2+3 x_0+3 \ms^2 x_0-4 x_0^2)
 \; , \\
g_1^{(9,10)}&=&
\frac{1}{\sqrt{x_0^2-\ms^2}} \frac{16}{9}(9 \ms^2+23 \ms^4
-18 \ms^2 x_0-18 x_0^2-52 \ms^2 x_0^2+36 x_0^3 + 20 x_0^4)
 \; , \\
g_2^{(9,10)}&=&
\frac{1}{\sqrt{x_0^2-\ms^2}} \frac{16}{3}(3 \ms^2+ 23 \ms^4-3 x_0 
-21 \ms^2 x_0 -6 x_0^2-52 \ms^2 x_0^2+36 x_0^3+20 x_0^4)
 \; , \\
g_0^{(7)}&=&
\sqrt{x_0^2-\ms^2}\frac{64}{3}(10 \ms^2+10 \ms^4-3 x_0-18 \ms^2 x_0
-3 \ms^4 x_0+ 2 x_0^2+2 \ms^2 x_0^2) 
 \; , \\
g_1^{(7)}&=&
\frac{1}{\sqrt{x_0^2-\ms^2}} \frac{1}{(x_0-\frac{1}{2}(1+\ms^2))^2}
\frac{-8}{9}(9 \ms^2+ 34 \ms^4+104 \ms^6+110 \ms^8+31 \ms^{10} \nonumber \\
&-&132 \ms^4 x_0
- 312 \ms^6 x_0 
- 180 \ms^8 x_0-18 x_0^2-170 \ms^2 x_0^2-58 \ms^4 x_0^2+74 \ms^6 x_0^2-20 \ms^8 x_0^2 \nonumber \\ 
&+& 72 x_0^3+564 \ms^2 x_0^3 + 576 \ms^4 x_0^3
+ 228 \ms^6 x_0^3-116 x_0^4-676 \ms^2 x_0^4-436 \ms^4 x_0^4-20 \ms^6 x_0^4
\nonumber \\ 
&+& 72 x_0^5+240 \ms^2 x_0^5+ 24 \ms^4 x_0^5)
 \; , \\
g_2^{(7)}&=&
 \frac{1}{\sqrt{x_0^2-\ms^2}} \frac{1}{x_0-\frac{1}{2}(1+\ms^2)}
\frac{16}{3}(27 \ms^2+ 93 \ms^4+97 \ms^6 + 31 \ms^8-3 x_0-63 \ms^2 x_0
\nonumber \\
&-&189 \ms^4 x_0 - 129 \ms^6 x_0-18 x_0^2 -108 \ms^2 x_0^2 -62 \ms^4 x_0^2-20 \ms^6 x_0^2+ 72 x_0^3+ 324 \ms^2 x_0^3 \nonumber \\
&+& 180 \ms^4 x_0^3-60 x_0^4
- 152 \ms^2 x_0^4-20 \ms^4 x_0^4)
 \; , \\
g_0^{(7,9)}&=&
\sqrt{x_0^2-\ms^2} 128 (-2 \ms^2+x_0+\ms^2 x_0)
 \; , \\
g_1^{(7,9)}&=&
\frac{1}{\sqrt{x_0^2-\ms^2}} 64 (\ms^2 +3 \ms^4 +2 \ms^2 x_0-2 x_0^2 -
4 \ms^2 x_0^2)
 \; , \\
g_2^{(7,9)}&=&
\frac{1}{\sqrt{x_0^2-\ms^2}} 64 (5 \ms^2+9 \ms^4-x_0 + 5 \ms^2 x_0-6 x_0^2
-12 \ms^2 x_0^2)
 \; , \\
h_1^{(9,10)}&=&
\frac{32}{9} \sqrt{x_0^2-\ms^2}
(-12 \ms^2-6 \ms^4+9 x_0+19 \ms^2 x_0+3 x_0^2+15 \ms^2 x_0^2-28 x_0^3) 
 \; , \\
h_2^{(9,10)}&=&
\frac{32}{3} \sqrt{x_0^2-\ms^2}
(-6 \ms^4+ 3 x_0+5 \ms^2 x_0+ 3 x_0^2+15 \ms^2 x_0^2-20 x_0^3)
 \; , \\
h_1^{(7,9)}&=&
\frac{128}{3} \sqrt{x_0^2-\ms^2}
(-8 \ms^2-2 \ms^4+3 x_0-3 \ms^2 x_0+5 x_0^2+5 \ms^2 x_0^2)
 \; , \\
h_2^{(7,9)}&=&
\frac{128}{3} \sqrt{x_0^2-\ms^2}
(-4 \ms^2-6 \ms^4+3 x_0-15 \ms^2 x_0+7 x_0^2+15 \ms^2 x_0^2)
 \; , \\
k_1^{(9,10)}&=&
\frac{64}{9} \sqrt{x_0^2-\ms^2}^3
(2 \ms^2-3 x_0-3 \ms^2 x_0+ 4 x_0^2) 
 \; , \\
k_1^{(7,9)}&=&
\frac{-256}{3} \sqrt{x_0^2-\ms^2}^3
(-2 \ms^2+ x_0+ \ms^2 x_0)
 \; .
\label{eq:coefffunc}
\end{eqnarray}

\section{Auxiliary Functions $f_{\delta}(\lo,\lt),f_{\delta'}(\lo,\lt)$
 \label{app:auxfunc}}
The auxiliary functions given below are the coefficients of the singular 
terms in the derivation of the leading 
$(1/m_b^2)$ corrections to the hadron energy spectrum in \bxsll, given in
eq.~(\ref{singlediff}).
 \begin{eqnarray}
 f_{\delta}(\lo,\lt)&=&{\cal B}_0 
 \left\{ 
\left[\frac{2}{9} (1-\ms^2)^3 (5-\ms^2) \lo
\right. \right. 
\nonumber \\
&+& \left. \frac{2}{3} (1-\ms^2)^3 (-1+5 \ms^2) \lt 
\right] \left( |C_9^{\mbox{eff}}|^2 + |C_{10}|^2 \right)
\nonumber \\
&+& \left[\frac{1}{9}(1+ 12 \ml^2-88 \ml^4-4 \ms^2-36 \ml^2 \ms^2-736 \ml^4 \ms^2+ 5 \ms^4+ 24 \ml^2 \ms^4+ 720 \ml^4 \ms^4
\right. \nonumber \\
&+&24 \ml^2 \ms^6+ 160 \ml^4 \ms^6-5 \ms^8 -36 \ml^2 \ms^8 -56 \ml^4 \ms^8 )\frac{\lo}{\ml^2} 
+ \frac{4}{3} (-1+\ms^2) (-3
 \nonumber \\
&+& \left. 14\ml^2-2 \ms^2+166 \ml^2 \ms^2+8 \ms^4+154 \ml^2 \ms^4+2 \ms^6+50 \ml^2 \ms^6-5 \ms^8) \lt
\right] \frac{|C_7^{\mbox{eff}}|^2}{\ml^2}
\nonumber \\
&+&   \left[\frac{8}{3} (1-\ms^2)^3 (7+\ms^2) \lo 
+ \frac{8}{3} (1-\ms^2)^3 (13+15 \ms^2 ) \lt
\right]
 Re(C_9^{\mbox{eff}}) \, C_7^{\mbox{eff}} \nonumber \\
&+& \left. \lo (-1+\ms^2)^5 (\frac{2}{9} \frac{d |C_9^{\mbox{eff}}|^2}{d\s_0} 
+ \frac{8}{3} \frac{d Re(C_9^{\mbox{eff}})}{d\s_0}  \, C_7^{\mbox{eff}})
 \right\}.
\end{eqnarray}
\begin{eqnarray}
 f_{\delta'}(\lo,\lt)&=&{\cal B}_0 \lo 
 \left\{ 
    \frac{1}{9} (1-\ms^2)^5 \left( |C_9^{\mbox{eff}}|^2 + |C_{10}|^2 \right)
\right. \nonumber \\
&+& \frac{2}{9} (-1+\ms^2)^3 (-1+14 \ml^2+\ms^2+52 \ml^2 \ms^2+\ms^4
+14 \ml^2 \ms^4 -\ms^6)
 \frac{|C_7^{\mbox{eff}}|^2}{\ml^2} \nonumber \\
&+& \left.  \frac{4}{3} (1-\ms^2)^5
 Re(C_9^{\mbox{eff}}) \, C_7^{\mbox{eff}}  \right\}.
\end{eqnarray}

\section{The Functions $\alpha_i, \beta_i, \gamma_i, \delta_i$ 
\label{app:moments}}
The functions entering in the definition of the hadron moments 
in eq.~(\ref{momentexp}) are given in this appendix. Note that the 
functions $\alpha_i^{(n,m)}$ and $\beta_i^{(n,m)}$ multiply the
Wilson coefficients $|C_7^{\mbox{eff}}|^2$ and $C_{10}^2$, respectively.
The functions $\gamma_i^{(n,m)},\delta_i^{(n,m)}$ result from the Wilson 
coefficients $C_7^{\mbox{eff}} Re (C_9^{\mbox{eff}}),|C_9^{{\mbox{eff}}}|^2$,
respectively. They cannot be given in a closed form since
$C_9^{{\mbox{eff}}}$ is an implicit function of
$x_0$. 

{\bf \underline{ The functions $\mathbf \alpha_i^{(n,m)}$}}
\begin{eqnarray}
\alpha_0^{(0,0)}\! \! \! &=& \! \! \!
\frac{16}{9}(-8-26\ms^2+18 \ms^4+22 \ms^6-11 \ms^8)+ 
   \frac{32}{3}(-1+\ms^2)^3(1+\ms^2) \ln(4 {\hat m_l}^2)  \nonumber \\
&+& 
   \frac{64}{3} \ms^4 (-9-2 \ms^2+\ms^4) \ln(\ms)  \; ,\\
\alpha_1^{(0,0)}\! \! \! &=& \! \! \!\frac{1}{2} \alpha_0^{(0,0)} \; , \\
\alpha_2^{(0,0)}\! \! \! &=& \! \! \!\frac{8}{3}(-4+38 \ms^2-42 \ms^4-26 \ms^6-15 \ms^8) + 
   16 (-1+\ms^2)^2(3+8 \ms^2+5 \ms^4) \ln(4 {\hat m_l}^2)\nonumber \\
&+& 
   32 \ms^2(-8-17\ms^2-2\ms^4+5\ms^6) \ln (\ms)  \; ,\\
\alpha_0^{(0,1)}\! \! \! &=& \! \! \!\frac{2}{9}(-41-49 \ms^2+256 \ms^4-128 \ms^6-43 \ms^8) + 
   \frac{16}{3}(-1+\ms^2)^3 (1+\ms^2)^2 \ln(4 {\hat m_l}^2) \nonumber \\
&+& 
   \frac{16}{3} \ms^4 (3-\ms^2-2 \ms^4) \ln(\ms)  \; , \\
\alpha_1^{(0,1)}\! \! \! &=& \! \! \!\alpha_1^{(0,0)}  \; ,\\
\alpha_2^{(0,1)}\! \! \! &=& \! \! \!\frac{4}{9}(21+167 \ms^2+128 \ms^4-276 \ms^6-319 \ms^8)
+ \frac{16}{3}(-1+\ms^2)^2 (3+14 \ms^2+21 \ms^4 \nonumber \\ 
&+&10 \ms^6) \ln(4 {\hat m_l}^2)
+ \frac{32}{3} \ms^2
      (3-24 \ms^2-18 \ms^4+\ms^6) \ln(\ms) \; , \\
\alpha_0^{(0,2)}\! \! \! &=& \! \! \!\frac{2}{45}(-119-144 \ms^2+45 \ms^4+320 \ms^6+45 \ms^8) + 
   \frac{8}{3}(-1+\ms^4)^3 \ln(4 {\hat m_l}^2)  \nonumber \\
& -& 
   \frac{16}{3} \ms^6 (8+3 \ms^2) \ln(\ms) \; , \\
\alpha_1^{(0,2)}\! \! \! &=& \! \! \!\frac{1}{27}(-127-278\ms^2+1075\ms^4-800 \ms^6+49\ms^8)
 +\frac{4}{9}(1-\ms^2)^3 (-7-17\ms^2\nonumber \\
&-&5\ms^4+5\ms^6) \ln(4 {\hat m_l}^2)
+ \frac{8}{9} \ms^4 (18-38 \ms^2-13 \ms^4) \ln(\ms)  \; , \\
\alpha_2^{(0,2)}\! \! \! &=& \! \! \!\frac{1}{9}(27-46 \ms^2+1681 \ms^4-688 \ms^6-1189 \ms^8)
+ \frac{4}{3}(-1+\ms^4)^2 (3+20 \ms^2+25 \ms^4)
\ln(4 {\hat m_l}^2)\nonumber \\
&-& \frac{8}{3} \ms^4 (18+54 \ms^2+47 \ms^4) \ln(\ms) \; , \\
\alpha_0^{(1,0)}\! \! \! &=& \! \! \!0 \;  ,\\
\alpha_1^{(1,0)}\! \! \! &=& \! \! \!\frac{2}{9}(-23-159 \ms^2-112 \ms^4+304 \ms^6-45 \ms^8)
- \frac{16}{3} (-1+\ms^2)^4 (1+\ms^2)\ln(4 {\hat m_l}^2)\nonumber \\
& +& 
  \frac{16}{3} \ms^4 (-39-7 \ms^2+6 \ms^4)\ln(\ms) \; , \\
\alpha_2^{(1,0)}\! \! \! &=& \! \! \!\frac{2}{9}(-93-469 \ms^2+704 \ms^4-127 \ms^8) + 
\frac{16}{3}(-1+\ms^2)^3 (3+8 \ms^2+5 \ms^4) \ln(4 {\hat m_l}^2)\nonumber \\ 
&-& 
   \frac{112}{3} \ms^4 (3+3 \ms^2+2 \ms^4)\ln(\ms) \; ,   \\
\alpha_0^{(1,1)}\! \! \! &=& \! \! \!0 \; ,  \\
\alpha_1^{(1,1)}\! \! \! &=& \! \! \!\frac{2}{27}(-4-131 \ms^2+307 \ms^4-416 \ms^6+178 \ms^8)
-\frac{8}{9}(-1+\ms^2)^4 (1+6 \ms^2+5 \ms^4)\ln(4 {\hat m_l}^2)\nonumber \\ 
&+& \frac{16}{9} \ms^4 (9-35 \ms^2-7 \ms^4) \ln(\ms) \; ,  \\
\alpha_2^{(1,1)}\! \! \! &=& \! \! \!\frac{2}{9}(-60-185 \ms^2+173 \ms^4+160 \ms^6+70 \ms^8)
+\frac{8}{3}(-1+\ms^2)^3 (1+\ms^2)^2 (3+5 \ms^2)\ln(4 {\hat m_l}^2)\nonumber \\
&+& \frac{16}{3} \ms^4 (3-21 \ms^2-13 \ms^4)\ln(\ms)  \; , \\
\alpha_0^{(2,0)}\! \! \! &=& \! \! \!0  \; , \\
\alpha_1^{(2,0)}\! \! \! &=& \! \! \!\frac{8}{135}(119-176 \ms^2-1085 \ms^4+400 \ms^6+835 \ms^8)
+\frac{32}{9}(1-\ms^2)^5 (1+\ms^2) \ln(4 {\hat m_l}^2)\nonumber \\
&-& 
  \frac{64}{9} \ms^6 (28 + 5 \ms^2)\ln(\ms)  \; , \\
\alpha_2^{(2,0)}\! \! \! &=& \! \! \!0 \, .
\end{eqnarray}
{\bf \underline{ The functions $\mathbf \beta_i^{(n,m)}$}}
\begin{eqnarray}
\beta_0^{(0,0)}\! \! \! &=& \! \! \!
 \frac{2}{3} (1-8 \ms^2+8 \ms^6-\ms^8-24 \ms^4 \ln(\ms)) \; ,\\
\beta_1^{(0,0)}\! \! \! &=& \! \! \!\frac{1}{2}\beta_0^{(0,0)} \; , \\ 
\beta_2^{(0,0)}\! \! \! &=& \! \! \!-3+8 \ms^2-24 \ms^4+24 \ms^6-5 \ms^8- 
       24 \ms^4 \ln(\ms) \; , \\
\beta_0^{(0,1)}\! \! \! \! &=& \! \! \! \! \frac{1}{30}(7- \! 25 \ms^2+ \! 160 \ms^4-160 \ms^6+25 \ms^8 -  \!
       7 \ms^{10} +  \! 120 \ms^4 \ln(\ms) + 120 \ms^6 \ln(\ms)) \; ,\\
\beta_1^{(0,1)}\! \! \! &=& \! \! \!\beta_1^{(0,0)} \; , \\
\beta_2^{(0,1)}\! \! \! &=& \! \! \!\frac{1}{3} \ms^2 (7-20 \ms^2+20 \ms^6-7 \ms^8+24
 \ln(\ms) - 
       48 \ms^2 \ln(\ms)) \; ,  \\
\beta_0^{(0,2)}\! \! \! &=& \! \! \!\frac{2}{45}(2-3 \ms^2-30 \ms^4+30 \ms^8+3\ms^{10}-2\ms^{12} 
-        120 \ms^6\ln(\ms) ) \; , \\
\beta_1^{(0,2)}\! \! \! &=& \! \! \!\frac{1}{270}(43-135 \ms^2+1260 \ms^4-1440 \ms^6+405 \ms^8 - 
       153 \ms^{10} + 20 \ms^{12}+1080 \ms^4 \ln(\ms)  \nonumber \\
&+& 840 \ms^6 \ln(\ms)) \; , \\
\beta_2^{(0,2)}\! \! \! &=& \! \! \!\frac{1}{90}(13-315 \ms^2+1500 \ms^4-1560 \ms^6+315 \ms^8 + 
147 \ms^{10}-100 \ms^{12}+360 \ms^4\ln(\ms) 
{\mbox{\hspace{0.8cm}}} \nonumber \\
 &+& 840 \ms^6 \ln(\ms) ) \; , \\
\beta_0^{(1,0)}\! \! \! &=& \! \! \!0 \; , \\
\beta_1^{(1,0)}\! \! \! \! &=& \! \! \! \! \frac{1}{30}(13-\! 135 \ms^2-\! 160 \ms^4 \! + \! 320 \ms^6-\! 45 \ms^8+
7 \ms^{10}\! -\! 600 \ms^4 \ln(\ms \! )-\! 120 \ms^6 \ln(\ms)) \; , \\
\beta_2^{(1,0)}\! \! \! &=& \! \! \!\frac{1}{6}(3-9 \ms^2+16 \ms^4-48 \ms^6+45 \ms^8-7 \ms^{10}+
24 \ms^4  \ln(\ms)-72 \ms^6  \ln(\ms)) \; , \\
\beta_0^{(1,1)}\! \! \! &=& \! \! \!0 \; , \\
\beta_1^{(1,1)}\! \! \! &=& \! \! \!\frac{1}{270}(23-45\ms^2+1080\ms^4-1440 \ms^6+585\ms^8-243
\ms^{10}+40 \ms^{12}+1080 \ms^4 \ln(\ms) \nonumber \\
&+&600 \ms^6 \ln(\ms) ) \; , \\
\beta_2^{(1,1)}\! \! \! &=& \! \! \!\frac{1}{90}(13+45 \ms^2-120 \ms^4-45 \ms^8+147 \ms^{10}-40 \ms^{12}+360 \ms^4\ln(\ms) \nonumber \\
&-&600 \ms^6 \ln(\ms) ) \; , \\
\beta_0^{(2,0)}\! \! \! &=& \! \! \!0 \; , \\
\beta_1^{(2,0)}\! \! \! &=& \! \! \!\frac{16}{135}(-1+9\ms^2-45 \ms^4+45 \ms^8-9 \ms^{10}+\ms^{12}-120 \ms^6\ln(\ms)) \; , \\
\beta_2^{(2,0)}\! \! \! &=& \! \! \!0 \;  .
\end{eqnarray}
{\bf \underline{ The functions $\mathbf \gamma_i^{(n,m)}$}}

Note that in the expressions given below 
$C_9^{\mbox{eff}}\equiv C_9^{\mbox{eff}}(\s=1-2 x_0+\ms^2)$.
The lower and upper limits of the $x_0$-integrals are:
$x_0^{min}=\ms$ and $x_0^{max}=\frac{1}{2}(1+\ms^2-4 \ml^2)$.
\begin{eqnarray}
\gamma_0^{(0,0)}\! \! \! &=& \! \! \!128 \int_{x_0^{min}}^{x_0^{max}} d x_0
\sqrt{x_0^2-\ms^2}(-2 \ms^2+x_0+\ms^2 x_0)
{\it{Re}}(C_9^{{\mbox{eff}}}) \; , \\
\gamma_1^{(0,0)}\! \! \! &=& \! \! \!\frac{1}{2} \gamma_0^{(0,0)} \; , \\ 
\gamma_2^{(0,0)}\! \! \! &=& \! \! \!\int_{x_0^{min}}^{x_0^{max}} d x_0
\frac{64}{\sqrt{x_0^2-\ms^2}} (4 \ms^2+14 \ms^4-x_0-\ms^2 x_0-12 \ms^4 x_0
-4 x_0^2-22 \ms^2 x_0^2+7 x_0^3 
%{\mbox{\hspace{2.6cm}}}
\nonumber \\
&+&15 \ms^2 x_0^3)
{\it{Re}}(C_9^{{\mbox{eff}}}) \; , \\
\gamma_0^{(0,1)}\! \! \! &=& \! \! \!128 \int_{x_0^{min}}^{x_0^{max}} d x_0
 x_0 \sqrt{x_0^2-\ms^2} (-2 \ms^2+x_0+\ms^2 x_0)
{\it{Re}}(C_9^{{\mbox{eff}}}) \; , \\
\gamma_1^{(0,1)}\! \! \! &=& \! \! \! \gamma_1^{(0,0)} \; , \\
\gamma_2^{(0,1)}\! \! \! &=& \! \! \!\frac{64}{3} \int_{x_0^{min}}^{x_0^{max}} d x_0
\frac{1}{\sqrt{x_0^2-\ms^2}}(4 \ms^4+6 \ms^6+9  \ms^2 x_0 +57 \ms^4 x_0-
3 x_0^2-14 \ms^2 x_0^2-57 \ms^4 x_0^2\nonumber \\
&-&9 x_0^3-81 \ms^2 x_0^3+28 x_0^4+60 \ms^2 x_0^4)
{\it{Re}}(C_9^{{\mbox{eff}}}) \; , \\
\gamma_0^{(0,2)}\! \! \! &=& \! \! \!128 \int_{x_0^{min}}^{x_0^{max}} d x_0
x_0^2 \sqrt{x_0^2-\ms^2} (-2 \ms^2+x_0+\ms^2 x_0)
{\it{Re}}(C_9^{{\mbox{eff}}}) \; , \\
\gamma_1^{(0,2)}\! \! \! &=& \! \! \!\frac{64}{3} \int_{x_0^{min}}^{x_0^{max}} d x_0
\sqrt{x_0^2-\ms^2} (-4 \ms^4-10 \ms^2 x_0+2 \ms^4 x_0+6 x_0^2 +16 \ms^2 x_0^2
-5 x_0^3 \nonumber \\
&-&5 \ms^2 x_0^3)
{\it{Re}}(C_9^{{\mbox{eff}}}) \; , \\
\gamma_2^{(0,2)}\! \! \! &=& \! \! \!\frac{64}{3} \int_{x_0^{min}}^{x_0^{max}} d x_0
\frac{x_0}{\sqrt{x_0^2-\ms^2}} (8 \ms^4+12 \ms^6+6 \ms^2 x_0+72 \ms^4 x_0-3 x_0^2-25 \ms^2 x_0^2-78 \ms^4 x_0^2 \nonumber \\
&-&6 x_0^3-96 \ms^2 x_0^3+35 x_0^4+75 \ms^2 x_0^4)
{\it{Re}}(C_9^{{\mbox{eff}}}) \; , \\
\gamma_0^{(1,0)}\! \! \! &=& \! \! \!0 \; , \\
\gamma_1^{(1,0)}\! \! \! &=& \! \! \! \!128 \int_{x_0^{min}}^{x_0^{max}} \!
\! d x_0
\frac{(x_0-1)}{\sqrt{x_0^2-\ms^2}} (-2 \ms^4+ \ms^2 x_0+\ms^4 x_0+2 \ms^2 x_0^2- \! x_0^3-\ms^2 x_0^3){\it{Re}}(C_9^{{\mbox{eff}}}) \; , \\
\gamma_2^{(1,0)}\! \! \! &=& \! \! \! \! \frac{128}{3} 
\int_{x_0^{min}}^{x_0^{max}} \! \!  d x_0 \sqrt{x_0^2-\ms^2}
(-4 \ms^2 - \! 6 \ms^4+3 x_0-\! 15 \ms^2 x_0+7 x_0^2+ \! 15 \ms^2 x_0^2)
{\it{Re}}(C_9^{{\mbox{eff}}}) \; , \\
\gamma_0^{(1,1)}\! \! \! &=& \! \! \!0 \; , \\
\gamma_1^{(1,1)}\! \! \! &=& \! \! \!\frac{128}{3} 
\int_{x_0^{min}}^{x_0^{max}}  d x_0 \frac{1}{\sqrt{x_0^2-\ms^2}}
(4 \ms^6+4 \ms^4 x_0-2 \ms^6 x_0 -3 \ms^2 x_0^2 -17 \ms^4 x_0^2+\ms^2 x_0^3+7 \ms^4 x_0^3 \nonumber \\
&+& 3 x_0^4+13 \ms^2 x_0^4-5 x_0^5-5 \ms^2 x_0^5)
{\it{Re}}(C_9^{{\mbox{eff}}}) \; , \\
\gamma_2^{(1,1)}\! \! \! &=& \! \! \! \!\frac{128}{3}  \!
\int_{x_0^{min}}^{x_0^{max}} \! \!  d x_0 x_0 \sqrt{x_0^2-\ms^2}
(-4 \ms^2 \! -\! 6 \ms^4 \! + \! 3 x_0\! -\! 15 \ms^2 x_0+ \! 7 x_0^2+\! 15 \ms^2 x_0^2)
{\it{Re}}(C_9^{{\mbox{eff}}}) \; , \\
\gamma_0^{(2,0)}\! \! \! &=& \! \! \!0 \; , \\
\gamma_1^{(2,0)}\! \! \! &=& \! \! \! \!\frac{512}{3} \int_{x_0^{min}}^{x_0^{max}} \! \!
d x_0 \sqrt{x_0^2-\ms^2}
(-2 \ms^4 + \ms^2 x_0 + \ms^4 x_0+2 \ms^2 x_0^2-x_0^3 
- \ms^2 x_0^3) {\it{Re}}(C_9^{{\mbox{eff}}}) \; , \\
\gamma_2^{(2,0)}\! \! \! &=& \! \! \!0 \; .
\end{eqnarray}
{\bf \underline{ The functions $\mathbf \delta_i^{(n,m)}$}}
\begin{eqnarray}
\delta_0^{(0,0)}\! \! \! &=& \! \! \!\frac{32}{3} \int_{x_0^{min}}^{x_0^{max}} d x_0
\sqrt{x_0^2-\ms^2} (-2 \ms^2+3 x_0+3 \ms^2 x_0-4 x_0^2)
|C_9^{{\mbox{eff}}}|^2 \; , \\
\delta_1^{(0,0)}\! \! \! &=& \! \! \!\frac{1}{2}\delta_0^{(0,0)} \; , \\ 
\delta_2^{(0,0)}\! \! \! &=& \! \! \! \int_{x_0^{min}}^{x_0^{max}} d x_0
\frac{16}{\sqrt{x_0^2-\ms^2}}(6 \ms^4-x_0-9 \ms^2 x_0-12 \ms^4 x_0+6 \ms^2 x_0^2+15 x_0^3+15 \ms^2 x_0^3 
%{\mbox{\hspace{3.0cm}}} 
\nonumber \\
&-& 20 x_0^4)
|C_9^{{\mbox{eff}}}|^2 \; , \\
\delta_0^{(0,1)}\! \! \! &=& \! \! \!\frac{32}{3} \int_{x_0^{min}}^{x_0^{max}} d x_0
x_0 \sqrt{x_0^2-\ms^2}  (-2 \ms^2+3 x_0+3 \ms^2 x_0-4 x_0^2)
|C_9^{{\mbox{eff}}}|^2 \; , \\
\delta_1^{(0,1)}\! \! \! &=& \! \! \!\delta_1^{(0,0)}  \; , \\
\delta_2^{(0,1)}\! \! \! &=& \! \! \!\frac{16}{3} \int_{x_0^{min}}^{x_0^{max}} d x_0
\frac{1}{\sqrt{x_0^2-\ms^2}} (6 \ms^6-3 \ms^2 x_0+13 \ms^4 x_0-3 x_0^2-30 \ms^2 x_0^2-57 \ms^4 x_0^2 +3 x_0^3 \nonumber \\
&+&43 \ms^2 x_0^3+48 x_0^4+60 \ms^2 x_0^4-80 x_0^5)
|C_9^{{\mbox{eff}}}|^2 \; , \\
\delta_0^{(0,2)}\! \! \! &=& \! \! \!\frac{32}{3} \int_{x_0^{min}}^{x_0^{max}} d x_0
 x_0^2 \sqrt{x_0^2-\ms^2} (-2 \ms^2+3 x_0+3 \ms^2 x_0-4 x_0^2)
|C_9^{{\mbox{eff}}}|^2 \; , \\
\delta_1^{(0,2)}\! \! \! &=& \! \! \!\frac{16}{9} \int_{x_0^{min}}^{x_0^{max}} d x_0
\sqrt{x_0^2-\ms^2} (-4 \ms^4-6 \ms^2 x_0+6 \ms^4 x_0+18 x_0^2+20 \ms^2 x_0^2-39 x_0^3-15 \ms^2 x_0^3 \nonumber \\
&+& 20 x_0^4)
|C_9^{{\mbox{eff}}}|^2 \; , \\
\delta_2^{(0,2)}\! \! \! &=& \! \! \!\frac{16}{3} \int_{x_0^{min}}^{x_0^{max}} d x_0
\frac{x_0}{\sqrt{x_0^2-\ms^2}}(12 \ms^6-6 \ms^2 x_0+8 \ms^4 x_0-3 x_0^2-33 \ms^2 x_0^2-78 \ms^4 x_0^2+6 x_0^3 \nonumber \\
&+&68 \ms^2 x_0^3+51 x_0^4+75 \ms^2 x_0^4-
100 x_0^5)
|C_9^{{\mbox{eff}}}|^2 \; , \\
\delta_0^{(1,0)}\! \! \! &=& \! \! \!0 \; , \\
\delta_1^{(1,0)}\! \! \! &=& \! \! \!\frac{32}{3} \int_{x_0^{min}}^{x_0^{max}} d x_0
\frac{(x_0-1)}{\sqrt{x_0^2-\ms^2}} (-2 \ms^4+ 3\ms^2 x_0+3\ms^4 x_0-2 \ms^2 x_0^2-3 x_0^3-3\ms^2 x_0^3 \nonumber \\
&+&4 x_0^4) |C_9^{{\mbox{eff}}}|^2 \; , \\
\delta_2^{(1,0)}\! \! \! &=& \! \! \!\frac{32}{3} \int_{x_0^{min}}^{x_0^{max}} d x_0
\sqrt{x_0^2-\ms^2} (-6 \ms^4+3 x_0+5 \ms^2 x_0+3 x_0^2+15 \ms^2 x_0^2-20 x_0^3)
|C_9^{{\mbox{eff}}}|^2 \; , \\
\delta_0^{(1,1)}\! \! \! &=& \! \! \! 0 \; , \\
\delta_1^{(1,1)}\! \! \! &=& \! \! \!\frac{32}{9} \int_{x_0^{min}}^{x_0^{max}} d x_0
\frac{1}{\sqrt{x_0^2-\ms^2}} (4 \ms^6-6 \ms^6 x_0-9 \ms^2 x_0^2-15 \ms^4 x_0^2+27 \ms^2 x_0^3+ 21 \ms^4 x_0^3 \nonumber \\
&+&9 x_0^4-9 \ms^2 x_0^4-27 x_0^5-15 \ms^2 x_0^5+20 x_0^6)
|C_9^{{\mbox{eff}}}|^2 \; , \\
\delta_2^{(1,1)}\! \! \! &=& \! \! \!  \!\frac{32}{3} \int_{x_0^{min}}^{x_0^{max}} d x_0 x_0  \!  \! 
\sqrt{x_0^2-\ms^2} (-6 \ms^4+3 x_0+5 \ms^2 x_0+3 x_0^2+15 \ms^2 x_0^2- \! 
20 x_0^3)|C_9^{{\mbox{eff}}}|^2 \; , \\
\delta_0^{(2,0)}\! \! \! &=& \! \! \!0 \; , \\
\delta_1^{(2,0)}\! \! \! &=& \! \! \!\frac{128}{9} \int_{x_0^{min}}^{x_0^{max}} 
d x_0 \sqrt{x_0^2-\ms^2}
(-2 \ms^4 +3 \ms^2 x_0 +3 \ms^4 x_0-2 \ms^2 x_0^2-3 x_0^3 \nonumber \\
&-& 3 \ms^2 x_0^3+
4 x_0^4) |C_9^{{\mbox{eff}}}|^2 \; , \\
\delta_2^{(2,0)}\! \! \! &=& \! \! \!0 \; .
\end{eqnarray}
\section{Lowest Hadronic Moments (Parton Level) \label{app:lowmoments}}

\begin{eqnarray}
\langle x_0\rangle {{\cal B}\over {\cal B}_0} &=&
\frac{2}{9 m_B^2}(-41 m_B^2-49 m_s^2-
24 (m_B^2- m_s^2) \ln(4 \frac{m_l^2}{m_B^2})){C_7^{\mbox{eff}}}^2+
\frac{1}{30 m_B^2} (7 m_B^2-25 m_s^2)C_{10}^2 \nonumber \\
&+&
\int_{m_s/m_B}^{\frac{1}{2}(1+m_s^2/m_B^2)} d x_0 \frac{64}{m_B^2} x_0
(-m_s^2-4 m_s^2 x_0+2 m_B^2 x_0^2+2 m_s^2 x_0^2)
{\it{Re}}(C_9^{{\mbox{eff}}})C_7^{{\mbox{eff}}} \nonumber \\
&+&
\int_{m_s/m_B}^{\frac{1}{2}(1+m_s^2/m_B^2)} d x_0
\frac{16}{3 m_B^2} x_0(-3 m_s^2+6 m_B^2 x_0^2+6 m_s^2 x_0^2-8 m_B^2 x_0^3)
|C_9^{{\mbox{eff}}}|^2\nonumber \\
&+&\frac{\alpha_s}{\pi} A^{(0,1)} C_9^2+ 
\frac{-32}{3} {C_7^{\mbox{eff}}}^2 \frac{\bar{\Lambda}}{m_B}+
\frac{-16}{3}{C_7^{\mbox{eff}}}^2\frac{\bar{\Lambda}^2}{m_B^2}
+ \left[ \frac{-16}{9} (1+3\ln(4 \frac{m_l^2}{m_B^2})){C_7^{\mbox{eff}}}^2+
 \frac{C_{10}^2}{3}  \right. \nonumber \\
&+& \left.
\int_{0}^{\frac{1}{2}} d x_0 
(64 x_0^2{\it{Re}}(C_9^{{\mbox{eff}}})C_7^{{\mbox{eff}}}
+\frac{16}{3} (3-4 x_0)x_0^2 |C_9^{{\mbox{eff}}}|^2 ) \right] 
\frac{\lambda_1}{m_B^2}
\nonumber \\
&+& 
 \left[ \frac{4}{3} (19+12 \ln(4 \frac{m_l^2}{m_B^2})){C_7^{\mbox{eff}}}^2
+\int_{0}^{\frac{1}{2}} d x_0
(\frac{64}{3} x_0(-3-9 x_0+28 x_0^2){\it{Re}}
(C_9^{{\mbox{eff}}})C_7^{{\mbox{eff}}}
 \right. \nonumber \\
&+& \left.
\frac{16}{3} x_0 (-3+3 x_0+48 x_0^2-80 x_0^3)|C_9^{{\mbox{eff}}}|^2 ) \right] 
\frac{\lambda_2}{m_B^2} \, ,
\end{eqnarray}
\begin{eqnarray}
\langle x_0^2\rangle {{\cal B}\over {\cal B}_0} \! \! \! \! &=& \! \! \! \!
\frac{2}{45 m_B^{12}}(-119 m_B^{12}-144 m_B^{10} m_s^2-
60 (m_B^{12}- m_s^{12}) \ln(4 \frac{m_l^2}{m_B^2})){C_7^{\mbox{eff}}}^2 
\! \! \! + \! \! \frac{2}{45 m_B^2} (2 m_B^2-3 m_s^2)C_{10}^2 \nonumber \\
&+&
\int_{m_s/m_B}^{\frac{1}{2}(1+m_s^2/m_B^2)} d x_0 \frac{64}{m_B^2} x_0^2
(-m_s^2-4 m_s^2 x_0+2 m_B^2 x_0^2+2 m_s^2 x_0^2)
{\it{Re}}(C_9^{{\mbox{eff}}})C_7^{{\mbox{eff}}} \nonumber \\
&+&
\int_{m_s/m_B}^{\frac{1}{2}(1+m_s^2/m_B^2)} d x_0
\frac{16}{3 m_B^2} x_0^2(-3 m_s^2+6 m_B^2 x_0^2+6 m_s^2 x_0^2-8 m_B^2 x_0^3)
|C_9^{{\mbox{eff}}}|^2\nonumber \\
&+&\frac{\alpha_s}{\pi} A^{(0,2)} C_9^2+ 
\frac{-16}{3} {C_7^{\mbox{eff}}}^2 \frac{\bar{\Lambda}}{m_B}+
\frac{-8}{3}{C_7^{\mbox{eff}}}^2\frac{\bar{\Lambda}^2}{m_B^2}
+ \left[ \frac{-1}{27} (55+84\ln(4 \frac{m_l^2}{m_B^2})){C_7^{\mbox{eff}}}^2+
 43 \frac{C_{10}^2}{270}  \right. \nonumber \\
&+& \left.
\int_{0}^{\frac{1}{2}} d x_0 
(\frac{64}{3} (6-5 x_0) x_0^3{\it{Re}}(C_9^{{\mbox{eff}}})C_7^{{\mbox{eff}}}
+\frac{16}{9} (18-39 x_0+20 x_0^2)x_0^3 |C_9^{{\mbox{eff}}}|^2 ) \right] 
\frac{\lambda_1}{m_B^2}
\nonumber \\
&+& 
 \left[ (11+4 \ln(4 \frac{m_l^2}{m_B^2})){C_7^{\mbox{eff}}}^2
  +13 \frac{C_{10}^2}{90} 
+ \int_{0}^{\frac{1}{2}} d x_0
(\frac{64}{3} x_0^2(-3-6 x_0+35 x_0^2){\it{Re}}(C_9^{{\mbox{eff}}})C_7^{{\mbox{eff}}} \right. \nonumber \\
&+& \left. 
\frac{16}{3} x_0^2 (-3+6 x_0+51 x_0^2-100 x_0^3)|C_9^{{\mbox{eff}}}|^2 )
\right] \frac{\lambda_2}{m_B^2} \, ,
\end{eqnarray}
\begin{eqnarray}
\langle x_0 (\s_0-\ms^2) \rangle {{\cal B}\over {\cal B}_0} &=&
\frac{\alpha_s}{\pi} A^{(1,1)} C_9^2
+ \left[ \frac{-8}{27} (1+3\ln(4 \frac{m_l^2}{m_B^2})){C_7^{\mbox{eff}}}^2+
 23 \frac{C_{10}^2}{270}  \right. \nonumber \\
&+& \! \! \! \left. \! \! \!
\int_{0}^{\frac{1}{2}} d x_0 
(\frac{128}{3} (3-5 x_0) x_0^3{\it{Re}}(C_9^{{\mbox{eff}}})C_7^{{\mbox{eff}}}
+\frac{32}{9} (9-27 x_0+20 x_0^2)x_0^3 |C_9^{{\mbox{eff}}}|^2 ) \right] 
\frac{\lambda_1}{m_B^2}
\nonumber \\
&+& 
 \left[ \frac{-8}{3}(5+3 \ln(4 \frac{m_l^2}{m_B^2})){C_7^{\mbox{eff}}}^2
  +13 \frac{C_{10}^2}{90} \right.\\
&+& \left.
\int_{0}^{\frac{1}{2}} d x_0
(\frac{128}{3} x_0^3(3+7 x_0){\it{Re}}(C_9^{{\mbox{eff}}})C_7^{{\mbox{eff}}}
+\frac{32}{3} x_0^3 (3+3 x_0-20 x_0^2)|C_9^{{\mbox{eff}}}|^2 ) \right] 
\frac{\lambda_2}{m_B^2}
\nonumber   \, ,
\end{eqnarray}
\begin{eqnarray}
\langle \s_0-\ms^2 \rangle {{\cal B}\over {\cal B}_0} &=&
\frac{\alpha_s}{\pi} A^{(1,0)} C_9^2
+ \left[ \frac{-2}{9} (23+24 \ln(4 \frac{m_l^2}{m_B^2})){C_7^{\mbox{eff}}}^2+
 13 \frac{C_{10}^2}{30}  \right. \nonumber \\
&+& \left.
\int_{0}^{\frac{1}{2}} d x_0 
(128 (1- x_0) x_0^2{\it{Re}}(C_9^{{\mbox{eff}}})C_7^{{\mbox{eff}}}
+\frac{32}{3} (3-7 x_0+4 x_0^2)x_0^2 |C_9^{{\mbox{eff}}}|^2 ) \right] 
\frac{\lambda_1}{m_B^2}
\nonumber \\
&+& 
 \left[ \frac{-2}{3}(31+24 \ln(4 \frac{m_l^2}{m_B^2})){C_7^{\mbox{eff}}}^2
  +\frac{C_{10}^2}{2} \right.\\
&+& \left.
\int_{0}^{\frac{1}{2}} d x_0
(\frac{128}{3} x_0^2 (3+7 x_0){\it{Re}}(C_9^{{\mbox{eff}}})C_7^{{\mbox{eff}}}
+\frac{32}{3} x_0^2 (3+3 x_0-20 x_0^2)|C_9^{{\mbox{eff}}}|^2 ) \right] 
\frac{\lambda_2}{m_B^2}
\nonumber  \, ,
\end{eqnarray}
\begin{eqnarray}
\langle (\s_0-\ms^2)^2 \rangle {{\cal B}\over {\cal B}_0} &=&
\frac{\alpha_s}{\pi} A^{(2,0)} C_9^2
+ \left[ \frac{8}{135} (119+60 \ln(4 \frac{m_l^2}{m_B^2})){C_7^{\mbox{eff}}}^2
 -16 \frac{C_{10}^2}{135}  \right.  \\
&+& \left.
\int_{0}^{\frac{1}{2}} d x_0 
(\frac{-512}{3} x_0^4{\it{Re}}(C_9^{{\mbox{eff}}})C_7^{{\mbox{eff}}}
+\frac{128}{9} (-3+4 x_0)x_0^4 |C_9^{{\mbox{eff}}}|^2 ) \right] 
\frac{\lambda_1}{m_B^2}
\nonumber \, .
\end{eqnarray}

\end{appendix}

\end{document}